\pdfoutput=1
\RequirePackage{fixltx2e}
\documentclass[11pt,a4paper]{article}
\usepackage{jheppub}
\usepackage{microtype}
\usepackage{etoolbox,xcolor,graphicx}
\usepackage{array,booktabs,multirow}
\usepackage{amsmath,amssymb,amsthm,mathtools}
\usepackage[only, llbracket, rrbracket]{stmaryrd}
\usepackage{simplewick}
\usepackage{hyperref}

\usepackage{tikz}
\usetikzlibrary{automata}
\usetikzlibrary{arrows}
\usetikzlibrary{calc}
\usetikzlibrary{decorations.markings}
\usetikzlibrary{decorations.pathreplacing}
\usetikzlibrary{intersections}
\usetikzlibrary{positioning}
\usetikzlibrary{topaths}
\usetikzlibrary{shapes.geometric}
\usetikzlibrary{shapes.misc}
\tikzset{->-/.style = {
    decoration = {markings, mark = at position #1 with {\arrow{>}}},
    postaction = {decorate}}}
\tikzset{color-group/.style = {
    shape = circle,
    minimum size = 2.5ex,
    inner sep = .5ex,
    draw}}
\tikzset{flavor-group/.style = {
    shape = rectangle,
    minimum size = 2.5ex,
    inner sep = .5ex,
    draw}}
\tikzset{cf-group/.style = {
    shape = rounded rectangle,
    rounded rectangle right arc = none,
    draw}}
\tikzset{fc-group/.style = {
    shape = rounded rectangle,
    rounded rectangle left arc = none,
    draw}}
\tikzset{cross/.style={minimum width=1pt, path picture={
      \draw[black, very thick]
               (path picture bounding box.south east)
            -- (path picture bounding box.north west)
               (path picture bounding box.south west)
            -- (path picture bounding box.north east);
          }}}
\newcommand{\mathtikz}[2][]
  {\ensuremath{\vcenter{\hbox{%
          \begin{tikzpicture}[#1]#2\end{tikzpicture}}}}}
\newcommand{\quiver}[2][]
  {\mathtikz[semithick,node distance=3em,#1]{#2}}
\hypersetup{unicode}
\hypersetup{linktoc = all}
\hypersetup{pdfborderstyle={/S/U/W 0.5}}
\hypersetup{bookmarksnumbered}
\DeclareMathOperator{\id}{id}
\DeclareMathOperator*{\res}{res}
\DeclareMathOperator{\Tr}{Tr}
\let\Re\relax\let\Im\relax
\DeclareMathOperator{\Re}{Re}
\DeclareMathOperator{\Im}{Im}
\DeclarePairedDelimiter{\abs}{\lvert}{\rvert}
\DeclarePairedDelimiter{\vev}{\langle}{\rangle}

\newcommand{\mathsemiclap}[2]
  {\mspace{-#1\thinmuskip}#2\mspace{-#1\thinmuskip}}

\newcommand{\SQED}         {{\text{SQED}}}
\newcommand{\SQCD}         {{\text{SQCD}}}
\newcommand{\SQCDA}        {{\text{SQCDA}}}
\newcommand{\classical}    {{\text{cl}}}
\newcommand{\oneloop}      {{\text{\(1\)l}}}
\newcommand{\vectormultiplet} {{\text{v.m.}}}
\newcommand{\chiralmultiplet} {{\text{c.m.}}}
\newcommand{\vortex}       {{\text{v}}}
\newcommand{\antivortex}   {{\bar{\vortex}}}
\newcommand{\free}         {{\text{free}}}
\newcommand{\bare}         {{\text{bare}}}
\newcommand{\enriched}     {{\text{enr}}}
\newcommand{\renormalized} {{\text{ren}}}
\newcommand{\Gf}           {{G_{\!f}}}
\newcommand{\Nsusy}        {{\mathcal{N}}}
\newcommand{\Nc}           {{N}}
\newcommand{\Nf}{\texorpdfstring{N_{\!f}}{Nf}}
\newcommand{\anti}[1]{\texorpdfstring{{\widetilde{#1}}}{#1'}}
\newcommand{\Q}    {{\mathcal{Q}}}
\newcommand{\Daux} {{\mathrm{D}}}
\newcommand{\Faux} {{\mathrm{F}}}
\newcommand{\bbZ}  {{\mathbb{Z}}}
\newcommand{\bbR}  {{\mathbb{R}}}
\newcommand{\bbC}  {{\mathbb{C}}}
\newcommand{\dd}[2][]{\mathop{\mathrm{d}#1#2}}
\providecommand{\llbracket}{[[}
\providecommand{\rrbracket}{]]}
\newcommand{\intset}[2]{\llbracket #1, #2\rrbracket}
\newcommand{\Weyl} {{\mathcal{W}}}
\newcommand{\repr} {{\mathcal{R}}}
\newcommand{\lie}[1]{{\mathfrak{#1}}}
\newcommand{\nop}[1]{\mathopen{}\mathclose{{:}#1{:}}}
\newcommand{\I}    {{\mathrm{i}}}
\newcommand{\Hypergeometric}[4][F]{%
  #1\left(\begin{smallmatrix}#2\\#3\end{smallmatrix}\middle|#4\right)}
\newcommand{\Fblock}[4]{%
  \mathinner{\Fblockname^{\text{(#1)}}_{#2}%
    \ifstrempty{#3}{}{\!\!\begin{bmatrix}#3\end{bmatrix}\!}%
    \ifstrempty{#4}{}{\!(#4)}}}
\newcommand{\Fblockname}{{\mathcal{F}}}
\newcommand{\Braiding}[3]{%
  \mathbf{B}^{#1}_{#2}%
  \ifstrempty{#3}{}{\!\begin{bmatrix}#3\end{bmatrix}\!}}
\newcommand{\Fusionname}{{\mathbf{F}}}
\newcommand{\dimToda}{\Delta}
\newcommand{\conj}{C}
\newcommand{\dual}{D}
\newcommand{\sch}{{\text{(s)}}}

\newcommand{\uch}{{\text{(u)}}}
\newcommand{\phiToda}{\varphi}
\newcommand{\quark}{q}
\newcommand{\biquark}{\phi}
\newcommand{\hyperquark}{\Phi}

\newcommand{\BLF}[1]{\relax\ifmmode\expandafter\text\fi{\textcolor{olive}{[BLF:} #1\textcolor{olive}{]}}}
\newcommand{\JG}[1]{\relax\ifmmode\expandafter\text\fi {\textcolor{blue}{[JG: #1]}}}

\date{July 2014}
\title{\Large M2-brane surface operators and  gauge theory dualities in Toda}
\hypersetup{pdftitle={M2-brane surface operators and  gauge theory dualities in Toda}}
\author[a]{Jaume Gomis}
\author[a,b]{and Bruno Le Floch}
\affiliation[a]{Perimeter Institute for Theoretical Physics,\\
  Waterloo, Ontario, N2L 2Y5, Canada}
\affiliation[b]{Laboratoire de Physique Th\'eorique de
  l'\'Ecole Normale Sup\'erieure,%
  \footnote{Unit\'e mixte (UMR 8549) du CNRS et de l'ENS, Paris.}\\
  Paris, 75005, France}
\emailAdd{jgomis@perimeterinstitute.ca}
\emailAdd{blefloch@princeton.edu}
\abstract{%
  We give a microscopic two dimensional ${\cal N}=(2,2)$ gauge theory description of arbitrary
  M2-branes ending on $\Nf$ M5-branes wrapping a punctured Riemann surface.  These realize surface
  operators in four dimensional ${\cal N}=2$ field theories.  We show that the expectation value
  of these surface operators on the sphere is captured by a Toda CFT correlation function in the
  presence of an additional degenerate vertex operator labelled by a representation ${\cal R}$ of
  $SU(\Nf)$, which also labels M2-branes ending on M5-branes.  We prove that symmetries of Toda
  CFT correlators provide a geometric realization of dualities between two dimensional gauge
  theories, including ${\cal N}=(2,2)$ analogues of Seiberg and Kutasov--Schwimmer dualities.  As
  a bonus, we find new explicit conformal blocks, braiding matrices, and fusion rules in Toda CFT\@.}
\keywords{Supersymmetry and Duality, Conformal and W Symmetry, M-Theory, Extended Supersymmetry}
\arxivnumber{1407.1852}

\begin{document}

\maketitle

\flushbottom

\bibliographystyle{JHEP}

\clearpage

\numberwithin{equation}{section}
\section{Introduction and Conclusions}
\label{sec:introduction}

The traditional order parameters for the phases of four dimensional gauge theories are the
Wilson~\cite{Wilson:1974sk} and 't~Hooft~\cite{Hooft:1977hy} operators.  In recent years, the
construction of nonlocal surface operators~\cite{Gukov-Witten-surface}, which insert probe
strings, have enlarged the space of order parameters of gauge theories.  Indeed, surface operators
can distinguish phases which are otherwise indistinguishable using the Wilson--'t~Hooft
criteria~\cite{Gukov:2013zka}.

A surface operator can be defined either by specifying a codimension-two singularity for the
elementary fields or by coupling a two dimensional field theory to the bulk four dimensional
one~\cite{Gukov-Witten-surface}.  The couplings between bulk and defect degrees of freedom can
result in rich dynamics for the combined system, arising from the synergy of two dimensional and
four dimensional strong coupling dynamics.  For a sample of early references on surface operators
see~\cite{Gomis-bubbling-surface,Gukov:2007ck,Witten:2007td,Buchbinder:2007ar,Beasley:2008dc,Gukov:2008sn,Dru-surface,Gaiotto:2009fs}.

Surface operators also play a fundamental role in the six dimensional $\Nsusy=(2,0)$
supersymmetric field theory living on the worldvolume of a collection of $\Nf$~coincident and flat
M5-branes.  A class of surface operators in this theory are labeled by a representation $\repr$ of~$A_{\Nf-1}$
and admit an M-theory realization as a collection of M2-branes ending on the M5-branes along the
domain of support of the surface operator.

\pagebreak[0]\bigskip

In this paper we give a microscopic two dimensional gauge theory description of all such surface
operators when the M5-branes wrap a punctured Riemann surface~$C$~\cite{Gaiotto:2009we}. This
realizes a surface operator in a four dimensional $\Nsusy=2$ gauge theory.
\begin{center}
  \vspace{-1.5ex}
  \begin{tabular}{lrrrrrrrr}
    & & & & & \multicolumn{2}{c}{\lower10pt\hbox{$\overbracket{\qquad}^C$}} & \\
    \toprule
    M5 \ & 0 & 1 & 2 & 3 & 4 & 5 &   \\
    M2 \ & 0 & 1 &   &   &   &   & 6 \\
    \bottomrule
  \end{tabular}
\end{center}
The surface operator associated to a collection of M2-branes labeled by a representation $\repr$
of~$A_{\Nf-1}$ corresponds to the following two dimensional $\Nsusy=(2,2)$ gauge theory
\vspace{-1.5ex}
\begin{equation}\label{quiver}
  \repr  \quad \longleftrightarrow \quad
  \quiver[color-group/.append style={minimum size=6.5ex, node distance=8ex}]{%
    \node (N1)  [color-group]                    {$\Nc_1$};
    \node (dots)[left of=N1]                     {$\cdots$};
    \node (Nn-1)[color-group, left of=dots]      {$\Nc_{n-1}$};
    \node (Nn)  [color-group, left of=Nn-1]      {$\Nc_n$};
    \node (f)   [left of=Nn]                     {};
    \node (Nf)  [flavor-group, above=0ex of f]   {$\Nf$};
    \node (Nf') [flavor-group, below=0ex of f]   {$\Nf$};
    \draw[->-=.55] (Nf) -- (Nn);
    \draw[->-=.55] (Nn) -- (Nf');
    \draw[->-=.55] (Nn)   to [bend left=30] (Nn-1);
    \draw[->-=.55] (Nn-1) to [bend left=30] (dots);
    \draw[->-=.55] (dots) to [bend left=30] (N1);
    \draw[->-=.55] (N1)   to [bend left=30] (dots);
    \draw[->-=.55] (dots) to [bend left=30] (Nn-1);
    \draw[->-=.55] (Nn-1) to [bend left=30] (Nn);
    \draw[->-=.5] (N1)    to [distance=5ex, in=60, out=120, loop] ();
    \draw[->-=.5] (Nn-1)  to [distance=5ex, in=60, out=120, loop] ();
  }
\end{equation}
coupled to the bulk theory.  A cubic superpotential couples each adjoint chiral multiplet to the
neighboring bifundamental chiral multiplets. The FI parameters associated to $U(\Nc_j)$ for
$j<n$ vanish.  The ranks~\(\Nc_j\) encode the representation \(\repr\) whose Young diagram
\vspace{-2ex}
\begin{equation}
  \mathtikz[semithick,x=1em,y=1em]{
    \foreach \X/\Y in {2/0,2/-1,2/-2,2/-3,1/-4}  \draw ( 0,\Y) -- (\X,\Y);
    \foreach \X/\Y in {0/-4,1/-4,2/-3}       \draw (\X,0) -- (\X,\Y);
    \node          at (3,-1)          {$\cdots$};
    \foreach \X/\Y in {4/-2,5/-2,6/-1}       \draw (\X, 0) -- (\X,\Y);
    \foreach \X/\Y in {6/0,6/-1,5/-2}        \draw ( 4,\Y) -- (\X,\Y);
    \node [anchor=base] at (0.5,-4.9) {$\scriptstyle\Nc_n-\Nc_{n-1}$};
    \node [anchor=base west] at (1,-3.9) {$\scriptstyle\Nc_{n-1}-\Nc_{n-2}$};
    \node [anchor=base] at (4.5,-2.9) {$\scriptstyle\Nc_2-\Nc_1$};
    \node [anchor=base west] at (5,-1.9) {$\scriptstyle\Nc_1$};
    \draw (0,2pt) -- (0,6pt) -- (6,6pt) node [midway,yshift=6pt] {$n$} -- (6,2pt);
  }
\end{equation}
has \(n\)~columns with \(\Nc_n-\Nc_{n-1} \geq \Nc_{n-1}-\Nc_{n-2} \geq \cdots \geq \Nc_2-\Nc_1
\geq \Nc_1 \geq 0\) boxes.\footnote{The highest weight of $\repr$ is \(\Omega=\sum_{j=1}^n
  \omega_{\Nc_j-\Nc_{j-1}}\) in terms of the fundamental weights~\(\omega_K\) of~\(A_{\Nf-1}\).}

Advances in the computation of supersymmetric partition functions of four dimensional $\Nsusy=2$
gauge theories on the squashed four-sphere $S^4_b$~\cite{Pestun:2007rz,Hama:2012bg} have resulted
in exact formulas for the expectation value of Wilson~\cite{Pestun:2007rz} and 't~Hooft
operators~\cite{Gomis:2011pf} as functions of the gauge couplings and masses of the
hypermultiplets.  The gauge theory computation of the expectation value of surface operators
supported on a squashed $S^2\subset S^4_b$ are not yet available.  However, recent results in the
exact computation of the two-sphere partition function of $\Nsusy=(2,2)$ supersymmetric field
theories~\cite{Benini:2012ui,Doroud:2012xw,Gomis:2012wy,Doroud:2013pka}, when suitably coupled to
those in~\cite{Pestun:2007rz,Hama:2012bg}, provide a concrete avenue of investigation of the
expectation value of half-BPS surface operators in four dimensional $\Nsusy=2$ theories on~$S^4_b$
using Feynman path integrals.

For the four dimensional $\Nsusy=2$ theories obtained by wrapping M5-branes on punctured Riemann
surfaces, also known as class~S theories~\cite{Gaiotto:2009we}, the $S^4_b$~partition function
in~\cite{Pestun:2007rz,Hama:2012bg} admits an elegant representation~\cite{AGT} (see
also~\cite{Wyllard:2009hg}) in terms of two dimensional Toda CFT correlation functions.  In the
correspondence between four dimensional $\Nsusy=2$ theories and Toda CFT, the expectation value of
Wilson and 't~Hooft operators on $S^4_b$ are realized as Toda CFT correlators in the presence of
loops operators and topological webs~\cite{Alday:2009fs,Drukker:2009id,Drukker:2010jp} (see
also~\cite{Passerini:2010pr,Gomis:2010kv,Bullimore:2013xsa}).  Degenerate vertex operators in
$A_{\Nf-1}$~Toda CFT are conjectured to realize the insertion of a supersymmetric surface
operator~\cite{Alday:2009fs} (see
also~\cite{Dimofte:2010tz,Taki:2010bj,Bonelli:2011fq,Bonelli:2011wx,Doroud:2012xw}).

\pagebreak[0]\bigskip

In this paper we identify the two dimensional $\Nsusy=(2,2)$ gauge theory that realizes an
arbitrary degenerate operator in Toda CFT, which in turn corresponds to an arbitrary M2-brane
configuration ending on wrapped M5-branes.\footnote{%
  Another class of surface operators can be realized by M5-branes, and are labeled by a
  partition~$\rho$ of~$\Nf$.  It was conjectured in~\cite{Braverman:2010ef} that the instanton
  partition function of four dimensional $\Nsusy=2$ $SU(\Nf)$ SYM in the presence of such an
  M5-defect labeled by~$\rho$ is the norm of a Whittaker vector in the $W$-algebra~$W_\rho$.  Some
  checks of this conjecture and generalizations have appeared
  in~\cite{Alday:2010vg,Kozcaz:2010yp,Wyllard:2010rp,Wyllard:2010vi,Tachikawa:2011dz,Kanno:2011fw,Belavin:2012qh,Tan:2013tq}.
  We propose that the surface operator associated to an M5-defect labeled by~$\rho$, with
  $\Nf=K_1+\cdots+K_n$, corresponds to coupling the bulk $\Nsusy=2$ superconformal field theory to
  the two dimensional $\Nsusy=(2,2)$ gauge theory
  \quiver[node distance=2.5em, color-group/.append style = {inner sep = .3ex}]{%
    \node (N1)  [color-group]                    {$\scriptscriptstyle\Nc_1$};
    \node (N2)  [color-group, left of=N1]        {$\scriptscriptstyle\Nc_2$};
    \node (dots)[left of=N2]                     {\smash{$\ldots$}};
    \node (Nn-1)[color-group, left=1.5ex of dots]{$\scriptscriptstyle\Nc_{n-1}$};
    \node (f)   [left of=Nn-1] {};
    \node (Nf)  [flavor-group, above=-0.5ex of f] {$\scriptscriptstyle\Nf$};
    \node (Nf') [flavor-group, below=-0.5ex of f] {$\scriptscriptstyle\Nf$};
    \draw[->-=.6] (Nf)   -- (Nn-1);
    \draw[->-=.6] (Nn-1) -- (Nf');
    \draw[->-=.6] (N1)   to (N2);
    \draw[->-=.6] (N2)   to (dots);
    \draw[->-=.6] (dots) to (Nn-1);
  }
  with \(\Nc_j=K_1+\cdots+K_j\): vacua of this theory yield a monodromy for
  the 4d gauge field, which breaks \(SU(\Nf)\) to \(S[U(K_1)\times\cdots\times U(K_n)]\).
  Surface operators labeled by a Young diagram have appeared in~\cite{Gadde:2013dda}.}
A degenerate operator with Toda momentum $\alpha=-b\Omega$, where $\Omega$~is the highest weight
vector of a representation $\repr(\Omega)$ of~$A_{\Nf-1}$, corresponds to the quiver gauge
theory~\eqref{quiver}. The complexified FI/theta parameter associated to the $U(\Nc_n)$ gauge group
encodes the position of the degenerate puncture (the other FI parameters must vanish in this
correspondence).  The surface operator is supported on an \(S^2\subset S^4_b\) invariant under the
\(U(1)\times U(1)\) isometries of~\(S^4_b\).\footnote{Degenerate operators with momentum
  $\alpha=-\Omega/b$ correspond to the same quiver gauge theory but now supported on the other
  \(U(1)\times U(1)\) invariant \(S^2\subset S^4_b\).  The most general degenerate momentum
  $\alpha=-b\Omega-\Omega'/b$ corresponds to the insertion of the associated surface operators on
  both~\(S^2\)'s, but with a non-trivial coupling at their intersection points, namely the poles
  of~\(S^4_b\).}

The quiver gauge theory~\eqref{quiver} can be used to construct a surface operator in any four
dimensional $\Nsusy=2$ gauge theory that contains an $SU(\Nf)\times SU(\Nf)\times U(1)$ flavour or
gauge symmetry group.  This is the flavour symmetry of the chiral multiplets charged only under
the $U(N_n)$ gauge group factor in~\eqref{quiver}. A surface operator is constructed by
identifying the common $SU(\Nf)\times SU(\Nf)\times U(1)$ symmetry groups of the four dimensional
and two dimensional theories.

The simplest four dimensional $\Nsusy=2$ class~S theory in which we can include a surface defect
is the theory of $\Nf^2$  hypermultiplets.  This theory is realized by wrapping
$\Nf$~M5-branes on a trinion with two full and one simple puncture. We explicitly show that the
partition function of this theory in the presence of the surface operator labeled by a
representation $\repr(\Omega)$ is given by the Toda four-point function\footnote{%
  The four point function in~\eqref{4pt} contains full \mathtikz{\fill (0,0) circle [radius=2pt];
    \draw (0,0) circle (4pt);}, simple \mathtikz{\fill (0,0) circle [radius=2pt];}, and degenerate
  \mathtikz{\node [cross] at (0,0) {};} punctures.}  obtained by adding to the trinion a
degenerate field with momentum $\alpha=-b\Omega$
\begin{equation}\label{4pt}
  Z_{S^2\subset S^4_b}^{\repr(\Omega)}
  = \mathtikz{
      \draw (0,0) ellipse (1.4 and .9);
      \fill (-.7,-.4) circle [radius=2pt];
      \fill (-.7, .4) circle [radius=2pt];
      \fill ( .7,-.4) circle [radius=2pt];
      \draw (-.7,-.4) circle (4pt);
      \draw ( .7,-.4) circle (4pt);
      \node [cross] at (.7,.4) {};
      \node at (.35,.4) {$\Omega$};
    }
  \,.
\end{equation}
The two dimensional quiver gauge theory~\eqref{quiver} is coupled to the four dimensional field
theory by (weakly) gauging the $SU(\Nf)\times SU(\Nf)\times U(1)$ flavour symmetry associated to
the trinion.  The coupling can also be described by a cubic superpotential between the bulk
hypermultiplets and the fundamental and antifundamental chiral multiplets on the defect.  The
combined 4d/2d quiver diagram describing the insertion of the surface operator in this four
dimensional theory is
\begin{equation}
  \quiver[color-group/.append style={minimum size=6.5ex, node distance=8ex}]{%
    \node (N1)  [color-group]                     {$\Nc_1$};
    \node (dots)[left of=N1]                      {$\cdots$};
    \node (Nn-1)[color-group, left of=dots]       {$\Nc_{n-1}$};
    \node (Nn)  [color-group, left of=Nn-1]       {$\Nc_n$};
    \node (f)   [left=4ex of Nn]                  {};
    \node (Nf)  [flavor-group, above=1ex of f]    {$\Nf$};
    \node (Nf') [flavor-group, below=1ex of f]    {$\Nf$};
    \draw (Nf) -- (Nf');
    \draw[->-=.55] (Nf) -- (Nn);
    \draw[->-=.55] (Nn) -- (Nf');
    \draw[->-=.55] (Nn)   to [bend left=30] (Nn-1);
    \draw[->-=.55] (Nn-1) to [bend left=30] (dots);
    \draw[->-=.55] (dots) to [bend left=30] (N1);
    \draw[->-=.55] (N1)   to [bend left=30] (dots);
    \draw[->-=.55] (dots) to [bend left=30] (Nn-1);
    \draw[->-=.55] (Nn-1) to [bend left=30] (Nn);
    \draw[->-=.5]  (N1)   to [distance=5ex, in=60, out=120, loop] ();
    \draw[->-=.5]  (Nn-1) to [distance=5ex, in=60, out=120, loop] ();
    \node (top-corner) at ($(Nf.north east)+(0.1,5ex)$) {};
    \node (bottom-corner) at ($(Nf'.south west)+(-0.3,-0.3)$) {};
    \node (4d)  [below left = -1pt of top-corner] {4d};
    \node (2d)  [below right = -1pt of top-corner] {2d};
    \draw [color=gray, dashed, rounded corners]
          (top-corner) rectangle (bottom-corner);
    \draw [color=gray, dashed, rounded corners]
          (top-corner) rectangle ($(bottom-corner -| N1.east)+(0.3,0)$);
  }
  \,.
\end{equation}

This construction can be enriched by allowing one (or both) of the \(SU(\Nf)\) flavour symmetry
groups of~\eqref{quiver} to be coupled to one (or two) \(SU(\Nf)\) gauge group factors of a four
dimensional theory.  An interesting theory where such surface operators can be inserted is four
dimensional $\Nsusy=2$ superconformal SQCD\@. The SQCD quiver description
\begin{equation}
  \quiver[x=1em,y=1em]{
    \node (c)           at (0,0)     [color-group]  {$\scriptstyle SU(\Nf)$};
    \coordinate (left)  at (-3,0);
    \node (left-1)      at (-4,1.73)  [flavor-group] {$\scriptstyle U(1)$};
    \node (left-N)      at (-4,-1.73) [flavor-group] {$\scriptstyle SU(\Nf)$};
    \coordinate (right) at (3,0);
    \node (right-1)     at (4,1.73)   [flavor-group] {$\scriptstyle U(1)$};
    \node (right-N)     at (4,-1.73)  [flavor-group] {$\scriptstyle SU(\Nf)$};
    \foreach \side in {left,right}
      { \foreach \I in {c,\side-1,\side-N}   \draw (\side) -- (\I); }
  }
\end{equation}
makes an \(U(\Nf)^2\) flavour symmetry manifest.  Both sides of the quiver represent a
hypermultiplet transforming in the bifundamental representation of the \(SU(\Nf)\) gauge group and
a \(U(\Nf)\) flavour group.  The two dimensional gauge theory~\eqref{quiver} can now be coupled to
SQCD by identifying the two dimensional flavour symmetry with the \(U(\Nf)\) flavour symmetry of
either of these hypermultiplets and the \(SU(\Nf)\) gauge group. The two resulting surface
operators in SQCD are realized by the following 4d/2d quiver diagrams
(we introduce the hybrid node \mathtikz{\node [cf-group] {\phantom{x}};} to denote a four
dimensional gauge group which gauges a two dimensional flavour symmetry):
\begin{equation}\label{SQCDs}
  \newcommand{\surfacequiver}[1][]{%
    \quiver{%
      \node (Nf)  [flavor-group]                   {$\Nf$};
      \node (Nf') [cf-group, below of=Nf]      {$\Nf$};
      \node (Nf'')[flavor-group, below of=Nf'] {$\Nf$};
      \node (c)  [right of=Nf#1] {};
      \node (c') [right of=Nf'#1] {};
      \node (Nn) at ($(c)!0.5!(c')$) [color-group] {$\Nc_n$};
      \node (dots)[right of=Nn]                    {$\cdots$};
      \node (N1)  [color-group, right of=dots]     {$\Nc_1$};
      \draw (Nf) -- (Nf');
      \draw (Nf') -- (Nf'');
      \draw[->-=.55] (Nf#1) -- (Nn);
      \draw[->-=.55] (Nn) -- (Nf'#1);
      \draw[->-=.55] (Nn)   to [bend left=30] (dots);
      \draw[->-=.55] (dots) to [bend left=30] (N1);
      \draw[->-=.55] (N1)   to [bend left=30] (dots);
      \draw[->-=.55] (dots) to [bend left=30] (Nn);
      \draw[->-=.5]  (N1)   to [distance=5ex, in=60, out=120, loop] ();
      \node (top-corner) at ($(Nf.north east)+(0.1,5ex)$) {};
      \node (bottom-corner) at ($(Nf''.south west)+(-0.3,-0.3)$) {};
      \node (4d)  [below left = -1pt of top-corner] {4d};
      \node (2d)  [below right = -1pt of top-corner] {2d};
      \draw [color=gray, dashed, rounded corners]
            (top-corner) rectangle (bottom-corner);
      \draw [color=gray, dashed, rounded corners]
            (top-corner) rectangle ($(bottom-corner -| N1.east)+(0.3,0)$);
    }}
  \surfacequiver
  \quad\text{and}\quad
  \surfacequiver[']
  \,.
\end{equation}

The correspondence we propose between these surface operators and Toda CFT correlators predicts a
duality between the two coupled 4d/2d theories in~\eqref{SQCDs}, since \(SU(\Nf)\) SQCD is the
theory on $\Nf$~M5-branes wrapping a sphere with two full and two simple punctures.  The weakly
coupled regime of SQCD corresponds to a pants decomposition where the two simple punctures belong
to distinct trinions, which are joined by a thin tube.  In this framework, coupling the two
dimensional theory~\eqref{quiver} to either of the two hypermultiplets in SQCD correspond to
inserting a degenerate operator with momentum \(\alpha=-b\Omega\) in either trinion.  The
partition functions of the two surface operators in SQCD are thus both realized as the same
five-point function of two full, two simple, and an additional degenerate puncture:
\begin{equation}
  Z[\eqref{SQCDs}]
  = \mathtikz{
    \draw (0,0) ellipse (1.4 and .9);
    \fill (-.7,-.4) circle [radius=2pt];
    \fill (-.7, .4) circle [radius=2pt];
    \fill ( .7,-.4) circle [radius=2pt];
    \fill ( .7, .4) circle [radius=2pt];
    \draw (-.7,-.4) circle (4pt);
    \draw ( .7,-.4) circle (4pt);
    \node [cross] at (0,0) {};
    \node at (0,.35) {$\Omega$};
  } \,.
\end{equation}
In this language, the two 4d/2d quiver diagrams~\eqref{SQCDs} correspond to two different
degeneration limits of the five-point function.  It is important to note that this
``node-hopping'' duality of the 4d/2d theory is distinct from the usual S-duality of four
dimensional $\Nsusy=2$ SQCD\@.  The node-hopping duality was first observed in the superconformal
index of some 4d/2d theories in~\cite{Gadde:2013dda}, whose 4d/2d quiver notation we have
borrowed.  The superconformal index with surface operators has been considered
in~\cite{Nakayama:2011pa,Gaiotto:2012xa,Alday:2013kda,Bullimore:2014nla}.

More generally, the surface operator~\eqref{quiver} can be inserted in an arbitrary class~S theory
whenever the corresponding Riemann surface has at least one simple puncture.\footnote{Inserting
  multiple degenerate punctures near distinct simple punctures corresponds to including multiple
  surface operators built using distinct $SU(\Nf)\times SU(\Nf)\times U(1)$ groups of the four
  dimensional theory.  In a pants decomposition where the degenerate punctures are all inserted
  near the same simple puncture, the surface operator describes a single two dimensional gauge
  theory coupled through a given $SU(\Nf)\times SU(\Nf)\times U(1)$ symmetry group.}  The
generalized S-duality symmetry groupoid of a class~S theory, which is realized as the
Moore-Seiberg groupoid of the punctured Riemann surface, is enriched in the presence of surface
operators.  The addition of a degenerate puncture to the Riemann surface allows for further pants
decomposition of the enriched Riemann surface, and thereby more duality transformations, that go
beyond the dualities of the purely four dimensional theory.  The node-hopping
duality~\eqref{SQCDs} provides an example of a new duality of the 4d/2d system.

\pagebreak[0]\bigskip

In the second part of the paper we ``geometrize'' dualities of two dimensional $\Nsusy=(2,2)$
quiver gauge theories in terms of symmetries of Toda CFT correlation functions.  The quiver gauge
theories we consider are
\begin{equation}\label{gquiver}
  \quiver{%
    \node (N1)  [color-group]                    {$\Nc_1$};
    \node (N2)  [color-group, left of=N1]        {$\Nc_2$};
    \node (dots)[left of=N2]                     {$\cdots$};
    \node (Nn)  [color-group, left of=dots]      {$\Nc_n$};
    \node (f)   [left of=Nn]                     {};
    \node (Nf)  [flavor-group, above=0ex of f]   {$\Nf$};
    \node (Nf') [flavor-group, below=0ex of f]   {$\Nf$};
    \draw[->-=.55] (Nf) -- (Nn);
    \draw[->-=.55] (Nn) -- (Nf');
    \draw[->-=.55] (Nn)   to [bend left=30] (dots);
    \draw[->-=.55] (dots) to [bend left=30] (N2);
    \draw[->-=.55] (N2)   to [bend left=30] (N1);
    \draw[->-=.55] (N1)   to [bend left=30] (N2);
    \draw[->-=.55] (N2)   to [bend left=30] (dots);
    \draw[->-=.55] (dots) to [bend left=30] (Nn);
    \draw[->-=.5, densely dashed] (N1) to [distance=5ex, in=60, out=120, loop] ();
    \draw[->-=.5, densely dashed] (N2) to [distance=5ex, in=60, out=120, loop] ();
    \draw[->-=.5, densely dashed] (Nn) to [distance=5ex, in=60, out=120, loop] ();
  }
\end{equation}
where an adjoint chiral multiplet can be added to any gauge group factor.  Each adjoint chiral
multiplet is coupled to the neighboring bifundamental chiral multiplets through a cubic
superpotential, while nodes without an adjoint chiral multiplet have a quartic superpotential for
the neighboring bifundamental chiral multiplets.
Finally, the $\Nf$ fundamental and antifundamental chiral multiplets have no superpotential coupling.

We show that surface operators obtained by coupling these two dimensional
gauge theories~\eqref{gquiver} to class~S theories have a Toda CFT realization.  The quiver with $n$~gauge nodes
corresponds to the insertion of $n$~degenerate fields labeled by either symmetric or antisymmetric
representations of~$A_{\Nf-1}$.  The $n$~complexified FI parameters encode the position of the
$n$~degenerate punctures.  We now build the representations labeling degenerate punctures
recursively from the matter content of~\eqref{gquiver}.  If the $U(\Nc_n)$ factor has an adjoint
chiral multiplet, then the representation carried by the $n$-th puncture is of symmetric type, and
otherwise of antisymmetric type.  Then sequentially for each gauge group factor $U(\Nc_j)$ from
$j=n-1$ to~$1$, the $j$-th puncture is labeled by a representation of the same type as the
$(j+1)$-th puncture if there is an adjoint chiral multiplet, and otherwise by a representation of
the other type.  The Young diagram labeling the $j$-th puncture has $\Nc_j-\Nc_{j-1}$ boxes for
$1\leq j\leq n$, where $\Nc_0=0$.  See Table~\ref{tab:defects} for useful special cases and
Figure~\ref{fig:defects-example} for a concrete example.  The sphere partition function of the
surface operator inserted by~\eqref{gquiver} in the trinion theory of free hypermultiplets is the
Toda CFT correlator%
\begin{equation}\label{n+3pt}
  Z_{S^2\subset S^4_b}^{\eqref{gquiver}} =
  \mathtikz{
    \draw (0,0) ellipse (1.7 and .9);
    \fill (-.8,-.4) circle [radius=2pt];
    \fill (-.8, .4) circle [radius=2pt];
    \fill ( .8,-.4) circle [radius=2pt];
    \draw (-.8,-.4) circle (4pt);
    \draw ( .8,-.4) circle (4pt);
    \node [cross] at (.8,.4) {};
    \node at (.8,.05) {$\Omega\mathrlap{_1}$};
    \node [cross] at (-.3,.4) {};
    \node at (-.3,.05) {$\Omega\mathrlap{_n}$};
    \node at (.25,.4) {$\cdots$};
  }
  \rule[-7ex]{0pt}{14ex}
  \,.
\end{equation}
We also identify the gauge theory corresponding to multiple degenerate punctures labeled by
arbitrary representations of~\(A_{\Nf-1}\).

\begin{table}[pt]\centering
  \caption{\label{tab:defects}%
    Correspondence between surface operators defined by \(\Nsusy=(2,2)\) gauge
    theories and degenerate vertex operators labeled by representations of~\(A_{\Nf-1}\).
    The positions of degenerate operators are controlled by a combination~\(\hat{z}\) of
    FI and theta parameters for each gauge group \(U(\Nc)\), which differs from
    \(z=e^{-2\pi\xi+\I\vartheta}\) by a sign: \((-1)^{n_f}\)~if the group has an adjoint chiral
    multiplet and otherwise~\((-1)^{n_f+\Nc-1}\), where \(n_f\)~is the number of fundamental
    chiral multiplets for that group.}
  \medskip
  \newcommand{\myref}[2]{\hbox to 6em{\hyperref[#1]{(\ref*{#2}) \hfill p.~\pageref*{#1}}}}
  \begin{tabular}{m{4.5cm}lm{8em}l}
    \toprule
    2d Gauge theory & Field content & Representation & Equation
    \\\midrule
    SQED
    & \quiver{%
      \node (c)   [color-group]              {\vphantom{$\Nc$}$1$};
      \node (Nf)  [flavor-group, left of=c]  {$\Nf$};
      \node (Nf') [flavor-group, right of=c] {$\Nf$};
      \draw[->-=.55] (Nf) -- (c);
      \draw[->-=.55] (c) -- (Nf');
    }
    & Fundamental
    & \myref{sec:SQED}{SQED-matching}
    \\
    SQCD
    & \quiver{%
      \node (c)   [color-group]              {\vphantom{$\Nc$}$\Nc$};
      \node (Nf)  [flavor-group, left of=c]  {$\Nf$};
      \node (Nf') [flavor-group, right of=c] {$\Nf$};
      \draw[->-=.55] (Nf) -- (c);
      \draw[->-=.55] (c) -- (Nf');
    }
    & Antisymmetric
    & \myref{sec:SQCD}{SQCD-matching}
    \\
    SQCDA
    & \multirow{3}{*}{%
      \quiver{%
        \node (c)   [color-group]              {\vphantom{$\Nc$}$\Nc$};
        \node (Nf)  [flavor-group, left of=c]  {$\Nf$};
        \node (Nf') [flavor-group, right of=c] {$\Nf$};
        \draw[->-=.55] (Nf) -- (c);
        \draw[->-=.55] (c) -- (Nf');
        \draw[->-=.5] (c) to [in=60, out=120, distance=5ex, loop] ();
      }}
    & Symmetric
    & \myref{sec:SQCDA}{SQCDA-matching}
    \\
    \quad with \(W=\sum_t \anti{\quark}_t X^{l_t} \quark_t\)
    &
    & Two symmetrics
    & \myref{sec:SQCDAW}{SQCDWA-simple-degenerates}
    \\
    \quad with \(W=\Tr X^{l+1}\)
    &
    & Quasi-rectangular
    & \myref{SQCDAW-TrXl}{SQCDAW-matching}
    \\\midrule\addlinespace[-.4pt]
    \(\prod_j U(\Nc_j)\) quiver\newline with some adjoints
    & \hspace{-2em}%
    \quiver[node distance=2em, color-group/.append style = {inner sep = .3ex}]{%
      \node (N1)  [color-group]                    {$\scriptscriptstyle\Nc_1$};
      \node (dots)[left of=N1]                     {\smash{$\ldots$}};
      \node (Nn-1)[color-group, left=1ex of dots]  {$\scriptscriptstyle\Nc_{n-1}$};
      \node (Nn)  [color-group, left=1ex of Nn-1]  {$\scriptscriptstyle\Nc_n$};
      \node (f)   [left of=Nn] {};
      \node (Nf)  [flavor-group, above=0ex of f] {$\scriptscriptstyle\Nf$};
      \node (Nf') [flavor-group, below=0ex of f] {$\scriptscriptstyle\Nf$};
      \draw[->-=.55] (Nf) -- (Nn);
      \draw[->-=.55] (Nn) -- (Nf');
      \draw[->-=.55] (Nn)   to [bend left=30] (Nn-1);
      \draw[->-=.55] (Nn-1) to [bend left=30] (dots);
      \draw[->-=.55] (dots) to [bend left=30] (N1);
      \draw[->-=.55] (N1)   to [bend left=30] (dots);
      \draw[->-=.55] (dots) to [bend left=30] (Nn-1);
      \draw[->-=.55] (Nn-1) to [bend left=30] (Nn);
      \draw[->-=.5, densely dashed, distance=3ex] (N1)   to [in=60, out=120, loop] ();
      \draw[->-=.5, densely dashed, distance=3ex] (Nn-1) to [in=60, out=120, loop] ();
      \draw[->-=.5, densely dashed, distance=3ex] (Nn)   to [in=60, out=120, loop] ();
    }
    & Antisymmetrics\newline and symmetrics
    & \myref{sec:Quivers}{Quivers-matching}
    \\
    \(\prod_jU(\Nc_j)\) quiver\newline \(\hat{z}_1=\cdots=\hat{z}_{n-1}=1\)
    & \hspace{-2em}%
    \quiver[node distance=2em, color-group/.append style = {inner sep = .3ex}]{%
      \node (N1)  [color-group]                    {$\scriptscriptstyle\Nc_1$};
      \node (dots)[left of=N1]                     {\smash{$\ldots$}};
      \node (Nn-1)[color-group, left=1ex of dots]  {$\scriptscriptstyle\Nc_{n-1}$};
      \node (Nn)  [color-group, left=1ex of Nn-1]  {$\scriptscriptstyle\Nc_n$};
      \node (f)   [left of=Nn] {};
      \node (Nf)  [flavor-group, above=0ex of f]   {$\scriptscriptstyle\Nf$};
      \node (Nf') [flavor-group, below=0ex of f]   {$\scriptscriptstyle\Nf$};
      \draw[->-=.55] (Nf) -- (Nn);
      \draw[->-=.55] (Nn) -- (Nf');
      \draw[->-=.55] (Nn)   to [bend left=30] (Nn-1);
      \draw[->-=.55] (Nn-1) to [bend left=30] (dots);
      \draw[->-=.55] (dots) to [bend left=30] (N1);
      \draw[->-=.55] (N1)   to [bend left=30] (dots);
      \draw[->-=.55] (dots) to [bend left=30] (Nn-1);
      \draw[->-=.55] (Nn-1) to [bend left=30] (Nn);
      \draw[->-=.5, distance=3ex]  (N1)   to [in=60, out=120, loop] ();
      \draw[->-=.5, distance=3ex]  (Nn-1) to [in=60, out=120, loop] ();
    }
    & Arbitrary
    & \myref{sec:Quivers-fuse}{Quivers-matching-fused}
    \\\bottomrule
  \end{tabular}
\end{table}

\begin{figure}[pt]
  \caption{\label{fig:defects-example}%
    Example of how multiple Toda CFT degenerate punctures map to a quiver gauge theory.}
  \medskip
  \begin{tabular}{p{0.97\textwidth}}
    \toprule
    The \(U(\Nc_1)\times\cdots\times U(\Nc_4)\) linear quiver given below has adjoint chiral
    multiplets for \(U(\Nc_1)\) and \(U(\Nc_4)\), hence two cubic superpotential terms coupling
    these to neighboring bifundamental multiplets.  It also has two quartic superpotential terms
    coupling bifundamentals charged under \(U(\Nc_2)\), and those charged under \(U(\Nc_3)\).
    The partition function of the surface operator inserted by coupling the theory to
    \(\Nf^2\)~hypermultiplets is captured by a Toda CFT correlator with two full punctures at
    \(0\) and~\(\infty\), one simple at~\(1\) and four degenerate punctures at
    \(x_4=\hat{z}_4\), \(x_3=\hat{z}_4\hat{z}_3\), \(x_2=\hat{z}_4\hat{z}_3\hat{z}_2\),
    \(x_1=\hat{z}_4\hat{z}_3\hat{z}_2\hat{z}_1\), where \(\hat{z}_4=(-1)^{\Nc_3+\Nf}z_4\),
    \(\hat{z}_3=(-1)^{\Nc_2+\Nc_4+\Nc_3-1}z_3\), \(\hat{z}_2=(-1)^{\Nc_1+\Nc_3+\Nc_2-1}z_2\),
    \(\hat{z}_1=(-1)^{\Nc_2}z_1\).
    \begin{equation*}
      Z_{S^2\subset S^4_b} \left[
        \:\:
        \raise5pt\hbox{%
        \quiver{%
          \node (N1)  [color-group]                    {$\Nc_1$};
          \node (N2)  [color-group, left of=N1]        {$\Nc_2$};
          \node (N3)  [color-group, left of=N2]        {$\Nc_3$};
          \node (N4)  [color-group, left of=N3]        {$\Nc_4$};
          \node (f)   [left of=N4]                     {};
          \node (Nf)  [flavor-group, above=0ex of f]   {$\Nf$};
          \node (Nf') [flavor-group, below=0ex of f]   {$\Nf$};
          \draw[->-=.55] (Nf) -- (N4);
          \draw[->-=.55] (N4) -- (Nf');
          \draw[->-=.55] (N4)   to [bend left=30] (N3);
          \draw[->-=.55] (N3)   to [bend left=30] (N2);
          \draw[->-=.55] (N2)   to [bend left=30] (N1);
          \draw[->-=.55] (N1)   to [bend left=30] (N2);
          \draw[->-=.55] (N2)   to [bend left=30] (N3);
          \draw[->-=.55] (N3)   to [bend left=30] (N4);
          \draw[->-=.5]  (N1)   to [distance=5ex, in=60, out=120, loop] ();
          \draw[->-=.5]  (N4)   to [distance=5ex, in=60, out=120, loop] ();
        }}
      \right]
      =
      \mathtikz{
        \draw (0,0) ellipse (1.9 and .8);
        \fill (-1.5,-.1) circle [radius=2pt];
        \fill (-1.0,-.1) circle [radius=2pt];
        \fill ( 1.5,-.1) circle [radius=2pt];
        \draw (-1.5,-.1) circle (4pt);
        \draw ( 1.5,-.1) circle (4pt);
        \node [cross] at (-.5,-.1) {};
        \node at (-.5,-.4) {$x_4$};
        \draw (-.65,.2) rectangle (-.35,.3);
        \node [cross] at (  0,-.1) {};
        \node at (  0,-.4) {$x_3$};
        \draw (-.05,.1) rectangle ( .05,.4);
        \node [cross] at ( .5,-.1) {};
        \node at ( .5,-.4) {$x_2$};
        \draw ( .35,.2) rectangle ( .65,.3);
        \node [cross] at (1.0,-.1) {};
        \node at (1.0,-.4) {$x_1$};
        \draw ( .85,.2) rectangle (1.15,.3);
      }
      \,.
    \end{equation*}
    The degenerate punctures are labelled by the \((\Nc_4-\Nc_3)\)-th symmetric, the
    \((\Nc_3-\Nc_2)\)-th antisymmetric, the \((\Nc_2-\Nc_1)\)-th symmetric, and the \(\Nc_1\)-th
    symmetric representations, depicted by cartoons of their Young diagrams.  Whenever two
    neighboring punctures have a different type of representation  the corresponding gauge theory
    node has no adjoint, while neighbors of the same type yield an adjoint.  The end node
    \(U(\Nc_4)\) is special and has an adjoint because the first puncture is symmetric.
    \\\bottomrule
  \end{tabular}
\end{figure}

\begin{table}[pt]\centering
  \caption{\label{tab:dualities}%
    Dualities of \(\Nsusy=(2,2)\) quiver gauge theories realized as symmetries in the Toda CFT\@.
    Chiral multiplets are denoted by \(\quark_t\)~(fundamentals),
    \(\anti{\quark}_t\)~(antifundamentals), and \(X\)~(adjoint).  Each has a twisted
    mass~\(\mathrm{m}\) and an \(R\)-charge~\(\mathrm{q}\).  FI and theta parameters combine into
    \(z=e^{-2\pi\xi+\I\vartheta}\) for each gauge group \(U(\Nc)\); denoting \(n_f\)~and~\(n_a\) the
    numbers of fundamental and of adjoint chiral multiplets, we also define
    \(\hat{z}=(-1)^{n_f+(n_a-1)(\Nc-1)}z\).
    For Seiberg and  Kutasov--Schwimmer dualities, the magnetic theory contains extra free chiral
    multiplets whose charges are identical to those of mesons in the electric theory.
    We assume \(\anti{\Nf}<\Nf\).}
  \medskip
  \newcommand{\myref}[1]{\unskip\space\hfil\mbox{\hyperref[#1]{p.~\pageref*{#1}}}}
  \arraycolsep=0pt\relax
  \begin{tabular}{m{4.49em}l@{\hspace{-.5em}}c@{}m{13.05em}@{\hspace{.5em}}m{9.9em}}
    \toprule
    Duality & Quiver & \(W\) & Dual parameters & Toda symmetry
    \\\midrule
    Seiberg
    & \quiver[node distance=2em, color-group/.append style = {inner sep = .3ex}]{%
        \node (f) {};
        \node (Nf)  [flavor-group, above=0ex of f] {$\scriptscriptstyle\Nf$};
        \node (Nf') [flavor-group, below=0ex of f] {$\scriptscriptstyle\anti{\Nf}$};
        \node (N)   [color-group, right of=f] {$\scriptscriptstyle\Nc$};
        \draw[->-=.55] (Nf) -- (N);
        \draw[->-=.55] (N) -- (Nf');
      }%
    & \(0\)
    & \(\Nc^\dual=\Nf-\Nc\), \(z^\dual=(-1)^{\anti{\Nf}}z\),\newline
    \(\mathrm{q}^\dual=1-\mathrm{q}\), \(\mathrm{m}^\dual=-\mathrm{m}\)
    & Conjugation \myref{sec:Seiberg}\linebreak
    \((-b\omega_\Nc)^\conj=-b\omega_{\Nc^\dual}\)
    \\\addlinespace[-.4pt]
    \((2,2)^*\)-like
    & \quiver[node distance=2em, color-group/.append style = {inner sep = .3ex}]{%
        \node (f) {};
        \node (Nf)  [flavor-group, above=0ex of f] {$\scriptscriptstyle\Nf$};
        \node (Nf') [flavor-group, below=0ex of f] {$\scriptscriptstyle\Nf$};
        \node (N)   [color-group, right of=f] {$\scriptscriptstyle\Nc$};
        \draw[->-=.55] (Nf) -- (N);
        \draw[->-=.55] (N) -- (Nf');
        \draw[->-=.5] (N)   to [in=60, out=120, loop] ();
      }%
    &\(\sum_t \anti{\quark}_t X^{l_t} \quark_t\)
    & \(\Nc^\dual=\sum_t l_t - \Nc\), \(z^\dual=z^{-1}\),\newline
    \(\mathrm{q}^\dual=\mathrm{q}\), \(\mathrm{m}^\dual=\mathrm{m}\)
    & Crossing \myref{sec:SeibergW}\linebreak
    simple \(\to\) degenerate
    \\\addlinespace
    Kutasov--\newline\rlap{Schwimmer}
    & \quiver[node distance=2em, color-group/.append style = {inner sep = .3ex}]{%
        \node (f) {};
        \node (Nf)  [flavor-group, above=0ex of f] {$\scriptscriptstyle\Nf$};
        \node (Nf') [flavor-group, below=0ex of f] {$\scriptscriptstyle\anti{\Nf}$};
        \node (N)   [color-group, right of=f] {$\scriptscriptstyle\Nc$};
        \draw[->-=.55] (Nf) -- (N);
        \draw[->-=.55] (N) -- (Nf');
        \draw[->-=.5] (N)   to [in=60, out=120, loop] ();
      }%
    & \(\Tr X^{l+1}\)
    & \(\Nc^\dual=l\Nf - \Nc\),\newline
    \(z^\dual=(-1)^{\Nf-\anti{\Nf}}z\),\newline
    \(\mathrm{q}^\dual=\frac{2}{l+1}-\mathrm{q}\), \(\mathrm{m}^\dual=-\mathrm{m}\)
    & Conjugation \myref{sec:Kutasov}\linebreak
    \((-\Nc bh_1)^\conj\equiv -\Nc^\dual bh_1\)
    \\\midrule\addlinespace[-.4pt]
    Quiver
    &
    \multicolumn{2}{c}{%
      \hspace{-2em}%
      \quiver[node distance=2em, color-group/.append style = {inner sep = .3ex}]{%
        \node (N1)  [color-group]                    {$\scriptscriptstyle\Nc_1$};
        \node (dots)[left of=N1]                     {\smash{$\ldots$}};
        \node (Nn-1)[color-group, left=1ex of dots]  {$\scriptscriptstyle\Nc_{n-1}$};
        \node (Nn)  [color-group, left=1ex of Nn-1]  {$\scriptscriptstyle\Nc_n$};
        \node (f)   [left of=Nn] {};
        \node (Nf)  [flavor-group, above=0ex of f] {$\scriptscriptstyle\Nf$};
        \node (Nf') [flavor-group, below=0ex of f] {$\scriptscriptstyle\Nf$};
        \draw[->-=.55] (Nf) -- (Nn);
        \draw[->-=.55] (Nn) -- (Nf');
        \draw[->-=.55] (Nn)   to [bend left=30] (Nn-1);
        \draw[->-=.55] (Nn-1) to [bend left=30] (dots);
        \draw[->-=.55] (dots) to [bend left=30] (N1);
        \draw[->-=.55] (N1)   to [bend left=30] (dots);
        \draw[->-=.55] (dots) to [bend left=30] (Nn-1);
        \draw[->-=.55] (Nn-1) to [bend left=30] (Nn);
        \draw[->-=.5, densely dashed, distance=3ex] (N1)   to [in=60, out=120, loop] ();
        \draw[->-=.5, densely dashed, distance=3ex] (Nn-1) to [in=60, out=120, loop] ();
        \draw[->-=.5, densely dashed, distance=3ex] (Nn)   to [in=60, out=120, loop] ();
      }%
    }
    & \(\Nc_j^\dual=\Nc_{j-1}+\Nc_{j+1}-\Nc_j\)\newline
    \(\hat{z}_j^\dual = \hat{z}_j^{-1}\), \(\hat{z}_{j\pm 1}^\dual = \hat{z}_j \hat{z}_{j\pm 1}\)
    & Crossing \myref{sec:SeibergQ-perm}\linebreak
    \(\omega_{\Nc_j-\Nc_{j-1}} \leftrightarrow \omega_{\Nc_{j+1}-\Nc_j}\)
    \\\addlinespace[-.4pt]
    Quiver
    &
    \multicolumn{2}{c}{%
      \hspace{-2em}%
      \quiver[node distance=2em, color-group/.append style = {inner sep = .3ex}]{%
        \node (N1)  [color-group]                    {$\scriptscriptstyle\Nc_1$};
        \node (dots)[left of=N1]                     {\smash{$\ldots$}};
        \node (Nn-1)[color-group, left=1ex of dots]  {$\scriptscriptstyle\Nc_{n-1}$};
        \node (Nn)  [color-group, left=1ex of Nn-1]  {$\scriptscriptstyle\Nc_n$};
        \node (f)   [left of=Nn] {};
        \node (Nf)  [flavor-group, above=0ex of f]   {$\scriptscriptstyle\Nf$};
        \node (Nf') [flavor-group, below=0ex of f]   {$\scriptscriptstyle\Nf$};
        \draw[->-=.55] (Nf) -- (Nn);
        \draw[->-=.55] (Nn) -- (Nf');
        \draw[->-=.55] (Nn)   to [bend left=30] (Nn-1);
        \draw[->-=.55] (Nn-1) to [bend left=30] (dots);
        \draw[->-=.55] (dots) to [bend left=30] (N1);
        \draw[->-=.55] (N1)   to [bend left=30] (dots);
        \draw[->-=.55] (dots) to [bend left=30] (Nn-1);
        \draw[->-=.55] (Nn-1) to [bend left=30] (Nn);
        \draw[->-=.5, distance=3ex]  (N1)   to [in=60, out=120, loop] ();
        \draw[->-=.5, distance=3ex]  (Nn-1) to [in=60, out=120, loop] ();
      }%
    }
    & \(\Nc_j^\dual=j\Nf-\Nc_j\) \(\forall j\)\newline
    \(\mathrm{q}^\dual=1-\mathrm{q}\), \(\mathrm{m}^\dual=-\mathrm{m}\)
    & Conjugation \myref{sec:SeibergQ-group}\linebreak
    \(\omega_{\Nc_j-\Nc_{j-1}}^\conj = \omega_{\Nc_j^\dual-\Nc_{j-1}^\dual}\)
    \\\bottomrule
  \end{tabular}
\end{table}

\begin{table}[pt]\centering
  \caption{\label{tab:moves}%
    The effect of a few Toda CFT moves on the corresponding 4d/2d gauge theory.  Besides the
    symmetry under changing trinion decomposition, Toda CFT correlators are also invariant under
    conjugation of all momenta.  Full punctures are drawn as solid lines, simple punctures as
    dashed lines, and degenerate punctures as dotted lines.  References are to papers describing
    the gauge theory duality and to papers giving its relation to Toda CFT.}
  \medskip
  \tikzset{x=1.5em,y=1.5ex}
  \tikzset{simple/.style={densely dashed}}
  \tikzset{degenerate/.style={densely dotted}}
  \begin{tabular}{r@{${}\leftrightarrow{}$}l >{$\:\Longleftrightarrow\quad$}lm{17.2em}}
    \toprule
    \multicolumn{2}{c}{Toda CFT move}
    && Gauge theory duality
    \\\midrule
    % _|_|_
    \mathtikz[thick]{
      \draw[->-=.55] (0,0) -- (1,0);
      \draw[->-=.55] (3,0) -- (2,0);
      \draw          (2,0) -- (1,0);
      \draw[->-=.55, simple] (1,3) -- (1,0);
      \draw[->-=.55, simple] (2,3) -- (2,0);
    }
    &
    % _X_
    \mathtikz[thick]{
      \draw[->-=.55] (0,0) -- (1,0);
      \draw[->-=.55] (3,0) -- (2,0);
      \draw          (2,0) -- (1,0);
      \draw[->-=.63, simple, name path=L1] (1,3) to [bend left=30] (2,0);
      \path[name path=L2] (2,3) to [bend right=30] (1,0);
      \fill[white, name intersections={of=L1 and L2}]
        (intersection-1) circle (2pt);
      \draw[->-=.63, simple] (2,3) to [bend right=30] (1,0);
    }
    && 4d generalized S-duality~\cite{Gaiotto:2009we};~\cite{AGT}
    \\\addlinespace
    % _V_|_ with 1 degenerate
    \mathtikz[thick]{
      \draw[->-=.55] (0,0) -- (1,0);
      \draw[->-=.55] (3,0) -- (2,0);
      \draw          (2,0) -- (1,0);
      \draw[->-=.7, simple] (1,3) -- (1,0);
      \draw[->-=.55, degenerate] (1.5,3) -- (1.1,0);
      \draw[->-=.55, simple] (2,3) -- (2,0);
    }
    &
    % _|_V_ with 1 degenerate
    \mathtikz[thick]{
      \draw[->-=.55] (0,0) -- (1,0);
      \draw[->-=.55] (3,0) -- (2,0);
      \draw          (2,0) -- (1,0);
      \draw[->-=.55, simple] (1,3) -- (1,0);
      \draw[->-=.55, simple] (2,3) -- (2,0);
      \draw[->-=.55, degenerate] (1.5,3) -- (1.9,0);
    }
    && 4d/2d node-hopping~\cite{Gadde:2013dda}
    \\\addlinespace
    % _|_|_ with 1 degenerate
    \mathtikz[thick]{
      \draw[->-=.55] (0,0) -- (1,0);
      \draw[->-=.55] (3,0) -- (2,0);
      \draw          (2,0) -- (1,0);
      \draw[->-=.55, simple] (1,3) -- (1,0);
      \draw[->-=.55, degenerate] (2,3) -- (2,0);
    }
    &
    % _X_ with 1 degenerate
    \mathtikz[thick]{
      \draw[->-=.55] (0,0) -- (1,0);
      \draw[->-=.55] (3,0) -- (2,0);
      \draw          (2,0) -- (1,0);
      \draw[->-=.63, simple, name path=L1] (1,3) to [bend left=30] (2,0);
      \path[name path=L2] (2,3) to [bend right=30] (1,0);
      \fill[white, name intersections={of=L1 and L2}]
        (intersection-1) circle (2pt);
      \draw[->-=.63, degenerate] (2,3) to [bend right=30] (1,0);
    }
    && 2d flop transition~\cite{Witten:1993yc};~\cite{Doroud:2012xw}
    \\\addlinespace
    % _|_|_ with 1 degenerate
    \mathtikz[thick]{
      \draw[->-=.55] (0,0) -- (1,0);
      \draw[->-=.55] (3,0) -- (2,0);
      \draw          (2,0) -- (1,0);
      \draw[->-=.55, simple] (1,3) -- (1,0);
      \draw[->-=.55, degenerate] (2,3) -- (2,0);
    }
    &
    % _|_|_ with 1 degenerate
    $\Biggl(
    \mathtikz[thick]{
      \draw[->-=.55] (0,0) -- (1,0);
      \draw[->-=.55] (3,0) -- (2,0);
      \draw          (2,0) -- (1,0);
      \draw[->-=.55, simple] (1,3) -- (1,0);
      \draw[->-=.55, degenerate] (2,3) -- (2,0);
    }
    \Biggr)^\conj$
    && 2d Seiberg duality~\cite{Benini:2012ui}: our Section~\ref{sec:Seiberg}\newline
    Kutasov--Schwimmer: our Section~\ref{sec:SeibergW}
    \\
    % _|_|_ with 2 degenerate
    \mathtikz[thick]{
      \draw[->-=.55] (0,0) -- (1,0);
      \draw[->-=.55] (3,0) -- (2,0);
      \draw          (2,0) -- (1,0);
      \draw[->-=.55, degenerate] (1,3) -- (1,0);
      \draw[->-=.55, degenerate] (2,3) -- (2,0);
    }
    &
    % _X_ with 2 degenerate
    \mathtikz[thick]{
      \draw[->-=.55] (0,0) -- (1,0);
      \draw[->-=.55] (3,0) -- (2,0);
      \draw          (2,0) -- (1,0);
      \draw[->-=.63, degenerate, name path=L1] (1,3) to [bend left=30] (2,0);
      \path[name path=L2] (2,3) to [bend right=30] (1,0);
      \fill[white, name intersections={of=L1 and L2}]
        (intersection-1) circle (2pt);
      \draw[->-=.63, degenerate] (2,3) to [bend right=30] (1,0);
    }
    && 2d Seiberg and \((2,2)^*\) dualities\newline for quivers~\cite{Benini:2014mia}: our Section~\ref{sec:SeibergQ}
    \\\bottomrule
  \end{tabular}
\end{table}

We consider several dualities in two dimensional $\Nsusy=(2,2)$ theories, realized as symmetries
of the corresponding Toda CFT degenerate operators.  As described below (see also
Table~\ref{tab:dualities}), some dualities correspond to the crossing symmetry exchanging two
degenerate operators, while others correspond to conjugating all Toda CFT momenta, under which the
fundamental weights of $A_{\Nf-1}$ transform as $(\omega_\Nc)^\conj = \omega_{\Nf-\Nc}$.  We also
establish these dualities through explicit evaluation of the exact two-sphere partition function
\cite{Benini:2012ui,Doroud:2012xw} of dual theories.  This completes the dictionary between
symmetries of Toda CFT correlators and dualities of 4d/2d gauge theories (see
Table~\ref{tab:moves}).

\pagebreak[0]\bigskip

The detailed description of the rest of the paper follows. Section~\ref{sec:partition} is devoted
to the correspondence between surface operators labeled by two dimensional quiver gauge theories
and Toda CFT degenerate operators.  We derive the identification by coupling the two dimensional
theories to the trinion theory of free hypermultiplets, as this choice of a free four dimensional
theory lets us concentrate on the two dimensional theories.  The \(S^2\subset S^4_b\) partition
function of these surface operators corresponds to Toda CFT correlators involving one simple, two
full, and additional degenerate operators.

After describing our gauge theory setup, and recalling explicit expressions for the \(S^4_b\)~and
\(S^2\) contributions, we proceed to expand \(S^2\)~partition functions in various limits and
compare them with Toda CFT results.  First, we review the case of SQED in some detail in
Section~\ref{sec:SQED}: this \(U(1)\) gauge theory corresponds to the insertion of the simplest
Toda CFT degenerate vertex operator, labeled by the fundamental representation
of~\(A_{\Nf-1}\)~\cite{Doroud:2012xw}.  Then, we move on to \(U(\Nc)\) SQCD in
Section~\ref{sec:SQCD}, which corresponds to inserting a degenerate operator labeled by an
antisymmetric representation of~\(A_{\Nf-1}\).  Using new braiding matrices derived in
Appendix~\ref{app:braiding}, we prove that the Toda CFT correlator and the partition function of
the 4d/2d theory are equal.  We then describe in Section~\ref{sec:irreg} how one can decouple some
free hypermultiplets from the four dimensional theory and chiral multiplets from the two
dimensional theory: the procedure translates to a collision limit where two Toda CFT vertex
operators combine into an irregular puncture (see also Appendix~\ref{app:irregular}).  In
Section~\ref{sec:SQCDA}, we add adjoint matter to SQCD to get SQCDA, and find that it corresponds
to a degenerate operator labeled by a symmetric representation.  We then consider SQCDA with
different superpotentials in Section~\ref{sec:SQCDAW} and give their Toda CFT interpretation.
Finally, in Section~\ref{sec:Quivers}, we show that the previous results arise as special cases of
surface operators described by the quivers~\eqref{gquiver}, which correspond to the insertion of
several (symmetric and antisymmetric) degenerate operators.
We briefly discuss a brane diagram interpretation of the dictionary.
By fusing representations, we deduce in Section~\ref{sec:Quivers-fuse} which surface operator corresponds to an arbitrary degenerate
operator.  All cases are summarized in Table~\ref{tab:defects}.

Section~\ref{sec:Dualities} describes dualities of two dimensional \(\Nsusy=(2,2)\) gauge theories
which can be obtained as manifest Toda CFT symmetries.  The dualities relate the IR limits of
these theories, and we probe them by comparing \(S^2\)~partition functions of proposed duals.
The contribution of free hypermultiplets to the partition function of the 4d/2d theory
plays little role.  We find several Seiberg-like dualities (generalizing the duality found by Hori
and Tong~\cite{Hori:2006dk}) relating theories with similar matter content but different gauge
groups (see Table~\ref{tab:dualities}).  The dualities are most clearly seen through the matching
with the Toda CFT, but we also show directly in Appendix~\ref{app:Proof} that the
\(S^2\)~partition functions of dual theories are equal.\footnote{This was shown previously for
  SQCD with \(\Nf\)~fundamental and \(\anti{\Nf}\leq\Nf-2\) antifundamental chiral
  multiplets~\cite{Benini:2012ui}, and generalized very recently to arbitrary~\(\anti{\Nf}\)
  in~\cite{Benini:2014mia}.  Our proofs follow the same logic but also apply to theories with an
  adjoint chiral multiplet and a superpotential.}

We start in Section~\ref{sec:Seiberg} with the two dimensional analogue of Seiberg
duality~\cite{Seiberg:1994pq}, between $\Nsusy=(2,2)$ \(U(\Nc)\) SQCD with \(\Nf\)~flavours, and
\(U(\Nf-\Nc)\) SQCD with \(\Nf\)~flavours.  The corresponding Toda CFT correlators are simply
related by conjugating all momenta.  This operation provides us with the precise map of
parameters: \(\Nc^\dual=\Nf-\Nc\), \(z^\dual=(-1)^{\Nf}z\), and \(m^\dual = \I/2 - m\) for the complexified
twisted masses of every chiral multiplet.\footnote{By coupling the flavour symmetry to a
  constant background vector multiplet, chiral multiplets can be
  given twisted masses and \(R\)-charges, which combine into a complex parameter~\(m\) for each
  chiral multiplet.}  In addition to fundamental and antifundamental chiral multiplets, the
\(U(\Nf-\Nc)\) theory involves a free chiral multiplet transforming in the bifundamental
representation of the flavour symmetry group $S[U(\Nf)\times U(\Nf)]$.  These free chiral
multiplets couple to the charged multiplets through a cubic superpotential, which must have total
\(R\)-charge~\(2\) (complexified twisted mass~\(\I\)) to be supersymmetric.  As was also observed
recently in~\cite{Benini:2014mia}, the theories differ by a shift in the FI parameter associated to
the \(U(1)\) flavour symmetry.  In Section~\ref{sec:Seiberg-irreg}, we deduce Seiberg duality relations between theories with
\(\Nf\)~fundamental and \(\anti{\Nf}<\Nf\) antifundamental chiral multiplets (with \(z^\dual=(-1)^{\anti{\Nf}}z\)).  For this, we let
some of the twisted masses of antifundamental multiplets go to infinity and take into account the
renormalization of the FI parameter: this limit precisely corresponds to merging the Toda CFT
operators inserted at \(\infty\) and~\(1\) into an irregular puncture~\cite{Gaiotto:2012sf}.

We then move on in Section~\ref{sec:SeibergW} to dualities of \(U(\Nc)\) SQCDA, which has
fundamental, antifundamental, and adjoint chiral multiplets.  Without further restriction, the
theory features no duality.  We find two choices of superpotentials for which the theory has a
dual description: both dualities appear to be new in two dimensional \(\Nsusy=(2,2)\) theories.

In Section~\ref{sec:SeibergW-cross}, we consider SQCDA with the superpotential
\begin{equation}\label{intro-supo-star}
  W = \sum_{t=1}^{\Nf} \anti{\quark}_t X^{l_t} \quark_t \,,
\end{equation}
described by a choice of \(\Nf\)~integers \(l_t \geq 0\), where \(\quark_t\), \(\anti{\quark}_t\)
and~\(X\) are the fundamental, antifundamental, and adjoint chiral multiplets.  The theory is a
simple generalization of \(\Nsusy=(2,2)^*\) SQCD\@.\footnote{\(\Nsusy=(2,2)^*\) SQCD is the mass
  deformation of the \(\Nsusy=(4,4)\) theory of a \(U(\Nc)\) vector multiplet coupled to
  \(\Nf\)~fundamental hypermultiplets.  Its cubic superpotential \(W=\sum_t \anti{\quark}_t X
  \quark_t\) corresponds to taking all \(l_t=1\).}  The constraint on \(R\)-charges due to the
superpotential translates to a very natural constraint in the Toda CFT language.  Namely, the
momentum labeling the simple puncture gets fine-tuned to become a degenerate operator, labeled
by a symmetric representation of~\(A_{\Nf-1}\).  The crossing symmetry exchanging these two
degenerate vertex operators thus provides us with a duality between two dimensional SQCDA theories
with the superpotential~\eqref{intro-supo-star}.  The \(U(\Nc^\dual)\) dual theory features the
same chiral multiplets and superpotential as the \(U(\Nc)\) theory, with identical complexified
twisted masses, \(\Nc^\dual = \sum_t l_t - \Nc\), and \(z^\dual = z^{-1}\).

In Section~\ref{sec:Kutasov}, we consider SQCDA with the superpotential
\begin{equation}\label{intro-supo-Kutasov}
  W = \Tr X^{l+1}
\end{equation}
for some integer \(l\geq 0\), where \(X\)~is the adjoint chiral multiplet.  We find a direct
analogue of the four dimensional Kutasov--Schwimmer duality \cite{Kutasov:1995ve,Kutasov:1995np}.
It turns out that given the superpotential constraint, conjugation maps the (symmetric) degenerate
operator describing \(U(\Nc)\) SQCDA to the degenerate operator describing \(U(\Nc^\dual)\) SQCDA\@.
The dual gauge theory has \(\Nc^\dual = l\Nf-\Nc\), \(z^\dual = z\), \(m_t^\dual = m_X - m_t\),
\(\anti{m}_t^\dual = m_X - \anti{m}_t\), and \(m_X^\dual = m_X\).  As in four
dimensions~\cite{Kutasov:1995ve,Kutasov:1995np}, the dual theory features $l$~additional free
chiral multiplets in the bifundamental representation of $S[U(\Nf)\times U(\Nf)]$, which
correspond to mesons of the electric theory.  As for SQCD, the limit where twisted masses of some
chiral multiplets are very large yields similar dualities between theories with a different number
of fundamental and antifundamental chiral multiplets.

Lastly, we describe dualities of quiver gauge theories in Section~\ref{sec:SeibergQ}.  We consider
the \(U(\Nc_1)\times\cdots\times U(\Nc_n)\) quiver theories~\eqref{gquiver} which correspond in
the Toda CFT to the insertion of \(n\)~degenerate vertex operators.  Dualities of another type of
\(\Nsusy=(2,2)\) quiver gauge theories were considered recently in~\cite{Benini:2014mia}.

In Section~\ref{sec:SeibergQ-perm} we apply Seiberg duality or the \(\Nsusy=(2,2)^*\) duality
(depending on the presence or absence of an adjoint chiral multiplet) to gauge group factors
\(U(\Nc_j)\) with \(j<n\).  We show that the duality translates to the exchange of degenerate
punctures numbered \(j\) and \(j+1\) in the Toda CFT\@.  Each permutation of the \(n\)~degenerate
punctures is thus realized as a combination of such Seiberg and \(\Nsusy=(2,2)^*\) dualities.

Based on this geometric realization of dualities for \(j<n\), we construct in
Section~\ref{sec:SeibergQ-group} the full set of dual theories obtained through Seiberg and
\(\Nsusy=(2,2)^*\) dualities acting on any gauge group.
We find no Toda CFT description of the duality acting on \(U(\Nc_n)\), except when all degenerate
vertex operators are labeled by antisymmetric representations of~\(A_{\Nf-1}\).  Then, conjugating
all Toda CFT momenta yields a different set of degenerate operators of the same type, and it turns
out that the corresponding dual gauge theories are related by a sequence of Seiberg and
\(\Nsusy=(2,2)^*\) dualities on all nodes.  A particular case is the quiver~\eqref{quiver}
which corresponds to a single degenerate vertex operator labeled by an arbitrary
representation~\(\repr\): applying the same sequence of Seiberg and \(\Nsusy=(2,2)^*\)
dualities corresponds to conjugating~\(\repr\) and all Toda CFT momenta.
This result concludes the description of dualities of two dimensional \(\Nsusy=(2,2)\)
gauge theories which correspond to manifest symmetries of the Toda CFT\@.

Many new Toda CFT results are presented in Appendix~\ref{app:Toda}.  We describe notations and the
normalization of vertex operators (Appendix~\ref{app:Toda-basics}), compare one-loop determinants
and three-point functions (Appendix~\ref{app:1loop-3pt}), work out new braiding matrices
(Appendix~\ref{app:braiding}), give new fusion rules (Appendix~\ref{app:fusion}), deduce new
conformal blocks from the correspondence (Appendix~\ref{app:blocks}), and collide vertex operators
to build irregular punctures of the \(W_{\Nf}\)~algebra (Appendix~\ref{app:irregular}).  Finally,
Appendix~\ref{app:Proof} features analytic proofs that vortex partition of dual theories are
equal, for Seiberg duality (Appendix~\ref{app:Proof-SQCD}), and for dualities of SQCD with an
adjoint (Appendix~\ref{app:Proof-SQCDAW}).

\numberwithin{equation}{section}
\section{Surface Operators as Toda Degenerate Operators}
\label{sec:partition}

In this section, we consider half-BPS surface operators obtained by
coupling two dimensional \(\Nsusy=(2,2)\) gauge theories to four
dimensional \(\Nsusy=2\) theories of class~S\@.  We enrich the dictionary
between class~S theories and Riemann surfaces by identifying surface
operators which correspond to the insertion of arbitrary degenerate
punctures.

To make the two dimensional features most visible, we restrict ourselves
to surface operators in the simplest class~S theory, which is the theory
on \(\Nf\)~M5-branes wrapping a sphere with two full and one simple
puncture, namely the theory of \(\Nf^2\)~free
hypermultiplets~\(\hyperquark^{\text{4d}}\).  The M-theory description
makes an \(SU(\Nf)\times SU(\Nf)\times U(1)\) flavour symmetry manifest,
and the hypermultiplets transform in the trifundamental representation
of this group.  All two dimensional theories we study contain
\(\Nf\)~fundamental chiral multiplets~\(\quark\) and
\(\Nf\)~antifundamental chiral multiplets~\(\anti{\quark}\) of a
\(U(\Nc)\) gauge group factor.  The 4d/2d coupling takes the form of a
superpotential term \(\sum_{s,t} \anti{\quark}_t \quark_s \bigl(
\hyperquark_{st}^{\text{4d}}|_{\text{2d}} \bigr)\) in two dimensions,
which identifies the flavour symmetries \(S[U(\Nf)\times U(\Nf)]\) of
these chiral multiplets\footnote{The full flavour symmetry of the two
  dimensional quiver gauge theories we consider also contains a \(U(1)\)
  factor, under which adjoint chiral multiplets have charge~\(\pm 2\)
  and bifundamental chiral multiplets have charge~\(\mp 1\).} and of the
hypermultiplets.  To write the superpotential term explicitly, the four
dimensional \(\Nsusy=2\) hypermultiplets should be decomposed into two
dimensional \(\Nsusy=(2,2)\) components.  Coupling the common
flavour group to a constant background vector multiplet then gives twisted masses to the two dimensional chiral
multiplets and masses to the four dimensional hypermultiplets, related
by~\eqref{part-mst}.

For definiteness, we place the four dimensional theory on a squashed
sphere~\(S^4_b\)
\begin{equation}\label{S4b}
  \frac{x_0^2}{r^2} + \frac{x_1^2+x_2^2}{\ell^2}
  + \frac{x_3^2+x_4^2}{\tilde{\ell}^2} = 1
\end{equation}
where \(b^2 = \ell/\tilde{\ell}\), and we place surface operators at
\(x_3=x_4=0\), hence on the squashed two-sphere\footnote{Inserting the
  surface operators at \(x_1=x_2=0\) instead would exchange
  \(\ell\leftrightarrow\tilde{\ell}\): we would find degenerate
  operators with momenta~\(-\frac{1}{b}\Omega\) instead of~\(-b\Omega\),
  where \(\Omega\) is a highest weight of~\(A_{\Nf-1}\).}
\begin{equation}
  \frac{x_0^2}{r^2} + \frac{x_1^2+x_2^2}{\ell^2} = 1 \,.
\end{equation}
The full partition function of the 4d/2d theory is then the product
\begin{equation}
  Z_{S^2\subset S^4_b} = Z_{S^4_b}^\free Z_{S^2}
\end{equation}
of the partition functions of the free hypermultiplets
on~\(S^4_b\)~\cite{Hama:2012bg} and of the two dimensional gauge theory
on the squashed
two-sphere~\cite{Benini:2012ui,Doroud:2012xw,Gomis:2012wy}.  The two
factors do not dependent on~\(r\), but only on the equatorial radii
\(\ell\) and~\(\tilde{\ell}\).

The \(S^4_b\)~partition function of a single free hypermultiplet of
mass~\(\mathrm{m}\) only depends on the dimensionless mass\footnote{In
  our correspondence \(m\) also has an imaginary part, which is linked
  to the \(U(1)\) \(R\)-charges of the two dimensional chiral
  multiplets.} \(m = \sqrt{\ell\tilde{\ell}} \, \mathrm{m}\).  It
reads~\cite{Hama:2012bg}\footnote{The sign of~\(m\) is irrelevant since
  the Upsilon function~\eqref{Upsilon-shift} obeys
  \(\Upsilon(b+\frac{1}{b}-x)=\Upsilon(x)\).}
\begin{equation}\label{ZS4-free}
  Z_{S^4_b}^\free (\mathrm{m})
  = \frac{1}{\Upsilon(\frac{b}{2} + \frac{1}{2b} - \I m)} \,.
\end{equation}
The \(S^4_b\)~partition function of the four dimensional theory is the
product of \(\Nf^2\)~such inverses of Upsilon functions.  The
complexified masses~\(m_{st}\) of the \(\Nf^2\)~hypermultiplets in this
class~S theory arise from coupling to a background vector multiplet
the \(S[U(\Nf)\times U(\Nf)]\)
flavour subgroup which is made manifest in the description
as M5-branes wrapping a trinion.  With such masses, the
\(S^4_b\)~partition function is then equal to a Toda CFT correlator with
one simple and two full punctures.  Inserting one or more degenerate
punctures in the correlator corresponds to including the associated
surface operator in the theory of \(\Nf^2\)~hypermultiplets: for given
degenerate punctures, we will find the gauge theory description of the
associated surface operator by comparing the enriched Toda CFT
correlator with the partition function of the 4d/2d theory on
\(S^2\subset S^4_b\).

The second contribution to the partition function of the \(S^2\subset
S^4_b\) system is the partition function of the two dimensional theory.
We recall now the data defining an \(\Nsusy=(2,2)\) theory of vector and
chiral multiplets, and expressions for its partition function
on~\(S^2\).  Besides the gauge group~\(G\) (throughout the paper,
\(G=U(\Nc)\) or a product of such factors) and the representation~\(R\)
of~\(G\) in which matter multiplets transform, the \(S^2\)~partition
function depends on a (real) twisted mass~\(\mathrm{m}\) and a \(U(1)\)
\(R\)-charge~\(\mathrm{q}\) for each chiral multiplet, that is, for each
irreducible factor in~\(R\).  Those are conveniently combined as the
dimensionless complexified twisted mass
\begin{equation}
  m = \ell \mathrm{m} + \frac{\I\mathrm{q}}{2} \,,
\end{equation}
where \(\ell\)~is the equatorial radius of the squashed~\(S^2\).
Furthermore, for each \(U(1)\) factor of~\(G\), an FI parameter~\(\xi\)
and a theta angle~\(\vartheta\) can be turned on.  It will be practical
to consider the complex combination
\begin{equation}\label{z-xi-theta}
  z = e^{-2\pi\xi+\I\vartheta}
\end{equation}
for each \(U(1)\) gauge group factor.  Unless stated otherwise, the
parameters~\(m\) and~\(z\) are generic.  We also assume that
\(R\)-charges are small and positive, \(0<\Re(-2\I m)<1\), and otherwise
define the partition function by analytic continuation.

For a choice of supercharge~\(\Q\) in the supersymmetry algebra, and of
a \(\Q\)-exact deformation term \(\Q V\) such that \(\Q^2 V=0\),
supersymmetric localization reduces the partition function to an
integral over saddle points of \(\Q V\).  When \(\Q V\) is chosen
appropriately, in particular with a positive semidefinite bosonic part,
the integral is finite dimensional and more tractable than
the original path integral.

One choice of deformation term leads to an expression of the partition
function as an integral over the Coulomb
branch~\cite{Benini:2012ui,Doroud:2012xw} (\cite{Hori:2013ika,Honda:2013uca}~corrected a sign):\footnote{Our normalization
  differs by \((2\pi)^{\dim\lie{h}}\) from~\cite{Doroud:2012xw} as this
  will simplify the expression of dualities.}
\begin{equation}\label{Z-Coulomb}
  Z = \frac{1}{\Weyl}
  \sum_{B\in\lie{h}_\bbZ} \int_{\lie{h}}
  \frac{\dd{\sigma}}{(2\pi)^{\dim\lie{h}}}
  Z_\classical
  \prod_{\alpha>0}\biggl[(-1)^{\alpha B}\biggl((\alpha \sigma)^2
    +\frac{(\alpha B)^2}{4}\biggr)\biggr]
  \prod_{w\in R} \biggl[
    \frac{\Gamma(- w(\I m + \I\sigma + \frac{B}{2}))}
    {\Gamma(1+ w(\I m + \I\sigma - \frac{B}{2}))}
  \biggr] \,.
\end{equation}
Here, \(\Weyl\)~is the order of the Weyl group of~\(G\), the sum is
restricted to GNO quantized fluxes~\(B\in\lie{h}\), and the integral
over the lowest component~\(\sigma\) of the vector multiplet ranges in
the Cartan algebra~\(\lie{h}\) of~\(G\).
The vector
multiplet one-loop determinant is a product over all positive
roots~\(\alpha\) of~\(G\), and the chiral multiplet one-loop
determinant, a product over all weights~\(w\) of~\(R\), involves the
complexified twisted mass \(w\cdot m\) of the irreducible factor
of~\(R\) containing~\(w\).\footnote{Roots and weights are linear forms
  on~\(\lie{h}\), and we use the notation \(\alpha\sigma =
  \alpha(\sigma) \in \bbR\).}
When \(G=U(\Nc_1)\times\cdots\times
U(\Nc_n)\), the classical contribution is
\begin{equation}
  Z_\classical(\sigma,B,z,\bar{z})
  = \prod_{i=1}^{n} \left[
  z_i^{\Tr\left(\I\sigma_i+\frac{B_i}{2}\right)}
  \bar{z}_i^{\Tr\left(\I\sigma_i-\frac{B_i}{2}\right)} \right]
  = \prod_{i=1}^{n} e^{-4\pi\xi_i\Tr(\I\sigma_i)+\I\vartheta_i\Tr(B_i)} \,,
\end{equation}
and it is invariant under any \(\vartheta_i\to\vartheta_i+2\pi\) since \(B_i\) are
\(\Nc_i\times\Nc_i\) (diagonal) matrices of integers.  The vector multiplet sign
simply shifts \(\vartheta_i\to\vartheta_i+(\Nc_i-1)\pi\).

A different choice of deformation
term~\cite{Benini:2012ui,Doroud:2012xw} localizes the path integral to
the Higgs branch of the theory rather than its Coulomb branch, yielding
a finite sum
\begin{align}\label{Z-Higgs}
  Z = \sum_{\mathsemiclap{7}{v\in\{\text{Higgs vacua}\}}}
  (z\bar{z})^{\I v}
  \res_{\sigma=v}\Biggl[
    \prod_{\alpha} (\I \alpha \sigma)
    \prod_{w\in R} \gamma(-w(\I m + \I\sigma))
  \Biggr]
  Z_\vortex(v,m,z) Z_\antivortex(v,m,\bar{z})
\end{align}
which includes a vortex contribution~\(Z_\vortex\) depending
holomorphically on~\(z\) and an antivortex contribution depending
on~\(\bar{z}\).  Here, \(\gamma(x) = \frac{\Gamma(x)}{\Gamma(1 - x)}\),
and factors other than \(Z_\vortex\) and \(Z_\antivortex\) are obtained
as the residue at \(\sigma=v\) and \(B=0\) of the Coulomb branch
integrand.  Higgs branch vacua are defined as having non-zero vevs for
the lowest component~\(\phi\) of some chiral fields.  They are labeled
by solutions \((\sigma,\phi)\) of the \(\Daux\)-term equation \(\phi
\phi^\dagger = \xi\) and of \((\sigma + m)\phi = 0\), where \(\sigma\)
and~\(m\) act on~\(\phi\) through the action of~\(G\) and of the flavour
symmetry group~\(\Gf\).  The set of values of~\(\sigma\) for which the
\(\Daux\)-term equation has a solution depends on signs of the FI
parameters~\(\xi_j\) for each \(U(1)\) factor in~\(G\): each choice of
signs leads to a different expansion~\eqref{Z-Higgs}.  Even after
solving these equations, one must in principle evaluate~\(Z_\vortex\) as
the volume of a moduli space of vortices.  However, the Coulomb branch
representation provides a convenient short-cut: closing the
\(\dd{\sigma}\) integrals~\eqref{Z-Coulomb} towards
\(\sigma\to\pm\I\infty\) depending on the matter content and on signs of
FI parameters expresses the partition function as a sum over poles,
which is then rewritten as a finite sum of factorized
terms~\eqref{Z-Higgs}.  The manipulations are most easily done on
specific examples, as we will see, but work for an arbitrary gauge group
and matter representation (see~\cite[Appendix~F]{Doroud:2012xw}).

\pagebreak[0]\bigskip

In the coming sections we associate a two dimensional \(\Nsusy=(2,2)\)
gauge theory, hence a surface operator, to each choice of representation
\(\repr(\Omega)\) of~\(A_{\Nf-1}\).  We work out equalities of the form
\begin{equation}\label{part-typical-matching}
  Z_{S^2\subset S^4_b}^{(\Omega)}
  = A \abs{x}^{2\gamma_0} \abs{1-x}^{2\gamma_1}
  \vev*{\widehat{V}_{\alpha_\infty}(\infty) \widehat{V}_{\hat{m}}(1)
    \widehat{V}_{\alpha_0}(0) \widehat{V}_{-b\Omega}(x,\bar{x})}
\end{equation}
between the partition function on \(S^2\subset S^4_b\) of the 4d/2d
system associated to a given representation \(\repr(\Omega)\) and Toda
CFT correlators with two full punctures at \(0\) and~\(\infty\), one
simple at~\(1\), and one degenerate.\footnote{Toda CFT notations are
  reviewed in Appendix~\ref{app:Toda-basics}.  Vertex
  operators~\(\widehat{V}_\alpha\) are labeled by their
  momentum~\(\alpha\), a linear combination of the weights~\(h_s\)
  (\(1\leq s\leq\Nf\)) of the fundamental representation
  of~\(A_{\Nf-1}\).  They are primary operators for the \(W_{\Nf}\)~chiral
  algebra.  Generic momenta depend on \(\Nf-1\) parameters and label
  full punctures.  Semi-degenerate vertex operators, with momentum
  \(\varkappa h_1\) (or its conjugate \(-\varkappa h_{\Nf}\)), have null
  descendants under~\(W_{\Nf}\) and label simple punctures.  Degenerate
  vertex operators have momentum \(-b\Omega-\Omega'/b\) for a pair of
  highest weights \(\Omega\) and~\(\Omega'\) of representations
  of~\(A_{\Nf-1}\).}  The position~\(x\) of the degenerate puncture is
related to a complexified FI parameter~\(z\).  The two dimensional
theories we consider involve \(\Nf\)~fundamental and
\(\Nf\)~antifundamental chiral multiplets of a gauge group factor
\(U(\Nc_n)\), whose twisted masses we denote by \(m_t\)
and~\(\anti{m}_t\).

Let us first explain how the factor \(A\abs{x}^{2\gamma_0}
\abs{1-x}^{2\gamma_1}\) can be absorbed into the partition function
(specifically the \(S^2\)~contribution).  In the coming sections it
will be easier to manipulate explicit expressions of
partition functions and correlators, hence we will keep the factor
explicitly, with the understanding that it has no physical content.  In
terms of gauge theory data, it turns out that we can split
\begin{align}
  \gamma_0
  & = \gamma_0^\circ (\Omega,b)
  - \frac{\Nc_n}{\Nf} \sum_{t=1}^{\Nf} \I m_t \,,
  &
  \gamma_1
  & = \gamma_1^\circ(\Omega,b)
  + \frac{\Nc_n}{\Nf} \sum_{t=1}^{\Nf} (\I m_t+\I\anti{m}_t)
  \,,
\end{align}
and \(A = A^\circ(\Omega,b)\, b^{-2\Nc_n \sum_t (\I m_t+\I\anti{m}_t)}\), where
\(A^\circ\), \(\gamma_0^\circ\) and \(\gamma_1^\circ\) depend only
on~\(b\) and~\(\Omega\).  The factor decomposes as
\begin{equation}\label{fudge-factor}
  A \abs{x}^{2\gamma_0} \abs{1-x}^{2\gamma_1}
  =
  \Bigl[ A^\circ \abs{x}^{2\gamma_0^\circ} \abs{1-x}^{2\gamma_1^\circ} \Bigr]
  \Bigl[\abs{x}^{-2(\Nc_n/\Nf) \sum_t \I m_t}\Bigr]
  \biggl[\frac{\abs{1-x}^{2\Nc_n}}{b^{2\Nc_n\Nf}}
  \biggr]^{\sum_t (\I m_t+\I\anti{m}_t)/\Nf}
\end{equation}
and can be absorbed in the partition function through three different
mechanisms.  Firstly, the two-sphere partition function is subject to
certain ambiguities~\cite{Gomis:2012wy} (see
also~\cite{Closset:2014pda}).  These are captured by local supergravity
counterterms~\cite{Gerchkovitz:2014gta}.  A change in the renormalization
scheme changes the partition function by
\begin{equation}
 Z \rightarrow f(z) \bar{f}(\bar{z}) Z \,,
\end{equation}
where \(f\)~is a holomorphic function of the complexified FI
parameter~\(z\).  This lets us absorb the first factor
in~\eqref{fudge-factor} as a renormalization ambiguity of
\(Z_{S^2\subset S^4_b}\).  Secondly, shifting the \(U(1)\) gauge
scalar by a constant lets us multiply the partition function by any power of
\(\abs{z}^2=\abs{x}^2\) hence absorb the second factor
in~\eqref{fudge-factor}.  Finally, the third factor can be absorbed
through a complexified FI parameter \(z_{\text{fl}} = b^{2\Nc_n\Nf} /
(1-x)^{2\Nc_n}\) for the \(U(1)\) subgroup of the flavour group
\(S[U(\Nf)\times U(\Nf)]\) which acts on the fundamental and
antifundamental chiral multiplets.  Indeed, such an FI parameter
multiplies the partition function by \((z_{\text{fl}}
\bar{z}_{\text{fl}} )^{\I\sigma_{\text{fl}}}\), where
\(\sigma_{\text{fl}}\) is the bottom component of the background
vector multiplet coupled to the \(U(1)\) flavour symmetry, that is,
\(\sigma_{\text{fl}} = \sum_t (m_t+\anti{m}_t) / (2\Nf)\).

We are now ready to discuss how we will derive equalities of the
form~\eqref{part-typical-matching}, or more generally for a set of
highest weights~\(\Omega_j\) of~\(A_{\Nf-1}\):
\begin{equation}\label{part-general-matching}
  Z_{S^2\subset S^4_b}^{\{\Omega_j\}}
  = \vev*{\widehat{V}_{\alpha_\infty}(\infty) \widehat{V}_{\hat{m}}(1)
    \widehat{V}_{\alpha_0}(0)
    \prod_{j=1}^n \widehat{V}_{-b\Omega_j}(x_j,\bar{x}_j)}
\end{equation}
where \(\alpha_0\) and~\(\alpha_\infty\) are generic, \(\hat{m}\) is
semi-degenerate, and we have omitted the factors which can be absorbed
into the partition function.  The dictionary between gauge theory and
Toda CFT data identifies the momenta \(\alpha_0\), \(\alpha_\infty\),
and~\(\hat{m}\) to the three factors of the flavour symmetry group
\(SU(\Nf)\times SU(\Nf)\times U(1)\) acting on fundamental and
antifundamental chiral multiplets:
\begin{equation}\label{part-alpha}
  \begin{aligned}
    \alpha_0
    & = Q - \frac{1}{b} \sum_{s=1}^{\Nf} \I m_s h_s \,,
    & \qquad\qquad
    \hat{m}
    & = (\varkappa + \Nc b) h_1 \,,
    \\
    \alpha_\infty
    & = Q - \frac{1}{b} \sum_{s=1}^{\Nf} \I\anti{m}_s h_s \,,
    & \qquad\qquad
    \varkappa
    & = \frac{1}{b} \sum_{s=1}^{\Nf} (1 + \I m_s + \I\anti{m}_s) \,,
  \end{aligned}
\end{equation}
with Toda CFT notations given in Appendix~\ref{app:Toda-basics}.  The
degenerate operators encode the choice of gauge groups and matter
content of the gauge theory.

As explained below, we will start in each case by matching the
dependence of the \(S^2\)~partition function on FI parameters~\(z_j\)
with the dependence of Toda CFT correlators on the position of
degenerate operators~\(x_j\).  Once this is done, there remains a
universal relative factor between the \(S^2\)~partition function and the
Toda CFT correlator, which turns out to be a Toda CFT three-point
function of two generic and one semi-degenerate vertex
operators\footnote{Note the shift between \(\varkappa h_1\) in the
  three-point function~\eqref{part-3pt} and \(\hat{m}\) in the
  \((n+3)\)-point function~\eqref{part-general-matching}.  The insertion
  of degenerate operators near a simple puncture thus shifts the
  dictionary between the semi-degenerate momentum of the puncture and
  the corresponding hypermultiplet mass.  As a result, the node-hopping
  duality relates surface operators in four dimensional theories which
  differ by shifts in complexified masses of hypermultiplets.}
\begin{equation}\label{part-3pt}
  \widehat{C}(\alpha_0,\alpha_\infty,\varkappa h_1)
  = \prod_{s=1}^{\Nf} \prod_{t=1}^{\Nf} \frac{1}{\Upsilon\bigl(
    \frac{1}{b} (1+\I m_s+\I\anti{m}_t)\bigl)}
  \,.
\end{equation}
These Upsilon functions are precisely reproduced by the
\(S^4_b\)~partition function~\eqref{ZS4-free} of \(\Nf^2\)~free
hypermultiplets with (dimensionless) masses
\begin{equation}\label{part-mst}
  m_{st} = \I\frac{1-b^2}{2b}-\frac{1}{b}(m_s+\anti{m}_t) \,.
\end{equation}
The dimensionful masses \((\ell\tilde{\ell})^{\frac{-1}{2}}m_{st}\) and
twisted masses \(\ell^{-1}m_s\) and \(\ell^{-1}\anti{m}_t\) both
originate from coupling the common flavour symmetry group
\(SU(\Nf)\times SU(\Nf)\times U(1)\) to a background vector
multiplet, and indeed, the relation between
dimensionful masses has no relative factor of~\(b\):
\begin{equation}
  \biggl[\frac{m_{st}}{\sqrt{\ell\tilde{\ell}}} + \frac{\I}{2\ell} + \frac{\I}{2\tilde{\ell}}\biggr]
  + \frac{m_s+\anti{m}_t}{\ell}
  = \frac{\I}{\ell} \,.
\end{equation}
The masses~\(m_{st}\) can also be found by requiring that the two
dimensional superpotential \(\sum_{s,t} \anti{\quark}_t \quark_s
\hyperquark_{st}^{\text{4d}}|_{\text{2d}}\) is supersymmetric hence has
\(R\)-charge~\(2\) (complexified twisted mass~\(\I\)).  From this
perspective, the shift in the four dimensional masses likely arises from
mixing the \(U(1)_R\) symmetry with geometrical
symmetries.\footnote{After the first version of this article, the term
  $\I/(2\ell)+\I/(2\tilde{\ell})$ was explained in~\cite{Gomis:2016ljm}:
  the two dimensional $R$-charge of $\Phi^{\text{4d}}|_{\text{2d}}$ is
  $1+b^2$.}

In Section~\ref{sec:SQED} and Section~\ref{sec:SQCD}, we identify
degenerate vertex operators labeled by the fundamental (resp.\@
antisymmetric) representation of~\(A_{\Nf-1}\) to SQED (resp.\@ SQCD\@).
The Toda CFT correlator is a four-point function which depends on a
single cross-ratio~\(x\), while the two dimensional theory has a single
\(U(1)\) gauge group factor hence a single complexified FI parameter
\(z=e^{-2\pi\xi+\I\vartheta}\).  We prove as follows that the Toda
correlator is equal to the \(S^2\subset S^4_b\) partition function.
First, we write the Higgs branch expressions of the \(S^2\)~partition
function in the regions \(\xi>0\) and \(\xi<0\), that is, \(\abs{z}<1\)
and \(\abs{z}>1\).  The two expressions match with expansions of the
Toda CFT correlator in the s-channel \(\abs{x}<1\) and u-channel
\(\abs{x}>1\) as described in Table~\ref{tab:pieces}: the Higgs branch
vacua correspond to choices of internal momenta and we match the leading
powers of \(\abs{z}^2=\abs{x}^2\).  On the gauge theory side, the
exponents of~\(\abs{z}^2\) are read from the classical contribution,
while on the Toda CFT side the exponents of~\(\abs{x}^2\) are sums of
dimensions of vertex operators.  We then derive the braiding matrices
which relate s-channel and u-channel conformal blocks and show that they
are equal to the corresponding gauge theory data.  These braiding
matrices let us express the monodromy around~\(\infty\) as a matrix in
the basis of s-channel conformal blocks (the monodromy around~\(0\) is
diagonal in this basis).  Finally, we prove that the \(S^2\)~partition
function has only one branch point besides \(z=0\) and \(z=\infty\), and
identify gauge theory exponents with those in the t-channel \(x\to 1\)
of the Toda correlator.  Therefore the monodromy matrix around~\(1\) is
simply the inverse product of the monodromies around \(0\)
and~\(\infty\).  Since their monodromy matrices around all three branch
points coincide, the \(S^2\)~partition function and Toda CFT four-point
function must be equal up to a factor with no monodromy.  Since in
expansions around \(0\), \(1\) and~\(\infty\) the exponents match, the
factor has no pole on the sphere hence is a constant: it is precisely
given by the \(S^4_b\)~contribution~\eqref{part-3pt} of
\(\Nf^2\)~hypermultiplets.

\begin{table}\centering
  \caption{\label{tab:pieces}%
    Relation between parts of the Higgs branch decomposition of the \(S^2\)~partition function,
    and the s-channel decomposition of corresponding Toda CFT correlators.  Explicit expressions
    differ by \(A\abs{x}^{2\gamma_0}\abs{1-x}^{2\gamma_1}\), an ambiguity in~\(Z\).}
  \begin{tabular}{rll}
    \toprule
    & Gauge theory & Toda CFT
    \\\midrule
    Terms in the sum & Higgs vacua & Internal momenta \\
    Asymptotics at~\(0\) & Classical contribution~\((z\bar{z})^{\I v}\) & \((x\bar{x})^{\dimToda(\alpha_0-bh)-\dimToda(\alpha_0)-\dimToda(-b\omega)}\)\\
    Leading coefficient & One-loop determinant~\(Z_\oneloop\) & Three-point functions \\
    Holomorphic series & Vortex partition function~\(Z_\vortex\) & Conformal blocks (normalized)
    \\\bottomrule
  \end{tabular}
\end{table}

When the FI parameter~\(\xi\) is changed continuously from \(\xi<0\) to
\(\xi>0\), the two dimensional gauge theory experiences a flop
transition between vortices carried by fundamental matter and vortices
carried by antifundamental matter.  The flop transition is realized in
the Toda CFT as crossing symmetry from the s-channel to the u-channel~\cite{Doroud:2012xw}.
This geometric approach implies that the results for \(\xi<0\) and
\(\xi>0\) are related by analytic continuation.  There is no Higgs
branch expansion as \(\xi\to 0\): instead, we build a decomposition of
the Coulomb branch integral in this limit.  It would be interesting to
provide a gauge theory interpretation of this ``t-channel''
decomposition, and of the braiding matrices relating \(\xi>0\) and
\(\xi<0\) vortex partition functions.

In Section~\ref{sec:SQCDA}, we identify degenerate vertex operators
labeled by symmetric representations of~\(A_{\Nf-1}\) to SQCD with an
additional adjoint chiral multiplet (SQCDA\@).  The discussion is very
similar to the previous cases, but braiding matrices are not
available.\footnote{It is technically difficult to write down braiding
  matrices in this case.  On the gauge theory side, the Mellin--Barnes
  integral (used for SQED and SQCD to interpolate between
  \(\abs{z}\lessgtr 1\) expansions) is much more involved.  On the Toda
  CFT side, recursion relations for the braiding matrices contain many
  more terms than for the antisymmetric case.}  Instead, we check that
the leading coefficients and powers of \(\abs{x}^2\)
coincide, both in the s-channel and in the u-channel.  We then
check that the \(S^2\)~partition function has a branch point
corresponding to the t-channel, and that the leading powers of
\(\abs{1-x}^2\) coincide.  As before, the Toda CFT four-point function
is equal to the \(S^2\)~partition function up to a constant, which is
the \(S^4_b\)~partition function of \(\Nf^2\)~free hypermultiplets.

In Section~\ref{sec:Quivers} we identify the quiver gauge theory which
corresponds to sets of degenerate operators labeled by symmetric or
antisymmetric representations of~\(A_{\Nf-1}\).  The identification is
checked by comparing the expansion of the \(S^2\subset S^4_b\) partition
function and of the Toda CFT correlator in various limits.  Seiberg-like
dualities let us probe further limits: as seen in
Section~\ref{sec:SeibergQ-perm}, permutations of the \(n\)~degenerate
vertex operators relate dual gauge theories.  First, we equate exponents
and leading coefficients in the channel where degenerate punctures are
at \(0<\abs{x_1}<\cdots<\abs{x_n}<1\).  Thanks to dualities, exponents
and leading coefficients also match for all other orderings of the
\(n\)~degenerate punctures.  By symmetry, the gauge theory and Toda CFT
exponents and leading coefficients also match in all channels with
\(1<\abs{x_1},\ldots,\abs{x_n}<\infty\).  In each of the \(2(\Nc!)\)
channels the decompositions involve many exponents and factors, and
all match.  We then equate exponents which appear in any of the limits
 \(x_n\to 1\) or \(x_j \to x_{j+1}\), hence also in the limits
\(x_j\to 1\) or \(x_j\to x_k\) thanks to dualities.

Building upon the identification of the quiver which corresponds to the
insertion of any number of antisymmetric degenerate vertex operators, we
show in Section~\ref{sec:Quivers-fuse} that bringing all punctures
\(x_j=x\) to the same position yields a degenerate vertex operator
labeled by an arbitrary representation of~\(A_{\Nf-1}\): all other
terms in the fusion of antisymmetric degenerate vertex operators appear
with higher powers of some \(\abs{x_j-x_k}^2\) hence are suppressed.
This determines the quiver gauge theory corresponding to an arbitrary
degenerate vertex operator~\(\widehat{V}_{-b\Omega}\).

The surface operators we consider are constructed by coupling
\(\Nf\)~fundamental and \(\Nf\)~antifundamental chiral multiplets of an
\(\Nsusy=(2,2)\) theory to \(\Nf^2\)~hypermultiplets.  Making some
antifundamental chiral multiplets and some hypermultiplets massive
yields surface operators described by \(\Nsusy=(2,2)\) theories with
\(\Nf\)~fundamental and \(\anti{\Nf}<\Nf\) antifundamental chiral
multiplets, coupled to \(\Nf\anti{\Nf}\)~free hypermultiplets.  The
limit corresponds to a collision limit of the punctures
\(\widehat{V}_{\hat{m}}\) and~\(\widehat{V}_{\alpha_\infty}\)
in~\eqref{part-general-matching}, which builds an irregular puncture
(see Appendix~\ref{app:irregular} and for \(\Nf=2\)
see~\cite{Gaiotto:2012sf}).  We only study this limit for SQCD (see
Section~\ref{sec:irreg}), but the discussion applies to all our
surface operators.

\numberwithin{equation}{subsection}
\subsection{SQED and Toda Fundamental Degenerate}
\label{sec:SQED}

We review in this section the case of \(\Nsusy=(2,2)\) SQED on~\(S^2\),
namely a \(U(1)\) vector multiplet coupled to \(\Nf\)~fundamental and
\(\Nf\)~antifundamental chiral multiplets, whose twisted masses (plus
\(R\)-charges) we denote by \(m_s\) and~\(\anti{m}_s\) for \(1\leq
s\leq\Nf\).  It was shown~\cite{Doroud:2012xw} that the
\(S^2\)~partition function of SQED matches an \(A_{\Nf-1}\)~Toda CFT
four-point function, up to a constant.  We find that the constant
factor reproduces the \(S^4_b\)~partition function of \(\Nf^2\)~free
hypermultiplets with masses~\eqref{part-mst}, hence the Toda correlator
in fact captures the partition function of the surface operator inserted
in this free four dimensional theory.
The precise relation is\footnote{As explained
  below~\eqref{part-typical-matching}, the factor \(A
  \abs{x}^{2\gamma_0} \abs{1-x}^{2\gamma_1}\) can be absorbed into the
  partition function.  To compare gauge theory and Toda CFT results it
  is best to keep the factor explicitly.}
\begin{equation}\label{SQED-matching}
  Z_{S^2\subset S^4_b}^\SQED(m,\anti{m},z,\bar{z})
  = A \abs{x}^{2\gamma_0} \abs{1-x}^{2\gamma_1}
  \vev*{\widehat{V}_{\alpha_\infty}(\infty)\widehat{V}_{\hat{m}}(1)
    \widehat{V}_{-bh_1}(x,\bar{x})\widehat{V}_{\alpha_0}(0)} \,.
\end{equation}
The Toda CFT correlator (see Appendix~\ref{app:Toda-basics} for
conventions) features a degenerate field \(\widehat{V}_{-bh_1}\)
inserted at \(x = (-1)^{\Nf} z\) and labeled by the fundamental
representation \(\repr(h_1)\) of \(A_{\Nf-1}\), a semi-degenerate field
\(\widehat{V}_{\hat{m}}\) at~\(1\), and two generic fields
\(\widehat{V}_{\alpha_0}\) and~\(\widehat{V}_{\alpha_\infty}\).  Momenta
are related to twisted masses through
\begin{equation}
  \label{SQED-alpha}
  \begin{aligned}
    \alpha_0 & = Q - \frac{1}{b} \sum_{s=1}^{\Nf} \I m_s h_s \,,
    & \qquad\qquad
    \hat{m} & = (\varkappa+b) h_1 \,,
    \\
    \alpha_\infty & = Q - \frac{1}{b} \sum_{s=1}^{\Nf} \I\anti{m}_s h_s \,,
    & \qquad\qquad
    \varkappa & = \frac{1}{b} \sum_{s=1}^{\Nf} (1+\I m_s+\I\anti{m}_s) \,,
  \end{aligned}
\end{equation}
and the exponents and coefficient are
\begin{align}
  \label{SQED-gamma0}
  \gamma_0 & = - \frac{1}{\Nf} \sum_{s=1}^{\Nf} \I m_s
  - \frac{\Nf-1}{2}(b^2+1) \,,
  \\
  \label{SQED-gamma1}
  \gamma_1 & = - \frac{\Nf-1}{\Nf} b^2
  + \frac{1}{\Nf} \sum_{s=1}^{\Nf} (\I m_s+\I\anti{m}_s) \,,
  \\
  \label{SQED-A}
  A & = b^{\Nf(1+b^2)-b^2-2b\varkappa} \,.
\end{align}
Permuting the~\(m_s\), or the~\(\anti{m}_s\), does not affect the
partition function.  This is reproduced in the Toda CFT by the
invariance of \(\widehat{V}_{\alpha_0}\)
and~\(\widehat{V}_{\alpha_\infty}\) under Weyl transformations (the
normalization is chosen to cancel reflection amplitudes).  The similarity
between \(\alpha_0\) and~\(\alpha_\infty\) is also expected, as swapping
them and changing \(x\) to its inverse amounts in gauge theory to charge
conjugation, which swaps \(m_s\) and~\(\anti{m}_s\), and changes \(z\)
to its inverse.  Under this transformation, \(Z_{S^2}\) is invariant,
while the Toda correlator receives a small shift controlled by the
dimension \(\dimToda(-bh_1)\) of the degenerate insertion: this shift is
absorbed by the factor \(\abs{x}^{2\gamma_0} \abs{1-x}^{2\gamma_1}\).

In~\cite{Doroud:2012xw}, the equality was shown directly thanks to known
expressions~\cite{Fateev:2007ab} for the \(W_{\Nf}\)~conformal block
involved.  The approach does not generalize, because conformal blocks
with higher degenerate insertions were not previously
known.\footnote{We derive such explicit conformal blocks from our matchings in Appendix~\ref{app:blocks}.}
Instead, we prove the
correspondence for SQED (treated here) and SQCD (see
Section~\ref{sec:SQCD}) by comparing monodromy matrices around branch
points.  In the main text, we find expansions around all branch points
and compare leading terms, as this is enough to fix uniquely the
dictionary between gauge theory and Toda CFT parameters.  In
Appendix~\ref{app:braiding} we derive the braiding matrices relating
s-channel and u-channel expansions of the Toda CFT correlator, and their
gauge theory analogues.  The braiding matrices match.  From this we
deduce the matching of monodromy matrices around all branch points,
expressed in a single basis, and not only of their eigenvalues compared
in the main text.  These results suffice to prove that the partition
function and the correlator are equal.

To prepare for the somewhat technical computations ahead, we first go
through the various steps here in the well-controlled case of SQED and
Toda CFT fundamental degenerate fields.  The expansions near \(z=0\) and
\(z=\infty\) follow~\cite{Doroud:2012xw} closely, while the expansion
near \(z=(-1)^{\Nf}\) is new.  All three play an important role in later
sections.

\subsubsection{Expanding the SQED Partition Function}

The Coulomb branch expression for the partition function of SQED is
\begin{equation}\label{SQED-Z-Coulomb}
  Z_{S^2}^\SQED =
  \sum_{B\in\bbZ} \int_{\bbR} \frac{\dd{\sigma}}{2\pi}
  z^{\I\sigma+\frac{B}{2}} \bar{z}^{\I\sigma-\frac{B}{2}}
  \prod_{s=1}^{\Nf} \left[
    \frac{\Gamma(-\I m_s - \I\sigma - \frac{B}{2})}
      {\Gamma(1+\I m_s + \I\sigma - \frac{B}{2})}
    \frac{\Gamma(-\I\anti{m}_s + \I\sigma + \frac{B}{2})}
      {\Gamma(1+\I\anti{m}_s - \I\sigma + \frac{B}{2})}
  \right] \,.
\end{equation}
As we will see shortly, the contour of integration for~\(\sigma\) can be
closed in the lower or upper half plane depending on whether
\(\abs{z}\lessgtr 1\), leading to distinct expressions of~\(Z\) as a sum
over poles lying in either half plane.  We will match the resulting
expressions with the s-channel and u-channel decompositions of the Toda
CFT four-point correlator.

To find out which half-plane the contour should enclose, we study the
asymptotic behavior of the integrand.  First, rewrite the ratios of
Gamma functions so that the numerator and denominator have no common
poles,
\begin{equation}
  \frac{\Gamma(-\upsilon\pm \frac{B}{2})}
  {\Gamma(1+\upsilon\pm \frac{B}{2})}
  = (-1)^{\frac{B\mp\abs{B}}{2}}
  \frac{\Gamma(-\upsilon+\frac{\abs{B}}{2})}
  {\Gamma(1+\upsilon+\frac{\abs{B}}{2})} \,,
\end{equation}
and absorb the resulting sign \((-1)^{\Nf B}\) by introducing
\begin{equation}
  x = (-1)^{\Nf} z \,, \qquad \bar{x} = (-1)^{\Nf} \bar{z} \,.
\end{equation}
Thanks to \(\frac{\Gamma(\upsilon+a)}{\Gamma(\upsilon+b)}\sim
\upsilon^{a-b}\), valid when \(\abs{\upsilon}\to\infty\) away from
the negative real axis, the integrand is
\begin{equation}\label{SQED-integrand}
  \begin{aligned}
    & x^{\I\sigma+\frac{B}{2}} \bar{x}^{\I\sigma-\frac{B}{2}}
    \prod_{s=1}^{\Nf} \left[
      \frac{\Gamma(-\I m_s - \I\sigma + \frac{\abs{B}}{2})}
        {\Gamma(1+\I m_s + \I\sigma + \frac{\abs{B}}{2})}
      \frac{\Gamma(-\I\anti{m}_s + \I\sigma + \frac{\abs{B}}{2})}
        {\Gamma(1+\I\anti{m}_s - \I\sigma + \frac{\abs{B}}{2})}
    \right]
    \\
    & \sim x^{\I\sigma+\frac{B}{2}} \bar{x}^{\I\sigma-\frac{B}{2}}
    \Bigl(\sigma^2 + \frac{B^2}{4}
    \Bigr)^{-\sum_{s=1}^{\Nf} (1+\I m_s+\I\anti{m}_s)}
  \end{aligned}
\end{equation}
as \(\abs{\I\sigma\pm\frac{B}{2}}\to\infty\).

As long as we keep \(\sigma\in\bbR\) on the integration contour, the
factor \(x^{\I\sigma+\frac{B}{2}} \bar{x}^{\I\sigma-\frac{B}{2}}\) is
simply a phase.  If \(\abs{x} = \abs{z} < 1\), this factor
decays exponentially towards \(\sigma\to -\I\infty\), hence the contour
of integration can be closed in this half-plane.  On the other hand, for
\(\abs{x} = \abs{z} > 1\), the integrand decays
exponentially in the \(\sigma\to\I\infty\) half-plane.

The integrand~\eqref{SQED-integrand} has poles whenever one of \(- \I
m_p - \I\sigma + \frac{\abs{B}}{2}\) or \(- \I\anti{m}_p + \I\sigma +
\frac{\abs{B}}{2}\) is a non-positive integer, that is, at
\begin{equation}\label{SQED-poles}
  \I\sigma = - \I m_p + k + \frac{\abs{B}}{2}
  \quad \text{or} \quad
  \I\anti{m}_p - k - \frac{\abs{B}}{2}
\end{equation}
for a fundamental or antifundamental flavour~\(1\leq p\leq\Nf\) and an
integer \(k\geq 0\).  Since \(R\)-charges are positive, \(-\I m_p\) has
a positive real part and \(\I\anti{m}_p\) a negative real part, hence
the poles of the fundamental multiplets' one-loop determinants lie in
the half-plane towards \(\sigma\to-\I\infty\), while the other half
plane contains those of antifundamental multiplets.

Let us focus on the case \(\abs{x} = \abs{z} < 1\).  We then sum
residues of the integrand of~\eqref{SQED-Z-Coulomb} over
poles~\eqref{SQED-poles} where \(\I\sigma\) has a positive real part.
This yields
\begin{equation}
  \begin{aligned}
    Z
    & = \sum_{p=1}^{\Nf} \sum_{k\geq 0} \sum_{B\in\bbZ} \Biggl\{
    z^{-\I m_p + k + \frac{\abs{B}+B}{2}}
    \bar{z}^{-\I m_p + k + \frac{\abs{B} - B}{2}}
    \\
    & \qquad \cdot
    \sideset{}{'}\prod_{s=1}^{\Nf} \Biggl[
      \frac{\Gamma(-\I m_s+\I m_p-k-\frac{\abs{B}+B}{2})}
        {\Gamma(1+\I m_s-\I m_p+k+\frac{\abs{B}-B}{2})}
      \frac{\Gamma(-\I\anti{m}_s-\I m_p+k+\frac{\abs{B}+B}{2})}
        {\Gamma(1+\I\anti{m}_s+\I m_p-k-\frac{\abs{B}-B}{2})}
    \Biggr] \Biggr\} \,,
  \end{aligned}
\end{equation}
where the singular factor \(\Gamma(-k-\frac{\abs{B}+B}{2})\) appearing for
\(s=p\) should be replaced by its residue \((-1)^{k+\frac{\abs{B}+B}{2}}
/\, \Gamma(1+k+\frac{\abs{B}+B}{2})\).  Note that \(k\) and~\(B\) appear
as the combinations \(k^\pm=k+\frac{\abs{B}\pm B}{2}\) only, and that
the sums over \(k\geq 0\) and \(B\in\bbZ\) are equivalent to sums over
\(k^+\geq 0\) and \(k^-\geq 0\).  Hence,
\begin{equation}
  Z
  = \sum_{p=1}^{\Nf} \sum_{k^\pm\geq 0}
  (z\bar{z})^{-\I m_p} z^{k^+} \bar{z}^{k^-}
  \sideset{}{'}\prod_{s=1}^{\Nf} \biggl[
    \frac{\Gamma(-\I m_s+\I m_p-k^+)}
      {\Gamma(1+\I m_s-\I m_p+k^-)}
    \frac{\Gamma(-\I\anti{m}_s-\I m_p+k^+)}
      {\Gamma(1+\I\anti{m}_s+\I m_p-k^-)}
  \biggr] \,,
\end{equation}
with the same caveat as above, namely,
\(\Gamma(-k^+)\to(-1)^{k^+}/\Gamma(1+k^+)\).  Since each Gamma function
argument depends on only one of \(k^+\) and~\(k^-\), the contribution
from each flavour~\(p\) factorizes as the product of two series in
(positive) powers of~\(z\) and of~\(\bar{z}\).  We extract from the
series a normalization factor (the value at \(k^\pm = 0\)), by writing
the Gamma functions in terms of Pochhammer symbols \((a)_n =
\frac{\Gamma(a+n)}{\Gamma(a)}\) and of
\(\gamma(x)=\frac{\Gamma(x)}{\Gamma(1-x)}\),
\begin{equation}
  \frac{\Gamma(-\I m_s+\I m_p-k^+)}{\Gamma(1+\I m_s-\I m_p+k^-)}
  =
  \frac{(-1)^{k^+}\gamma(-\I m_s+\I m_p)}
    {(1+\I m_s-\I m_p)_{k^+}(1+\I m_s-\I m_p)_{k^-}} \,.
\end{equation}

We deduce the partition function for \(\abs{x}=\abs{z}<1\)
in terms of ``s-channel'' vortex partition functions
\begin{gather}\label{SQED-Zs}
  Z = \sum_{p=1}^{\Nf} \Biggl\{
  (x\bar{x})^{-\I m_p}
  \frac{\prod_{s\neq p}^{\Nf}\gamma(-\I m_s+\I m_p)}
    {\prod_{s=1}^{\Nf}\gamma(1+\I\anti{m}_s+\I m_p)}
  f^\sch_p(m,\anti{m},x) f^\sch_p(m,\anti{m},\bar{x}) \Biggl\} \,,
  \\\label{SQED-fs}
  f^\sch_p(m,\anti{m},x)
  = \sum_{k\geq 0} x^k \prod_{s=1}^{\Nf}
  \frac{(-\I\anti{m}_s-\I m_p)_k}{(1+\I m_s-\I m_p)_k}
  =
  \Hypergeometric
    {(-\I\anti{m}_s-\I m_p), \, 1\leq s\leq\Nf}
    {(1+\I m_s-\I m_p), \, s\neq p}
    {x} \,.
\end{gather}
The \(f^\sch_p\) are hypergeometric functions, related later on to
s-channel conformal blocks in the Toda CFT\@.  Similar computations for
\(\abs{x}=\abs{z}>1\) convert the sum over poles at
\(\I\sigma=\I\anti{m}_p-\cdots\) to a factorized form, related to the
u-channel decomposition of a Toda CFT correlator,
\begin{gather}\label{SQED-Zu}
  Z = \sum_{p=1}^{\Nf} \Biggl\{
  (x\bar{x})^{\I\anti{m}_p}
  \frac{\prod_{s\neq p}^{\Nf}\gamma(-\I\anti{m}_s+\I\anti{m}_p)}
    {\prod_{s=1}^{\Nf}\gamma(1+\I m_s+\I\anti{m}_p)}
  f^\uch_p(m,\anti{m},x) f^\uch_p(m,\anti{m},\bar{x}) \Biggr\} \,,
  \\\label{SQED-fu}
  f^\uch_p(m,\anti{m},x)
  = \sum_{k\geq 0} x^{-k} \prod_{s=1}^{\Nf}
  \frac{(-\I m_s-\I\anti{m}_p)_k}{(1+\I\anti{m}_s-\I\anti{m}_p)_k}
  =
  \Hypergeometric
    {(-\I m_s-\I\anti{m}_p), \, 1\leq s\leq\Nf}
    {(1+\I\anti{m}_s-\I\anti{m}_p), \, s\neq p}
    {\frac{1}{x}} \,.
\end{gather}

The factorized results \eqref{SQED-Zs} and~\eqref{SQED-Zu} reproduce the
general form~\eqref{Z-Higgs}
\begin{equation}
  Z = \sum_{\text{vacua}}
  \res\Bigl[ Z_\classical(\sigma,0,z,\bar{z}) Z_\oneloop(m,\sigma,0) \Bigr]
  Z_{\vortex,p}(m,z) Z_{\antivortex,p}(m,\bar{z})
\end{equation}
obtained when localizing to the Higgs branch of the theory for positive
and for negative FI parameter~\(\xi\), respectively.
Indeed, Higgs branch vacua are labeled by solutions of
\begin{equation}
  \sum_{s=1}^{\Nf} \Bigl( \abs{\quark_s}^2
    - \abs{\anti{\quark}_s}^2 \Bigr)
  = \xi = -\frac{1}{2\pi} \ln\abs{z}
\end{equation}
and \((\sigma+m_s)\quark_s = 0 = (\sigma-\anti{m}_s)\anti{\quark}_s\)
for all~\(s\).  For \(\abs{z}<1\), that is, \(\xi>0\), at least one of
the positively charged fields \(\quark_s\) is non-zero, thus
\(\sigma=-m_s\).  For \(\abs{z}>1\), that is, \(\xi<0\), one of the
negatively charged fields is non-zero, and \(\sigma=\anti{m}_s\).  One
easily checks that evaluating the classical contribution, and the
residue of the one-loop contribution \big(which is the integrand
of the Coulomb branch representation~\eqref{SQED-Z-Coulomb}\big) at
those values of~\(\sigma\) and at \(B=0\) yields the relevant factors in
\eqref{SQED-Zs} and~\eqref{SQED-Zu}.  The hypergeometric functions
\(f^\sch_p\) and~\(f^\uch_p\) obtained from factorization also match with
known vortex and anti-vortex partition functions
(see~\cite{Benini:2012ui,Doroud:2012xw}).  For more general theories,
factorization always yields explicit expressions for the vortex
partition functions, while earlier methods soon become intractable.

The s-channel factors in~\eqref{SQED-Zs} also have a Mellin--Barnes
integral representation
\begin{equation}\label{SQED-fs-integral}
  \begin{aligned}
    & (-x)^{-\I m_p} f^\sch_p(x)
    \\
    & =
    \prod_{s=1}^{\Nf}\biggl[
      \frac{\Gamma(1+\I m_s-\I m_p)}{\Gamma(-\I\anti{m}_s-\I m_p)}
    \biggr]
    \int_{-\I\infty}^{\I\infty} \frac{\dd{\kappa}}{2\pi\I}
    \frac{\prod_{s=1}^{\Nf}\Gamma(-\I\anti{m}_s+\kappa)}
      {\prod_{s\neq p}^{\Nf}\Gamma(1+\I m_s+\kappa)}
    \Gamma(-\kappa-\I m_p) (-x)^{\kappa}
  \end{aligned}
\end{equation}
which converges for \(\abs{\arg(-x)}<\pi\), that is, away from the
positive real axis.  On the other hand, the s-~and u-channel expansions
found above imply that the partition function has branch points at \(0\)
and~\(\infty\), but is otherwise smooth away from the unit circle.
Hence, the partition function can only have branch points at
\(x\in\{0,1,\infty\}\).

We have already given expansions near \(0\) and~\(\infty\), so we now
focus on powers of \(\abs{1-x}^2\) as \(x\to 1\).  The Higgs branch
localization has no analogue at \(x=1\), because the FI parameter
\(\xi=-\frac{1}{2\pi}\ln\abs{z}\) vanishes and the manifold of solutions
of \(\sum_{s=1}^{\Nf} \bigl(\abs{\quark_s}^2 -
\abs{\anti{\quark}_s}^2\bigr) = \xi\) experiences a flop transition.
Instead, we find an explicit decomposition starting from the Coulomb
branch integral.

\label{page:SQED-x-to-1}%
As \(x\to 1\), split the Coulomb branch
representation~\eqref{SQED-Z-Coulomb} into the two regions,
\(\abs{\I\sigma+\frac{B}{2}} \lessgtr \abs{\ln x}^{-1}\).  In the first,
\(x^{\I\sigma+\frac{B}{2}}\bar{x}^{\I\sigma-\frac{B}{2}}\) is given by a
convergent series in integer powers of \(\ln x\) and \(\ln\bar{x}\)
thanks to
\begin{equation}
  x^{\I\sigma+\frac{B}{2}}
  = \sum_{k\geq 0} \frac{(\I\sigma+\frac{B}{2})^k}{k!} (\ln x)^k \,.
\end{equation}
In the second, the product of Gamma functions in the integrand can be
approximated as~\eqref{SQED-integrand} through Stirling's approximation,
and the sum over~\(B\) can be replaced by a continuous integral, leading
to a contribution
\begin{equation}\label{SQED-large-sigma}
  \int \dd{B} \frac{\dd{\sigma}}{2\pi}
  e^{(\I\sigma+\frac{B}{2})\ln x} e^{(\I\sigma-\frac{B}{2})\ln\bar{x}}
  \Bigl(\sigma^2 + \frac{B^2}{4}\Bigr)^{-\Sigma}
  =
  \frac{1}{\pi}
  \int \dd{\rho} \rho \dd{\theta}
  e^{2\I\rho\cos\theta \abs{\ln x}}
  \rho^{-2\Sigma}
  \,,
\end{equation}
where \(\Sigma = \sum_{s=1}^{\Nf} (1+\I m_s+\I\anti{m}_s)\) and we applied
the change of variables \(\rho e^{i\theta} \abs{\ln x} =
(\sigma-\I\frac{B}{2}) \ln x\).  Rescaling then \(\rho\) by \(\abs{\ln
  x}\), we find that the contribution behaves as
\begin{equation}\label{SQED-one-pow}
  \abs{\ln x}^{2\Sigma-2}
  \sim \abs{1-x}^{2\bigl[-1+\sum_{s=1}^{\Nf}(1+\I m_s+\I\anti{m}_s)\bigr]} \,,
\end{equation}
as \(x\to 1\), multiplied by a series in powers of \((1-x)\) and
\((1-\bar{x})\).  We thus find
\begin{equation}\label{SQED-Zt}
  Z = \abs{1-x}^0 G(1-x,1-\bar{x})
  + \abs{1-x}^{2\bigl[-1+\sum_{s=1}^{\Nf}(1+\I m_s+\I\anti{m}_s)\bigr]}
  H(1-x,1-\bar{x})
\end{equation}
for some series \(G\) and~\(H\) in positive integer powers of \(1-x\)
and \(1-\bar{x}\).  Since the \(\Nf\)~terms of the Higgs branch
expansions around \(x=0\) and~\(\infty\) are linearly independent, the
series \(G\) and~\(H\) cannot both factorize.  When studying the gauge
theory analogue of the braiding matrix relating the s-~and u-channel
expansions in Appendix~\ref{app:braiding}, we find that \(H\)~factorizes
as \(h(1-x)\bar{h}(1-\bar{x})\), while \(G\) is a sum of \(\Nf-1\) such
factorized terms, with no preferred choice of splitting.  We can expect
the factorization of~\(H\) because in the limit
\(\abs{\I\sigma\pm\frac{B}{2}}\to\infty\) the
integrand~\eqref{SQED-large-sigma} factorizes into functions of
\(\I\sigma\pm\frac{B}{2}\).

\subsubsection{Matching Parameters for SQED}

We wish to equate the expansions of~\(Z\) obtained so far with an
\(A_{\Nf-1}\)~Toda CFT correlator.  Since the \(S^2\)~partition function
has branch points at \((-1)^{\Nf}z\in\bigl\{0,1,\infty\bigr\}\), and
factorizes when expanded around each of those points, the Toda
correlator must be a four-point function with insertions at \(0\),
\(1\), \(\infty\), and \(x = (-1)^{\Nf} z\).  The expansions near branch
points have finitely many terms, hence the operator inserted at~\(x\)
must be a degenerate operator~\(\widehat{V}_{-b\omega}\) (labeled by
the highest weight~\(\omega\) of a representation
\(\repr(\omega)\) of~\(A_{\Nf-1}\)), and the correlator has the
form
\begin{equation}
  \vev*{\widehat{V}_{\alpha_\infty}(\infty)\widehat{V}_{\hat{m}}(1)
    \widehat{V}_{-b\omega}(x,\bar{x})\widehat{V}_{\alpha_0}(0)} \,.
\end{equation}
The number of internal momenta allowed by the fusion rule for
\(\widehat{V}_{-b\omega}\) with a generic operator is equal to the
dimension of \(\repr(\omega)\), hence \(\repr(\omega)\) must be the
fundamental or antifundamental representation, to match the number of
terms in \eqref{SQED-Zs} and~\eqref{SQED-Zu}.  Without loss of
generality (we can at this point conjugate all momenta), we choose the
operator~\(\widehat{V}_{-bh_1}\), where \(h_1\) is the highest weight of
the fundamental representation.  The momenta \(\alpha_0\), \(\hat{m}\)
and~\(\alpha_\infty\) can then be obtained by comparing dimensions of
Toda CFT operators with the powers of~\(\abs{x}^2\) and of
\(\abs{1-x}^2\) appearing in the expansions of~\(Z\) around \(x=0\),
\(x=1\), and \(x=\infty\).

The s-channel decomposition of the Toda correlator is a sum over
internal momenta \(\alpha_0-bh_p\) labeling \(W_{\Nf}\)~primary
operators:
\begin{equation}\label{SQED-conf-s}
  \begin{aligned}
    & \vev*{\widehat{V}_{\alpha_\infty}(\infty)\widehat{V}_{\hat{m}}(1)
      \widehat{V}_{-bh_1}(x,\bar{x})\widehat{V}_{\alpha_0}(0)}
    \\
    & =
    \sum_{p=1}^{\Nf}
    \widehat{C}(\alpha_\infty,\hat{m},\alpha_0-bh_p)
    \widehat{C}_{-bh_1,\alpha_0}^{\alpha_0-bh_p}
    \Fblock{s}{\alpha_0-bh_p}{\hat{m}&-bh_1\\\alpha_\infty&\alpha_0}{x}
    \Fblock{s}{\alpha_0-bh_p}{\hat{m}&-bh_1\\\alpha_\infty&\alpha_0}{\bar{x}} \,,
  \end{aligned}
\end{equation}
where \(\widehat{C}\)~denote three-point functions and
\(\Fblock{s}{\alpha_0-bh_p}{}{x}\)~are \(W_{\Nf}\)~conformal blocks.
Conformal invariance fixes \(\Fblock{s}{\alpha_0-bh_p}{}{x} =
x^{\dimToda(\alpha_0-bh_p) - \dimToda(\alpha_0) - \dimToda(-bh_1)}
(1+\cdots)\), with a series \((1+\cdots)\) in positive integer powers
of~\(x\).  We compute
\begin{equation}
  \dimToda(\alpha_0-bh_p) - \dimToda(\alpha_0) - \dimToda(-bh_1)
  = b \vev{\alpha_0-Q,h_p} + \frac{\Nf-1}{2} (b^2+1) \,.
\end{equation}
This should be compared with the powers \(x^{-\I m_p}\) appearing
in~\eqref{SQED-Zs}.  Since the weights \(h_p\) sum to zero,
\(\sum_p\vev{\alpha-Q,h_p}=0\), and we must allow for an overall shift
by~\(x^{\gamma_0}\) between the partition function and the correlator.
Power matching then dictates
\begin{equation}
  b \vev{\alpha_0-Q,h_p} + \frac{\Nf-1}{2} (b^2+1) + \gamma_0
  = -\I m_p \,,
\end{equation}
up to permutations, from which we deduce \(\alpha_0\) and~\(\gamma_0\)
given in \eqref{SQED-alpha} and~\eqref{SQED-gamma0}.  Permuting
the~\(m_p\) is equivalent to permuting the components of \(\alpha_0-Q\), a
Weyl reflection under which the primary
operator~\(\widehat{V}_{\alpha_0}\) is invariant.

Next, the u-channel decomposition is a sum over the internal momenta
\(\alpha_\infty-bh_p\).  Conformal invariance fixes
\(\Fblock{u}{\alpha_\infty-bh_p}{}{x} = x^{\dimToda(\alpha_\infty) -
  \dimToda(\alpha_\infty-bh_p) - \dimToda(-bh_1)} (1+\cdots)\), with a
series \((1+\cdots)\) in negative integer powers of~\(x\).  We compute
\begin{equation}
  \dimToda(\alpha_\infty) - \dimToda(\alpha_\infty-bh_p) - \dimToda(-bh_1)
  = - b \vev{\alpha_\infty-Q,h_p} + \frac{\Nf-1}{2} (b^2+1)
  + \frac{\Nf-1}{\Nf} b^2 \,,
\end{equation}
which should be compared with \(x^{\I\anti{m}_p-\gamma_0}\).  Once more,
we must allow for an overall ambiguity: besides \(x^{\gamma_0}\), the
only other factor that can appear is \((1-x)^{\gamma_1}\), since the
Toda correlator is only singular at \(0\), \(1\), and~\(\infty\).  This
factor does not alter powers at \(x=0\), and the power matching at
\(x=\infty\) reads
\begin{equation}
  - b \vev{\alpha_\infty-Q,h_p} + \frac{\Nf-1}{2} (b^2+1)
  + \frac{\Nf-1}{\Nf} b^2 + \gamma_0 + \gamma_1 = \I\anti{m}_p \,,
\end{equation}
up to permutations: this fixes \(\alpha_\infty\) and~\(\gamma_1\) to
\eqref{SQED-alpha} and~\eqref{SQED-gamma1}.

Third, the expansion of~\(Z\) near \(x=1\) involves the leading powers
\((1-x)^0\) with multiplicity \(\Nf - 1\) and
\((1-x)^{-1+\sum_{p=1}^{\Nf}(1+\I m_p +\I\anti{m}_p)}\) with no
multiplicity.  On the Toda CFT side, the exponents that can appear in
the t-channel are
\begin{equation}\label{SQED-t-pow}
  \begin{aligned}
    & \dimToda(\alpha_1-bh_p) - \dimToda(\alpha_1) - \dimToda(-bh_1)
    + \gamma_1
    \\[-1ex]
    & = b \vev{\alpha_1-Q,h_p} + \frac{\Nf-1}{2} (b^2+1)
    - \frac{\Nf-1}{\Nf} b^2
    + \frac{1}{\Nf} \sum_{p=1}^{\Nf} (\I m_p+\I\anti{m}_p) \,.
  \end{aligned}
\end{equation}
If \(\alpha_1\) were generic, all shifts \(-bh_p\) would be allowed by
the fusion, but summing the powers~\eqref{SQED-t-pow} for \(1\leq
p\leq\Nf\) does not yield the similar gauge theory sum \(- 1 +
\sum_{p=1}^{\Nf}(1+\I m_p + \I\anti{m}_p)\).  Instead, we take \(\alpha_1
= \hat{m} = (\varkappa+b) h_1\) to be a semi-degenerate momentum (with a
shift by~\(b\) to simplify expressions), so that the fusion rule only
allows shifts to \(\hat{m} - bh_2\) and \(\hat{m} - bh_1\).  Setting the
exponent for a shift \(\hat{m} - bh_2\) to~\(0\) fixes~\(\varkappa\)
to~\eqref{SQED-alpha}, and the second power matches (setting
\(\hat{m}-bh_1\) to~\(0\) instead would fail to match the second power).
The \(SU(\Nf)\times SU(\Nf)\times U(1)\) flavour symmetry of the gauge
theory is reproduced by the two generic and one semi-degenerate
operators in the correlator, allowing us to package the twisted masses
of fundamental chiral multiplets into~\(\alpha_0\), those of
antifundamental multiplets into~\(\alpha_\infty\), and the axial mass
into~\(\hat{m}\).

Finally, the overall constant~\(A\) is fixed in
Appendix~\ref{app:1loop-3pt} by comparing gauge theory one-loop
determinants and Toda three-point functions: for \(A\)~given
by~\eqref{SQED-A},
\begin{equation}
  Z_{S^4_b}^\free
  \frac{\prod_{s\neq p}^{\Nf} \gamma(\I m_p-\I m_s)}
    {\prod_{t=1}^{\Nf} \gamma(1+\I m_p+\I\anti{m}_t)}
  = A \widehat{C}(\alpha_\infty, (\varkappa+b) h_1, \alpha_0 - bh_p)
  \widehat{C}_{-bh_1,\alpha_0}^{\alpha_0-bh_p} \,.
\end{equation}
The same relation holds for u-channel constant factors (with an
identical value of~\(A\)), as we can obtain most readily thanks to the
invariance of~\(Z\) under \(m_p\leftrightarrow\anti{m}_p\) and
\(z\leftrightarrow\frac{1}{z}\) (gauge theory charge conjugation) and
equivalently of the Toda correlator (up to a shift in
exponents) under \(\alpha_0\leftrightarrow\alpha_\infty\) and
\(x\leftrightarrow\frac{1}{x}\).

We have thus fixed how gauge theory and Toda CFT parameters match.  One
way to prove the matching is to directly equate gauge theory factors
with conformal blocks as done in~\cite{Doroud:2012xw}, but this approach
does not generalize.  Instead, we show in Appendix~\ref{app:braiding}
that the matrix to change basis from s-channel factors \(x^{-\I
  m_p}f^\sch_p(x)\) to u-channel factors is identical to the appropriate
braiding matrix in the Toda CFT\@.  Since the eigenvalues of monodromies
around \(0\) and~\(\infty\) also match up to shifts by the~\(\gamma_i\)
as we just saw, the monodromy matrices themselves agree.  The last
monodromy matrix, around \(x=1\), thus also matches.  Therefore, the
partition function and the correlator differ by a factor with no
monodromy.  Since the precise exponents match, the relative factor is in
fact constant, and comparing constant coefficients establishes the
matching~\eqref{SQED-matching}.

\subsection{SQCD and Toda Antisymmetric Degenerate}
\label{sec:SQCD}

We now extend the matching to the case of \(\Nsusy=(2,2)\) SQCD, that
is, a \(U(\Nc)\) vector multiplet coupled to \(\Nf\)~fundamental and
\(\Nf\)~antifundamental chiral multiplets, with twisted masses (plus
\(R\)-charges) \(m_s\) and~\(\anti{m}_s\).  The partition function of
the \(S^2\)~surface operator defined by this theory coupled to
\(\Nf^2\)~hypermultiplets with masses~\eqref{part-mst} on~\(S^4_b\) is
captured by a Toda CFT four-point function with a degenerate operator
\(\widehat{V}_{-b\omega_\Nc}\) labeled by the \(\Nc\)-th antisymmetric
representation of \(A_{\Nf-1}\).  Explicitly, we prove that\footnote{As
  explained below~\eqref{part-typical-matching}, the factor \(A
  \abs{x}^{2\gamma_0} \abs{1-x}^{2\gamma_1}\) can be absorbed into the
  partition function.  To compare gauge theory and Toda CFT results it
  is best to keep the factor explicitly.}%
\begin{equation}\label{SQCD-matching}
  Z_{S^2\subset S^4_b}^{U(\Nc)\ \SQCD}(m,\anti{m},z,\bar{z})
  = A \abs{x}^{2\gamma_0} \abs{1-x}^{2\gamma_1}
  \vev*{\widehat{V}_{\alpha_\infty}(\infty)\widehat{V}_{\hat{m}}(1)
    \widehat{V}_{-b\omega_\Nc}(x,\bar{x})\widehat{V}_{\alpha_0}(0)}
\end{equation}
with \(x = (-1)^{\Nf+\Nc-1} z\), momenta
\begin{equation}
  \label{SQCD-alpha}
  \begin{aligned}
    \alpha_0 & = Q - \frac{1}{b} \sum_{s=1}^{\Nf} \I m_s h_s \,,
    & \qquad\qquad
    \hat{m} & = (\varkappa+\Nc b) h_1 \,,
    \\
    \alpha_\infty & = Q - \frac{1}{b} \sum_{s=1}^{\Nf} \I\anti{m}_s h_s \,,
    & \qquad\qquad
    \varkappa & = \frac{1}{b} \sum_{s=1}^{\Nf} (1+\I m_s+\I\anti{m}_s) \,,
  \end{aligned}
\end{equation}
and coefficients
\begin{align}
  \label{SQCD-gamma0}
  \gamma_0 & = - \frac{\Nc}{\Nf} \sum_{s=1}^{\Nf} \I m_s
  - \frac{\Nc(\Nf-\Nc)}{2}(b^2+1) \,,
  \\
  \label{SQCD-gamma1}
  \gamma_1 & = - \frac{\Nc(\Nf-\Nc)}{\Nf} b^2
  + \frac{\Nc}{\Nf} \sum_{s=1}^{\Nf} (\I m_s + \I\anti{m}_s) \,,
  \\
  \label{SQCD-A}
  A & = b^{\Nc\Nf(1+b^2)-\Nc^2 b^2-2\Nc b\varkappa}
  \,.
\end{align}
Setting \(\Nc=1\) in~\eqref{SQCD-matching} reproduces the SQED
matching~\eqref{SQED-matching}.  We recognize the same symmetries as
SQED\@.  Permuting twisted masses \(m_s\) or~\(\anti{m}_s\) amounts to a
Weyl transformation of \(\alpha_0\) or~\(\alpha_\infty\).  Gauge theory
charge conjugation, which swaps \(m_s\leftrightarrow\anti{m}_s\) and
\(z\leftrightarrow\frac{1}{z}\), corresponds to the conformal map
\((\infty,1,x,0)\to(0,1,\frac{1}{x},\infty)\), which exchanges
\(\alpha_0\leftrightarrow\alpha_\infty\) and
\(x\leftrightarrow\frac{1}{x}\) in the Toda CFT correlator.

We start the analysis from the Coulomb branch representation
\begin{equation}\label{SQCD-Z-Coulomb}
  \begin{aligned}
    & Z_{S^2}^\SQCD
    =
    \frac{1}{\Nc!}
    \sum_{B\in\bbZ^\Nc} \int_{\bbR^\Nc}
    \frac{\dd[^\Nc]{\sigma}}{(2\pi)^\Nc} \Biggl\{
    \bigl[(-1)^{\Nc-1} z\bigr]^{\Tr(\I\sigma+\frac{B}{2})}
    \bigl[(-1)^{\Nc-1} \bar{z}\bigr]^{\Tr(\I\sigma-\frac{B}{2})}
    \\
    &
    \cdot \prod_{i<j} \biggl[(\sigma_i-\sigma_j)^2+\frac{(B_i-B_j)^2}{4}\biggr]
    \prod_{j=1}^\Nc \prod_{s=1}^{\Nf} \biggl[
      \frac{\Gamma(-\I m_s - \I\sigma_j - \frac{B_j}{2})}
        {\Gamma(1+\I m_s + \I\sigma_j - \frac{B_j}{2})}
      \frac{\Gamma(-\I\anti{m}_s + \I\sigma_j + \frac{B_j}{2})}
        {\Gamma(1+\I\anti{m}_s - \I\sigma_j + \frac{B_j}{2})}
    \biggr] \Biggr\} \,.
  \end{aligned}
\end{equation}
The partition function can be studied in the same way as that of SQED,
by closing the integration contours towards either half-plane depending
on whether \(\abs{z} \lessgtr 1\), thus obtaining an s-channel and a
u-channel decompositions akin to \eqref{SQED-Zs} and~\eqref{SQED-Zu}.
Interestingly, there is a shortcut, as the SQCD partition function can
be expressed as a differential operator acting on the product of \(\Nc\)
copies of the SQED partition function:%
\begin{equation}\label{SQCD-Z-from-SQED}
  Z_{S^2}^\SQCD
  =
  \frac{1}{\Nc!} \Biggl[
  \prod_{i<j} \Bigl[- (z_i \partial_{z_i} - z_j \partial_{z_j})
  (\bar{z}_i \partial_{\bar{z}_i} - \bar{z}_j \partial_{\bar{z}_j}) \Bigr]
  \prod_{j=1}^\Nc Z_{S^2}^\SQED(m,\anti{m},z_j,\bar{z}_j)
  \Biggr]_{\substack{z_j=(-1)^{\Nc-1}z\\\bar{z}_j=(-1)^{\Nc-1}\bar{z}}} \!\!
  \,.
\end{equation}
Since the differential operator cannot introduce branch points, the SQCD
partition function has the same branch points
\((-1)^{\Nc-1}z\in\{0,(-1)^{\Nf},\infty\}\) as the SQED partition function, and we
switch to using the coordinate \(x = (-1)^{\Nf+\Nc-1} z\).

\subsubsection{Expanding the SQCD Partition Function}
\label{sec:SQCD-expand}

Using the s-channel decomposition~\eqref{SQED-Zs} of \(Z^\SQED\) in the
above yields a sum over flavours \(1\leq p_1,\ldots,p_\Nc\leq\Nf\).  The
summand factorizes, since both the differential operator and the terms
in~\(Z^\SQED\) are products of a holomorphic and an antiholomorphic
parts.  The holomorphic and the antiholomorphic factors are each totally
antisymmetric in the~\(p_j\), hence reducing the sum to \(1\leq
p_1<\cdots<p_\Nc\leq\Nf\).  Explicitly,
\begin{equation}\label{SQCD-Zs}
  Z =
  \mspace{-20mu}
  \sum_{1\leq p_1<\cdots<p_\Nc\leq\Nf} \Biggl[
  (x\bar{x})^{-\sum_{j=1}^\Nc \I m_{p_j}}
  \prod_{j=1}^\Nc
  \frac{\prod_{s\not\in\{p\}}^{\Nf}\gamma(-\I m_s+\I m_{p_j})}
    {\prod_{s=1}^{\Nf}\gamma(1+\I\anti{m}_s+\I m_{p_j})}
  f^\sch_{\{p\}}(x) f^\sch_{\{p\}}(\bar{x})
  \Biggr]
\end{equation}
where we have canceled \(\prod_{i\neq j} \gamma(-\I m_{p_i} + \I
m_{p_j}) = \prod_{i\neq j} (\I m_{p_i} - \I m_{p_j})^{-1}\) and defined
\begin{align}
  \label{SQCD-fs-from-SQED}
  f^\sch_{\{p\}}(x)
  & =
  \biggl[\prod_{i<j}
  \frac{-\I m_{p_i} + \I m_{p_j} + x_i \partial_{x_i} - x_j \partial_{x_j}}
    {-\I m_{p_i} + \I m_{p_j}}
  \prod_{j=1}^\Nc f^\sch_{p_j}(x_j) \biggr]_{x_j=x}
  \\
  \label{SQCD-fs}
  & = \!\!
  \sum_{k_1,\ldots,k_\Nc\geq 0}
  \frac{x^{\sum_{j=1}^\Nc k_j}}{\prod_{j=1}^\Nc k_j!}
  \frac{\prod_{j=1}^\Nc \prod_{s=1}^{\Nf} (-\I\anti{m}_s-\I m_{p_j})_{k_j}}
    {\prod_{i\neq j}^\Nc (\I m_{p_i}-\I m_{p_j}-k_i)_{k_j}
      \prod_{j=1}^\Nc \prod_{s\not\in\{p\}}^{\Nf} (1+\I m_s-\I m_{p_j})_{k_j}}
  \,,
\end{align}
a series in positive integer powers of~\(x\), with radius of
convergence~\(1\), and whose first term is normalized to be~\(1\).
Similarly, the u-channel expansion near \(x=\infty\) reads
\begin{equation}\label{SQCD-Zu}
  Z =
  \mspace{-20mu}
  \sum_{1\leq p_1<\cdots<p_\Nc\leq\Nf}
  \Biggl[
  (x\bar{x})^{\sum_{j=1}^\Nc \I\anti{m}_{p_j}}
  \frac{\prod_{j=1}^\Nc \prod_{s\not\in\{p\}}^{\Nf}
    \gamma(-\I\anti{m}_s+\I\anti{m}_{p_j})}
  {\prod_{j=1}^\Nc \prod_{s=1}^{\Nf}\gamma(1+\I m_s+\I\anti{m}_{p_j})}
  f^\uch_{\{p\}}(x)
  f^\uch_{\{p\}}(\bar{x})
  \Biggr]
\end{equation}
where
\begin{equation}\label{SQCD-fu}
  f^\uch_{\{p\}}(x)
  = \!\!
  \sum_{k_1,\ldots,k_\Nc\geq 0}
  \frac{x^{-\sum_{j=1}^\Nc k_j}}{\prod_{j=1}^\Nc k_j!}
  \frac{\prod_{j=1}^\Nc \prod_{s=1}^{\Nf} (-\I m_s-\I\anti{m}_{p_j})_{k_j}}
    {\prod_{i\neq j}^\Nc (\I\anti{m}_{p_i}-\I\anti{m}_{p_j}-k_i)_{k_j}
      \prod_{j=1}^\Nc \prod_{s\not\in\{p\}}^{\Nf}
      (1+\I\anti{m}_s-\I\anti{m}_{p_j})_{k_j}}
\end{equation}
are series in negative integer powers of~\(x\).

The s-~and u-channel decompositions above can also be obtained by
localizing to the Higgs branch of the theory, with a positive or a
negative FI parameter.  In this setting, they arise as sums over Higgs
branch vacua, labeled by solutions \((\sigma, \quark_s,
\anti{\quark}_s)\) of
\begin{align}
  \label{SQCD-D-term}
  \begin{aligned}
    (\sigma + m_s) \quark_s & = 0
    \\
    (- \sigma + \anti{m}_s) \anti{\quark}_s & = 0
  \end{aligned}
  &&
  \sum_{s=1}^{\Nf} (\quark_s \quark_s^\dagger
  - \anti{\quark}_s^\dagger \anti{\quark}_s)
  = \xi \id_\Nc \,,
\end{align}
up to gauge transformations.  In the region \(\abs{x} = \abs{z} < 1\),
that is, \(\xi > 0\), the \(\Daux\)-term equation~\eqref{SQCD-D-term}
can be rewritten as
\begin{equation}
  \sum_{s=1}^{\Nf} \quark_s \quark_s^\dagger
  = \xi \id_\Nc
  + \sum_{s=1}^{\Nf} \anti{\quark}_s^\dagger \anti{\quark}_s \,,
\end{equation}
which is positive definite, hence has full rank~\(\Nc\).  Therefore, the
non-zero vectors \(\quark_s\), which are eigenvectors of~\(\sigma\),
span~\(\bbC^{\Nf}\).  The eigenvalues of~\(\sigma\) are thus completely
fixed to be \(-m_{p_j}\) for a choice of \(\Nc\)~distinct
flavours~\(p_j\).  On the contrary, for \(\abs{x} = \abs{z} > 1\), that
is, \(\xi < 0\), the antifundamental chiral fields \(\anti{\quark}_s\)
span~\(\bbC^{\Nf}\), and \(\sigma\) has eigenvalues \(\anti{m}_{p_j}\).
The classical and one-loop contributions derived for each of those vacua
is equal to those appearing in \eqref{SQCD-Zs} and~\eqref{SQCD-Zu}.
More tediously, one checks that the vortex partition functions are
indeed given by \(f^\sch_{\{p\}}(x)\) and~\(f^\uch_{\{p\}}(x)\).

Once more, the t-channel is the most troublesome.  We know
from~\eqref{SQED-Zt} the expansion of the SQED partition function near
\(x=1\), leading to
\begin{equation}
  Z^\SQED
  = G(1-x,1-\bar{x})
  + \abs{1-x}^{2(\gamma-1)}
  h(1-x)\bar{h}(1-\bar{x}) \,,
\end{equation}
where
\begin{equation}
  \gamma = \sum_{s=1}^{\Nf}(1+\I m_s+\I\anti{m}_s) \,.
\end{equation}
The functions \(G\) and \(h\bar{h}\) are series in positive integer
powers of \(1-x\) and \(1-\bar{x}\), and \(G\)~does not factorize
because the eigenvalue~\(1\) of the monodromy has multiplicity
\(\Nf-1\).  Plug this t-channel expansion into~\eqref{SQCD-Z-from-SQED}:
\begin{equation}
  \begin{aligned}
    Z^\SQCD(z,\bar{z})
    & =
    \frac{1}{\Nc!} \Biggl[
    \prod_{i<j} \Bigl[- (x_i \partial_{x_i} - x_j \partial_{x_j})
    (\bar{x}_i \partial_{\bar{x}_i} - \bar{x}_j \partial_{\bar{x}_j}) \Bigr]
    \\
    & \quad \cdot \prod_{j=1}^\Nc
    \Bigl\{G(1-x_j,1-\bar{x}_j)
    + \abs{1-x_j}^{2\gamma-2} h(1-x_j) \bar{h}(1-\bar{x}_j)\Bigr\}
    \Biggr]_{\substack{x_j=x\\\bar{x}_j=\bar{x}}} \,.
  \end{aligned}
\end{equation}
Among the \(2^\Nc\) terms in the product of SQED partition functions,
any which contains the factor \(\abs{1-x_j}^{2\gamma-2} h(1-x_j)
\bar{h}(1-\bar{x}_j)\) for two indices \(i\) and~\(j\) is annihilated by
\(x_i\partial_{x_i}-x_j\partial_{x_j}\), hence does not contribute.  The
annihilation does not take place when \(G(1-x_j,1-\bar{x}_j)\) appears
twice, as it relies on separating the holomorphic and antiholomorphic
parts.  Thus, \(1 + \Nc\) terms remain, and we can replace the product
by
\begin{equation}
  \prod_{j=1}^\Nc G(1-x_j,1-\bar{x}_j)
  + \sum_{j=1}^\Nc \abs{1-x_j}^{2\gamma-2} h(1-x_j) \bar{h}(1-\bar{x}_j)
  \prod_{i\neq j}^\Nc G(1-x_i,1-\bar{x}_i) \,.
\end{equation}
Derivatives acting on \(G\), \(h\) and~\(\bar{h}\) yield other series in
positive integer powers of \(1-x_j\) and~\(1-\bar{x}_j\), hence for the
purpose of finding exponents for \(\abs{1-x}^2\) we only need to keep
track of \(\abs{1-x_j}^{2\gamma-2}\).  At most \((\Nc - 1)\)
\(x_j\)~derivatives can affect it, hence the SQCD partition function
takes the form
\begin{equation}\label{SQCD-tch-expo}
  Z^\SQCD(z,\bar{z}) = G'(1-x,1-\bar{x})
  + \abs{1-x}^{2(\gamma-\Nc)} H'(1-x,1-\bar{x}) \,,
\end{equation}
for some series \(G'\) and~\(H'\).  The two terms correspond to
eigenvalues \(1\) and \(e^{2\pi\I(\gamma-\Nc)}\) of the monodromy around
\(x=1\).  We find out the multiplicities with which the powers appear by
doing a finer expansion: split \(G(1-x_j, 1-\bar{x}_j) =
\sum_{i=1}^{\Nf-1} g_i(1-x_j) \bar{g}_i(1-\bar{x}_j)\) non-canonically.
Antisymmetry restricts the sum of \(\Nf^\Nc\) terms to
\(\binom{\Nf}{\Nc}\), each of which is a product of \(\Nc\) distinct
terms of \(Z^\SQED\) among \(h\bar{h}\) and the \(g_i\bar{g}_i\).  The
exponent for a given combination is \(2(\gamma-\Nc)\)~if \(h\bar{h}\)
appears, and \(0\)~otherwise.  The multiplicity of \(\abs{1-x}^{0}\) is
thus \(\binom{\Nf-1}{\Nc}\), and that of \(\abs{1-x}^{2(\gamma-\Nc)}\)
is~\(\binom{\Nf-1}{\Nc-1}\).

\subsubsection{Matching Parameters for SQCD}

We are at last ready to match SQCD and Toda CFT parameters.  The
partition function depends on a single parameter~\(x\) encoded as the
position of a puncture, hence we expect a four-point function on the
Toda side.  The s-channel and u-channel decompositions involve
\(\binom{\Nf}{\Nc}\) terms, hence the Toda degenerate operator is
labeled by the \(\Nc\)-th antisymmetric representation
\(\repr(\omega_\Nc)\) of \(A_{\Nf-1}\), which has the correct
dimension.  The highest weight of this representation is~\(\omega_\Nc =
h_1 + \cdots + h_\Nc\), and its weights are \(h_{\{p\}} = h_{p_1} +
\cdots + h_{p_\Nc}\), labeled by \(\Nc\)-element sets \(1\leq
p_1<\cdots<p_\Nc\leq\Nf\).

The s-channel Toda exponents
\begin{equation}
  \begin{aligned}
    & \dimToda(\alpha_0-bh_{\{p\}})
    - \dimToda(\alpha_0) - \dimToda(-b\omega_\Nc) + \gamma_0
    \\
    & = b \sum_{j=1}^\Nc \vev{\alpha_0-Q, h_{p_j}}
    + \frac{\Nc(\Nf-\Nc)}{2}(b^2+1)
    + \gamma_0
  \end{aligned}
\end{equation}
must be equal to \(-\sum_{j=1}^\Nc \I m_{p_j}\) from gauge theory (up to
permutations): this constraint fixes \(\alpha_0\) and~\(\gamma_0\) as
given in \eqref{SQCD-alpha} and~\eqref{SQCD-gamma0}.  Matching powers
in the u-channel,
\begin{align}
  \sum_{j=1}^\Nc \I\anti{m}_{p_j}
  & = \dimToda(\alpha_\infty)
  - \dimToda(\alpha_\infty-bh_{\{p\}})
  - \dimToda(-b\omega_\Nc)
  + \gamma_0 + \gamma_1
  \\\nonumber
  & = - b \sum_{j=1}^\Nc \vev{\alpha_\infty-Q,h_{p_j}}
  + \frac{\Nc(\Nf-\Nc)}{2}(b^2+1)
  + \frac{\Nc(\Nf-\Nc)}{\Nf} b^2
  + \gamma_0 + \gamma_1
\end{align}
fixes \(\alpha_\infty\) and~\(\gamma_1\).

We finally match powers in the t-channel.  From our SQED experience, we
expect the momentum at~\(1\) to be the semi-degenerate
\(\hat{m}=(\varkappa+\Nc b)h_1\) (the shift by \(\Nc b\) simplifies
expressions).  We compute the exponents
\begin{equation}
  \begin{aligned}
    & \dimToda\bigl((\varkappa+\Nc b)h_1-bh_{\{p\}}\bigr)
    - \dimToda\bigl((\varkappa+\Nc b)h_1\bigr)
    - \dimToda(-b\omega_\Nc) + \gamma_1
    \\
    & =
    (b\varkappa+\Nc b^2)
    \vev{h_1,h_{\{p\}}}
    + \frac{\Nc}{\Nf} \biggl[\sum_{s=1}^{\Nf} (1+\I m_s + \I\anti{m}_s)
    +  \Nc b^2 \biggr]
    + (1+b^2) \sum_{j=1}^\Nc (p_j-j-1) \,,
  \end{aligned}
\end{equation}
where \(\vev{h_1,h_{\{p\}}} = \delta_{1\in\{p\}} - \Nc/\Nf\).  Two
different sets~\(\{p\}\) must reproduce the gauge theory exponents
\(0\) and \(-\Nc+\sum_{s=1}^{\Nf}(1+\I m_s+\I\anti{m}_s)\).  One set
must contain~\(1\) and the other not, since the exponents would
otherwise only differ by an integer multiple of \(1+b^2\): this fixes
\(\varkappa=\pm\sum_{s=1}^{\Nf}(1+\I m_s+\I\anti{m}_s)+n(b+\frac{1}{b})\) for some integer~\(n\).
Comparing the coefficients of \(\sum_{s=1}^{\Nf} (1+\I m_s+\I\anti{m}_s)\)
selects the positive sign, and also implies that the exponent~\(0\) corresponds to
a case where \(1\not\in\{p\}\) while the other exponent has
\(1\in\{p'\}\).  Comparing the coefficients of \(b^2+1\), the Toda CFT
and gauge exponents match if
\begin{equation}
  - \frac{\Nc}{\Nf} n + \sum_{j=1} (p_j - j - 1) = 0
  \quad\text{and}\quad
  \frac{\Nf-\Nc}{\Nf} n + \sum_{j=1} (p'_j - j - 1) = -\Nc
\end{equation}
for the choices of \(\{p\}\) and~\(\{p'\}\) corresponding to the two
exponents.  Since \(1\not\in\{p\}\), \(p_j\geq j+1\) and the first
relation implies \(n\leq 0\).  Since \(p'_j\geq j\), the second implies
\(n\geq 0\), and we conclude that \(\varkappa\) is given
by~\eqref{SQCD-alpha}, that \(\{p\}=\intset{2}{\Nc+1}\) and that
\(\{p'\}=\intset{1}{\Nc}\).\footnote{Here and later we denote integer intervals by \(\intset{a}{b}=[a,b]\cap\bbZ\).}  After we show independently that the
partition function and Toda correlator are equal, we deduce that the
fusion of \(\widehat{V}_{-b\omega_\Nc}\) with
\(\widehat{V}_{\varkappa'h_1}\) allows the momenta \(\varkappa'h_1 -
b\omega_\Nc\) and \(\varkappa'h_1 + bh_1 - b\omega_{\Nc+1}\).  This is
consistent with the case \(\varkappa' = -kb\) for which the
semi-degenerate insertion becomes a degenerate field labeled by the
\(k\)-th symmetric representation: the tensor product of this
representation with the \(\Nc\)-th antisymmetric splits as a sum of two
irreducible representations of \(A_{\Nf-1}\), with highest weights
\(kh_1+\omega_\Nc\) and \((k-1)h_1+\omega_{\Nc+1}\).  We discuss such
fusion rules further in Appendix~\ref{app:fusion}.

Last, we fix the constant~\(A\).  We check in
Appendix~\ref{app:1loop-3pt} that the one-loop determinant and the
three-point functions appearing in the s-channel decompositions
of~\(Z^\SQCD\) and of the Toda correlator match, for \(A\)~given
in~\eqref{SQCD-A}:
\begin{equation}
  Z_{S^4_b}^\free
  \frac{\prod_{s\not\in\{p\}}^{\Nf} \prod_{t\in\{p\}} \gamma(\I m_t - \I m_s)}
    {\prod_{s=1}^{\Nf} \prod_{t\in\{p\}} \gamma(1 + \I m_t + \I\anti{m}_s)}
  =
  A \widehat{C}(\alpha_\infty, (\varkappa+\Nc b)h_1, \alpha_0-bh_{\{p\}})
  \widehat{C}_{-b\omega_\Nc,\alpha_0}^{\alpha_0-bh_{\{p\}}}
  \,.
\end{equation}

Having settled the dictionary above, we know that gauge theory and Toda
CFT monodromy matrices around each of \(0\), \(1\) and~\(\infty\) have
matching eigenvalues.  In Appendix~\ref{app:braiding}, we compute the
braiding matrix of \(\widehat{V}_{-b\omega_\Nc}\)
and~\(\widehat{V}_{\hat{m}}\) by combining the fusion of
\(\Nc\)~operators \(\widehat{V}_{-bh_1}\)
into~\(\widehat{V}_{-b\omega_\Nc}\) with the braiding matrices for each
individual \(\widehat{V}_{-bh_1}\) with~\(\widehat{V}_{\hat{m}}\).  The
result agrees with the analogue for SQCD, an antisymmetric combination
of the matrix for SQED, worked out in the same appendix.  Therefore, the
monodromy matrices around \(0\) and around~\(\infty\) are equal for SQCD
and the Toda CFT\@.  Monodromy matrices around~\(1\) then also match,
hence the Toda CFT correlator and gauge theory partition function are
equal up to a factor with no monodromy, which is constant since the
precise exponents at \(0\), \(1\) and~\(\infty\) match.  The constant
factors work out, thereby concluding the proof of the
matching~\eqref{SQCD-matching}.

\subsubsection{Decoupled Multiplets and Irregular Puncture}
\label{sec:irreg}

In this section, we give large twisted masses to \(\Nf-\anti{\Nf}\) of
the \(\Nf\)~antifundamental chiral multiplets of the SQCD surface
operator, hence to \(\Nf(\Nf-\anti{\Nf})\) of the four dimensional
hypermultiplets.  The massive multiplets decouple, and we obtain in this
limit~\eqref{irreg-Z-as-SQCD} a surface operator described by a
\(U(\Nc)\) vector multiplet, \(\Nf\)~fundamental and \(\anti{\Nf}<\Nf\)
antifundamental chiral multiplets, coupled to the remaining
\(\Nf\anti{\Nf}\) free hypermultiplets.  On the Toda CFT side of the
matching~\eqref{SQCD-matching}, the limit amounts to building a Toda CFT
irregular puncture from the collision of two vertex operators.  We give
the precise matching~\eqref{irreg-matching} in the case
\(\anti{\Nf}=\Nf-1\), and claim that further limits for
\(\anti{\Nf}\leq\Nf-2\) also lead to well-defined irregular punctures.

In a two dimensional \(\Nsusy=(2,2)\) gauge theory, whenever the total
charge \(Q=\sum_i Q_i\) of all chiral multiplets under a given \(U(1)\)
gauge group factor is non-zero (in our case, \(Q=\Nf-\anti{\Nf}>0\)),
the corresponding FI parameter runs logarithmically, and the theta angle
is shifted.  An ultraviolet cutoff can be introduced supersymmetrically
by enriching the theory with a single ``spectator'' chiral multiplet of
large twisted mass\footnote{The dimensionful cutoff is \(\Lambda/\ell\)
  in terms of the equatorial radius~\(\ell\) of the squashed two-sphere.}
\(\Lambda\in\bbR\) and \(U(1)\) charge \(-Q\), or with
\(Q\)~antifundamental spectator chiral multiplets of twisted
masses~\(\Lambda\).  We take the latter approach, as the resulting
enriched theory is simply SQCD with \(\Nf\)~fundamental and
\(\Nf\)~antifundamental chiral multiplets.  Each spectator chiral
multiplet brings a one-loop contribution
\begin{equation}
  \prod_{j=1}^\Nc \frac{\Gamma(-\I\Lambda+\I\sigma_j+\frac{B_j}{2})}
    {\Gamma(1+\I\Lambda-\I\sigma_j+\frac{B_j}{2})}
  \stackrel{\Lambda\to\infty}{\sim}
  \prod_{j=1}^\Nc \biggl(\frac{\Gamma(-\I\Lambda)}{\Gamma(1+\I\Lambda)}
  (-\I\Lambda)^{\I\sigma_j+B_j/2} (\I\Lambda)^{\I\sigma_j-B_j/2}\biggr)
\end{equation}
to the Coulomb branch expression for the enriched theory.  The original
partition function is thus a limit of the enriched partition function,
\begin{equation}\label{irreg-Z-as-SQCD}
  Z(m,\anti{m},z, \bar{z})
  = \lim_{\Lambda\to\infty} \Biggl[
  \frac{1}{\gamma(-\I\Lambda)^{\Nc(\Nf-\anti{\Nf})}}
  Z_\enriched
  \bigl(m,\{\anti{m},\Lambda\}, z_\bare, \bar{z}_\bare\bigr)
  \Biggr]
  \,,
\end{equation}
where the factor \(\gamma(-\I\Lambda)^{-\Nc(\Nf-\anti{\Nf})}\) has no
physical effect, and the bare parameter~\(z_\bare\) appearing in the
enriched theory is related to the renormalized~\(z=z_\renormalized\) (at
the scale~\(\ell\) given by the equatorial radius of the squashed
sphere) via
\begin{equation}
  z_\bare
  = \frac{z_\renormalized}{(-\I\Lambda)^{\Nf-\anti{\Nf}}}
  \quad \text{and} \quad
  \bar{z}_\bare
  = \frac{\bar{z}_\renormalized}{(\I\Lambda)^{\Nf-\anti{\Nf}}}
  \,.
\end{equation}
In particular, the FI parameter runs logarithmically, and the theta
angle is shifted:
\begin{equation}
  \xi_\renormalized
  = \xi_\bare - \frac{1}{2\pi} (\Nf-\anti{\Nf}) \ln\Lambda
  \quad \text{and} \quad
  \vartheta_\renormalized
  = \vartheta_\bare + \frac{\pi}{2} (\Nf-\anti{\Nf}) \,.
\end{equation}
Since the Coulomb branch representation involves an integral over
arbitrarily large values of~\(\sigma\pm\I\frac{B}{2}\), our derivation
of~\eqref{irreg-Z-as-SQCD} above is not rigorous.  However, one can
split the integral into a region \(\abs*{\sigma\pm\I\frac{B}{2}}\ll
\Lambda\) and its complement, and check that the contribution from large
\(\sigma\pm\I\frac{B}{2}\) becomes negligible as \(\Lambda\to\infty\).
It is more convenient to perform such steps on the Higgs branch
decomposition~\eqref{SQCD-Zs} of \(Z_\enriched\) near~\(0\).  Regardless
of the value of~\(z\), the series expansions of vortex partition
functions converges for \(\Lambda\)~large enough that \(\abs{z_\bare} =
\abs{z}/\Lambda^{\Nf-\anti{\Nf}} < 1\).  Then each term in the series
for the enriched theory converges to the appropriate term for the
\(\anti{\Nf}<\Nf\) theory.  Since the sum of terms with \(\sum_{j=1}^\Nc
k_j > K\) decreases exponentially with~\(K\) in both series,
\(Z_{\vortex,\enriched}(z_\bare) \to Z_\vortex(z)\).  Other factors work
out as for the Coulomb branch representation.

In the limit above, \(\Nf(\Nf-\anti{\Nf})\) of the \(\Nf^2\) free
hypermultiplets on~\(S^4_b\) become infinitely massive, and the
corresponding factors must be removed from the enriched partition
function to retain a finite result.  The partition function of the
surface operator with \(\anti{\Nf}<\Nf\) in a theory of
\(\Nf\anti{\Nf}\) free hypermultiplets of masses~\eqref{part-mst} is
thus the limit
\begin{equation}\label{irreg-ZZ-lim}
  Z_{S^2\subset S^4_b}^{U(\Nc)} (z, \bar{z})
  = \lim_{\Lambda\to\infty} \Biggl[
  \biggl(
  \frac{\prod_{s=1}^{\Nf}\Upsilon\bigl(
    \frac{1}{b}(1+\I m_s+\I\Lambda)\bigr)}
    {\gamma(-\I\Lambda)^\Nc}
  \biggr)^{\Nf-\anti{\Nf}}
  Z_{S^2\subset S^4_b,\enriched}^{U(\Nc)}
  \bigl(z_\bare, \bar{z}_\bare \bigr)
  \Biggr]
  \,.
\end{equation}

We now provide a Toda CFT interpretation of the limit for
\(\Nf-\anti{\Nf}=1\).  For simplicity, label antifundamental multiplets
of the enriched theory starting with the spectator multiplet, so that
\(\anti{m}_1 = \Lambda \to \infty\).  Replace the partition function of
the enriched defect in~\eqref{irreg-ZZ-lim} by its corresponding Toda
CFT four-point function through the matching~\eqref{SQCD-matching}.
After a conformal transformation which maps
\((\infty,1,x/(-\I\Lambda),0)\) to \((0,x/(-\I\Lambda),1,\infty)\),
\begin{align}
  \nonumber
  & Z_{S^2\subset S^4_b}^{U(\Nc)} (z, \bar{z})
  = \lim_{\Lambda\to\infty} \Biggl[
  A_\enriched
  \abs*{\frac{x}{\Lambda}}^{
    2\gamma_{0,\enriched}
    -2\dimToda(\alpha_0)
    -2\dimToda(-b\omega_\Nc)
    +2\dimToda(\alpha_\infty)
    +2\dimToda(\hat{m})}
  \abs*{1-\frac{x}{-\I\Lambda}}^{2\gamma_{1,\enriched}}
  \\\label{irreg-ZZ-VVVV}
  & \quad \cdot
  \frac{\prod_{s=1}^{\Nf}
    \Upsilon\bigl(\tfrac{1}{b}(1+\I m_s+\I\Lambda)\bigr)}
    {\gamma(-\I\Lambda)^\Nc}
  \vev*{
    \widehat{V}_{\alpha_0}(\infty)
    \widehat{V}_{-b\omega_\Nc}(1)
    \widehat{V}_{\hat{m}}\biggl(\frac{x}{-\I\Lambda},
      \frac{\bar{x}}{\I\Lambda}\biggr)
    \widehat{V}_{\alpha_\infty}(0)
  }
  \Biggr]
\end{align}
with \(x = (-1)^{\Nf+\Nc-1} z\), and parameters \(\alpha_0\),
\(\hat{m}=(\varkappa+\Nc b)h_1\), \(\alpha_\infty\), \(A_\enriched\),
\(\gamma_{0,\enriched}\) and~\(\gamma_{1,\enriched}\) given
below~\eqref{SQCD-matching}.  As \(\Lambda\to\infty\),
\(\gamma_{1,\enriched} \sim \frac{\Nc}{\Nf} \I\Lambda\), thus
\(\abs*{1-x/(-\I\Lambda)}^{2\gamma_{1,\enriched}} \to
e^{(\Nc/\Nf)(x+\bar{x})}\).

In the same limit, the punctures \(\widehat{V}_{\hat{m}}\)
and~\(\widehat{V}_{\alpha_\infty}\) collide, with momenta growing as the
inverse of the distance, keeping a constant sum \(c_0+\Nc bh_1 =
(\varkappa+\Nc b)h_1 + \alpha_\infty\) given in~\eqref{irreg-alpha}.  We
study such collision limits in Appendix~\ref{app:irregular} and
define~\eqref{irregular-normalized-V}
\begin{equation}\label{irreg-V}
  \begin{aligned}
    & \widehat{\mathbb{V}}_{c_0+\Nc bh_1;-(x/b)h_1,(\bar{x}/b)h_1}(0)
    = \lim_{\Lambda\to\infty} \biggl[
    \Upsilon\bigl(\varkappa+\Nc b+\vev{Q-c_0-\Nc bh_1,h_1}\bigr)^{\Nf}
    \\
    & \quad \cdot
      \left[\frac{\Lambda}{b}\right]^{\vev{Q,Q}-2\dimToda(c_0+\Nc bh_1)}
      \abs*{\frac{x}{\Lambda}}^{2\vev{(\varkappa+\Nc b)h_1, c_0 - \varkappa h_1}}
      \widehat{V}_{(\varkappa+\Nc b)h_1}
      \biggl(\frac{x}{-\I\Lambda}, \frac{\bar{x}}{\I\Lambda}\biggr)
      \widehat{V}_{c_0-\varkappa h_1}(0)
    \biggr]_{\varkappa \sim \I\Lambda/b} \,.
  \end{aligned}
\end{equation}
The Upsilon functions and gamma functions in \eqref{irreg-ZZ-VVVV}
and~\eqref{irreg-V} can be recast in the same form through the
asymptotics \eqref{Upsilon-asymptotic-prod},
\eqref{Upsilon-asymptotic-ratio}, and
\(\gamma(1+\I\Lambda+a)\sim\gamma(1+\I\Lambda)\Lambda^{2a}\).  Let
\(\I\underline{m}=\frac{1}{\Nf}\sum_{s=1}^{\Nf}\I m_s\).  Then,
\begin{equation}\label{irreg-Upsilon}
  \begin{aligned}
    & \gamma(1+\I\Lambda)^\Nc
    \prod_{s=1}^{\Nf}\Upsilon\bigl(\tfrac{1}{b}(1+\I\Lambda+\I m_s)\bigr)
    =
    \gamma(1+\I\Lambda)^\Nc
    \prod_{s=1}^{\Nf}
    \Upsilon\bigl(\tfrac{1}{b}(1+\I\Lambda+\I\underline{m})
      +\vev{Q-\alpha_0,h_s}\bigr)
    \\
    & \quad \sim
    \gamma(1+\I\Lambda)^\Nc
    \Upsilon\bigl(\tfrac{1}{b}(1+\I\Lambda+\I\underline{m})\bigr)^{\Nf}
    [\Lambda/b]^{\vev{Q,Q}-2\dimToda(\alpha_0)}
    \\
    & \quad \sim
    \Upsilon\bigl(\tfrac{1}{b}(1+\I\Lambda+\I\underline{m})
    +b\Nc/\Nf\bigr)^{\Nf}
    b^{\Nc+2\Nc\I\Lambda}
    [\Lambda/b]^{\vev{Q,Q}-2\dimToda(\alpha_0)
      -2\Nc\I\underline{m}+\Nc(\Nf-\Nc)b^2/\Nf} \,.
  \end{aligned}
\end{equation}
The last Upsilon functions are precisely those appearing
in~\eqref{irreg-V}.  Plugging back into~\eqref{irreg-ZZ-VVVV}, all
powers of \(\Lambda\) and \(b^\Lambda\) cancel, and we can drop the
limit.

All in all, the partition function of a surface operator describing a
\(U(\Nc)\) vector multiplet with \(\Nf\)~fundamental and
\(\anti{\Nf}=\Nf-1\) antifundamental chiral multiplets, coupled to
\(\Nf(\Nf-1)\) hypermultiplets on~\(S^4_b\) is equal to a Toda CFT
correlator with an antisymmetric degenerate insertion and a rank~\(1\)
irregular puncture:\footnote{Following the argument
  below~\eqref{part-typical-matching}, the factor \(A
  \abs{x}^{2\gamma_0} e^{(\Nc/\Nf)(x+\bar{x})}\) can be absorbed in the
  \(S^2\)~partition function.  We keep the factor explicitly to compare
  gauge theory and Toda CFT results.}%
\begin{equation}\label{irreg-matching}
  Z_{S^2\subset S^4_b}^{U(\Nc),\Nf,\Nf-1}(z,\bar{z})
  =
  A \abs{x}^{2\gamma_0} e^{\frac{\Nc}{\Nf}(x+\bar{x})}
  \vev*{\widehat{V}_{\alpha_0}(\infty) \widehat{V}_{-b\omega_\Nc}(1)
    \widehat{\mathbb{V}}_{c_0+\Nc bh_1;c_1,\bar{c}_1}(0)} \,.
\end{equation}
As before, \(x=(-1)^{\Nf+\Nc-1} z\).  The irregular
puncture~\(\widehat{\mathbb{V}}\) is defined above and studied in
Appendix~\ref{app:irregular}.  The momenta \(c_0\), \(c_1\),
\(\bar{c}_1\), and~\(\alpha_0\) are
\begin{equation}
  \label{irreg-alpha}
  \begin{aligned}
    c_0 & = Q + \frac{1}{b}\sum_{s=1}^{\Nf}(1+\I m_s)h_1
    + \frac{1}{b} \sum_{s=2}^{\Nf} \I\anti{m}_s (h_1-h_s) \,,
    \\
    c_1 & = - \frac{x}{b} h_1 \,,
    \qquad
    \bar{c}_1 = \frac{\bar{x}}{b} h_1 \,,
    \qquad
    \alpha_0 = Q - \frac{1}{b} \sum_{s=1}^{\Nf} \I m_s h_s \,,
  \end{aligned}
\end{equation}
and the constant~\(A\) and exponent~\(\gamma_0\) are\footnote{Mapping
  \(\{0,1,\infty\}\) to \(\{\infty,x,0\}\) gives a closer analogue of
  the \(\anti{\Nf}=\Nf\) matching.  This replaces \(\gamma_0\) by the
  simpler \(\gamma_0 - \dimToda(c_0+\Nc bh_1) + \dimToda(\alpha_0) +
  \dimToda(-b\omega_\Nc) = - \frac{\Nc}{\Nf} \sum_{s=1}^{\Nf} \I m_s -
  \frac{\Nc(\Nf-\Nc)}{2}(b^2+1)\).  However, the transformation
  properties~\eqref{irregular-conformal-transfo} of rank~\(1\) irregular
  punctures would make the parameters \(c_1\) and~\(\bar{c}_1\)
  infinite.  The best convention to cancel this infinity is not clear.}
\begin{align}
  A & = b^{\Nc(\Nf-1)(b^2+1)+2\dimToda(\alpha_0)-2\dimToda(c_0)} \,,
  \\
  \gamma_0 & = \dimToda(c_0) - \dimToda(\alpha_0)
  - \Nc \sum_{s=1}^{\Nf} \I m_s - \Nc \sum_{s=2}^{\Nf} (1 + \I\anti{m}_s)
  - \frac{\Nc(\Nc-1)}{2} b^2
  \,.
\end{align}

As we have seen, it is natural from the gauge theory point of view to
decouple further antifundamental chiral multiplets by making them
massive.  Specifically, from~\eqref{irreg-ZZ-lim} we know that the
partition function of a surface operator described by a \(U(\Nc)\)
vector multiplet coupled to \(\Nf\)~fundamental and
\(\anti{\Nf}=\Nf-k\leq\Nf-2\) antifundamental chiral multiplets is a
limit of \(Z_{S^2\subset S^4_b}^{U(\Nc),\Nf,\Nf-1}
\bigl(z/(-\I\Lambda)^{k-1},\bar{z}/(\I\Lambda)^{k-1}\bigr)\) with
twisted masses \(\anti{m}_2=\cdots=\anti{m}_k=\Lambda\), multiplied by
some factor.  On the Toda CFT side of the
matching~\eqref{irreg-matching}, the limit amounts to taking
\(\vev{c_0,h_s}\sim\I\Lambda/b\) for \(2\leq s\leq k\) and letting
\(c_1\) and~\(\bar{c}_1\) decrease as~\(\Lambda^{-(k-1)}\).  Such a
limit does not fit in the framework described in
Appendix~\ref{app:irregular}, since the parameter~\(c_0\) blows up.
However, translating the gauge theory factors to the Toda CFT and
setting \(\Nc=0\) for simplicity, we find that the two-point function of
a generic vertex operator~\(\widehat{V}_{\alpha_0}\) with
\begin{equation}\label{irreg-Vlim}
  \biggl[
  \abs{\nu}^{2\dimToda(c_0)-\vev{Q,Q}}
  \prod_{t=2}^k \Upsilon\bigl(\vev{Q-c_0,h_t}\bigr)^{\Nf}
  \widehat{\mathbb{V}}_{c_0;-\nu h_1, \bar{\nu}h_1}
  \biggr]_{\begin{subarray}{l}\nu=x/[b(-\I\Lambda)^{k-1}]\\
      c_0\sim\frac{\I\Lambda}{b}(kh_1-\omega_k)\end{subarray}}
\end{equation}
remains finite as \(\Lambda\to\infty\).  This suggests that the
operator~\eqref{irreg-Vlim} itself has a limit.  Additionally, the
OPE~\eqref{irregular-T-OPE} of the stress-energy tensor with a
rank~\(1\) puncture includes a term
\(\dimToda(c_0)+\vev{c_1,\partial_{c_1}}\), and the normalization factor
\(\abs{\nu}^{2\dimToda(c_0)}\) ensures that the singular term
\(\dimToda(c_0)\) is absorbed in \(\vev{c_1,\partial_{c_1}}\).
Unfortunately, it is difficult to go further, as the OPE with higher
currents of the \(W_{\Nf}\)~algebra contain many singular terms, and
all must be carefully canceled by the choice of normalization before
taking the limit.

Having dissected the partition function of theories with fundamental and
antifundamental matter, we consider next theories with an adjoint chiral
multiplet.

\subsection{SQCDA and Toda Symmetric Degenerate}
\label{sec:SQCDA}

We focus in this section on \(\Nsusy=(2,2)\) SQCDA: a \(U(\Nc)\) vector
multiplet coupled to an adjoint chiral multiplet~\(X\) and
\(\Nf\)~fundamental and \(\Nf\)~antifundamental chiral multiplets.
Twisted masses (plus \(R\)-charges) are \(m_X\), \(m_s\),
and~\(\anti{m}_s\).  This theory, coupled to \(\Nf^2\)~hypermultiplets
with masses given by~\eqref{part-mst}, defines a surface operator.  We
equate the \(S^2\subset S^4_b\) partition function of the 4d/2d system
to a Toda CFT correlator with a degenerate field \(\widehat{V}_{-\Nc
  bh_1}\) labeled by the \(\Nc\)-th symmetric representation
of~\(A_{\Nf-1}\).  Namely, we check that\footnote{As explained
  below~\eqref{part-typical-matching}, the factor \(A
  \abs{x}^{2\gamma_0} \abs{1-x}^{2\gamma_1}\) can be absorbed into the
  partition function.  To compare gauge theory and Toda CFT results it
  is best to keep the factor explicitly.}
\begin{equation}\label{SQCDA-matching}
  Z_{S^2 \subset S^4_b}^{U(\Nc)\ \SQCDA}(m,\anti{m},m_X,z,\bar{z})
  = A \abs{y}^{2\gamma_0} \abs{1-y}^{2\gamma_1}
  \vev*{\widehat{V}_{\alpha_\infty}(\infty)\widehat{V}_{\hat{m}}(1)
    \widehat{V}_{-\Nc bh_1}(y,\bar{y})\widehat{V}_{\alpha_0}(0)}
\end{equation}
with \(y = (-1)^{\Nf} z\) and\footnote{The full flavour group of
  SQCDA is \(U(1) \times S[U(\Nf)\times U(\Nf)]\), where the factors act
  on the adjoint, fundamental, and antifundamental chiral multiplets.
  The relation \(b^2=\I m_X\) identifies the first \(U(1)\) flavour
  symmetry with rotations transverse to the surface operator.}
\(b^2 = \I m_X\), momenta
\begin{equation}
  \label{SQCDA-alpha}
  \begin{aligned}
    \alpha_0 & = Q - \frac{1}{b} \sum_{s=1}^{\Nf} \I m_s h_s \,,
    & \qquad\qquad
    \hat{m} & = (\varkappa+\Nc b) h_1 \,,
    \\
    \alpha_\infty & = Q - \frac{1}{b} \sum_{s=1}^{\Nf} \I\anti{m}_s h_s \,,
    & \qquad\qquad
    \varkappa & = \frac{1}{b} \sum_{s=1}^{\Nf} (1+\I m_s+\I\anti{m}_s) \,,
  \end{aligned}
\end{equation}
and coefficients
\begin{align}
  \label{SQCDA-gamma0}
  \gamma_0
  & = - \frac{\Nc}{\Nf} \sum_{s=1}^{\Nf} \I m_s
  - \frac{\Nc(\Nf-1)}{2} (b^2+1)
  - \frac{\Nc(\Nc-1)}{2} b^2 \,,
  \\
  \label{SQCDA-gamma1}
  \gamma_1 & = - \frac{\Nc(\Nf-\Nc)}{\Nf} b^2
  + \frac{\Nc}{\Nf} \sum_{s=1}^{\Nf} (\I m_s + \I\anti{m}_s) \,,
  \\
  \label{SQCDA-A}
  A & = b^{\Nc\Nf(1+b^2)-\Nc^2 b^2-2\Nc b\varkappa}
  \prod_{\nu=1}^\Nc \gamma(-\nu b^2) \,.
\end{align}
We recognize the same symmetries as for SQED and SQCD, under
permutations of the \(m_s\) or the~\(\anti{m}_s\), and under \(z\leftrightarrow
\frac{1}{z}\) and exchanging those two sets of masses.  Setting \(\Nc=1\)
reproduces the matching~\eqref{SQED-matching} of SQED, but \(A\)~has an
additional factor of~\(\gamma(-b^2)=\gamma(-\I m_X)\): this is the
one-loop determinant of the adjoint chiral multiplet, which decouples in
an abelian theory.

Given the geometrical origin of the deformation parameter, one has
\(b^2>0\).  On the other hand, the \(S^2\)~partition function is defined
with positive \(R\)-charges \(\Re(-2\I m)\).  The two requirements are
incompatible with \(b^2=\I m_X\), hence one of those two parameters must
be continued beyond its usual range.  For now, we analytically continue
the \(R\)-charge: it is easier because the partition function depends
holomorphically on~\(\I m_X\), as deduced from explicit expressions.
However, we will encounter in Section~\ref{sec:SQCDAW} a setting where
\(b^2=\I m_X\) is fixed to a real negative value.  Given that the
Upsilon function which appears in~\(Z_{S^4_b}^\free\) and in Toda
correlators cannot be continued to negative~\(b^2\), we will have to
first recast the relation~\eqref{SQCDA-matching} in the form \(Z_{S^2} =
\vev{\cdots} / Z_{S^4_b}^\free\) for the analytic continuation in~\(b\)
to make sense.

Once more, we fix the dictionary and demonstrate the equality by
comparing exponents in the s-, t- and u-channels.  The equality of Toda
CFT three-point functions and gauge theory one-loop determinants (for
the s-~and u-channels) is checked in Appendix~\ref{app:1loop-3pt}, and
the expression of~\(A\) is found there.

The Coulomb branch representation reads
\begin{equation}\label{SQCDA-Z-Coulomb}
  \begin{aligned}
    & Z^\SQCDA =
    \frac{1}{\Nc!}
    \sum_{B\in\bbZ^\Nc} \int_{\bbR^\Nc}
    \frac{\dd[^\Nc]{\sigma}}{(2\pi)^\Nc} \Biggl\{
    \bigl[(-1)^{\Nc-1} z\bigr]^{\Tr(\I\sigma+\frac{B}{2})}
    \bigl[(-1)^{\Nc-1} \bar{z}\bigr]^{\Tr(\I\sigma-\frac{B}{2})}
    \\
    & \quad \cdot
    \prod_{i<j} \biggl[(\sigma_i-\sigma_j)^2+\frac{(B_i-B_j)^2}{4}\biggr]
    \prod_{j=1}^\Nc \prod_{s=1}^{\Nf} \biggl[
      \frac{\Gamma(-\I m_s - \I\sigma_j - \frac{B_j}{2})}
        {\Gamma(1+\I m_s + \I\sigma_j - \frac{B_j}{2})}
      \frac{\Gamma(-\I\anti{m}_s + \I\sigma_j + \frac{B_j}{2})}
        {\Gamma(1+\I\anti{m}_s - \I\sigma_j + \frac{B_j}{2})}
    \biggr]
    \\
    & \quad \cdot
    \prod_{i=1}^\Nc \prod_{j=1}^\Nc \biggl[
    \frac{\Gamma(-\I m_X -\I\sigma_i+\I\sigma_j -\frac{B_i-B_j}{2})}
      {\Gamma(1+\I m_X +\I\sigma_i-\I\sigma_j - \frac{B_i-B_j}{2})}
    \biggr]
    \Biggr\} \,.
  \end{aligned}
\end{equation}
We will expand this partition function around the points \(0\),
\((-1)^{\Nf}\) and~\(\infty\), where, as a function of~\(z\), it
has branch points.  This follows the path we traced for SQED: the
behaviors near \(z=0\) and~\(\infty\) are probed by closing integration
contours towards \(\pm\I\infty\).  The partition function is then
expressed as a sum over poles of the integrand, which are characterized
up to integers by the set of Gamma functions which are singular for
those values of~\(\I\sigma\).  The behavior near \((-1)^{\Nf}\)
is found by splitting the Coulomb branch integral depending on whether
each \(\abs{\sigma_j\pm\frac{\I B_j}{2}}\lessgtr \ln\abs{z}\).

\subsubsection{Expanding the SQCDA Partition Function}

We start with the s-channel expansion for \(\abs{z} < 1\).  Ignoring for
a moment the magnetic flux~\(B\), and integer shifts due to the infinite
set of poles of the Gamma function, we find that poles enclosed by the
contour must be such that each \(\I\sigma_j\) is either \(- \I m_s\) for
some flavour~\(s\), or \(\I\sigma_i - \I m_X\) for some other
color~\(i\).  As in the case of SQCD, the vector multiplet one-loop
determinant enforces \(\I\sigma_i \neq \I\sigma_j\) for any two distinct
colors, hence \(\{\I\sigma_j\}\) is \(\{-\I m_s-\mu\I m_X \mid 1\leq
s\leq\Nf , 0\leq\mu<n_s\}\) for some choice of integers~\(n_s\) with
\(n_1+\cdots+n_{\Nf} = \Nc\).  It is convenient to label colors with
indices \((s,\mu)\) instead of \(j\in\intset{1}{\Nc}\), and denote
\(I=\{(s,\mu)\}\).  The sums over~\(B\) and over poles of Gamma
functions introduce shifts, in the form of sums over \(2\Nc\) integers
\(k^\pm_{s\mu} \geq 0\), and poles are%
\begin{equation}\label{SQCDA-poles}
  \I\sigma_{s\mu} \pm \frac{B_{s\mu}}{2}
  = - \I m_s - \mu \I m_X + k^\pm_{s\mu}
\end{equation}
for \((s,\mu)\in I\).  The partition function can then be recast as a
sum over residues at those values of \(\I\sigma \pm \frac{B}{2}\).  It
turns out that the residues vanish unless \(k^\pm_{s\mu} \leq
k^\pm_{s(\mu+1)}\) for every \((s,\mu)\in I\) and sign~\(\pm\): this
indicates that~\eqref{SQCDA-poles} also labels some points which are not
poles; thankfully, the residue formula is robust against such
overcounting.

Since every factor in the Coulomb branch formula depends only on \(\I
\sigma + \frac{B}{2}\) hence on~\(k^+\), or on \(\I \sigma -
\frac{B}{2}\) hence on~\(k^-\), the series over~\(k^+\) and
over~\(k^-\) decouple, and~\(Z^\SQCDA\) splits into a sum of factorized
terms labeled by the choice of~\(\{n_s\}\),
\begin{align}\label{SQCDA-Zs}
  Z_{S^2}^\SQCDA
  & = \mspace{-18mu} \sum_{n_1+\cdots+n_{\Nf}=\Nc} \mspace{-18mu}
  (z\bar{z})^{\sum_{(s,\mu)\in I} (- \I m_s - \mu \I m_X)}
  Z_{\oneloop,\{n\}} Z_{\vortex,\{n\}}\bigl[(-1)^{\Nf}z\bigr] Z_{\vortex,\{n\}}\bigl[(-1)^{\Nf}\bar{z}\bigr]
  \\
  \label{SQCDA-Z1l}
  Z_{\oneloop,\{n\}}
  & = \prod_{(s,\mu)\in I} \prod_{t=1}^{\Nf}
  \frac{\gamma(-\I m_t - n_t \I m_X + \I m_s + \mu \I m_X)}
    {\gamma(1+\I\anti{m}_t+\I m_s + \mu \I m_X)} \,,
  \\
  \label{SQCDA-Zv}
  Z_{\vortex,\{n\}}(y)
  & =
  \begin{aligned}[t]
    & \sum_{k:I\to\bbZ_{\geq 0}}
    \prod_{(s,\mu) \in I} \biggl[
    y^{k_{s\mu}}
    \prod_{t=1}^{\Nf}
    \frac
    {(-\I\anti{m}_t - \I m_s - \mu \I m_X)_{k_{s\mu}}}
    {(1+\I m_t - \I m_s + (n_t - \mu) \I m_X)_{k_{s\mu}}}
    \\
    & \cdot
    \frac
    {\prod_{t=1}^{\Nf} (1 + \I m_t - \I m_s + (n_t - \mu) \I m_X
      + k_{s\mu} - k_{t(n_t - 1)})_{k_{t(n_t - 1)}}}
    {\prod_{(t,\nu)\in I}
      (1 + \I m_t - \I m_s + (\nu - \mu) \I m_X
      + k_{s\mu} - k_{t\nu})_{k_{t\nu} - k_{t(\nu - 1)}}}
    \biggr]
  \end{aligned}
\end{align}
where we define \(k_{t,-1}=0\) for convenience.  Carrying through the
same procedure for \(\abs{z}>1\) yields a u-channel decomposition
similar to the s-channel decomposition~\eqref{SQCDA-Zs}, with
\(m_s\leftrightarrow\anti{m}_s\), \(z\to z^{-1}\) and
\(\bar{z}\to\bar{z}^{-1}\).

Having found powers of \(\abs{z}\) in the s-channel and u-channel
decompositions of~\(Z^\SQCDA\), we now expand the Coulomb branch
integral in the t-channel.  The first step is to use the identity
\(\frac{\Gamma(-\I a - B / 2)}{\Gamma(1 + \I a - B / 2)} = (-1)^B
\frac{\Gamma(-\I a + B / 2)}{\Gamma(1 + \I a + B / 2)}\) on the one-loop
determinants of fundamental chiral multiplets, and on half of the Gamma
functions stemming from the adjoint chiral multiplet, and absorb the
resulting signs into
\begin{equation}
  y = (-1)^{\Nf} z \qquad \text{and} \qquad
  \bar{y} = (-1)^{\Nf} \bar{z} \,.
\end{equation}
The integrand resulting from this operation can be recast as
\begin{equation}
  \begin{aligned}
    &
    y^{\Tr(\I\sigma+\frac{B}{2})} \bar{y}^{\Tr(\I\sigma-\frac{B}{2})}
    \prod_{j=1}^\Nc \prod_{s=1}^{\Nf} \biggl[
      \frac{\Gamma(-\I m_s - \I\sigma_j + \frac{B_j}{2})}
        {\Gamma(1+\I\anti{m}_s - \I\sigma_j + \frac{B_j}{2})}
      \frac{\Gamma(-\I\anti{m}_s + \I\sigma_j + \frac{B_j}{2})}
        {\Gamma(1+\I m_s + \I\sigma_j + \frac{B_j}{2})}
    \biggr]
    \\
    & \quad \cdot
    \gamma(-\I m_X)^\Nc
    \prod_{\pm}
    \prod_{i<j}^\Nc \biggl[
    \frac{\bigl(\pm(\I\sigma_i-\I\sigma_j)+\frac{B_i-B_j}{2}\bigr)
      \Gamma(-\I m_X\pm(\I\sigma_i-\I\sigma_j)+\frac{B_i-B_j}{2})}
      {\Gamma(1 + \I m_X\pm(\I\sigma_i-\I\sigma_j)+\frac{B_i-B_j}{2})}
    \biggr]
  \end{aligned}
\end{equation}
by writing the vector multiplet one-loop determinant as a product of
\(\pm(\I\sigma_i-\I\sigma_j)+\frac{B_i-B_j}{2}\).  We now split the sums
and integrals in the same way as for SQED on
page~\pageref{page:SQED-x-to-1}, one pair \((\sigma_j,B_j)\) at a time.
For \(\abs{\I\sigma_j+\frac{B_j}{2}} < \abs{\ln y}^{-1}\), we expand the
classical contribution \(y^{\I\sigma_j+\frac{B_j}{2}}
\bar{y}^{\I\sigma_j-\frac{B_j}{2}}\) as a series in \(\ln y\) and
\(\ln\bar{y}\); the integral and sum only contributes a constant factor.
For \(\abs{\I\sigma_j+\frac{B_j}{2}} > \abs{\ln y}^{-1}\), the sum
over~\(B_j\) is well approximated by an integral, and we expand the
Gamma functions which involve this particular combination as a power of
\(\abs{\I\sigma_j+\frac{B_j}{2}}\) times a power series in
\((\I\sigma_j\pm\frac{B_j}{2})^{-1}\).  Rescaling
\(\I\sigma_j+\frac{B_j}{2}\) by \(\ln y\) makes the classical
contribution independent of~\(y\), and extracting a power of \(\abs{\ln
  y}\) leaves a series in \(\ln y\) and \(\ln\bar{y}\) as the sole
dependence in~\(y\).  After performing this procedure for all pairs
\((\sigma_j, B_j)\), we obtain \(2^\Nc\) contributions, labeled by the
set \(K\subseteq\{1,\ldots,\Nc\}\) of colors~\(j\) such that
\(\abs{\I\sigma_j+\frac{B_j}{2}} > \abs{\ln y}^{-1}\) is large.  The
contribution for a given set~\(K\) behaves as
\begin{equation}\label{SQCDA-t-powers}
  Z_K \sim
  \abs{1-y}^{-2k+2k\sum_{s=1}^{\Nf}(1+\I m_s+\I\anti{m}_s)+2k[2\Nc-k-1]\I m_X}
  \,,
\end{equation}
multiplied by a constant and by a series in powers of \(1-y\) and
\(1-\bar{y}\), where \(k = \#K\) is the number of elements in~\(K\) and
we used \((\ln y)^\alpha = (1-y)^\alpha \cdot (\text{series})\).  There
are \(\Nc+1\) distinct exponents, corresponding to values
\(k\in\intset{0}{\Nc}\).  This approach does not seem amenable to
finding multiplicities attached to each power of \(1-y\), hence we will
not be able to probe that aspect of the correspondence.

\subsubsection{Matching Parameters for SQCDA}

We are now ready to match the gauge theory data to Toda CFT data.  The
s-~and u-channel decompositions of~\(Z^\SQCDA\) have
\begin{equation}
  \binom{\Nf+\Nc-1}{\Nc}
  = \dim\bigl(\repr(\Nc h_1)\bigr)
\end{equation}
terms, which is the dimension of the \(\Nc\)-th symmetric representation
\(\repr(\Nc h_1)\) of \(A_{\Nf-1}\), with highest weight \(\Nc
h_1\).  Thus, in analogy with SQCD, we expect \(Z^\SQCDA\) to match a
Toda four-point correlation function involving the degenerate operator
\(\widehat{V}_{-\Nc bh_1}\).  The fusion rule then allows shifts of
generic momenta by \(-bh = -b\sum_{s=1}^{\Nf} n_s h_s\) for a choice of
integers \(n_1+\cdots+n_{\Nf} = \Nc\).  We thus wish to match the
s-channel exponents
\begin{equation}
  \dimToda(\alpha_0-bh)
  - \dimToda(\alpha_0) - \dimToda(-\Nc b h_1) + \gamma_0
  = - \sum_{s=1}^{\Nf}
  \Bigl[ n_s \I m_s + \frac{n_s(n_s-1)}{2} \I m_X \Bigr] \,.
\end{equation}
This equality holds if \(\I m_X = b^2\), and \(\alpha_0\)
and~\(\gamma_0\) are as given in \eqref{SQCDA-alpha}
and~\eqref{SQCDA-gamma0}.  The u-channel powers are similar,
\begin{equation}
  \dimToda(\alpha_\infty)
  - \dimToda(\alpha_\infty-bh)
  - \dimToda(-\Nc bh_1)
  + \gamma_0 + \gamma_1
  = \sum_{s=1}^{\Nf}
  \Bigl[ n_s \I\anti{m}_s + \frac{n_s(n_s-1)}{2} \I m_X \Bigr] \,,
\end{equation}
and the equality holds for values of \(\alpha_\infty\) and~\(\gamma_1\)
in \eqref{SQCDA-alpha} and~\eqref{SQCDA-gamma1}.

We find in Appendix~\ref{app:fusion} that the fusion of \((\varkappa+\Nc
b)h_1\) with \(-\Nc b h_1\) allows the t-channel internal momenta
\((\varkappa+nb)h_1 - nbh_2\) for \(0\leq n\leq\Nc\).  This fusion
rule~\eqref{fusion-kap-sym} provides the powers of \(1-y\) for the
t-channel of the Toda correlator, and power matching then requires
\begin{equation}\label{SQCDA-match-1}
  \begin{aligned}
    & \dimToda((\varkappa+nb)h_1 - nbh_2)
    - \dimToda((\varkappa+\Nc b)h_1) - \dimToda(-\Nc bh_1)
    + \gamma_1
    \\
    & \quad = k \Bigl[ \sum_{s=1}^{\Nf}(\I m_s + \I\anti{m}_s)
    + (\Nf - 1)
    + (2 \Nc - k - 1) \I m_X \Bigr]
    \,.
  \end{aligned}
\end{equation}
The exponents are equal if \(n = \Nc - k\), and \(\varkappa\)~is as
given in~\eqref{SQCDA-alpha}.

Finally, as checked in Appendix~\ref{app:1loop-3pt}, the Toda CFT
three-point functions which appear in the s-channel decomposition of the
correlator reproduce the corresponding one-loop determinants
in~\eqref{SQCDA-Zs}, provided \(A\)~is as given in~\eqref{SQCDA-A}.  For
any given~\(\Nc\), the techniques of Appendix~\ref{app:braiding} can
yield the Toda CFT braiding matrix of \(\widehat{V}_{-\Nc bh_1}\)
with~\(\widehat{V}_{\hat{m}}\).  However, we did not find a closed form
of those matrices or their gauge theory analogues to provide a proof of
the matching~\eqref{SQCDA-matching}.

\subsubsection{Adding a Superpotential to SQCDA}
\label{sec:SQCDAW}

We now discuss the effect of adding to SQCDA a superpotential term of
the form \(W=\sum_{t=1}^{\Nf} \anti{\quark}_t X^{l_t} \quark_t\) or \(W =
\Tr X^{l+1}\), where \(\quark_t\), \(\anti{\quark}_t\), and~\(X\) denote
the fundamental, antifundamental, and adjoint chiral multiplets, and
\(l_t\) and~\(l\) are non-negative integers.

The deformation term which localizes to the Higgs branch of the theory
with no superpotential can still be used in the presence of a
superpotential, and it yields the same decomposition into vortex and
anti-vortex partition functions.  Hence, the only effect of the
superpotential on the partition function is to constrain the
(complexified) twisted masses of chiral multiplets.  On the other hand,
the superpotential term is in fact \(\Q\)-exact for the choice of
localization supercharge~\(\Q\), thus one can include it into the
deformation term.  This lifts some vacua of the deformation term through
\(\Faux\)-term constraints, thus removes some terms from the sum over
Higgs branch vacua.  The two deformation terms must yield equal results
for the partition function.  Therefore, the terms forbidden by
\(\Faux\)-term constraints must vanish in the larger sum: they must have
zero one-loop determinant.  As a result, we can either solve
\(\Daux\)-term and \(\Faux\)-term equations to find vacua of the
enhanced deformation term, or remove vacua of the original deformation
term whose one-loop determinant vanishes when imposing the
superpotential constraint on \(R\)-charges.

First, we focus on a generalization of the superpotential
\(\anti{\quark} X \quark\) of \(\Nsusy=(2,2)^*\)
SQCD,\footnote{\(\Nsusy=(2,2)^*\) SQCD is the mass deformation of
  \(\Nsusy=(4,4)\) SQCD\@.}
\begin{equation}
  W = \sum_{t=1}^{\Nf} \anti{\quark}_t X^{l_t} \quark_t \,,
\end{equation}
where \(l_t\geq 0\) is an integer for each flavour \(1\leq t\leq\Nf\).
We let \(L = \sum_{t=1}^{\Nf} l_t\).  The superpotential must have a total
\(R\)-charge of~\(2\) and a vanishing twisted mass, hence
\(\I\anti{m}_t+l_t\I m_X+\I m_t=-1\) for each \(1\leq t\leq\Nf\).  The
one-loop determinant~\eqref{SQCDA-Z1l} then contains a vanishing factor
\(1/\gamma(1+\I\anti{m}_t+\I m_t+l_t\I m_X)=0\) whenever any
\(n_t>l_t\), thus those terms do not contribute to the partition
function.  An equivalent point of view is that the corresponding Higgs
branch vacua have \(X^{n_t-1} \quark_t \neq 0\) and are forbidden
by the \(\Faux\)-term equation \(X^{l_t} \quark_t = 0\).  Terms in the
Higgs branch representation of the partition function are thus labeled
by integers \(0\leq n_t\leq l_t\) with \(\sum_{t=1}^{\Nf} n_t = \Nc\).
Note that \(n_t\leq l_t\) implies \(\Nc\leq L\), analogous to the
condition \(\Nc\leq\Nf\) for SQCD\@.

The constraint on (complexified) twisted masses translates to a
constraint on the momenta of operators appearing in the corresponding
Toda CFT correlator.  The semi-degenerate operator becomes degenerate:
\begin{equation}\label{SQCDWA-simple-degenerates}
  \hat{m}
  = \biggl[ \frac{1}{b} \sum_{t=1}^{\Nf} (1+\I m_t+\I\anti{m}_t) + \Nc b
  \biggr] h_1
  = - (L - \Nc) b h_1 \,,
\end{equation}
where we used \(\I m_X = b^2\).  Thus, the outgoing momentum \(2Q -
\alpha_\infty\) must take the form \(\alpha_0 - bh - bh'\), where \(h =
\sum_t n_t h_t\) is a weight of \(\repr(\Nc h_1)\) and \(h' =
\sum_t n_t' h_t\) is a weight of \(\repr((L-\Nc)h_1)\).
The superpotential ensures that this is the case:
\begin{equation}
  2Q-\alpha_\infty
  = Q + \frac{1}{b} \sum_{t=1}^{\Nf} \I\anti{m}_t h_t
  = Q - \frac{1}{b} \sum_{t=1}^{\Nf} (\I m_t + l_t b^2 + 1)h_t
  = \alpha_0 - b \sum_{t=1}^{\Nf} l_t h_t \,.
\end{equation}
The conformal block decomposition contains one term for each way of
splitting \(\sum_t l_t h_t\) into a sum \(h+h'\) of weights of
\(\repr(\Nc h_1)\) and \(\repr((L-\Nc)h_1)\), that is, each set of
integers \(0\leq n_t\leq l_t\) with \(\sum_t n_t = \Nc\).  In
Section~\ref{sec:SeibergW-cross}, we note that the vertex operators
\(\widehat{V}_{-(L-\Nc)bh_1}\) and~\(\widehat{V}_{-\Nc bh_1}\) have the
same form with \(\Nc \leftrightarrow L-\Nc\), and deduce a duality
between theories with gauge groups \(U(\Nc)\) and \(U(L-\Nc)\).  This
duality reduces when all \(l_t=1\) to an \(\Nsusy=(2,2)^*\) analogue of
Seiberg duality.

\label{SQCDAW-TrXl}%
Our second example of superpotential only involves the adjoint chiral
multiplet, and constrains its complexified twisted mass:
\begin{equation}
  W = \Tr X^{l+1} \,, \qquad b^2 = \I m_X = \frac{-1}{l+1}
\end{equation}
for some \(l\geq 1\).  The superpotential constraint sets \(b\)~to an
imaginary value, for which \(S^4_b\)~does not make sense.  Instead of a
surface operator on \(S^2\subset S^4_b\) we must thus manipulate the two
dimensional theory on~\(S^2\) only.  Correspondingly, the
matching~\eqref{SQCDAW-matching} with the Toda CFT is written in the
form \(Z_{S^2} = \bigl[ \vev{\cdots} / Z_{S^4_b}^\free
\bigr]_{b^2=-1/(l+1)}\), where the right-hand side is analytically
continued after taking the ratio.\footnote{The central charge \(c=(\Nf-1)\bigl[1+\Nf(\Nf+1)(b^2+2+b^{-2})\bigr] = -
  (\Nf-1)(\Nf l-1)(\Nf l+l+1)/(l+1)\) is negative for the value \(b^2=-1/(l+1)\) we
  consider.}

For \(\I m_X=\frac{-1}{l+1}\), the one-loop
determinant~\eqref{SQCDA-Z1l} vanishes whenever any \(n_s > l\): the
numerator factor for \(t=s\) and \(\mu = n_s-l-1\) is \(\gamma(\I m_s
-\I m_s + (n_s-l-1-n_s) \I m_X) = \gamma(1) = 0\).  Equivalently, Higgs
branch vacua have \(X^{n_s-1} \quark_s \neq 0\) and are forbidden if
\(n_s>l\) by the \(\Faux\)-term equation \(X^l = 0\).  The
\(S^2\)~partition function in the presence of \(W = \Tr X^{l+1}\) is
thus a sum over choices of integers \(0\leq n_s\leq l\) with
\(\sum_{s=1}^{\Nf} n_s = \Nc\).

We see that introducing the superpotential \(W=\Tr X^{l+1}\) replaces
the sum over weights \(\sum_{s=1}^{\Nf} n_s h_s\) of the symmetric
representation \(\repr(\Nc h_1)\) by a sum over a restricted set
of weights, with \(0\leq n_s\leq l\).  Those are precisely the weights
of the representation with highest weight
\vspace{-2ex}
\begin{equation}\label{SQCDAW-weight}
  \omega_{\Nc,l} = l \omega_k + (\Nc - lk)h_{k+1}
  \quad\text{and Young diagram}\quad
  \mathtikz[semithick,x=1em,y=1em]{
    \foreach \X/\Y in {0/-3,1/-3,2/-3,3/-3,4/-3,5/-2,6/-2}
      \draw (\X, 0) -- (\X,\Y);
    \foreach \X/\Y in {6/0,6/-1,6/-2,4/-3}
      \draw ( 0,\Y) -- (\X,\Y);
    \draw [decorate,decoration={brace,amplitude=6pt},yshift=4pt]
      (0, 0) -- (6, 0) node [midway,yshift=14pt] {$l$};
    \draw [decorate,decoration={brace,amplitude=6pt},yshift=-4pt]
      (4,-3) -- (0,-3) node [midway,yshift=-14pt] {$\Nc-lk$};
    \draw [decorate,decoration={brace,amplitude=6pt},xshift=4pt]
      (6,0) -- (6,-2) node [midway,xshift=14pt] {$k$};
  }
  \,,
\end{equation}
where \(k\)~is defined by \(kl\leq\Nc<(k+1)l\).  The
``quasi-rectangular'' Young diagram is obtained by placing \(\Nc\)~boxes
into as many \(l\)-box rows as possible followed by a row with any
remaining box.  For \(l\geq\Nc\), none of the one-loop determinants
vanish, and the Young diagram is that of the \(\Nc\)-th symmetric
representation: this is the same as for SQCDA\@.  For \(l=1\), the Young
diagram becomes a column, hence we sum over weights of the \(\Nc\)-th
antisymmetric representation, as for SQCD with no adjoint:
correspondingly, the superpotential \(W=\Tr X^2\) lets us integrate out
the adjoint chiral multiplet.

From our experience with SQCD and SQCDA, we expect the sum over weights
of~\(\repr(\omega_{\Nc,l})\) to have a Toda CFT analogue involving the
degenerate operator~\(\widehat{V}_{-b\omega_{\Nc,l}}\).  This is
confirmed by the observation that the momenta \(-\Nc bh_1\) and
\(-b\omega_{\Nc,l}\) are Weyl conjugate when \(b^2 = \frac{-1}{l + 1}\)
since
\begin{equation}
  \begin{aligned}
    & \bigl\{\tfrac{1}{b} \vev{-\Nc bh_1 - Q, h_p}
    \bigm|
    1\leq p\leq\Nf\bigr\}
    \\
    & = \Bigl\{\frac{\Nc}{\Nf} + \frac{\Nf-1}{2} l - \Nc\Bigr\}
    \cup \Bigl\{\frac{\Nc}{\Nf} + \frac{\Nf-1}{2} l - kl
    \Bigm|
    1\leq k\leq\Nf-1\Bigr\}
    \\
    & = \bigl\{\tfrac{1}{b} \vev{-b\omega_{\Nc,l} - Q, h_p}
    \bigm|
    1\leq p\leq\Nf\bigr\} \,.
  \end{aligned}
\end{equation}
Therefore, \(\widehat{V}_{-\Nc bh_1}\)
and~\(\widehat{V}_{-b\omega_{\Nc,l}}\) are equal up to a scalar factor
for this value of~\(b^2\).  This assertion should be handled with care,
as the Toda CFT is ill defined for \(b^2<0\).

Trusting the assertion leads us to the proposal\footnote{As explained
  below~\eqref{part-typical-matching}, the factor \(A
  \abs{x}^{2\gamma_0} \abs{1-x}^{2\gamma_1}\) can be absorbed into the
  partition function.  To compare gauge theory and Toda CFT results it
  is best to keep the factor explicitly.}
\begin{equation}\label{SQCDAW-matching}
  \begin{aligned}
    & Z_{S^2}^{U(\Nc)\ \SQCDA,W=\Tr X^{l+1}}
    \biggl(m,\anti{m},m_X=\frac{\I}{l+1},z,\bar{z}\biggr)
    \\
    & \quad = A \abs{y}^{2\gamma_0} \abs{1-y}^{2\gamma_1}
    \left[
    \frac{\vev*{\widehat{V}_{\alpha_\infty}(\infty)
        \widehat{V}_{(\varkappa+\Nc b)h_1}(1)
        \widehat{V}_{-b\omega_{\Nc,l}}(y,\bar{y})
        \widehat{V}_{\alpha_0}(0)}}
    {\vev*{\widehat{V}_{\alpha_\infty}(\infty)
        \widehat{V}_{\varkappa h_1}(1)
        \widehat{V}_{\alpha_0}(0)}}
    \right]_{b^2\to\frac{-1}{l+1}}
  \end{aligned}
\end{equation}
for some~\(A\), and with other parameters given below the SQCDA
matching~\eqref{SQCDA-matching}.  Importantly, we have moved the
\(S^4_b\)~partition function of \(\Nsusy=2\) free hypermultiplets to the
right-hand side (in the form of a Toda CFT three-point function), and we
only set \(b^2=\frac{-1}{l+1}\) after evaluating the ratio of Toda CFT
correlators.  We can thus expect Upsilon functions in the numerator and
denominator to cancel, leaving a product of gamma functions which can be
analytically continued to \(b^2=\frac{-1}{l+1}\) and should reproduce
one-loop determinants in the left-hand side.

When \(l\geq\Nc\), \eqref{SQCDAW-matching}~is simply the SQCDA
matching~\eqref{SQCDA-matching} at \(\I m_X = b^2 = \frac{-1}{l+1}\),
with the same value of~\(A\).  When \(l=1\), we expect the claim to
reproduce the SQCD result~\eqref{SQCD-matching}, and indeed the SQCDA
parameters which appear in~\eqref{SQCDAW-matching} are equal for \(\I
m_X = b^2 = \frac{-1}{2}\) to the corresponding SQCD parameters, with
the exception of~\(A\).

It is difficult to find~\(A\) in general, because three-point functions
involving \(\widehat{V}_{-b\omega_{\Nc,l}}\) take complicated forms for
\(1<l<\Nc\).  Using \cite[equations (1.53) and~(1.56)]{Fateev:2007ab},
we tested the proposal~\eqref{SQCDAW-matching} for \(\Nf=\Nc=3\) and
\(l=2\), which corresponds to the adjoint representation of \(SU(3)\).
Three-point functions \(\widehat{C}_{-b(h_1-h_3),\alpha}^{\alpha-bh}\)
associated to non-zero weights~\(h\) of the adjoint representation are
ratios of Gamma functions.  When \(b^2=\frac{-1}{l+1}=\frac{-1}{3}\),
they yield the expected one-loop determinants up to a factor
\(A = b^{9-6b\varkappa} \gamma(1/3)\).  For general~\(b\), the
three-point function~\(\widehat{C}_{-b(h_1-h_3),\alpha}^{\alpha}\)
associated to the zero weight is expressed in terms of hypergeometric
functions evaluated at~\(1\), but at the point \(b^2=\frac{-1}{3}\) the
value agrees numerically with the Gamma functions expected from gauge
theory:
\begin{align}
  \widehat{C}_{-b(h_1-h_{\Nf}),Q-\I a}^{Q-\I a}
  & = \sum_{p=1}^{\Nf}
  \prod_{s\neq p}^{\Nf} \biggl(\frac{\gamma(b\vev{\I a,h_p-h_s})}{\gamma(1+b^2+b\vev{\I a,h_p-h_s})}\biggr)
  \Hypergeometric
    {-b^2+b\vev{\I a,h_s-h_p}, \, 1\leq s\leq \Nf}
    {1+b\vev{\I a,h_s-h_p}, \, s\neq p}
    {1}^2
  \\
  & \xrightarrow[\text{numerically}]{\Nf=3,b^2\to-1/3}
  \frac{\gamma(-b^2)}{\gamma(-2b^2)} \prod_{1\leq s\neq t\leq 3} \gamma(b\vev{\I a,h_s-h_t}-b^2) \,.
\end{align}

More generally, a Toda CFT four-point function with a fully degenerate
vertex operator other than \(\widehat{V}_{-b\omega_\Nc}\)
or~\(\widehat{V}_{-\Nc bh_1}\) (and the usual two generic and one
semi-degenerate vertex operators) cannot coincide with the partition
function of a surface operator described by a single \(\Nsusy=(2,2)\)
\(U(\Nc)\) vector multiplet coupled to some chiral multiplets, except
for special values of~\(b\) as is the case here.  Indeed, as described
by Fateev and Litvinov~\cite{Fateev:2007ab}, the Toda three-point
function \(\widehat{C}_{-b\omega,\alpha}^{\alpha-bh}\) only takes the
form of a ratio of Gamma functions if the weight~\(h\) appears with no
multiplicity in~\(\repr(\omega)\).  Since one-loop determinants are
always such ratios, they can only reproduce Toda CFT three-point
functions for general~\(b\) if weights have no multiplicities.

However, higher degenerate fields can be obtained by considering the
collision limit of simpler degenerate fields.  For instance, the
three-point function \(\widehat{C}_{-b(h_1-h_3),\alpha}^{\alpha}\)
mentioned above is equal to a four-point function involving a
fundamental and an antifundamental degenerate fields, in the limit where
the two punctures collide.  In the next section, we match Toda CFT
correlators involving more than one (symmetric or antisymmetric)
degenerate vertex operator with \(S^2\)~partition functions of quiver
gauge theories.  Colliding antisymmetric degenerate operators, we obtain
expressions for Toda CFT correlators of arbitrary degenerate
operators~\(\widehat{V}_{-b\Omega}\) with two generic and one
semi-degenerate vertex operators, for any~\(b\).

\subsection{Quivers and Multiple Toda Degenerates}
\label{sec:Quivers}

We have focused so far on surface operators described by \(U(\Nc)\)
gauge theories, which have a single FI parameter.  Those correspond to
Toda CFT four-point functions, which involve a single anharmonic
ratio~\(x\).  Here, we equate the partition function of surface
operators described by certain \(U(\Nc_1)\times\cdots\times U(\Nc_n)\)
quiver gauge theories and \((n+3)\)-point functions
with \(n\)~symmetric or antisymmetric degenerate operators.  In
detail,\footnote{Following the arguments
  below~\eqref{part-typical-matching}, the factor \(A a(x)
  a(\bar{x})\) can be absorbed into the partition function.  To
  compare gauge theory and Toda CFT results it is best to keep the
  factor explicitly.}%
\begin{equation}\label{Quivers-matching}
  Z_{S^2\subset S^4_b}^{\prod_j U(\Nc_j), W_\eta}\bigl(m,z,\bar{z}\bigr)
  = A a(x) a(\bar{x})
  \vev*{\widehat{V}_{\alpha_\infty}(\infty)
    \widehat{V}_{\hat{m}}(1)
    \prod_{j=1}^n
    \widehat{V}_{-b\Omega(K_j,\epsilon_j)}(x_j,\bar{x}_j)
    \widehat{V}_{\alpha_0}(0)} \,.
\end{equation}
The matching gives a detailed description of the moduli space
parametrized by the~\(z_j\).  We describe notations below, then
consider several limits to fix all parameters of the matching in
Section~\ref{sec:Quivers-fix}.  Fine-tuning FI parameters such that
degenerate punctures collide on the Toda CFT side, we deduce in
Section~\ref{sec:Quivers-fuse} the microscopic description of the
surface operator which corresponds to arbitrary degenerate punctures
in the Toda CFT\@.  Brane diagrams (see Figure~\ref{fig:branes}) clarify
some aspects of the correspondence.

\begin{figure}[pt]
  \caption{\label{fig:branes}%
    A 4d/2d quiver, its corresponding brane diagram, and Toda CFT correlator.}%
  \medskip
  \begin{tabular}{p{0.97\textwidth}}
    \toprule
    \centerline{$%
      \quiver{%
        \node (N1)  [color-group]                     {$3$};
        \node (N2)  [color-group, left of=N1]         {$5$};
        \node (Nn)  [color-group, left of=N2]         {$7$};
        \node (f)   [left=4ex of Nn]                  {};
        \node (Nf)  [flavor-group, above=1ex of f]    {$\Nf$};
        \node (Nf') [flavor-group, below=1ex of f]    {$\Nf$};
        \draw (Nf) -- (Nf');
        \draw[->-=.55] (Nf) -- (Nn);
        \draw[->-=.55] (Nn) -- (Nf');
        \draw[->-=.55] (Nn) to [bend left=30] (N2);
        \draw[->-=.55] (N2) to [bend left=30] (N1);
        \draw[->-=.55] (N1) to [bend left=30] (N2);
        \draw[->-=.55] (N2) to [bend left=30] (Nn);
        \draw[->-=.5]  (N1) to [distance=5ex, in=60, out=120, loop] ();
        \draw[->-=.5]  (Nn) to [distance=5ex, in=60, out=120, loop] ();
        \node (top-corner) at ($(Nf.north east)+(0.1,5ex)$) {};
        \node (bottom-corner) at ($(Nf'.south west)+(-0.3,-0.3)$) {};
        \node (4d)  [below left = -1pt of top-corner] {4d};
        \node (2d)  [below right = -1pt of top-corner] {2d};
        \draw [color=gray, dashed, rounded corners]
              (top-corner) rectangle (bottom-corner);
        \draw [color=gray, dashed, rounded corners]
              (top-corner) rectangle ($(bottom-corner -| N1.east)+(0.3,0)$);
      }
      \quad
      \mathtikz[thick,x=1em,y=1em]{
        %D4
        \node at (0,1.7) {D4};
        \foreach \Y in {-1.4,-0.4,0.6,1}
          \draw (-1,\Y) -- (2,\Y);
        \foreach \Y in {-1,-0.2,0.4,1.4}
          \draw (2,\Y) -- (5,\Y);
        \draw [decorate,decoration={brace,amplitude=6pt},xshift=-.1em]
          (-1, -1.8) -- (-1, 1.4) node [midway,xshift=-1.2em] {$\Nf$};
        \draw [decorate,decoration={brace,amplitude=6pt},xshift=.1em]
          (5, 1.8) -- (5, -1.4) node [midway,xshift=1.2em] {$\Nf$};
        %NS5
        \node at (2,3.1)  {NS5};
        \draw (2,-4.9) -- (2,2.4);
        %\node at (5.1,-2.2) {NS5};
        \draw (4,-5.9) -- (4,-2.5);
        %NS5'
        \node at (8.4,-5.5)  {NS5'};
        \draw (5,-5.7) -- (7,-3.7);
        \draw (6,-6.2) -- (8,-4.2);
        \draw [decorate,decoration={brace,amplitude=6pt},yshift=.1em]
          (3.7,-2.2) -- (8.4,-4.4) node [midway,xshift=1em,yshift=1em] {$n=3$};
        %D2
        \node at (3,-5.3) {D2};
        \draw (2,-2.5) -- (7.1,-5.05);
        \draw (2,-2.71) -- (6.96,-5.19);
        \draw (2,-2.92) -- (6.82,-5.33);
        \draw (2,-3.19) -- (5.64,-5.01);
        \draw (2,-3.4) -- (5.5,-5.15);
        \draw (2,-3.69) -- (4,-4.69);
        \draw (2,-3.9) -- (4,-4.9);
      }
      \quad
      \mathtikz{
        \draw (-0.25,0) ellipse (1.65 and .8);
        \fill (-1.5,-.1) circle [radius=2pt];
        \fill (-1.0,-.1) circle [radius=2pt];
        \fill ( 1.0,-.1) circle [radius=2pt];
        \draw (-1.5,-.1) circle (4pt);
        \draw ( 1.0,-.1) circle (4pt);
        \node [cross] at (-.5,-.1) {};
        \draw (-.6,.2) rectangle (-.4,.3);
        \draw (-.5,.2) -- (-.5,.3);
        \node [cross] at (0,-.1) {};
        \draw (-.05,.15) rectangle (.05,.35);
        \draw (-.05,.25) -- (.05,.25);
        \node [cross] at (.5,-.1) {};
        \draw (.45,.1) rectangle (.55,.4);
        \draw (.45,.2) -- (.55,.2);
        \draw (.45,.3) -- (.55,.3);
      }
    $}%
    \setlength{\parindent}{20pt}%
    \smallskip
    \noindent
    $\Nf$~semi-infinite D4~branes ending on each side of a single
    NS5~brane engineer at low energies the theory of $\Nf^2$~free
    hypermultiplets on their four-dimensional intersection.  Adding
    D2~branes stretched between the NS5~brane and $n$~additional
    NS5~branes inserts a surface operator with support on the boundary
    of the added D2~branes.  Rotating some NS5~branes (rotated branes
    are denoted as~NS5' and are all parallel) alters the surface
    operator, which is then precisely the one discussed in the main
    text.

    The ranks $\Nc_n \geq \cdots \geq \Nc_1$ are the numbers of
    D2~branes between consecutive NS5/NS5'~branes.  When these are
    parallel (both~NS5 or both~NS5'), the corresponding $U(\Nc_j)$ group
    has an adjoint chiral multiplet ($\eta_j=+1$), otherwise not
    ($\eta_j=-1$).  Equivalently, the $j$-th brane is an~NS5 if
    $\epsilon_j = \prod_{i=j}^n \eta_i$ is~$1$ and otherwise it is
    an~NS5'.  The Toda CFT data appears by turning on FI~parameters, as
    this separates the NS5/NS5'~branes along the D4~brane direction.
    Then $K_j=(\Nc_j-\Nc_{j-1})$ D2~branes stretch between the original
    NS5~brane and the $j$-th NS5/NS5'~brane, corresponding to the
    $K_j$-th symmetric (or antisymmetric if $\epsilon_j=-1$) degenerate
    operator.

    We will see in Section~\ref{sec:SeibergQ} that permuting the
    $(\epsilon_j,K_j)$ or equivalently the NS5/NS5'~branes is a
    (Seiberg-like) duality of the surface operator.
    \\\bottomrule
  \end{tabular}
\end{figure}

The surface operator depends on a choice of \(n\)~signs \(\eta_j = \pm
1\) and integer parameters \(\Nc_n \geq \cdots \geq \Nc_1 \geq 0\).
It also depends on \(n\)~FI and theta parameters combined as
\begin{equation}\label{Quivers-zhat}
  z_j = e^{-2\pi\xi_j+\I\vartheta_j}
  \quad \text{and} \quad
  \hat{z}_j = (-1)^{\Nc_{j-1}+\Nc_{j+1}} z_j
\end{equation}
for \(1\leq j\leq n\), where \(\Nc_0=0\), \(\Nc_{n+1}=\Nf\), and the
sign is chosen for later convenience.  The operator is defined by the
\(\Nsusy=(2,2)\) quiver
\begin{equation}\label{Quivers-quiver}
  \quiver{%
    \node (Nn)  [color-group]                    {$\Nc_n$};
    \node (dots)[right of=Nn]                    {$\cdots$};
    \node (N1)  [color-group, right of=dots]     {$\Nc_1$};
    \node (f)   [left of=Nn]                     {};
    \node (Nf)  [flavor-group, above=1ex of f]   {$\Nf$};
    \node (Nf') [flavor-group, below=1ex of f]   {$\Nf$};
    \draw[->-=.55] (Nf) -- (Nn);
    \draw[->-=.55] (Nn) -- (Nf');
    \draw[->-=.55] (Nn)   to [bend left=30] (dots);
    \draw[->-=.55] (dots) to [bend left=30] (N1);
    \draw[->-=.55] (N1)   to [bend left=30] (dots);
    \draw[->-=.55] (dots) to [bend left=30] (Nn);
    \draw[->-=.5] (Nn) to [distance=5ex, in=60, out=120, loop] ();
    \draw[->-=.5] (N1) to [distance=5ex, in=60, out=120, loop] ();
  }
\end{equation}
which describes a \(U(\Nc_1)\times\cdots\times U(\Nc_n)\) vector
multiplet coupled to various chiral multiplets.
First, \(\Nf\)~fundamentals~\(\quark_t\)
and \(\Nf\)~antifundamentals~\(\anti{\quark}_t\) of \(U(\Nc_n)\).
Next, for each \(1\leq j\leq n-1\), one pair of bifundamentals
of \(U(\Nc_j)\times U(\Nc_{j+1})\): \(\biquark_{j(j+1)}\)~in the
representation \(\Nc_j\otimes\overline{\Nc}_{j+1}\)
and \(\biquark_{(j+1)j}\)~in the
representation \(\overline{\Nc}_j\otimes\Nc_{j+1}\).  Finally, for
each \(1\leq j\leq n\), one adjoint~\(X_j\).  The (complexified)
twisted masses \(m_t\), \(\anti{m}_t\), \(m_{j(j+1)}\), \(m_{(j+1)j}\)
and~\(m_{jj}\) of these fields are constrained by a superpotential
coupling~\(W_\eta\).

The superpotential has the following terms,
\begin{equation}\label{Quivers-superpotential}
  \begin{cases}
    \Tr \bigl( X_j^2 \bigr)
    & \text{for \(1\leq j\leq n\) if \(\eta_j = -1\)} \,,
    \\
    \Tr \bigl( \biquark_{j(j+1)} \biquark_{(j+1)j}
    \biquark_{j(j-1)} \biquark_{(j-1)j} \bigr)
    & \text{for \(1<j<n\) if \(\eta_j = -1\)} \,,
    \\
    \Tr \bigl( X_j \biquark_{j(j+1)} \biquark_{(j+1)j} \bigr)
    & \text{for \(1\leq j<n\) if \(\eta_j = 1\)} \,,
    \\
    \Tr \bigl( X_j \biquark_{j(j-1)} \biquark_{(j-1)j} \bigr)
    & \text{for \(1<j\leq n\) if \(\eta_j = 1\)} \,.
  \end{cases}
\end{equation}
In other words the adjoint multiplets of nodes with $\eta_j=1$ have a
cubic coupling to neighboring bifundamental multiplets, while nodes
with $\eta_j=-1$ entail a quartic coupling of neighboring
bifundamental multiplets.  The $\Tr(X_j^2)$ term for $\eta_j=-1$ gives
a mass to the adjoint multiplet~$X_j$, hence the
theory~\eqref{Quivers-quiver} is equivalent in the low-energy to the
analogous theory~\eqref{gquiver} from the introduction, which omits
these~$X_j$.  Here, we include adjoint multiplets for all nodes to
simplify signs in the definition~\eqref{Quivers-zhat}
of~\(\hat{z}_j\).  Indeed, integrating out~\(X_j\) when \(\eta_j =
-1\) shifts the corresponding theta angle \(z_j\to(-1)^{\Nc_j-1}z_j\),
hence in order to keep \(\hat{z}_j\) fixed one should complicate
the definition~\eqref{Quivers-zhat} of~\(\hat{z}_j\) by including
the sign~\((-1)^{\Nc_j-1}\).

The superpotential \(W_\eta\) must have \(R\)-charge~\(2\) (twisted
mass~\(\I\)) to be supersymmetric.  This fixes twisted masses of
bifundamental and adjoint multiplets in terms of the signs~\(\eta\) and
a single continuous parameter,\footnote{The full flavour symmetry of the
  two dimensional theory is \(S[U(\Nf)\times U(\Nf)]\times U(1)\), where
  the first factor acts on fundamental and antifundamental chiral
  multiplets.  Under the \(U(1)\) factor, the adjoint chiral
  multiplet~\(X_j\) has charge \(\epsilon_j+\epsilon_{j+1}\) and the
  bifundamental multiplets \(\biquark_{(j-1)j}\)
  and~\(\biquark_{j(j-1)}\) have charge \(-\epsilon_j\), where
  \(\epsilon_j = \prod_{i=j}^n \eta_i\).} which will match with~\(b^2\)
in the Toda CFT\@.  To ease the comparison with the Toda CFT correlator,
we define signs \(\epsilon_j = \prod_{i=j}^n \eta_i\) for \(1\leq j\leq
n+1\) and find
\begin{equation}\label{Quivers-masses}
  \begin{aligned}
    \I m_{jj}
    & = \begin{cases}
      -1-b^2 & \text{if \(\epsilon_{j+1} = \epsilon_j = -1\)} \,, \\
      -1/2   & \text{if \(\epsilon_{j+1} \neq \epsilon_j\)} \,, \\
      b^2    & \text{if \(\epsilon_{j+1} = \epsilon_j = +1\)} \,,
    \end{cases}
    \\
    \I m_{(j-1)j} + \I m_{j(j-1)}
    & = \begin{cases}
      b^2      & \text{if \(\epsilon_j = -1\)} \,, \\
      -1 - b^2 & \text{if \(\epsilon_j = +1\)} \,.
    \end{cases}
  \end{aligned}
\end{equation}
Equivalently, \(W_\eta\) could be defined as containing all gauge
invariant combinations of the fields which have total \(R\)-charge~\(2\)
(twisted mass~\(\I\)), given the mass assignment~\eqref{Quivers-masses}.
As always, the twisted masses and \(R\)-charges of fundamental and
antifundamental chiral multiplets are unconstrained.

On the other hand, the Toda CFT \((n+3)\)-point function involves two
generic and one semi-degenerate vertex operators
\(\widehat{V}_{\alpha_\infty}(\infty)\), \(\widehat{V}_{\hat{m}}(1)\),
and \(\widehat{V}_{\alpha_0}(0)\) with momenta
\begin{equation}\label{Quivers-alpha}
  \begin{aligned}
    \alpha_0
    & = Q - \frac{1}{b} \sum_{s=1}^{\Nf} \I m_s h_s \,,
    & \qquad\qquad
    \hat{m}
    & = (\varkappa + \Nc_n b) h_1 \,,
    \\
    \alpha_\infty
    & = Q - \frac{1}{b} \sum_{s=1}^{\Nf} \I\anti{m}_s h_s \,,
    & \qquad\qquad
    \varkappa
    & = \frac{1}{b} \sum_{s=1}^{\Nf} (1 + \I m_s + \I\anti{m}_s) \,,
  \end{aligned}
\end{equation}
which coincide with those of earlier sections.  It also involves
\(n\)~fully degenerate vertex operators
\(\widehat{V}_{-b\Omega(K_j,\epsilon_j)}(x_j,\bar{x}_j)\) at
\begin{equation}
  x_j = \prod_{i=j}^n \hat{z}_i
  \quad \text{for \(1\leq j\leq n\)} \,.
\end{equation}
Each degenerate operator is labeled by the highest weight
\(\Omega(K,+1) = Kh_1\) of a symmetric representation or \(\Omega(K,-1)
= \omega_K\) of an antisymmetric representation of~\(A_{\Nf-1}\),
depending on the signs \(\epsilon_j = \prod_{i=j}^n \eta_i\) and the integers
\begin{equation}
  K_1 = \Nc_1 \,, \quad \text{and} \quad
  K_j = \Nc_j - \Nc_{j-1} \quad \text{for \(1<j\leq n\)} \,.
\end{equation}
Finally, the factors \(A\) and~\(a\) are
\begin{align}
  \label{Quivers-A}
  A & = b^{\Nc_n\Nf(1+b^2)-\Nc_n^2 b^2-2\Nc_n b\varkappa}
  \prod_{j\mid\epsilon_j=+1} \prod_{1\leq\nu\leq K_j} \gamma(-\nu b^2) \,,
  \\
  \label{Quivers-a}
  a(x) a(\bar{x})
  & = \prod_{j=1}^n \abs{x_j}^{2\beta_j}
  \prod_{j=1}^n \abs{1-x_j}^{2\gamma_j}
  \prod_{i<j}^n \abs{x_j-x_i}^{2\gamma_{ij}} \,,
\end{align}
with the exponents
\begin{align}
  \label{Quivers-beta}
  \beta_j & =
  \dimToda\bigl(-b\Omega(K_j,\epsilon_j)\bigr)
  -\frac{K_j}{\Nf} \sum_{t=1}^{\Nf} \I m_t
  + \frac{K_j(\Nf-K_j)}{2\Nf} b^2
  - \Nc_{j-1} \I m_{j(j-1)} - K_j \sum_{i=j+1}^n \I m_{(i-1)i} \,,
  \\
  \label{Quivers-gamma1}
  \gamma_j & = (-1-b^2)K_j + b(\varkappa+\Nc_n b) K_j / \Nf \,,
  \\
  \label{Quivers-gamma2}
  \gamma_{ij} & = \begin{cases}
    b^2 K_i - b^2 K_i K_j / \Nf & \text{if \(\epsilon_j=-1\)} \,, \\
    (-1-b^2) K_i - b^2 K_i K_j / \Nf & \text{if \(\epsilon_j=+1\)} \,,
  \end{cases}
\end{align}
for \(1\leq i<j\leq n\).  When \(n=1\), the
matching~\eqref{Quivers-matching} reproduces the known cases of SQCD
(\(\eta_1=-1\)) and SQCD with an adjoint (\(\eta_1=1\)).  Also, for
\(n>1\) setting \(\Nc_1=0\) reduces the matching to the case \(n\to
n-1\).

As a preliminary check of the equality~\eqref{Quivers-matching}, we can
recognize a few symmetries.  Permuting the flavours of fundamental
quarks~\(\quark_t\), hence their twisted masses~\(m_t\), does not alter
the partition function.  This is translated on the Toda CFT side into a
Weyl transformation of the momentum~\(\alpha_0\), which permutes the
\(\vev{\alpha_0-Q,h_t}\).  Similarly, permuting the~\(\anti{m}_t\)
amounts to a Weyl transformation of~\(\alpha_\infty\).  Next, performing
charge conjugation on all gauge group factors maps \(\hat{z}_j \to
\hat{z}_j^{-1}\), \(m_t \leftrightarrow \anti{m}_t\), and \(m_{j(j+1)}
\leftrightarrow m_{(j+1)j}\): this corresponds on the Toda CFT side to
the conformal map \(x\to x^{-1}\), which swaps
\(\alpha_0\leftrightarrow\alpha_\infty\) and maps \(x_j\to x_j^{-1}\).
The transformation of \(a(x)a(\bar{x})\) compensates exactly the
conformal factor \(\abs{x_j}^{-4\dimToda(-b\Omega(K_j,\epsilon_j))}\)
for each~\(j\).
Finally, shifting the twisted masses of bifundamentals while keeping the
sums \(m_{j(j+1)} + m_{(j+1)j}\) fixed amounts to redefining gauge
multiplet scalars by a constant shift,
whose sole effect on the partition function is in
overall powers of \(\abs{x_j}^2\): on the Toda CFT side
of~\eqref{Quivers-matching}, only the exponents~\(\beta_j\) change.

\subsubsection{Matching Parameters for Quivers}
\label{sec:Quivers-fix}

We first expand the partition function and the correlator in the
s-channel, that is, the region where \(0<\abs{x_1}<\cdots
<\abs{x_n}<1\) or equivalently where all FI parameters are
positive: \(\abs{\hat{z}_j}<1\).  We map vacua of the gauge theory to
choices of internal momenta in the correlator.  The classical and
one-loop contributions match as expected with the exponents and
three-point functions, while the vortex partition functions give
predictions for Toda CFT conformal blocks (see
Appendix~\ref{app:blocks}).  This check fixes \(\{K_j,\epsilon_j\}\),
the momentum~\(\alpha_0\), the overall constant factor~\(A\) and the
exponents \(\beta_j+\sum_{i<j} \gamma_{ij}\).  The momentum
\(\alpha_\infty\)~is fixed by the symmetry under charge conjugation
discussed earlier.  Then, we justify the relation between the gauge
theory data \(\{\eta_j,\hat{z}_j\}\) and the Toda CFT data
\(\{\epsilon_j,x_j\}\) by counting distinct exponents in the limit where
two neighboring punctures collide.  Comparing the exponents only fixes
the momentum~\(\hat{m}\) and the exponents \(\gamma_n\) and
\(\gamma_{(j-1)j}\).  The remaining exponents \(\gamma_j\)
and~\(\gamma_{ij}\) are fixed thanks to Seiberg dualities which
translates in this setting to permutations of the \(n\)~punctures (see
Section~\ref{sec:SeibergQ}).

It is easiest to find Higgs branch vacua of the gauge theory by solving
the \(\Daux\)-term and \(\Faux\)-term equations, assuming as before that
fundamental chiral multiplets have generic twisted masses~\(m_s\).
The derivation goes as follows.  Diagonalize
all~\(\sigma_j\).  Introduce \(\I\sigma_{n+1} = \operatorname{diag}(-\I m_1,\ldots,-\I
m_{\Nf})\), \(\Nc_{n+1}=\Nf\), and \(\Nc_0=0\) to simplify the discussion.
Integrate out all \(X_j\) which have twisted mass \(m_{jj}=\I/2\), that is, \(\eta_j=-1\).  The
\(\Daux\)-term equations (for \(\abs{\hat{z}_j}<1\)) impose that the
images of \(X_j\), \(\biquark_{j(j+1)}\) and \(\biquark_{j(j-1)}\) span
\(\bbC^{\Nc_j}\), hence all eigenvalues of~\(\sigma_j\) are constrained
to be equal to another eigenvalue of~\(\sigma_j\) or of~\(\sigma_{j\pm
  1}\), minus a twisted mass.  All eigenvalues of~\(\I\sigma_j\) thus
take the form \(\I\sigma_{j,\mathrm{a}} = -\I m_s -
\sum_{i=j+1}^n \I m_{(i-1)i} + \mu(1+b^2) - \nu b^2\) where
\(\mu,\nu\in\bbZ_{\geq 0}\).  Using the \(\Faux\)-term constraint, one
can then bound the multiplicity of such an eigenvalue by the
multiplicity of the eigenvalue \(\I\sigma_{j,\mathrm{a}}-\I m_{jk}\) of
\(\I\sigma_k\), for \(k\in\{j,j\pm 1\}\) (only \(k\in\{j\pm 1\}\) if
\(X_j\)~was integrated out).  Since each eigenvalue \(-\I m_s\) of
\(\I\sigma_{n+1}\) has multiplicity~\(1\), we deduce by induction on
\(n+1-j\), \(\mu\), and~\(\nu\) that all eigenvalues have
multiplicity~\(1\).  The statement is in fact stronger: for any
eigenvalue \(\I\sigma_{j,\mathrm{a}}\) of \(\I\sigma_j\), and for
\(k\in\{j,j\pm 1\}\) (or only \(k\in\{j\pm 1\}\)), \(\I\sigma_{j,\mathrm{a}}-\I
m_{jk}\) is an eigenvalue of \(\I\sigma_k\), and the relevant component
of \(\biquark_{jk}\) is non-zero.
Solving the \(\Faux\)-term
constraints is then a combinatorical problem whose details depend
on the superpotential~\(W_\eta\).

At the end of the day, one finds that vacua obey
\begin{equation}\label{Quivers-spec-sigmaj}
  \I\sigma_j = \operatorname{diag}\biggl(
  - \I m_s - \sum_{i=j+1}^n \bigl( \I m_{(i-1)i} \bigr) - \nu b^2
  \biggm| 0\leq \nu<n_s^j , 1 \leq s\leq\Nf \biggr)
\end{equation}
for \(1\leq j\leq n\), where \(n_s^j\geq 0\) are integers such that
\(\sum_{s=1}^{\Nf} n_s^j = \Nc_j\) and
\begin{equation}
  \begin{cases}
    n_s^{j-1} \leq n_s^j \leq n_s^{j-1} + 1 & \text{if \(\epsilon_j=-1\)} \,,
    \\
    n_s^{j-1} \leq n_s^j & \text{if \(\epsilon_j=+1\)} \,,
  \end{cases}
\end{equation}
where \(n_s^0 = 0\).  These conditions are equivalent to requiring
that for each \(1\leq j\leq n\) the difference \(h_{[n^j]} -
h_{[n^{j-1}]} = \sum_{s=1}^{\Nf} (n_s^j - n_s^{j-1}) h_s\) is a weight of the symmetric or antisymmetric representation
\(\repr(\Omega(K_j,\epsilon_j))\) of rank \(K_j=\Nc_j-\Nc_{j-1}\).  The \(S^2\)~partition function is
then a sum
\begin{equation}
  Z_{S^2} = \sum_{\{n_s^j\}} Z_\classical Z_\oneloop Z_\vortex Z_\antivortex
\end{equation}
over choices of~\(\{n_s^j\}\) consistent with the constraints above.
Terms of this sum are in a natural bijection with terms of the s-channel
decomposition of the Toda CFT correlator in~\eqref{Quivers-matching}:
the internal momenta are \(\alpha_0-bh_{[n^j]}\) for \(1\leq j\leq n\).
Thus, counting terms fixes the degenerate momenta
\(-b\Omega(K_j,\epsilon_j)\) in terms of the \(\Nc_j\) and~\(\eta_j\).

Since the Higgs branch and Coulomb branch representations of
\(S^2\)~partition functions coincide, \(Z_\classical Z_\oneloop\) is the
residue at the pole~\eqref{Quivers-spec-sigmaj} of the Coulomb branch
integrand, and \(Z_\vortex Z_\antivortex\) is the additional
contribution from poles for which \(\I\sigma_j^\pm\)
is~\eqref{Quivers-spec-sigmaj} plus integers.  We find in particular
that the classical contribution reproduces the powers of~\(x_j\)
expected from the Toda CFT up to shifts by \(\beta_j + \sum_{i=1}^{j-1}
\gamma_{ij}\),
\begin{equation}
  \begin{aligned}
    Z_\classical
    & = \prod_{j=1}^n \abs{z_j}^{2\Tr\I\sigma_j}
    = \prod_{j=1}^n \abs{z_j}^{2\bigl[
      - \sum_{s=1}^{\Nf} (n_s^j \I m_s)
      - \Nc_j \sum_{i=j+1}^n(\I m_{(i-1)i})
      - \sum_{s=1}^{\Nf} \sum_{\nu=0}^{n_s^j-1} \nu b^2\bigr]}
    \\
    & = \prod_{j=1}^n \abs{x_j}^{2\bigl[
      \beta_j + \sum_{i=1}^{j-1} ( \gamma_{ij} )
      + \dimToda(\alpha_0-bh_{[n^j]})
      - \dimToda(\alpha_0-bh_{[n^{j-1}]})
      - \dimToda(-b\Omega(K_j,\epsilon_j))
      \bigr]} \,,
  \end{aligned}
\end{equation}
provided that \(\alpha_0\) is as given in~\eqref{Quivers-alpha}, and
\(\beta_j+\sum_{i=1}^{j-1}\gamma_{ij}\) as in \eqref{Quivers-beta}
and~\eqref{Quivers-gamma2}.  By symmetry, \(\alpha_\infty\) is as given
in~\eqref{Quivers-alpha}.  Similarly, a tedious calculation shows that
for each term the one-loop determinant~\(Z_\oneloop\) matches with the
product of Toda CFT three-point functions, up to
precisely the constant~\(A\) given in~\eqref{Quivers-A}.

Next, let us probe the collision of two neighboring punctures, starting
again from the s-channel \(0<\abs{x_1}<\cdots<\abs{x_n}<1\).  The Coulomb
branch representation of the \(S^2\)~partition function of interest has
the form
\begin{equation}
  Z_{S^2}^{\prod_j U(\Nc_j), W_\eta}
  = \prod_{j=1}^n \biggl[
  \frac{1}{\Nc_j!} \sum_{B_j\in\bbZ^{\Nc_j}}
  \int \frac{\dd[^{\Nc_j}]{\sigma_j}}{(2\pi)^{\Nc_j}}
  \biggr]
  \prod_{j=1}^n \Bigl[ z_j^{\Tr\I\sigma_j^+} \bar{z}_j^{\Tr\I\sigma_j^-} \Bigr]
  Z_{\oneloop,\vectormultiplet}
  Z_{\oneloop,\chiralmultiplet}
\end{equation}
where \(\I\sigma_j^\pm = \I\sigma_j \pm B_j/2\), \(Z_{\oneloop,\vectormultiplet}\) is
the one-loop determinant of vector multiplets (a Vandermonde factor and a sign),
and~\(Z_{\oneloop,\chiralmultiplet}\) is the one-loop determinant of chiral
multiplets (a product of Gamma functions).  Collecting all factors which
depend on~\(\sigma_k^\pm\) for a given \(k<n\) yields the integral
\begin{equation}\label{Quivers-Zk}
  \begin{aligned}
    & Z_k
    = \mspace{-9mu} \sum_{B_k\in\bbZ^{\Nc_k}}
    \int
    \frac{\dd[^{\Nc_k}]{\sigma_k}}
    {\Nc_k! (2\pi)^{\Nc_k}}
    \bigl[(-1)^{\Nc_k-1} z_k\bigr]^{\Tr\I\sigma_k^+}
    \bigl[(-1)^{\Nc_k-1} \bar{z}_k\bigr]^{\Tr\I\sigma_k^-}
    \prod_{i<j}^{\Nc_k} \prod_\pm \bigl(\sigma_{ki}^\pm-\sigma_{kj}^\pm\bigr)
    \prod_{i=1}^{\Nc_k}\Biggl[
    \\
    &
    \prod_{j=1}^{\Nc_k}
    \frac{\Gamma(-\I m_{kk}-\I\sigma_{ki}^++\I\sigma_{kj}^+)}
      {\Gamma(1+\I m_{kk}+\I\sigma_{ki}^--\I\sigma_{kj}^-)}
    \prod_{\substack{l\in\{k\pm 1\}\\1\leq j\leq\Nc_l}} \biggl[
    \frac{\Gamma(-\I m_{kl}+\I\sigma_{lj}^+-\I\sigma_{ki}^+)}
      {\Gamma(1+\I m_{kl}-\I\sigma_{lj}^-+\I\sigma_{ki}^-)}
    \frac{\Gamma(-\I m_{lk}-\I\sigma_{lj}^++\I\sigma_{ki}^+)}
      {\Gamma(1+\I m_{lk}+\I\sigma_{lj}^--\I\sigma_{ki}^-)}
    \biggr] \Biggr]
  \end{aligned}
\end{equation}
which resembles the \(S^2\)~partition function of SQCDA with \(\Nc_k\)
colors and \(\Nc_{k-1}+\Nc_{k+1}\) flavours, with twisted masses
\(m_{kl}-\sigma_{lj}\) and \(m_{lk}+\sigma_{lj}\).  The shifts of
\(\sigma_{lj}\) by \(\pm B_{lj}/2\) cannot be incorporated in such
twisted masses, as the ratios of Gamma functions involve both
\(\sigma_{lj}^+\) and~\(\sigma_{lj}^-\).

However, we can still apply the same techniques as in
Section~\ref{sec:SQCDA}, and close the \(\I\sigma_k\) integration
contours towards \(\pm\infty\) depending on whether \(\abs{z_k}\lessgtr
1\).  The sum over poles factorizes as in the case of SQCDA, and the
resulting vortex and antivortex partition functions are those of SQCDA
with twisted masses \(m_{kl}-\sigma_{lj}^+\) and
\(m_{lk}+\sigma_{lj}^+\) for vortices, and \(m_{kl}-\sigma_{lj}^-\) and
\(m_{lk}+\sigma_{lj}^-\) for antivortices.  As we saw in
Section~\ref{sec:SQCDA}, those vortex partition functions have branch
points when \(\hat{z}_k = (-1)^{\Nc_{k-1}+\Nc_{k+1}} z_k\) is
\(1\) or~\(\infty\).  We now prove that the powers of \(1-\hat{z}_k\)
which appear in an expansion of~\(Z_\vortex\) near \(\hat{z}_k=1\)
coincide with the powers of \(x_{k+1}-x_k\) obtained in the fusion of
the punctures at \(x_k\) and~\(x_{k+1}\).  This implies that \(x_k =
x_{k+1} \hat{z}_k\), as announced, and fixes \(\gamma_{k(k+1)}\).  To
proceed further, we need to distinguish the cases \(\eta_k=\pm 1\).

If \(\eta_k=-1\), then \(\I m_{kk} = -1/2\), and the adjoint chiral
multiplet one-loop determinant is simply a sign.  Thus, the vortex
partition functions are those of SQCD\@.  From~\eqref{SQCD-tch-expo}, the
exponents of \(1-\hat{z}_k\) which occur in an expansion near~\(1\) are
\(0\)~and
\begin{equation}\label{Quivers-SQCD-tch-expo}
  \begin{aligned}
    & \Nc_{k-1} (1 + \I m_{k(k-1)} + \I m_{(k-1)k})
    + \Nc_{k+1} (1 + \I m_{k(k+1)} + \I m_{(k+1)k}) - \Nc_k
    \\
    & = \begin{cases}
      - K_k (1 + b^2) - K_{k+1} b^2
      & \text{if \(\epsilon_k=-\epsilon_{k+1}=-1\)} \,,
      \\
      K_k b^2 + K_{k+1} (1 + b^2)
      & \text{if \(\epsilon_k=-\epsilon_{k+1}=+1\)} \,.
    \end{cases}
  \end{aligned}
\end{equation}
The analogous limit in the Toda CFT correlator is \(x_k\to x_{k+1}\).
The channel where the punctures at \(x_k\) and~\(x_{k+1}\) are fused
allows two internal momenta.  Indeed, \(\epsilon_k=-\epsilon_{k+1}\),
hence one of the punctures is labeled by a symmetric representation and
the other one by an antisymmetric representation.  The fusion of two
such representations is the direct sum of two irreducible
representations:
\begin{equation}
  \repr(K h_1) \otimes \repr(\omega_L)
  = \repr(K h_1+\omega_L) \oplus \repr((K-1)h_1+\omega_{L+1})
\end{equation}
assuming \(K,L\geq 1\).  The corresponding exponents of \(x_{k+1}-x_k\)
are
\begin{align}
  \dimToda(-Kbh_1-b\omega_L)
  - \dimToda(-Kbh_1) - \dimToda(-b\omega_L)
  & = - K b^2 + \frac{KL}{\Nf} b^2 \,,
  \\
  \dimToda(-(K-1)bh_1-b\omega_{L+1})
  - \dimToda(-Kbh_1) - \dimToda(-b\omega_L)
  & = L (1+b^2) + \frac{KL}{\Nf} b^2 \,,
\end{align}
and match with the gauge theory exponents up to precisely
\(\gamma_{k(k+1)}\) given in~\eqref{Quivers-gamma2}.  Indeed, if
\(\epsilon_k=-\epsilon_{k+1}=-1\), then \(K\) and~\(L\) above are
\(K_{k+1}\) and~\(K_k\), the first Toda CFT exponent corresponds to the
gauge theory exponent~\eqref{Quivers-SQCD-tch-expo}, and the second
to~\(0\).  If \(\epsilon_k=-\epsilon_{k+1}=1\), then \(K=K_k\) and
\(L=K_{k+1}\), the first Toda CFT exponent corresponds to~\(0\) and the
second to~\eqref{Quivers-SQCD-tch-expo}.

If instead \(\eta_k=+1\), then the adjoint chiral multiplet remains, and
the vortex partition functions involve more powers of \(1-\hat{z}_k\),
given in~\eqref{SQCDA-t-powers}.  Namely,
\begin{equation}
  \begin{aligned}
    & \bigl(1-\hat{z}_k\bigr)^{
      -\nu
      +\nu\Nc_{k-1}(1+\I m_{k(k-1)}+\I m_{(k-1)k})
      +\nu\Nc_{k+1}(1+\I m_{k(k+1)}+\I m_{(k+1)k})
      +\nu[2\Nc_k-\nu-1]\I m_{kk}}
    \\
    & \quad =
    \bigl(1-\hat{z}_k\bigr)^{
      -\nu(1+\I m_{kk})+\nu[K_k-K_{k+1}-\nu]\I m_{kk}}
  \end{aligned}
\end{equation}
for \(0\leq\nu\leq\Nc_k\).  The remaining~\(\sigma_j\) integrals
(\(j\neq k\)) do not affect these exponents.  From the derivation
of~\eqref{SQCDA-t-powers}, we know that the contribution for a
given~\(\nu\) comes from the region where \(\nu\)~components
\(\sigma_{k,\mathrm{a}}\) of~\(\sigma_k\) are large.  The corresponding
Gamma functions in the Coulomb branch integral are expanded as powers of
\(\I\sigma_{k,\mathrm{a}}^\pm\).  Afterwards, one can close contours of
all~\(\sigma_j\) for \(j<k\) as we have done to find the s-channel
expansion.  The Gamma functions which were expanded in powers of
\(\I\sigma_{k,\mathrm{a}}^\pm\) do not contribute poles, hence the sum
over poles is non-empty only if \(\Nc_k-\nu \geq \Nc_{k-1} \geq \cdots
\geq \Nc_1\).  As a result, \(\nu\leq \Nc_k-\Nc_{k-1} = K_k\).  Changing
variables to \(\mu=K_k-\nu\), we deduce
\begin{equation}\label{Quivers-SQCDA-tch-expo}
  Z = \abs{1-\hat{z}_k}^{2[-K_k(1+\I m_{kk})]}
  \sum_{\mu=0}^{K_k}
  \abs{1-\hat{z}_k}^{2[\mu(1+\I m_{kk})-(K_k-\mu)(K_{k+1}-\mu)\I m_{kk}]}
  \bigl(\text{series}\bigr)
\end{equation}
where \((\text{series})\) denote series in non-negative integer powers
of \(1-\hat{z}_k\) and \(\overline{1-\hat{z}_k}\).  In
Section~\ref{sec:SeibergQ}, we relate the \(S^2\)~partition function of
the quiver gauge theory we are studying to another such partition
function, with in particular \(K_k\leftrightarrow K_{k+1}\).  This
restricts the sum over~\(\mu\) to \(0\leq\mu\leq\min(K_k,K_{k+1})\).  On
the Toda CFT side, the limit is \(x_k\to x_{k+1}\), and we are
interested in the fusion of two degenerate punctures, labeled by two
symmetric or two antisymmetric representations since
\(\epsilon_k=\epsilon_{k+1}\).  Given that
\begin{equation}
  \begin{aligned}
    \repr(\omega_K) \otimes \repr(\omega_L)
    & = \bigoplus_{\mu=0}^{\min(K,L)}
    \repr\bigl(\omega_{K+L-\mu} + \omega_\mu\bigr) \,,
    \\
    \repr(K h_1) \otimes \repr(L h_1)
    & = \bigoplus_{\mu=0}^{\min(K,L)}
    \repr\bigl((K+L-\mu) h_1 + \mu h_2\bigr) \,,
  \end{aligned}
\end{equation}
the Toda CFT exponents of \(\abs{x_{k+1}-x_k}^2\) are
\begin{align}
  & \begin{aligned}[t]
    & \dimToda\bigl(-b\omega_{K+L-\mu}-b\omega_\mu\bigr)
    - \dimToda(-b\omega_K) - \dimToda(-b\omega_L)
    \\
    & \quad =
    \frac{KL}{\Nf} b^2
    - \mu b^2
    +(K-\mu) (L-\mu) (b^2+1)
    \qquad \text{if \(\epsilon_k=\epsilon_{k+1}=-1\)} \,,
  \end{aligned}
  \\
  & \begin{aligned}[t]
    & \dimToda\bigl(-(K+L-\mu)bh_1-\mu bh_2\bigr)
    - \dimToda(-Kbh_1) - \dimToda(-Lbh_1)
    \\
    & \quad =
    \frac{KL}{\Nf} b^2
    + \mu (b^2+1)
    - (K-\mu) (L-\mu) b^2
    \qquad \text{if \(\epsilon_k=\epsilon_{k+1}=+1\)} \,,
  \end{aligned}
\end{align}
where \(K\) and~\(L\) are \(K_k\) and~\(K_{k+1}\).  Again, the Toda CFT
exponents match with the gauge theory exponents up to precisely
\(\gamma_{k(k+1)}\) given in~\eqref{Quivers-gamma2}.

Note that matching the number of distinct powers of \(1-\hat{z}_k\) in
gauge theory with the number of internal momenta in the fusion of
punctures at \(x_{k-1}\) and~\(x_k\) is enough to fix the relation
between the signs \(\{\eta_j\}\) and~\(\{\epsilon_j\}\).  When the
adjoint~\(X_j\) can be integrated out (\(\eta_j=-1\)), the gauge theory
involves two exponents only, and correspondingly the two neighboring
punctures are labeled by different types of representations (one is
symmetric and the other antisymmetric), whose fusion has two terms.
When the adjoint~\(X_j\) remains (\(\eta_j=+1\)), the gauge theory
involves many exponents, and the two punctures have the same type, hence
a fusion with many terms.

The situation is very similar in the limit \(x_n = \hat{z}_n \to 1\).
The gauge theory involves two exponents if \(\eta_n = -1\), and
\(\Nc_n-\Nc_{n-1}\) if \(\eta_n = +1\).  On the Toda CFT side, the
fusion of the semidegenerate momentum~\(\hat{m}\) with the degenerate
\(-b\Omega(K_n,\epsilon_n)\) gives two momenta if \(\epsilon_n = -1\),
and \(K_n\)~if \(\epsilon_n = +1\).  Hence \(\epsilon_n = \eta_n\) and
\(K_n = \Nc_n-\Nc_{n-1}\).  Calculating the exponents and comparing them
fixes \(\hat{m}\) to~\eqref{Quivers-alpha} and \(\gamma_n\)
to~\eqref{Quivers-gamma1}.

All other exponents \(\gamma_{ij}\) and~\(\gamma_j\) are fixed thanks to
the identification of permutations of degenerate punctures with gauge
theory dualities found in Section~\ref{sec:SeibergQ-perm}.

\subsubsection{Arbitrary Toda Degenerates}
\label{sec:Quivers-fuse}

We now consider the matching~\eqref{Quivers-matching} in the case where
\(K_{j+1} \geq K_j\) for \(1\leq j\leq n-1\), and \(\epsilon_j=-1\) for
all \(1\leq j\leq n\), that is, \(\eta_n=-1\) and \(\eta_j=+1\) for all
\(1\leq j\leq n-1\).  In the course of fixing parameters for the
matching, we have found that the expansion near \(x_k = x_{k+1}\)
involves the \(\min(K_k,K_{k+1})=K_k\)
powers~\eqref{Quivers-SQCDA-tch-expo} of \(x_{k+1}-x_k = x_{k+1}
(1-\hat{z}_k)\), for \(1\leq k\leq n-1\).  Given our assumptions, these
exponents all have a non-negative real part (the vortex partition
functions contribute integer exponents \(\nu\geq 0\)):
\begin{equation}
  \Re\bigl((K_k-\mu)b^2 + (K_k-\mu)(K_{k+1}-\mu)(1+b^2) + \nu\bigr)
  \geq 0 \,.
\end{equation}
The real part vanishes if and only if \(\mu=K_k\) and \(\nu=0\).
As \(\hat{z}_k \to 1\), only the term with \(\mu=K_k\) and \(\nu=0\)
remains.  On the Toda CFT side, this limit selects the fusion%
\begin{equation}
  \repr(\omega_{K_{k+1}}) \otimes \repr(\omega_{K_k})
  \longrightarrow \repr(\omega_{K_{k+1}}+\omega_{K_k}) \,.
\end{equation}

We can carry this process further and take the fusion of arbitrarily
many antisymmetric degenerate operators.  For definiteness, let us send
\(x_k \to x_n\) for \(k\) going from \(n-1\) to~\(1\), in this order.  At a
given step \(x_k \to x_n\), the Littlewood--Richardson rule gives
\begin{equation}
  \repr(\Omega) \otimes \repr(\omega_{K_k})
  =
  \sideset{}{'}\bigoplus_{h\in\repr(\omega_{K_k})}
  \repr\bigl(\Omega+h\bigr)
\end{equation}
with a sum running over weights~\(h\) of \(\repr(\omega_{K_k})\) such
that \(\Omega+h\) is a dominant weight.  In our setting, \(\Omega =
\omega_{K_n}+\cdots+\omega_{K_{k+1}}\).  The power of \(x_n-x_k\) for a
weight~\(h\) is
\begin{equation}
  \dimToda(-b\Omega-bh)-\dimToda(-b\Omega)-\dimToda(-b\omega_{K_k})
  + \sum_{l=k+1}^n \gamma_{kl}
  =
  b \vev{Q,\omega_{K_k}-h} + b^2 \vev{\Omega,\omega_{K_k}-h} \,,
\end{equation}
which has a positive real part unless \(h=\omega_{K_k}\), in which case
it vanishes.  Thus, setting \(x_k=x_n\) selects precisely the fusion of
\(-b\Omega\) and \(-b\omega_{K_k}\) into \(-b\Omega-b\omega_{K_k}\).

Any dominant weight~\(\Omega\) is a sum of fundamental weights, hence
the four-point function of two generic and one semi-degenerate vertex
operators with an arbitrary degenerate vertex
operator~\(\widehat{V}_{-b\Omega}\) is equal to the partition function
of an \(S^2\)~surface operator built from a certain quiver on~\(S^4_b\),
with some fine-tuned FI parameters and theta angles.  Namely, decomposing
\(\Omega=\omega_{K_n}+\cdots+\omega_{K_1}\) with \(K_n\geq\cdots\geq
K_1\), we find
\begin{equation}\label{Quivers-matching-fused}
  Z_{S^2\subset S^4_b}^{\prod_k U(\Nc_k), W_\eta}\bigl(m,z,\bar{z}\bigr)
  = A a(x) a(\bar{x})
  \vev*{\widehat{V}_{\alpha_\infty}(\infty)
    \widehat{V}_{\hat{m}}(1)
    \widehat{V}_{-b\Omega}(x,\bar{x})
    \widehat{V}_{\alpha_0}(0)} \,,
\end{equation}
where\footnote{As explained below~\eqref{part-typical-matching}, the
  factor \(A a(x) a(\bar{x})\) can be absorbed into the partition
  function.}  \(\Nc_k = \sum_{j=1}^k K_j\) for \(1\leq k\leq n\),
\begin{equation}
  \begin{aligned}
    & \eta_n = -1 \quad\text{and}\quad
    \hat{z}_n = x \,,
    \\
    & \eta_k = +1 \quad\text{and}\quad
    \hat{z}_k = 1
    \quad \text{for \(1\leq k\leq n-1\)} \,,
  \end{aligned}
\end{equation}
and the momenta \(\alpha_0\), \(\alpha_\infty\), and \(\hat{m}\) are
given by~\eqref{Quivers-alpha}.  The factor
\begin{equation}
  a(x) a(\bar{x}) = \abs{x}^{2\beta} \abs{1-x}^{2\gamma}
\end{equation}
differs from~\eqref{Quivers-a} and has the exponents
\begin{align}
  \beta
  & = \vev{Q,-b\Omega}
  - \frac{\Nc_n}{\Nf} \sum_{t=1}^{\Nf} \I m_t
  - \sum_{j=1}^{n-1} \Nc_j b^2 \,,
  \\
  \gamma
  & = - b^2 \frac{\Nc_n(\Nf-\Nc_n)}{\Nf}
  + \frac{\Nc_n}{\Nf} \sum_t (\I m_t+\I\anti{m}_t) \,.
\end{align}
Finally, the overall constant~\(A\) is identical to the constant
in~\eqref{Quivers-matching}, given by~\eqref{Quivers-A}, because the
three-point functions
\(\widehat{C}_{-b\omega_K,-b\Omega}^{-b(\Omega+\omega_K)}\) are in
fact all equal to~\(1\).  Incidentally, in the case \(\Omega = \Nc
h_1\), the factor \(Aa(x)a(\bar{x})\) coincides with the factor we
found in the matching between the same Toda CFT correlator and the
SQCDA surface operator.  Thus, SQCDA and the
\(U(\Nc)\times\cdots\times U(1)\) theory which appears in this
matching have equal \(S^2\)~partition functions.  The relation between
these theories may run deeper.

Since the partition function in~\eqref{Quivers-matching-fused} is known
explicitly, the matching gives an explicit expression for the Toda CFT
four-point function of two full, one simple, and a degenerate
operator~\(\widehat{V}_{-b\Omega}\).  The Higgs branch expansion
of~\(Z\) provides conformal blocks as explicit series.  From the Coulomb
branch representation of~\(Z\) for \(\hat{m}=0\) one can extract
integral expressions for the three-point function of a degenerate
operator~\(\widehat{V}_{-b\Omega}\) with generic vertex operators.
These expressions typically involve fewer integrals than expressions
obtained form the Coulomb gas formalism, but we have not investigated
this direction further.

More generally, any Toda CFT \((p+3)\)-point function with two generic
and one semi-degenerate operators at \(0\), \(\infty\) and~\(1\), and
\(p\)~arbitrary degenerate
operators~\(\widehat{V}_{-b\Omega_l}(x_l,\bar{x}_l)\) is equal to the
partition function of a surface operator describing a certain quiver
gauge theory.  This matching directly derives from the
matching~\eqref{Quivers-matching}, with only antisymmetric degenerate
operators, and taking all but~\(p\) of the~\(\hat{z}\) equal to~\(1\).
Concretely, we express each highest weight as
\begin{equation}
  \Omega_l = \sum_{j=1}^{c_l} \omega_{K_{l,j}} \,,
\end{equation}
where \(c_l\)~is the number of columns in the Young diagram
of~\(\Omega_l\) and \(K_{l,c_l} \geq \cdots \geq K_{l,2} \geq K_{l,1}
\geq 0\) are the number of boxes in each column.  We then define an order on the pairs \(\bigl\{ (l,j) \bigm|
1\leq l\leq p, 1\leq j\leq c_l \bigr\}\) by \((k,i)\leq (l,j)\) if
\(k<l\) or if \(k=l\) and \(i\leq j\).  The gauge group is then
\begin{equation}
  \prod_{l=1}^p \prod_{j=1}^{c_l} U(\Nc_{l,j})
  \quad\text{where}\quad
  \Nc_{l,j} = \sum_{(k,i)\leq (l,j)} K_{k,i} \,.
\end{equation}
The matter content of the theory consists as usual of pairs of bifundamental
chiral multiplets between neighboring nodes, namely \((k,i)
\leftrightarrow (k,i+1)\) and \((k,c_k) \leftrightarrow (k+1,1)\), of an
adjoint chiral multiplet for every node except \(U(\Nc_{p,c_p})\), and of
\(\Nf\)~fundamental and \(\Nf\)~antifundamental chiral multiplets for
this last node \(U(\Nc_{p,c_p})\).  Complexified FI parameters
associated to each node \(U(\Nc_{l,j})\) are given by
\begin{equation}
  \hat{z}_{l,j} = \begin{cases}
    1 & \text{if \(1\leq j<c_l\)} \,, \\
    x_l / x_{l+1} & \text{if \(j = c_l\)} \,,
  \end{cases}
\end{equation}
where \(x_{p+1} = 1\).  Detailed factors can be read from the
matching~\eqref{Quivers-matching} using this gauge theory data.

All in all, we have identified the \(\Nsusy=(2,2)\) surface operator
corresponding to the insertion of an arbitrary set of degenerate vertex
operators in a Toda CFT three-point function.  It would be interesting
to calculate the expectation values of such surface operators in an
interacting four dimensional theory of class~S\@.

\section{Gauge Theory Dualities as Toda Symmetries}
\label{sec:Dualities}

In this section, we probe low-energy dualities between two dimensional
\(\Nsusy=(2,2)\) gauge theories through the correspondence of surface
operators with Toda CFT degenerate operators.  In
Section~\ref{sec:Seiberg}, we show that some pairs of \(\Nsusy=(2,2)\)
SQCD theories have equal partition functions on~\(S^2\), as predicted by
the Seiberg duality.  The equality is realized as the symmetry of Toda
CFT correlators under conjugation of momenta.  In
Section~\ref{sec:SeibergW}, we consider \(\Nsusy=(2,2)\) SQCDA theories
with superpotentials.  We focus first on a generalization of
\(\Nsusy=(2,2)^*\) SQCD, and find the analogue of Seiberg duality for
such theories, which amounts to crossing symmetry of a Toda CFT
correlator.  We then obtain the Kutasov--Schwimmer duality between
\(\Nsusy=(2,2)\) SQCDA theories with a \(W = \Tr X^{l+1}\)
superpotential as conjugation of momenta.  In Section~\ref{sec:SeibergQ}
finally, we describe the groupoid of Seiberg dualities for some quiver
gauge theories: some dualities correspond to permutations of degenerate
punctures on the Toda CFT side, and in one case to momentum conjugation.

We check all dualities by proving that the \(S^2\)~partition functions
of dual theories are equal up to simple ambiguous factors: besides the Toda CFT
approach, we provide direct proofs in Appendix~\ref{app:Proof}.  In all
cases, the factors can be absorbed in either one of the dual partition
functions through the ambiguities described
below~\eqref{part-typical-matching}, namely a renormalization scheme
ambiguity, a constant shift of gauge multiplet scalars, and a flavour FI parameter.

\subsection{Seiberg Duality as Momentum Conjugation}
\label{sec:Seiberg}

Seiberg duality relates theories with different gauge groups but the
same flavour symmetry.  In our two dimensional \(\Nsusy=(2,2)\) context,
it is expected that \(U(\Nc)\) SQCD with \(\Nf\)~fundamental and
\(\anti{\Nf}\leq\Nf\)~antifundamental chiral multiplets is dual to
\(U(\Nf-\Nc)\) SQCD with the same number of chiral multiplets and
\(\Nf\anti{\Nf}\) additional free mesons, for an appropriate choice of
twisted masses.  In the case \(\anti{\Nf}\leq\Nf-2\), the series giving
vortex partition functions were proven term by term to be equal
in~\cite{Benini:2012ui}, and the relation for \(S^2\)~partition
functions was deduced.  For \(\anti{\Nf}=\Nf-1\) or \(\anti{\Nf}=\Nf\),
vortex partition functions differ by a non-trivial factor, found
numerically in~\cite[Appendix~F]{Honda:2013uca}.  Our direct proof in
Appendix~\ref{app:Proof-SQCD} (similar to that of~\cite{Benini:2014mia}
found independently) is technical and by itself provides no insight on
the factor.  In contrast, the factor appears naturally in the proof we
give here via the Toda CFT\@.

We denote by \(m_s\) and~\(\anti{m}_s\) the twisted masses (with
\(R\)-charges) in the electric theory, and by \(m_s^\dual\)
and~\(\anti{m}_s^\dual\) those in the dual magnetic theory.  We shall prove
that
\begin{equation}\label{Seiberg-Z}
  Z_{S^2}^{U(\Nc),\Nf,\anti{\Nf}} (z,\bar{z},m,\anti{m})
  =
  a(z,\bar{z})
  \prod_{s=1}^{\Nf} \prod_{t=1}^{\anti{\Nf}}
  \Bigl[ \gamma(-\I m_s-\I\anti{m}_t) \Bigr]
  Z_{S^2}^{U(\Nc^\dual),\Nf,\anti{\Nf}} (z^\dual,\bar{z}^\dual,m^\dual,\anti{m}^\dual)
\end{equation}
where \(z\) and~\(z^\dual\) are renormalized values at the scale~\(\ell\) of
the sphere, and dual parameters are \(\Nc^\dual=\Nf-\Nc\), \(z^\dual =
(-1)^{\anti{\Nf}} z\), \(\bar{z}^\dual = (-1)^{\anti{\Nf}}
\bar{z}\), \(m_s^\dual = \frac{\I}{2} - m_s\), and \(\anti{m}_s^\dual =
\frac{\I}{2} - \anti{m}_s\).  The factor
\begin{equation}\label{Seiberg-a}
  a(z,\bar{z}) =
  \begin{cases}
    \abs{z}^{2\delta_0}
    & \text{if \(\anti{\Nf}\leq\Nf-2\)} \,, \\
    \abs{z}^{2\delta_0} e^{(-1)^{\Nf+\Nc-1} (z - \bar{z})}
    & \text{if \(\anti{\Nf}=\Nf-1\)} \,, \\
    \abs{z}^{2\delta_0} \abs*{1-(-1)^{\Nf+\Nc-1} z}^{2\delta_1}
    & \text{if \(\anti{\Nf}=\Nf\)}
  \end{cases}
\end{equation}
is given in terms of the exponents
\begin{align}\label{Seiberg-delta0}
  \delta_0 = \gamma_0 - \gamma_0^\dual
  & = - \frac{\Nf-\Nc}{2} - \sum_{s=1}^{\Nf} \I m_s \,,
  \\\label{Seiberg-delta1}
  \delta_1 = \gamma_1 - \gamma_1^\dual
  & = \Nf - \Nc + \sum_{s=1}^{\Nf} (\I m_s + \I\anti{m}_s) \,,
\end{align}
which we will obtain from the exponents~\(\gamma_i\) in the
matching~\eqref{SQCD-matching}, and their duals~\(\gamma_i^\dual\).  The
factor \(a(z,\bar{z})\) could be absorbed into~\(Z\) in three parts as
discussed below~\eqref{part-typical-matching}.  First, a renormalization
scheme ambiguity absorbs any factor independent of twisted masses.
Next, a global gauge transformation shifts the partition function by any
power of~\(\abs{z}^2\).  A last factor (present only for
\(\anti{\Nf}=\Nf\)) can be absorbed by turning on an FI parameter for
the \(U(1)\) flavour group under which fundamental and antifundamental
chiral multiplets have the same charge.

The product of gamma functions in~\eqref{Seiberg-Z} is the (one-loop
determinant) contribution from \(\Nf\anti{\Nf}\) free mesons with
twisted masses \(m_s + \anti{m}_t = \I - m^\dual_s - \anti{m}^\dual_t\).  These
twisted masses are equal to those of the mesons \(\anti{\quark}_t
\quark_s\), where \(\quark_s\) and~\(\anti{\quark}_t\) are fundamental
and antifundamental quarks of the electric theory.  In the magnetic
theory, the twisted masses derive from the superpotential coupling \(W =
\anti{\quark}^\dual M \quark^\dual\), which has total \(R\)-charge~\(2\), hence
total (complexified) twisted mass~\(\anti{m}^\dual_t + (\I - m^\dual_s -
\anti{m}^\dual_t) + m^\dual_s = \I\).

Applied twice, the duality~\eqref{Seiberg-Z} yields the original theory,
since parameters are mapped back to those of the electric theory.  An
immediate consistency check is thus%
\begin{equation}
  \gamma(-\I m_s-\I\anti{m}_t)
  \gamma(-\I m_s^\dual-\I\anti{m}_t^\dual)
  =
  \frac{\Gamma(-\I m_s-\I\anti{m}_t)}{\Gamma(1+\I m_s+\I\anti{m}_t)}
  \frac{\Gamma(1+\I m_s+\I\anti{m}_t)}{\Gamma(-\I m_s-\I\anti{m}_t)}
  = 1
\end{equation}
and that the \(a(z,\bar{z})\) factors cancel thanks to
\begin{gather}
  \delta_0^\dual
  = - \frac{\Nf-\Nc^\dual}{2} - \sum_{s=1}^{\Nf} \I m_s^\dual
  = - \frac{\Nc}{2} + \sum_{s=1}^{\Nf} \I m_s + \frac{\Nf}{2}
  = - \delta_0 \,,
  \\
  \delta_1^\dual
  = \Nf - \Nc^\dual + \sum_{s=1}^{\Nf} (\I m_s^\dual + \I\anti{m}_s^\dual)
  = \Nc - \sum_{s=1}^{\Nf} (\I m_s + \I\anti{m}_s) - \Nf
  = - \delta_1 \,,
\end{gather}
and, for \(\anti{\Nf}=\Nf-1\), \((-1)^{\Nf+\Nc^\dual-1} (z^\dual - \bar{z}^\dual) = - (-1)^{\Nf+\Nc-1}
(z - \bar{z})\).  A second consistency check, in the case
\(\anti{\Nf}=\Nf\), is the symmetry under charge conjugation
\(z\leftrightarrow 1/z\), \(\bar{z}\leftrightarrow 1/\bar{z}\), and \(\I
m_s\leftrightarrow\I\anti{m}_s\).  We find that \(\delta_1\)~is left
unchanged, and that \(\delta_0\) is mapped to \(-\delta_0-\delta_1\),
consistent with \(a(1/z,1/\bar{z}) = \abs{z}^{-2\delta_0-2\delta_1}
\abs{1-(-1)^{\Nf+\Nc-1} z}^{2\delta_1}\).

\subsubsection{Momentum Conjugation for \(\anti{\Nf}=\Nf\)}
\label{sec:Seiberg-Nf-Nf}

To derive the Seiberg duality relation~\eqref{Seiberg-Z} for
\(\anti{\Nf}=\Nf\), we rely on the matching~\eqref{SQCD-matching}
relating the \(S^2\)~partition function of \(U(\Nc)\) SQCD to a Toda CFT
four-point function:
\begin{equation}
  Z_{S^2\subset S^4_b}^{U(\Nc), \anti{\Nf}=\Nf}(m,\anti{m},z,\bar{z})
  = A \abs{x}^{2\gamma_0} \abs{1-x}^{2\gamma_1}
  \vev*{\widehat{V}_{\alpha_\infty}(\infty)\widehat{V}_{\hat{m}}(1)
    \widehat{V}_{-b\omega_\Nc}(x,\bar{x})\widehat{V}_{\alpha_0}(0)} \,.
\end{equation}
The four-point function features two generic operators
\(\widehat{V}_\alpha\), one semi-degenerate operator
\(\widehat{V}_{\hat{m}}\), and the degenerate
operator~\(\widehat{V}_{-b\omega_\Nc}\) inserted at \(x = (-1)^{\Nf+\Nc-1} z\)
and labeled by the highest weight~\(\omega_\Nc\) of the \(\Nc\)-th
antisymmetric representation of~\(A_{\Nf-1}\).  The relation between
gauge theory twisted masses \(m\) and~\(\anti{m}\), and Toda CFT momenta
\(\alpha_0\), \(\alpha_\infty\), and~\(\hat{m}\) is given in
Section~\ref{sec:SQCD}.

Toda CFT correlators are invariant under changing all momenta to their
conjugate, that is, applying the \(\bbC\)-linear transformation \(h_s
\to h_s^\conj = - h_{\Nf+1-s}\) which maps weights of a representation of
\(A_{\Nf-1}\) to weights of the conjugate representation.  This
transformation maps the degenerate momentum \(-b\omega_\Nc\)
to\footnote{The third step uses that the weights~\(h_s\) of the
  fundamental representation of~\(A_{\Nf-1}\) sum to zero.}
\begin{equation}
  (-b\omega_\Nc)^\conj
  = - \sum_{s=1}^\Nc b h_s^\conj
  = \sum_{\mathsemiclap{4}{s=\Nf-\Nc+1}}^{\Nf} b h_s
  = - \sum_{s=1}^{\Nf-\Nc} b h_s
  = - b \omega_{\Nf-\Nc} \,,
\end{equation}
which is precisely the degenerate momentum appearing in the description
of the Seiberg dual SQCD theory.  The semi-degenerate momentum \(\hat{m}
= (\varkappa + \Nc b) h_1\) becomes \(\hat{m}^\conj = - (\varkappa + \Nc
b) h_{\Nf}\), which is not along~\(h_1\).  However, the Weyl reflection
defined by the cyclic permutation of \(\intset{1}{\Nf}\) maps
\(\hat{m}^\conj\) to
\begin{equation}\label{Seiberg-hatmD}
  \left[\Nf \left(b+\frac{1}{b}\right)-\varkappa-\Nc b\right] h_1
  = (\varkappa^\dual+\Nc^\dual b) h_1
  = \hat{m}^\dual
  \,,
\end{equation}
where \(\varkappa^\dual = \Nf/b - \varkappa\).  Indeed,
\(\vev{\hat{m}^\conj - Q, h_s} = \vev{\hat{m}^\dual - Q, h_{s+1}}\) for
all \(1\leq s\leq\Nf-1\), and \(\vev{\hat{m}^\conj - Q, h_{\Nf}} =
\vev{\hat{m}^\dual - Q, h_1}\).  Finally, the generic momenta
\(\alpha_0\) and~\(\alpha_\infty\) remain generic after conjugation, and
we have
\begin{equation}
  \vev{\alpha^\conj - Q, h_p}
  = \vev{\alpha - Q^\conj, h_p^\conj}
  = - \vev{\alpha - Q, h_{\Nf+1-p}} \,,
\end{equation}
where we used that \(\vev{\alpha_1, \alpha_2} = \vev{\alpha_1^\conj,
  \alpha_2^\conj}\) and that \(Q = Q^\conj\).  A Weyl reflection then
permutes \(\vev{\alpha-Q, h_{\Nf+1-p}} \to \vev{\alpha-Q,h_p}\), hence
conjugation followed by this reflection simply changes \(\alpha\to
2Q-\alpha\).

We thus find that conjugation of all momenta (with subsequent Weyl
reflections) relates two correlators which correspond to SQCD with
\(U(\Nc)\) and \(U(\Nf-\Nc)\) gauge groups.  Converting the relation
between momenta to gauge theory variables, we find \(m^\dual_s =
\frac{\I}{2} - m_s\) and \(\anti{m}^\dual_s = \frac{\I}{2} - \anti{m}_s\),
as we claimed below~\eqref{Seiberg-Z}.\footnote{A global \(U(1)\) gauge
  transformation is identical to the flavour symmetry which shifts \(\I
  m_s\) and \(-\I\anti{m}_s\) by the same amount.  This has no physical
  effect: the Toda correlator is invariant, and the partition function
  is multiplied by a power of~\(\abs{z}^2\).  Dual twisted masses are
  only defined up to such a shift, which also alters~\(\delta_0\).}

In our normalization, both generic and non-degenerate operators are Weyl
reflection invariant, without reflection amplitudes.  The two Toda CFT
correlators are thus equal, and we divide the factor relating the
\(S^2\)~partition functions and Toda CFT correlator for one theory by
the factor for the other theory to find (for \(\anti{\Nf}=\Nf\))
\begin{equation}
  Z_{S^2}^{U(\Nc)} (z,\bar{z},m,\anti{m})
  =
  \frac{Z_{S^4_b}^{\free,\dual} A \,
    \abs{x}^{2\gamma_0} \abs*{1-x}^{2\gamma_1}}
  {Z_{S^4_b}^\free A^\dual \,
    \abs{x}^{2\gamma_0^\dual} \abs*{1-x}^{2\gamma_1^\dual}}
  Z_{S^2}^{U(\Nc^\dual)} (z^\dual,\bar{z}^\dual,m^\dual,\anti{m}^\dual) \,.
\end{equation}
We recognize the factor \(a(z,\bar{z}) =
\abs{z}^{2\gamma_0-2\gamma_0^\dual} \abs{1-(-1)^{\Nf+\Nc-1}
  z}^{2\gamma_1-2\gamma_1^\dual}\) announced in~\eqref{Seiberg-a}.  The
ratio \(A/A^\dual\) simplifies:
\begin{equation}
  \frac{A}{A^\dual}
  = \frac{b^{\Nc\Nf(1+b^2)-\Nc^2 b^2-2\Nc\sum_{s=1}^{\Nf} (1+\I m_s+\I\anti{m}_s)}}
    {b^{\Nc^\dual\Nf(1+b^2)-(\Nc^\dual)^2 b^2-2\Nc^\dual\sum_{s=1}^{\Nf} (1+\I m_s^\dual+\I\anti{m}_s^\dual)}}
  = b^{-\Nf\sum_{s=1}^{\Nf} (1+2\I m_s+2\I\anti{m}_s)} \,.
\end{equation}
The hypermultiplets masses~\eqref{part-mst} involved in the
\(S^4_b\)~partition functions~\eqref{ZS4-free} associated to the
electric and magnetic theories are
\begin{align}
  m_{st} & = \I\frac{1-b^2}{2b}-\frac{1}{b}(m_s+\anti{m}_t) \,, \\
  m_{st}^\dual & = \I\frac{1-b^2}{2b}-\frac{1}{b}(\I-m_s-\anti{m}_t)
  = -\I b - m_{st} \,,
\end{align}
thus the constant factor is
\begin{align}
  & \frac{Z_{S^4_b}^{\free,\dual} A}{Z_{S^4_b}^\free A^\dual}
  =
  \frac{A}{A^\dual}
  \prod_{s,t}
  \frac{\Upsilon(\frac{b}{2} + \frac{1}{2b} - \I m_{st})}
    {\Upsilon(\frac{b}{2} + \frac{1}{2b} - \I m_{st}^\dual)}
  =
  \frac{A}{A^\dual}
  \prod_{s,t}
  \frac{\Upsilon(\frac{b}{2} + \frac{1}{2b} + \I m_{st})}
    {\Upsilon(\frac{b}{2} + \frac{1}{2b} + \I m_{st} - b)}
  \\\nonumber
  & \qquad = b^{-\Nf\sum_{s=1}^{\Nf} (1+2\I m_s+2\I\anti{m}_s)}
  \prod_{s,t} \biggl[ b^{b^2 - 2b \I m_{st}}
  \gamma\biggl(\frac{1-b^2}{2} + b \I m_{st}\biggr) \biggr]
  = \prod_{s,t} \gamma(-\I m_s-\I\anti{m}_t) \,.
\end{align}
The one-loop determinants of free mesons appear here thanks to the shift
by~\(b\) in \(\I m_{st}^\dual = b-\I m_{st}\), which relies on the shift
between \(m_{st}\) and \(\frac{-1}{b}(m_s+\anti{m}_t)\)
in~\eqref{part-mst}.  We obtain this constant factor more directly for
any \(\anti{\Nf}\leq\Nf\) in the next section.

\subsubsection{Decoupling Multiplets Towards \(\anti{\Nf}<\Nf\)}
\label{sec:Seiberg-irreg}

We could find analogous Seiberg duality relations for
\(\anti{\Nf}<\Nf\) via the matching of Section~\ref{sec:irreg} with Toda
irregular punctures, but those cases are also easily accessed by taking
some twisted masses of antifundamental multiplets to be very large in
the \(\anti{\Nf}=\Nf\) duality.  The reverse process, which decreases
\(\Nf>\anti{\Nf}\) by giving some fundamental multiplets large twisted
masses, is more difficult, and must be significantly altered to reach
the case \(\Nf=\anti{\Nf}\) in Appendix~\ref{app:Proof-SQCD}.

Our starting point to prove~\eqref{Seiberg-Z} is the Higgs branch
decomposition of the \(S^2\)~partition function of
interest~\cite{Benini:2012ui,Doroud:2012xw}:
\begin{gather}\nonumber
  Z^{U(\Nc),\Nf,\anti{\Nf}}
  =
  \mspace{-21mu}
  \sum_{p_1<\cdots<p_\Nc}^{\Nf} \mspace{-21mu}
  (z\bar{z})^{-\sum_{j=1}^\Nc \I m_{p_j}}
  \, Z^{\Nf,\anti{\Nf}}_{\oneloop,\{p\}} \,
  f^{\sch,\Nf,\anti{\Nf}}_{\{p\}}\!\bigl((-1)^{\Nf+\Nc-1} z\bigr)
  f^{\sch,\Nf,\anti{\Nf}}_{\{p\}}\!\bigl((-1)^{\anti{\Nf}+\Nc-1} \bar{z}\bigr)
  \,,
  \\\label{Seiberg-Zs}
  Z^{\Nf,\anti{\Nf}}_{\oneloop,\{p\}}
  = \prod_{j=1}^\Nc
  \frac{\prod_{s\not\in\{p\}}^{\Nf}\gamma(-\I m_s+\I m_{p_j})}
    {\prod_{s=1}^{\anti{\Nf}}\gamma(1+\I\anti{m}_s+\I m_{p_j})}
  \,,\quad
  f^{\sch,\Nf,\anti{\Nf}}_{\{p\}}(x) =
  \sum_{k=0}^\infty \frac{x^k}{k!} f^{\sch,\Nf,\anti{\Nf}}_{\{p\},k}
  \,,
  \\\nonumber
  f^{\sch,\Nf,\anti{\Nf}}_{\{p\},k}
  = k! \!\!
  \sum_{k_1+\cdots+k_\Nc=k}
  \prod_{j=1}^\Nc\Biggl[
  \frac
    {\prod_{s=1}^{\anti{\Nf}}
      (-\I\anti{m}_s-\I m_{p_j})_{k_j}}
    {k_j! \prod_{i\neq j}^\Nc (\I m_{p_i}-\I m_{p_j}-k_i)_{k_j}
      \prod_{s\not\in\{p\}}^{\Nf} (1+\I m_s-\I m_{p_j})_{k_j}}
  \Biggr]
  \,,
\end{gather}
which generalizes~\eqref{SQCD-Zs} to \(\anti{\Nf}<\Nf\).  The series
\(f^{\sch,\Nf,\anti{\Nf}}_{\{p\}}(x)\) converge on the unit disc if
\(\anti{\Nf}=\Nf\), and on the whole complex plane if
\(\anti{\Nf}<\Nf\).  We shall equate the term of~\eqref{Seiberg-Zs}
labeled by~\(\{p\}\subseteq\intset{1}{\Nf}\) with the term labeled by
the complement~\(\{p\}^\complement\) for the dual theory.  The powers
of~\(\abs{z}^2\) match:
\begin{equation}
  -\sum_{j=1}^\Nc \I m_{p_j}
  =
  -\sum_{s=1}^{\Nf} \I m_s
  +\sum_{s\in\{p\}^\complement} \I m_s
  =
  - \sum_{s=1}^{\Nf} \I m_s
  - \frac{\Nf-\Nc}{2}
  - \sum_{s\in\{p\}^\complement}\I m_s^\dual
  =
  \delta_0
  - \sum_{s\in\{p\}^\complement}\I m_s^\dual \,.
\end{equation}
The constant is fixed as the ratio of one-loop
determinants~\(Z_\oneloop\)
\begin{equation}\label{Seiberg-1l}
  \frac{Z_{\oneloop,\{p\}}^{\Nf,\anti{\Nf}}}
    {Z_{\oneloop,\{p\}^\complement}^{\Nf,\anti{\Nf},\dual}}
  =
  \prod_{s=1}^{\anti{\Nf}}
  \frac
    {\prod_{t\in\{p\}^\complement} \gamma(1+\I\anti{m}_s^\dual+\I m_t^\dual)}
    {\prod_{t\in\{p\}} \gamma(1+\I\anti{m}_s+\I m_t)}
  = \prod_{s=1}^{\anti{\Nf}} \prod_{t=1}^{\Nf} \gamma(-\I\anti{m}_s-\I m_t) \,,
\end{equation}
which is independent of~\(\{p\}\).  There remains to match vortex
partition functions:
\begin{equation}\label{Seiberg-fs}
  f^{\sch,\Nf,\anti{\Nf}}_{\{p\}}(x)
  = a(x) f^{\sch,\Nf,\anti{\Nf},\dual}_{\{p\}^\complement}(x^\dual) \,,
\end{equation}
where
\begin{equation}
  a(x) = \begin{cases}
    (1-x)^{\delta_1} & \text{if \(\anti{\Nf}=\Nf\)} \,, \\
    e^x & \text{if \(\anti{\Nf}=\Nf-1\)} \,, \\
    1 & \text{if \(\anti{\Nf}\leq\Nf-2\)} \,,
  \end{cases}
\end{equation}
and \(x^\dual = (-1)^{\Nf+\Nc^\dual-1} z^{\dual} = (-1)^{\anti{\Nf}+\Nc-1} z = (-1)^{\Nf-\anti{\Nf}} x\).  From the case \(\anti{\Nf}=\Nf\)
studied in the previous section, we now derive the relations for
\(\anti{\Nf}<\Nf\) by taking a limit where \(\Nf-\anti{\Nf}\)
antifundamental chiral multiplets are given large twisted masses.  We
give a proof independent of the Toda CFT matching in
Appendix~\ref{app:Proof-SQCD}.

Let us expand~\(f^{\sch,\Nf,\anti{\Nf}}_{\{p\},k}\), for some
\(\anti{\Nf}\leq\Nf\), in the limit \(\anti{m}_{\anti{\Nf}} = \Lambda
\to +\infty\).  This relies on the asymptotic behavior \((\rho+a)_k
\sim \rho^k\) of Pochhammer symbols when \(\abs{\rho}\to\infty\):
\begin{equation}
  f^{\sch,\Nf,\anti{\Nf}}_{\{p\},k}
  \sim (-\I\Lambda)^k f^{\sch,\Nf,\anti{\Nf}-1}_{\{p\},k} \,.
\end{equation}
Summing over~\(k\geq 0\),
\begin{equation}
  f^{\sch,\Nf,\anti{\Nf}}_{\{p\}}\biggl(\frac{x}{-\I\Lambda}\biggr)
  \to f^{\sch,\Nf,\anti{\Nf}-1}_{\{p\}}(x) \,,
\end{equation}
as \(\Lambda \to \infty\), and for a fixed~\(x\).  We then apply this
limit to~\eqref{Seiberg-fs} for \(\anti{\Nf}=\Nf\) after changing \(x\to
x/(-\I\Lambda)\).  Since \(\delta_1\sim\I\Lambda\), we get
\(a(\anti{\Nf}=\Nf, \I x/\Lambda) = e^{\delta_1 \ln(1-\I x/\Lambda)}
\sim e^x\), which is the exponential factor for \(\anti{\Nf}=\Nf-1\).
In the limit where further twisted masses become very large while
keeping the appropriate combination \(-\I\Lambda x\) fixed, the
exponential factor becomes \(e^{x/(-\I\Lambda)} \to 1\), yielding
\(a(x)=1\) for \(\anti{\Nf}\leq\Nf-2\).  The relative sign between \(x\)
and~\(x^\dual\) is due to the sign difference \(\I\anti{m}^\dual \sim -
\I\anti{m}\) for each of the \(\Nf-\anti{\Nf}\) antifundamental
multiplets which we decouple.

This concludes the proof of the Seiberg duality
relation~\eqref{Seiberg-Z} for all \(\anti{\Nf}\leq\Nf\) as limits of
the case \(\anti{\Nf}=\Nf\), itself derived from the correspondence with
the Toda CFT\@.

\subsection{SQCDA Dualities: Crossing Symmetry and Conjugation}
\label{sec:SeibergW}

In this section, we find two new Seiberg-like dualities between pairs of
\(\Nsusy=(2,2)\) theories with adjoint matter and a superpotential.  The
first is realized in the Toda CFT as crossing symmetry, and contains as
a special case a duality between \(\Nsusy=(2,2)^*\) theories, proposed
in~\cite{Orlando:2010uu} for particular twisted masses, and recently
in~\cite{Benini:2014mia}.  The second is realized as conjugation
symmetry, and is a direct two dimensional analogue of the four
dimensional Kutasov--Schwimmer
duality~\cite{Kutasov:1995ve,Kutasov:1995np}.  We test both dualities by
comparing \(S^2\)~partition functions using the matching with Toda CFT
correlators.  We also provide direct proofs that the \(S^2\)~partition
functions of dual theories are equal, without relying on the Toda CFT\@.
Namely, classical and one-loop contributions are compared in the main
text, and vortex partition functions in Appendix~\ref{app:Proof-SQCDAW}.
It would be interesting to work out the mapping of chiral rings of dual
theories.

Each duality relates theories with \(U(\Nc)\) and \(U(\Nc^\dual)\) vector
multiplets coupled to one adjoint, \(\Nf\)~fundamental, and
\(\anti{\Nf}\)~antifundamental chiral multiplets.  We assume by symmetry
that \(\anti{\Nf}\leq\Nf\).  As for the Seiberg duality, the magnetic
theory includes additional free matter.  In the electric theory, chiral
multiplets are denoted by \(X\), \(\quark_t\), and~\(\anti{\quark}_t\),
and their (complexified) twisted masses by \(m_X\), \(m_t\),
and~\(\anti{m}_t\) respectively.  The FI parameters and theta angles
(renormalized, at the scale~\(\ell\) of the sphere) are combined as
usual into a complex parameter~\(z\).  We use the notations \(X^\dual\),
\(\quark_t^\dual\), \(\anti{\quark}_t^\dual\), \(m_X^\dual\), \(m_t^\dual\),
\(\anti{m}_t^\dual\), and~\(z^\dual\) for the magnetic theory.

Recall that when \(\anti{\Nf}=\Nf\) we have the matching
\begin{equation}
  Z_{S^2\subset S^4_b}^{U(\Nc)\ \SQCDA}(m,\anti{m},m_X,z,\bar{z})
  = A \abs{y}^{2\gamma_0} \abs{1-y}^{2\gamma_1}
  \vev*{\widehat{V}_{\alpha_\infty}(\infty)\widehat{V}_{\hat{m}}(1)
    \widehat{V}_{-\Nc bh_1}(y,\bar{y})\widehat{V}_{\alpha_0}(0)}
\end{equation}
for \(y=(-1)^{\Nf} z\), \(b^2=\I m_X\), with other parameters
given below~\eqref{SQCDA-matching}.  The four-point function can exhibit
two types of symmetries.  If the semi-degenerate momentum \(\hat{m} =
(\varkappa+\Nc b)h_1\) is in fact degenerate (\(\hat{m} = - \Nc^\dual
bh_1\)), then crossing symmetry exchanges the two degenerate operators
via \(\Nc\leftrightarrow\Nc^\dual\) and \(y\to y^{-1}\).  This yields the
duality in Section~\ref{sec:SeibergW-cross}.  On the other hand, it
turns out that for fined-tuned values \(\I m_X = b^2 = \frac{-1}{l+1}\)
the degenerate operator \(\widehat{V}_{-\Nc bh_1}\) is conjugate to
another degenerate operator, \(\widehat{V}_{-\Nc^\dual bh_1}\).  This leads
to the Kutasov--Schwimmer duality in Section~\ref{sec:Kutasov}, which we
then extend to \(\anti{\Nf}<\Nf\) as we did for the Seiberg duality.

\subsubsection{\texorpdfstring{\((2,2)^*\)}{(2,2)*}-like Duality as a Braiding Move}
\label{sec:SeibergW-cross}

Let us describe the first duality more precisely.  With notations as
above, the electric and magnetic theories are \(\Nsusy=(2,2)\) SQCDA
theories with \(\Nc\) and~\(\Nc^\dual\) colors and the same matter content
and superpotential
\begin{equation}
  W = \sum_{t=1}^{\Nf} \anti{\quark}_t X^{l_t} \quark_t
  \qquad \text{hence} \quad 1+\I m_t+\I\anti{m}_t+l_t\I m_X=0 \,,
\end{equation}
where \(l_t \geq 0\) are integers, and \(\anti{\Nf}=\Nf\).  We will find
that \(\Nc^\dual = L-\Nc\) with \(L = \sum_{t=1}^{\Nf} l_t\), twisted
masses are the same in the two theories, \(z^\dual = z^{-1}\), and
\(\bar{z}^\dual = \bar{z}^{-1}\).

In particular, when all \(l_t=1\) the theories are \(\Nsusy=(2,2)^*\)
SQCD with gauge groups \(U(\Nc)\) and \(U(\Nf-\Nc)\), and the duality is
an \(\Nsusy=(2,2)^*\) analogue of the \(\Nsusy=(2,2)\) Seiberg duality
found earlier.  In the special case \(m_X=\I/2\), the two dualities
agree after charge conjugation, which exchanges
\(m_s^\dual\leftrightarrow\anti{m}_s^\dual\) and maps \(z^\dual\to
(z^\dual)^{-1}\).  The agreement is expected since the superpotential
term \(W=\Tr X^2\) is then supersymmetric and \(X\)~can be integrated
out, shifting the theta angle by \((\Nc-1)\pi\) in the process: this
leads to a sign difference in the maps \(z\to z^\dual\) of the two
dualities.

We test the duality by proving that \(S^2\)~partition functions match:
\begin{equation}\label{SeibergW-Z}
  Z_{S^2}^{U(\Nc)} (z,\bar{z})
  =
  \abs{z}^{2\delta_0} \abs*{1-(-1)^{\Nf}z}^{2\delta_1}
  Z_{S^2}^{U(\Nc^\dual)} (z^\dual,\bar{z}^\dual)
  \quad \text{for } W = \sum_{t=1}^{\Nf} \anti{\quark}_t X^{l_t} \quark_t
\end{equation}
with dual parameters given above, and the exponents
\begin{align}
  \label{SeibergW-delta0}
  \delta_0 & = - (L-\Nc)(1+\I m_X)
  - \sum_{t=1}^{\Nf} \sum_{\nu=0}^{l_t-1} (\I m_t + \nu\I m_X) \,,
  \\\label{SeibergW-delta1}
  \delta_1 & = (L-2\Nc)(1+\I m_X) \,.
\end{align}
As discussed below~\eqref{Seiberg-a} for the Seiberg duality, the powers
of \(\abs{z}^2\) and \(\abs{1-(-1)^{\Nf}z}^2\) can be absorbed as ambiguities of
the \(S^2\)~partition function.\footnote{For \(\Nsusy=(2,2)^*\)
  theories, the power of \(1-(-1)^{\Nf}z\) relating dual vortex partition
  functions was found numerically by Honda and
  Okuda~\cite{Honda:2013uca}.}  The same consistency checks as for the
Seiberg duality apply.  Repeating the duality yields the original
parameters, and the factors cancel thanks to \(\delta_0^\dual =
\delta_0+\delta_1\) and \(\delta_1^\dual = -\delta_1\).  The relation is
also invariant under charge conjugation, which exchanges twisted masses
of fundamental and antifundamental chiral multiplets, since
\(\delta_1\)~is unchanged and \(\delta_0 \to -\delta_0-\delta_1\).  We
first derive dual parameters from the matching of SQCDA partition
functions to Toda CFT correlators.  For completeness, we then prove the
relation by comparing classical, one-loop and vortex contributions of
the two theories.

Recall the matching~\eqref{SQCDA-matching} between the partition
function of a sphere surface operator describing \(U(\Nc)\) SQCDA and a
Toda CFT correlator with the symmetric degenerate
operator~\(\widehat{V}_{-\Nc bh_1}\), a semi-degenerate
operator~\(\widehat{V}_{\hat{m}}\), and two generic operators.  We find
in Section~\ref{sec:SQCDAW} that the superpotential \(W = \sum_t
\anti{\quark}_t X^{l_t} \quark_t\) constrains twisted masses in such a
way that \(\hat{m} = -(L-\Nc)bh_1 = -\Nc^\dual bh_1\).  The
\(S^2\)~partition function of the electric theory we are studying is
thus
\begin{equation}
  Z_{S^2}^{U(\Nc)}(z,\bar{z})
  = \widetilde{A}
  \abs{y}^{2\gamma_0} \abs{1-y}^{2\gamma_1}
  \vev*{\widehat{V}_{\alpha_\infty}(\infty)\widehat{V}_{-\Nc^\dual bh_1}(1)
    \widehat{V}_{-\Nc bh_1}(y,\bar{y})\widehat{V}_{\alpha_0}(0)}
\end{equation}
where \(y = (-1)^{\Nf} z\), \(b^2 = \I m_X\), momenta and
exponents are given below~\eqref{SQCDA-matching}, and we have absorbed
in~\(\widetilde{A}\) the contributions from the
\(S^4_b\)~hypermultiplets and from the differing normalization of
semidegenerate and degenerate operators.  The Toda CFT correlator is
invariant under \(\Nc\to\Nc^\dual\), \(y\to y^\dual = y^{-1}\), and the
conformal map \((\infty,1,y^{-1},0) \to (\infty,y,1,0)\).  This implies
that
\begin{equation}
  Z_{S^2}^{U(\Nc)}(z,\bar{z})
  =
  \frac{\widetilde{A}\abs{y}^{2\gamma_0}\abs{1-y}^{2\gamma_1}}
    {\widetilde{A}^\dual\abs{y^\dual}^{2\gamma_0^\dual} \abs{1-y^\dual}^{2\gamma_1^\dual}}
  \abs{y}^{\dimToda(\alpha_\infty)-\dimToda(-\Nc^\dual bh_1)
    -\dimToda(-\Nc bh_1)-\dimToda(\alpha_0)}
  Z_{S^2}^{U(\Nc^\dual)}(z^\dual,\bar{z}^\dual) \,.
\end{equation}
We deduce the exponents \eqref{SeibergW-delta0}
and~\eqref{SeibergW-delta1} by computing \(\delta_1 = \gamma_1 -
\gamma_1^\dual\) and
\begin{equation}
  \delta_0 =
  \gamma_0 + \dimToda(\alpha_\infty)-\dimToda(-\Nc^\dual bh_1)
  -\dimToda(-\Nc bh_1)-\dimToda(\alpha_0)
  + \gamma_0^\dual + \gamma_1^\dual \,.
\end{equation}
We also obtain \(z^\dual = (-1)^{\Nf} y^\dual = (-1)^{\Nf} y^{-1} = z^{-1}\)
and \(\Nc^\dual = L-\Nc\) as announced.

There remains to fix the overall constant factor, since
\(\widetilde{A}/\widetilde{A}^\dual\) is difficult to evaluate
(\(\widetilde{A}\) and~\(\widetilde{A}^\dual\) are singular for our
choice of twisted masses).  This is done by comparing the s-channel
decomposition (as \(z\to 0\)) of the electric theory with the u-channel
decomposition (as \(z^\dual\to\infty\)) of the magnetic theory.  Recall
from Section~\ref{sec:SQCDAW} that the s-channel Higgs branch vacua of
the electric theory are labeled by ordered partitions \(\sum_{t=1}^{\Nf}
n_t = \Nc\) with \(0\leq n_t\leq l_t\).  The classical and one-loop
contributions~\eqref{SQCDA-Z1l} are
\begin{align}
  Z_{\classical,\{n_t\}}^\sch(z,\bar{z})
  & = (z\bar{z})^{\sum_{s=1}^{\Nf} \sum_{\mu=0}^{n_s-1} (- \I m_s - \mu \I m_X)} \,,
  \\
  Z_{\oneloop,\{n_t\}}^\sch
  & = \prod_{s=1}^{\Nf} \prod_{\mu=0}^{n_s-1} \prod_{t=1}^{\Nf}
  \frac{\gamma(\I m_s - \I m_t + (\mu - n_t) \I m_X)}
    {\gamma(\I m_s - \I m_t + (\mu - l_t) \I m_X)} \,.
\end{align}
Similarly, u-channel Higgs branch vacua of the magnetic theory are
labeled by partitions \(\sum_{t=1}^{\Nf} n_t^\dual = \Nc^\dual\) with
\(0\leq n_t^\dual\leq l_t\), and are in a natural bijection with those
of the electric theory through \(n_t^\dual = l_t - n_t\).  The classical
contributions match up to~\(\abs{z}^{2\delta_0}\):
\begin{equation}
  \sum_{s=1}^{\Nf} \sum_{\mu=0}^{n_s-1} (- \I m_s - \mu \I m_X)
  = \delta_0
  - \sum_{s=1}^{\Nf} \sum_{\mu=0}^{l_s-n_s-1} (\I\anti{m}_s + \mu \I m_X) \,.
\end{equation}
The one-loop contributions are equal, with no relative constant factor,
since
\begin{equation}
  Z_{\oneloop,\{n_t\}}^\sch
  = \prod_{s=1}^{\Nf} \prod_{\mu=0}^{n_s-1} \prod_{t=1}^{\Nf}
  \prod_{\nu=0}^{l_t-n_t-1}
  \frac{\gamma(1 + \I m_s + \I\anti{m}_t + (\mu + \nu + 1) \I m_X)}
    {\gamma(1 + \I m_s + \I\anti{m}_t + (\mu + \nu) \I m_X)}
\end{equation}
is invariant under \(m\leftrightarrow\anti{m}\) and \(n\to l-n\).  We
prove in Appendix~\ref{app:Proof-SQCDAW} that the vortex partition
functions match up to \((1-y)^{\delta_1}\).  This establishes the
duality relation~\eqref{SeibergW-Z}.

From the duality we can extract information about powers of
\(\abs{1-y}^2\) which appear in the expansion of~\(Z\) near \(y=1\).  In
the electric theory, the powers are given by~\eqref{SQCDA-t-powers},
valid for all SQCDA theories: replacing \(k\) by \(\Nc - k\) there,
\begin{equation}\label{SeibergW-t-powers}
  Z_{S^2}^{U(\Nc)} (z,\bar{z})
  =
  \abs{1-y}^{-2\Nc(1+\I m_X)}
  \sum_{k=0}^\Nc \biggl[
  \abs{1-y}^{2[k(1+\I m_X)-(\Nc-k)(\Nc^\dual-k)\I m_X]}
  (\text{series}) \biggr]
\end{equation}
for some series in non-negative powers of \((1-y)\) and \((1-\bar{y})\).
The magnetic theory has a similar expansion with
\(\Nc\leftrightarrow\Nc^\dual\).  Since the two must match, we deduce that
the expansion~\eqref{SeibergW-t-powers} holds, with a sum restricted to
\(0\leq k\leq\min(\Nc,\Nc^\dual)\).  This list of exponents is useful to
identify the correct relation between quiver gauge theories and
correlators in Section~\ref{sec:Quivers}.

\subsubsection{Kutasov--Schwimmer Duality as Momentum Conjugation}
\label{sec:Kutasov}

The Kutasov--Schwimmer duality~\cite{Kutasov:1995ve,Kutasov:1995np},
initially proposed between four dimensional theories, is similar to the
Seiberg duality, with an additional adjoint matter multiplet~\(X\)
subject to the superpotential coupling \(W=\Tr X^{l+1}\).  Through the
matching found in Section~\ref{sec:SQCDAW}, the duality is realized as
conjugation of momenta in the Toda CFT when \(\anti{\Nf}=\Nf\).
Theories with \(\anti{\Nf}<\Nf\) are obtained by decoupling chiral
multiplets.  For \(l=1\), integrating out~\(X\) reproduces the Seiberg
duality between SQCD theories.

The electric and magnetic theories are \(\Nsusy=(2,2)\) SQCDA theories
with gauge groups \(U(\Nc)\) and \(U(\Nc^\dual)\) and the superpotential
coupling
\begin{equation}
  W = \Tr X^{l+1}
  \qquad \text{hence} \quad \I m_X^\dual = \I m_X = \frac{-1}{l+1}
\end{equation}
for some integer \(l\geq 1\).  As we will see, \(z^\dual =
(-1)^{\Nf-\anti{\Nf}} z\), \(\bar{z}^\dual = (-1)^{\Nf-\anti{\Nf}}
\bar{z}\) (in terms of renormalized parameters at the scale~\(\ell\) of
the sphere), \(\Nc^\dual=l\Nf-\Nc\), \(m_t^\dual = m_X - m_t\), \(\anti{m}_t^\dual =
m_X - \anti{m}_t\), \(m_X^\dual = m_X\), and the magnetic theory also
features \(l\Nf\anti{\Nf}\) free mesons~\(M_{jst}^\dual\) with twisted
masses \(m_{jst}^\dual = m_s + \anti{m}_t + j m_X\) for \(0\leq j<l\),
\(1\leq s\leq\Nf\), \(1\leq t\leq\anti{\Nf}\).  We assume that
\(l\leq\Nc\leq l\Nf-l\).

We test the duality by comparing \(S^2\)~partition functions.  Namely,
we prove that
\begin{equation}
  \label{Kutasov-Z}
  Z_{S^2}^{U(\Nc),\Nf,\anti{\Nf}} (m;z,\bar{z})
  =
  a(z,\bar{z})
  \prod_{j,s,t} \gamma\bigl(-\I m_{jst}^\dual\bigr)
  Z_{S^2}^{U(\Nc^\dual),\Nf,\anti{\Nf}} (m^\dual;z^\dual,\bar{z}^\dual)
\end{equation}
with dual parameters given above.  The constant factor
in~\eqref{Kutasov-Z} is the one-loop determinant of free
mesons~\(M_{jst}^\dual\) whose twisted masses \(m_{jst}^\dual = m_s + \anti{m}_t
+ j m_X\) are fixed by the full superpotential coupling
\begin{equation}\label{Kutasov-W}
  W = \Tr\bigl[(X^\dual)^{l+1}\bigr]
  + \sum_{s=1}^{\Nf} \sum_{t=1}^{\anti{\Nf}} \sum_{j=0}^{l-1}
  M_{jst}^\dual\,\bigl[\anti{\quark}^\dual_t (X^\dual)^{l-1-j} \quark^\dual_s\bigr] \,.
\end{equation}
Relative coefficients are unimportant, as the superpotential only affects
the \(S^2\)~partition function by fixing complexified twisted masses.
The electric theory features mesons \(\anti{\quark}_t X^j \quark_s\)
which have the same twisted masses \(m_s + \anti{m}_t + j m_X\).  The
factor \(a(z,\bar{z})\) is
\begin{gather}
  \label{Kutasov-a}
  a(z,\bar{z}) =
  \begin{cases}
    \abs{z}^{2\delta_0}
    & \text{if \(\anti{\Nf}\leq\Nf-2\)} \,, \\
    \abs{z}^{2\delta_0} e^{l (-1)^{\Nf} (z - \bar{z})}
    & \text{if \(\anti{\Nf}=\Nf-1\)} \,, \\
    \abs{z}^{2\delta_0} \abs*{1-(-1)^{\Nf} z}^{2\delta_1}
    & \text{if \(\anti{\Nf}=\Nf\)} \,,
  \end{cases}
  \\
  \label{Kutasov-delta}
  \delta_0
  = - \frac{l\Nf}{2} + \frac{l\Nc}{l+1} - l \sum_{s=1}^{\Nf} \I m_s \,,
  \qquad
  \delta_1
  = l \Nf - \frac{2l\Nc}{l+1} + l \sum_{s=1}^{\Nf} (\I m_s + \I\anti{m}_s)
  \,.
\end{gather}
As discussed below~\eqref{Seiberg-a} for the Seiberg duality, this factor
can be absorbed as an ambiguity of the \(S^2\)~partition function.

The same consistency checks as for the Seiberg duality apply.  Repeating
the duality yields the original parameters, and the factors
\(a(z,\bar{z})\) and products of gamma functions cancel.  Charge
conjugation leaves the relation invariant in the case
\(\anti{\Nf}=\Nf\).

\medskip

Let us first derive~\eqref{Kutasov-Z} for \(\anti{\Nf}=\Nf\) from Toda
CFT conjugation.  Recall~\eqref{SQCDAW-matching}, which expresses the
partition functions of interest as \(b^2 \to \frac{-1}{l+1}\) limits of
Toda CFT four-point functions.  The relevant correlator is
\(\vev{\widehat{V}_{\alpha_\infty}(\infty)\widehat{V}_{\hat{m}}(1)
  \widehat{V}_{-b\omega_{\Nc,l}}(x,\bar{x})\widehat{V}_{\alpha_0}(0)}\).
Here, \(\omega_{\Nc,l} = l \omega_k + (\Nc - lk)h_{k+1}\) with
\(k\)~defined by \(kl\leq\Nc<(k+1)l\), and its conjugate weight is
\(\omega_{\Nc,l}^\conj = \omega_{\Nc^\dual,l}\) with \(\Nc^\dual=l\Nf-\Nc\).  As
for the Seiberg duality, we follow the conjugation of \(\hat{m} =
(\varkappa+\Nc b) h_1\) by a Weyl reflection to get a momentum
along~\(h_1\),
\begin{equation}
  \hat{m}^\dual
  = (\varkappa^\dual+\Nc^\dual b) h_1
  = \left[\Nf \left(b+\frac{1}{b}\right)-\varkappa-\Nc b\right] h_1
  \,.
\end{equation}
Thus, \(\varkappa^\dual = \frac{1}{b}\Nf\bigl(1-(l-1)b^2\bigr)-\varkappa\),
which is \(\frac{2}{b}\Nf(1+b^2) - \varkappa\) when
\(b^2=\frac{-1}{l+1}\).  Finally, the generic momenta \(\alpha_0\)
and~\(\alpha_\infty\) are mapped as \(\alpha \to 2Q-\alpha\) under a
conjugation followed by the maximal Weyl reflection.  Translating to
gauge theory parameters thanks to the dictionary~\eqref{SQCDA-alpha}
yields the values of \(\Nc^\dual\), \(m_X^\dual\), \(m_s^\dual\), and
\(\anti{m}_s^\dual\) claimed earlier.  The position \(y=(-1)^{\Nf}z\)
of the degenerate operator is not affected by conjugation, hence
\(y^\dual=y\) and \(z^\dual = z\).  The factor \(a(z,\bar{z})\)
given in~\eqref{Kutasov-a} is the ratio of factors \(\abs{y}^{2\gamma_0}
\abs{1-y}^{2\gamma_1}\) for the electric and magnetic theories.

Since the constant factor~\(A\) which appears in the
matching~\eqref{SQCDAW-matching} is not known, we cannot deduce the
presence of free mesons in the magnetic theory through conjugation of
momenta.  Instead, we use the Higgs branch
decomposition~\eqref{SQCDA-Zs}, which expresses both partition functions
as sums over choices of \(0\leq n_s\leq l\) with
\(n_1+\cdots+n_{\Nf}=\Nc\).  The classical contribution for the term
labeled by~\(\{n_s\}\) in the electric theory is
\(\abs{z}^{2\delta_0}\)~times the classical contribution for the term
labeled by~\(\{n_s^\dual=l-n_s\}\) in the magnetic theory.  We then compare
the one-loop determinants~\eqref{SQCDA-Z1l} for those vacua,
\begin{align}
  \nonumber
  \frac{Z_{\oneloop,\{n\}}}{Z^\dual_{\oneloop,\{l-n\}}}
  & =
  \prod_{s=1}^{\Nf} \prod_{t=1}^{\Nf} \Biggl[
  \prod_{\mu=0}^{n_s-1}
  \frac
    {\gamma(-\I m_t + \I m_s + \frac{n_t - \mu}{l+1})}
    {\gamma(1+\I\anti{m}_t+\I m_s - \frac{\mu}{l+1})}
  \prod_{\mu=0}^{l-n_s-1}
  \frac{\gamma(1+\I\anti{m}_t^\dual+\I m_s^\dual - \frac{\mu}{l+1})}
    {\gamma(-\I m_t^\dual + \I m_s^\dual + \frac{n_t^\dual - \mu}{l+1})}
  \Biggr]
  \\
  & =
  \prod_{s=1}^{\Nf} \prod_{t=1}^{\Nf}
  \prod_{j=0}^{l-1}
  \gamma\biggl(-\I\anti{m}_t-\I m_s + \frac{j}{l+1}\biggr) \,,
\end{align}
and find the one-loop determinants of mesons with twisted
masses~\(m_{jst}^\dual\) as announced.  Finally, we prove in
Appendix~\ref{app:Proof-SQCDAW} that vortex partition functions of dual
theories are equal up to the factor \((1-y)^{\delta_1}\), hence
establishing the relation~\eqref{Kutasov-Z} for \(\anti{\Nf}=\Nf\).

For completeness, we compare exponents which appear when expanding the
\(S^2\)~partition functions of the dual theories near \(y=1\).  Those
exponents are given by~\eqref{SQCDA-t-powers}: for a general
\(\Nsusy=(2,2)\) SQCDA theory with \(\Nc\)~colors there are \(\Nc+1\)
exponents labeled by an integer \(0\leq k\leq\Nc\).  The set of
exponents thus does not match for the \(U(\Nc)\) and \(U(\Nc^\dual)\)
theories we consider here.  In fact, it turns out that only the subset
labeled by \(0\leq k\leq l\) matches (assuming that \(l\leq\Nc\leq
l\Nf-l\)): the coefficients of all other exponents must thus vanish when
\(\I m_X=\frac{-1}{l+1}\).

\medskip

We now take the limit \(\I\anti{m}_t = \I\Lambda \to \pm\I\infty\) for
\(\Nf-\anti{\Nf}\) antifundamental flavours \(\anti{\Nf}<t\leq\Nf\).
The multiplets~\(\anti{\quark}_t\) decouple, the FI parameter is
renormalized, and we will be left with the Kutasov--Schwimmer
duality~\eqref{Kutasov-Z} for \(\anti{\Nf}<\Nf\).

The twisted mass~\(\Lambda\) appears in the Coulomb branch
expansion~\eqref{SQCDA-Z-Coulomb} through the one-loop determinant of
antifundamental chiral multiplets: for fixed \(\sigma\) and~\(B\)
\begin{equation}
  \prod_{j=1}^\Nc \prod_{t=\anti{\Nf}+1}^{\Nf}
  \frac{\Gamma(-\I\anti{m}_t + \I\sigma_j + \frac{B_j}{2})}
    {\Gamma(1+\I\anti{m}_t - \I\sigma_j + \frac{B_j}{2})}
  \sim
  \biggl[
  \gamma(-\I\Lambda)^\Nc
  (-\I\Lambda)^{\Tr(\I\sigma + \frac{B}{2})}
  (\I\Lambda)^{\Tr(\I\sigma - \frac{B}{2})}
  \biggr]^{\Nf-\anti{\Nf}} \,.
\end{equation}
The powers of \(\pm\I\Lambda\) combine with the classical contribution
\(z_\bare^{\Tr(\I\sigma+B/2)} \bar{z}_\bare^{\Tr(\I\sigma-B/2)}\), and
we get the integrand of the Coulomb branch representation for the theory
with \(\anti{\Nf}<\Nf\) and \(z = z_\renormalized =
(-\I\Lambda)^{\Nf-\anti{\Nf}} z_\bare\).  A careful treatment shows
that the limit \(\Lambda\to\pm\infty\) and the integration commute,
because the contribution for large values of \(\sigma\) and~\(B\) falls
off fast enough at infinity.  As mentioned for a similar limit of the
SQCD theory in Section~\ref{sec:irreg}, it is easier to work out this
convergence issue in the Higgs branch decomposition where terms decrease
exponentially in the vorticity~\(k\).  Either way yields
\begin{equation}\label{Kutasov-limit}
  \begin{aligned}
    & Z_{S^2}^{\Nf,\Nf} \biggl(
    \frac{z}{(-\I\Lambda)^{\Nf-\anti{\Nf}}},
    \frac{\bar{z}}{(\I\Lambda)^{\Nf-\anti{\Nf}}},
    \{m_s\},
    \{\anti{m}_s,\Lambda\}
    \biggr)
    \\
    & \quad \sim
    \gamma(-\I\Lambda)^{\Nc(\Nf-\anti{\Nf})}
    Z_{S^2}^{\Nf,\anti{\Nf}}
    \bigl(\{m_s\},\{\anti{m}_s\},z,\bar{z}\bigr)
    \,.
  \end{aligned}
\end{equation}

Given the form of~\eqref{Kutasov-limit}, the next step is to consider
the duality~\eqref{Kutasov-Z} with the replacement \(\anti{\Nf}\to\Nf\),
\(z\to z/(-\I\Lambda)^{\Nf-\anti{\Nf}}\) and
\(\bar{z}\to\bar{z}/(\I\Lambda)^{\Nf-\anti{\Nf}}\), in the limit where
\(\Lambda\to\pm\infty\).  The factor \(a(z,\bar{z}) =
\abs{z}^{2\delta_0} \abs{1-(-1)^{\Nf}z}^{2\delta_1}\) with
\(\delta_1\sim \I\Lambda l\) becomes
\begin{equation}\label{Kutasov-limit-a}
  \begin{aligned}
    & a_{\Nf,\Nf}\biggl(\frac{z}{(-\I\Lambda)^{\Nf-\anti{\Nf}}},
    \frac{\bar{z}}{(\I\Lambda)^{\Nf-\anti{\Nf}}}\biggr)
    \sim
    \Lambda^{-2(\Nf-\anti{\Nf})\delta_0}
    a_{\Nf,\anti{\Nf}}(z,\bar{z})
    \\
    & \qquad \sim
    \begin{cases}
      \Lambda^{-2(\Nf-\anti{\Nf})\delta_0}
      \abs{z}^{2\delta_0}
      & \text{if \(\anti{\Nf} \leq \Nf-2\)} \,,
      \\
      \Lambda^{-2(\Nf-\anti{\Nf})\delta_0}
      \abs{z}^{2\delta_0}
      e^{l (-1)^{\Nf} (z-\bar{z})}
      & \text{if \(\anti{\Nf} = \Nf-1\)} \,.
    \end{cases}
  \end{aligned}
\end{equation}
The gamma functions in~\eqref{Kutasov-Z} become those for
\(\anti{\Nf}<\Nf\), multiplied by
\begin{equation}\label{Kutasov-limit-gamma}
  \begin{aligned}
    & \prod_{s=1}^{\Nf} \prod_{j=0}^{l-1} \prod_{t=\anti{\Nf}+1}^{\Nf}
    \gamma\biggl(-\I\Lambda-\I m_s+\frac{j}{l+1}\biggr)
    \sim
    \prod_{s=1}^{\Nf} \prod_{j=0}^{l-1} \Bigl[
      \gamma(-\I\Lambda) \,
      \Lambda^{2(-\I m_s+\frac{j}{l+1})}
      \Bigr]^{\Nf-\anti{\Nf}}
    \\
    & \quad \sim
    \Bigl[
    \gamma(-\I\Lambda)^{l\Nf}
    \Lambda^{2(\delta_0 + \Nc^\dual \frac{l}{l+1})}
    \Bigr]^{\Nf-\anti{\Nf}}
    \sim
    \Bigl[
    \Lambda^{2\delta_0}
    \gamma(-\I\Lambda)^{\Nc}
    \gamma(-\I\Lambda^\dual)^{-\Nc^\dual}
    \Bigr]^{\Nf-\anti{\Nf}}
    \,,
  \end{aligned}
\end{equation}
where we used \(\gamma(\I x+a)\sim\gamma(\I x)\abs{x}^{2a}\) as
\(x\to\pm\infty\), and \(\Lambda^{\frac{2l}{l+1}}
\gamma\bigl(-\I\Lambda\bigr) \sim
\gamma\bigl(-\I\Lambda^\dual\bigr)^{-1}\).  Combining
\eqref{Kutasov-limit-a} and~\eqref{Kutasov-limit-gamma} with the power
of \(\gamma(-\I\Lambda)\) from~\eqref{Kutasov-limit} and the power of
\(\gamma(-\I\Lambda^\dual)\) for the dual theory establishes
the Kutasov--Schwimmer duality relation~\eqref{Kutasov-Z} for all
\(\anti{\Nf}\leq\Nf\).

\subsection{Dualities for Quivers}
\label{sec:SeibergQ}

We revisit here the \(\Nsusy=(2,2)\) quivers of Section~\ref{sec:Quivers} and
express some Seiberg and \(\Nsusy=(2,2)^*\) dualities as permutations of
Toda CFT punctures in Section~\ref{sec:SeibergQ-perm}.  This lets us
construct in Section~\ref{sec:SeibergQ-group} the full set of theories
obtained through Seiberg and \(\Nsusy=(2,2)^*\) dualities.  For a
particular choice of matter content, a certain combination of dualities
is realized as conjugation of momenta in the Toda CFT\@.

The gauge theories depend on ranks
\(\Nc_j\geq 0\), signs \(\eta_j\), and complexified FI parameters
\((\hat{z}_j,\overline{\hat{z}_j})\) for \(1\leq j\leq n\), as well as
twisted masses.  They are described by the quiver
\begin{equation}\label{SeibergQ-quiver}
  \quiver{%
    \node (Nn)  [color-group]                    {$\Nc_n$};
    \node (dots)[right of=Nn]                    {$\cdots$};
    \node (N1)  [color-group, right of=dots]     {$\Nc_1$};
    \node (f)   [left of=Nn]                     {};
    \node (Nf)  [flavor-group, above=1ex of f]   {$\Nf$};
    \node (Nf') [flavor-group, below=1ex of f]   {$\Nf$};
    \draw[->-=.55] (Nf) -- (Nn);
    \draw[->-=.55] (Nn) -- (Nf');
    \draw[->-=.55] (Nn)   to [bend left=30] (dots);
    \draw[->-=.55] (dots) to [bend left=30] (N1);
    \draw[->-=.55] (N1)   to [bend left=30] (dots);
    \draw[->-=.55] (dots) to [bend left=30] (Nn);
    \draw[->-=.5] (N1) to [distance=5ex, in=60, out=120, loop] ();
    \draw[->-=.5] (Nn) to [distance=5ex, in=60, out=120, loop] ();
  } \,.
\end{equation}
The theories consist of a \(U(\Nc_1)\times\cdots\times U(\Nc_n)\) vector
multiplet coupled to chiral multiplets which transform in the following
representations: \(\Nf\)~fundamentals and \(\Nf\)~antifundamentals of
\(U(\Nc_n)\), two bifundamentals of \(U(\Nc_j)\times U(\Nc_{j-1})\) for
each \(2\leq j\leq n\), and one adjoint of \(U(\Nc_j)\) for each \(1\leq
j\leq n\).  Let \(\epsilon_j = \prod_{i=j}^n \eta_i\).  The twisted
masses \(m_t\), \(\anti{m}_t\), \(m_{j(j-1)}\), \(m_{(j-1)j}\),
and~\(m_{jj}\) of those chiral multiplets obey~\eqref{Quivers-masses}, that is,
\begin{equation}
  \I m_{j(j-1)}+\I m_{(j-1)j} = - 1 - 2 q_j
  \quad \text{and} \quad
  \I m_{jj} = q_j + q_{j+1} \,,
\end{equation}
where \(q_j = b^2/2\) if \(\epsilon_j=1\) and \(q_j = -(1+b^2)/2\)
otherwise for some parameter~\(b^2\).  The
twisted masses are such that a given superpotential~\(W_\eta\) has
\(R\)-charge~\(2\) (twisted mass~\(\I\)).  Whenever \(\eta_j=-1\), the
superpotential term \(\Tr(X_j^2)\) lets us integrate out~\(X_j\).

We gave evidence in Section~\ref{sec:Quivers} that the partition
function on~\(S^4_b\) of the \(S^2\)~surface operator obtained by
coupling such a theory to free hypermultiplets is equal to a Toda CFT
\((n+3)\)-point function, namely the correlator of two generic, one
semi-degenerate, and \(n\)~degenerate vertex operators.  The momenta of
the first three operators encode the twisted masses \(m_t\)
and~\(\anti{m}_t\).  The degenerate operators are inserted at positions
\(x_j = \prod_{i=j}^n \hat{z}_i\), and have momenta \(-b\Omega_j =
-b\Omega(K_j,\epsilon_j)\), where \(K_j = \Nc_j-\Nc_{j-1}\),
\(\epsilon_j = \prod_{i=j}^n \eta_i\), and \(\Omega(K,+1)=Kh_1\) and
\(\Omega(K,-1)=\omega_K\).

Crossing symmetry of the Toda CFT correlator states that the labeling
of degenerate operators by integers \(1\leq j\leq n\) is irrelevant.
Therefore, the \(n!\)~gauge theories which correspond to each labeling
of the degenerate punctures should all have identical \(S^2\)~partition
functions, up to simple factors.  It turns out that each transposition
\(k\leftrightarrow k+1\) (for \(k<n\)) corresponds to a duality acting
on the node \(U(\Nc_k)\) of the quiver gauge theory: Seiberg duality
(see Section~\ref{sec:Seiberg}) if \(\eta_k=-1\), or the
\(\Nsusy=(2,2)^*\) duality (see Section~\ref{sec:SeibergW-cross}) if
\(\eta_k=+1\).  In Section~\ref{sec:SeibergQ-perm} we work out details
and make sure that transpositions correctly reproduce the mapping of
parameters for such dualities.  As a result, the groupoid generated by
Seiberg and \(\Nsusy=(2,2)^*\) dualities acting on nodes with \(k<n\) is
realized as permutations of punctures in the Toda CFT\@.

In Section~\ref{sec:SeibergQ-group}, we extend the groupoid by including
the action of Seiberg duality on the node \(U(\Nc_n)\) when it is
applicable (\(\eta_n=-1\)): the \(\Nsusy=(2,2)^*\) duality never
applies, since the \(\Nf\)~fundamental and antifundamental chiral multiplets are
not constrained by a superpotential.  The result of acting with Seiberg
duality on \(U(\Nc_n)\) is not a quiver of the same type, hence is not
given a Toda CFT interpretation in our work.  However, for a specific
choice of matter content which corresponds to the case where all
degenerate vertex operators are labeled by antisymmetric
representations of~\(A_{\Nf-1}\), applying Seiberg duality in turn to
all the nodes yields a quiver of the original form.  This combination of
dualities corresponds to conjugating Toda CFT momenta.

All our results extend to theories with any number
\(\anti{\Nf}\leq\Nf\) of antifundamental chiral multiplets following the
discussion for Seiberg duality of SQCD in
Section~\ref{sec:Seiberg-irreg}.  We focus on
\(\anti{\Nf}=\Nf\) because the matching between partition functions and
Toda CFT correlators was only derived in this case in
Section~\ref{sec:Quivers}: for \(\anti{\Nf}<\Nf\), the
correlator contains an irregular puncture as described for SQCD in
Section~\ref{sec:irreg}.

\subsubsection{Seiberg Dualities from Braiding Moves}
\label{sec:SeibergQ-perm}

We now prove that the action of Seiberg duality or the
\(\Nsusy=(2,2)^*\) duality (depending on~\(\eta_k\)) on the node
\(U(\Nc_k)\) translates to the transposition \((x_k,\epsilon_k,K_k)
\leftrightarrow (x_{k+1},\epsilon_{k+1},K_{k+1})\) of two degenerate
punctures, for \(1\leq k\leq n-1\).  Specifically, we show that the
\(S^2\)~partition functions of the theories described by the Toda CFT
data before and after the transposition are equal.  Most gauge theory
parameters describing the electric and magnetic theories are the same,
with the following changes: \(\eta_{k\pm 1}^\dual = \eta_{k\pm 1}
\eta_k\), \(\Nc_k^\dual = \Nc_{k+1} + \Nc_{k-1} - \Nc_k\),
\(\hat{z}_{k\pm 1}^\dual = \hat{z}_{k\pm 1} \hat{z}_k\) and
\(\hat{z}_k^\dual = \hat{z}_k^{-1}\).

The multiplets which interact with the \(U(\Nc_k)\) vector multiplet of
the electric theory are those of \(\Nsusy=(2,2)\) SQCDA with
\(\Nc_k\)~colors and \(\Nc_{k-1}+\Nc_{k+1}\) flavours.  If \(\eta_k=-1\),
then \(\I m_{kk} = -1/2\) and \(W_\eta\)~contains the term
\(\Tr(X_k^2)\) which lets us integrate out the adjoint chiral
multiplet~\(X_k\), leaving \(\Nsusy=(2,2)\) SQCD\@.  If instead
\(\eta_k=+1\), then \(\I m_{kk}+\I m_{k(k\pm 1)}+\I m_{(k\pm 1)k}=-1\)
and \(W_\eta\)~contains the terms \(\Tr(\biquark_{(k\pm 1)k} X_k
\biquark_{k(k\pm 1)})\): this is \(\Nsusy=(2,2)^*\) SQCD\@.
In both settings, the theory admits a dual description with \(\Nc_{k+1}
+ \Nc_{k-1} - \Nc_k\) colors, and some mesons if \(\eta_k=-1\) (see
sections \ref{sec:Seiberg} and~\ref{sec:SeibergW-cross}).  As we will
now see, parameters map precisely as expected from the Toda CFT\@.

In the Coulomb branch representation of the \(S^2\)~partition function
of the electric theory, we collect all factors which depend on the
scalar~\(\sigma_k\) of the \(U(\Nc_k)\) vector multiplet.  This yields
an integral~\(Z_k\)~\eqref{Quivers-Zk} very similar to the partition
function of \(\Nsusy=(2,2)\) SQCDA with \(\Nc_k\)~colors and
\(\Nc_{k-1}+\Nc_{k+1}\) flavours.  The usual contour techniques apply
and yield a factorized expression for~\(Z_k\) in the region
\(\abs{\hat{z}_k}<1\), namely
\begin{equation}\label{SeibergQ-Zk}
  Z_k
  = \sum_{\text{Higgs vacuum \(v^\pm\)}}
  \begin{aligned}[t]
    & \hat{z}_k^{\Tr\I v^+} \bar{\hat{z}}_k^{\Tr\I v^-}
    Z_{\oneloop,\{v^\pm\}}\bigl(\{m_{kl}-\sigma_{lj}^\pm\},
    \{m_{lk}+\sigma_{lj}^\pm\}\bigr)
    \\
    & \cdot
    Z_{\vortex,v^+}\bigl(\{m_{kl}-\sigma_{lj}^+\},
      \{m_{lk}+\sigma_{lj}^+\};\hat{z}_k\bigr)
    \\
    & \cdot
    Z_{\antivortex,v^-}\bigl(\{m_{kl}-\sigma_{lj}^-\},
      \{m_{lk}+\sigma_{lj}^-\};\bar{\hat{z}}_k\bigr) \,.
  \end{aligned}
\end{equation}
As discussed above, the superpotential~\(W_\eta\) reduces SQCDA to
\(\Nsusy=(2,2)\) SQCD or \(\Nsusy=(2,2)^*\) SQCD depending
on~\(\eta_k\).  In both cases, Higgs branch vacua are labeled by sets
of \(\Nc_k\) ``flavours'' among
\begin{equation}
  L_k = \bigl\{ (l,j) \bigm| l=k\pm 1, 1\leq j\leq\Nc_l \bigr\} \,,
\end{equation}
and the eigenvalues of~\(v^\pm\) for a given \(\Nc_k\)-element subset
\(E\subset L_k\) are
\begin{equation}
  v^\pm_{(l,j)} = -m_{kl} + \sigma_{lj}^\pm
  \quad \text{for \((l,j)\in E\)} \,.
\end{equation}
The vortex partition functions in~\eqref{SeibergQ-Zk} are those of the
relevant \(\Nsusy=(2,2)\) SQCD or \(\Nsusy=(2,2)^*\) SQCD theory with
\(\Nc_{k+1}+\Nc_{k-1}\) fundamental multiplets of twisted masses
\(\{m_{kl}-\sigma_{lj}^+\}\) and the same number of antifundamental
multiplets of twisted masses \(\{m_{lk}+\sigma_{lj}^+\}\), in the Higgs
branch vacuum~\(v^+\).  The antivortex partition functions are obtained
by replacing \(\sigma_{lj}^+\) by \(\sigma_{lj}^-\) and \(v^+\)
by~\(v^-\).  The one-loop determinant for the vacuum labeled by
\(E\subset L_k\) is
\begin{equation}\label{SeibergQ-Z1l}
  \begin{aligned}
    Z_{\oneloop,E}
    =
    \prod_{(l,j)\in E} \prod_{(l',j')\in L_k} \biggl[
    & \frac{\Gamma(-\I m_{kl'} - \delta_{l'j'\in E} \I m_{kk} + \I m_{kl}
      + \I\sigma_{l'j'}^+ - \I\sigma_{lj}^+)}
    {\Gamma(1+\I m_{kl'} + \delta_{l'j'\in E} \I m_{kk} - \I m_{kl}
      - \I\sigma_{l'j'}^- + \I\sigma_{lj}^-)}
    \\
    & \cdot
    \frac{\Gamma(- \I m_{l'k} - \I m_{kl}
      - \I\sigma_{l'j'}^+ + \I\sigma_{lj}^+)}
    {\Gamma(1 + \I m_{l'k} + \I m_{kl}
      + \I\sigma_{l'j'}^- - \I\sigma_{lj}^-)}
    \biggr] \,.
  \end{aligned}
\end{equation}
We now need to distinguish \(\eta_k=\pm 1\) because explicit expressions
differ.  We will bring the results together at the end of this section.

Focus first on the case \(\eta_k=+1\).  Since \(1 + \I m_{kk} + \I
m_{k(k\pm 1)}+\I m_{(k\pm 1)k} = 0\), the factors with \((l,j)\in E\)
and \((l',j')\in E\) in~\eqref{SeibergQ-Z1l} cancel.  The remaining
factors are invariant under the exchanges \(E\to E^\complement\) and
\(m_{kl}-\sigma_{lj}^\pm \leftrightarrow m_{lk}+\sigma_{lj}^\pm\).  As a
result, the one-loop determinant for the s-channel vacuum~\(E\) of the
\(U(\Nc_k)\) theory is equal to the one-loop determinant for the
u-channel vacuum~\(E^\complement\) of a theory with identical twisted
masses but \(\Nc_k^\dual = \# E^\complement = \Nc_{k-1}+\Nc_{k+1}-\Nc_k\)
colors.  As discussed in Section~\ref{sec:SeibergW-cross} and shown
directly in Appendix~\ref{app:Proof-SQCDAW}, the vortex partition
functions of the \(U(\Nc_k)\) theory in the s-channel vacuum~\(E\) and
of the \(U(\Nc_k^\dual)\) theory in the u-channel vacuum~\(E^\complement\)
are equal up to a factor~\eqref{Proof-SQCDAW-Z=Z}
\begin{equation}
  Z^{U(\Nc_k)}_{\vortex,E}(\hat{z}_k)
  = (1-\hat{z}_k)^{-\delta_1}
  Z^{U(\Nc_k^\dual)}_{\vortex,E^\complement} \bigl((\hat{z}_k^\dual)^{-1}\bigr)
\end{equation}
with \(\delta_1 = (\Nc_k-\Nc_k^\dual)(1+\I m_{kk})\), provided \(\hat{z}_k^\dual
= \hat{z}_k^{-1}\) as expected from the Toda CFT symmetry.  Finally, the
classical contribution transforms as follows:
\begin{equation}
  \prod_{(l,j)\in E} \hat{z}_k^{-\I m_{kl} + \I\sigma_{lj}^+}
  =
  \hat{z}_k^{-\delta_0 + \Tr \I\sigma_{k-1}^+ + \Tr \I\sigma_{k+1}^+}
  \prod_{(l,j)\in E^\complement} (\hat{z}_k^\dual)^{\I m_{lk} + \I\sigma_{lj}^+}
\end{equation}
with \(\delta_0 = \Nc_{k-1} \I m_{k(k-1)} + \Nc_{k+1} \I m_{k(k+1)} + (1 +
\I m_{kk}) \Nc_k^\dual\).  All in all,
\begin{equation}
  Z_k^{U(\Nc_k)}(z_k,\bar{z}_k)
  =
  \abs{\hat{z}_k}^{-2\delta_0}
  \abs{1-\hat{z}_k}^{-2\delta_1}
  \hat{z}_k^{\Tr \I\sigma_{k-1}^+ + \Tr \I\sigma_{k+1}^+}
  \bar{\hat{z}}_k^{\Tr \I\sigma_{k-1}^- + \Tr \I\sigma_{k+1}^-}
  Z_k^{U(\Nc_k^\dual)}(z_k^\dual,\bar{z}_k^\dual) \,.
\end{equation}
In the full \(S^2\)~partition function of the quiver theory, the powers
of \(\hat{z}_k\) and~\(\bar{\hat{z}}_k\) combine with the classical
contribution for the gauge group factors \(U(\Nc_{k\pm 1})\) and yield
\begin{equation}
  \abs{\hat{z}_k}^{-2\delta_0}
  \abs{1-\hat{z}_k}^{-2\delta_1}
  \prod_{l=k\pm 1}
  (\hat{z}_l \hat{z}_k)^{\Tr\I\sigma_l^+}
  \bigl(\overline{\hat{z}_l \hat{z}_k}\bigr)^{\Tr\I\sigma_l^-} \,.
\end{equation}
Therefore, the \(S^2\)~partition functions of the
\(U(\Nc_1)\times\cdots\times U(\Nc_n)\) theory and of the theory with
\(\Nc_k^\dual = \Nc_{k-1}+\Nc_{k+1}-\Nc_k\), \(\hat{z}_k^\dual =
\hat{z}_k^{-1}\), and \(\hat{z}_{k\pm 1}^\dual = \hat{z}_{k\pm 1}
\hat{z}_k\) are equal up to \(\abs{\hat{z}_k}^{-2\delta_0}
\abs{1-\hat{z}_k}^{-2\delta_1}\).  On the Toda CFT side, this factor is
due to differences in powers of \(\abs{x_k}^2\), \(\abs{x_{k+1}}^2\) and
\(\abs{x_{k+1}-x_k}^2\) which appear in the correspondences for the
electric and magnetic theories.  In gauge theory, the factor can be
absorbed into the partition function: since~\(\delta_1\) only depends
on~\(b\), the~\(\Nc_j\), and the matter content,
\(\abs{1-\hat{z}_k}^{-2\delta_1}\) is a renormalization scheme
ambiguity, while \(\abs{\hat{z}_k}^{-2\delta_0}\) can be absorbed by a
global \(U(1)\subset U(\Nc_k)\) gauge transformation.  Such ambiguities
are described below~\eqref{part-typical-matching}.

The case \(\eta_k=-1\) follows the same ideas, but is more intricate.
The Higgs branch decomposition~\eqref{SeibergQ-Z1l} involves vortex
partition functions of \(\Nsusy=(2,2)\) SQCD\@.  As for the previous case,
those are equal up to a power of \((1-\hat{z}_k)\) to vortex partition
functions of a dual theory with \(\Nc_k^\dual\)~colors and twisted
masses \(m^\dual = \I/2 - m\).  Explicitly,
\begin{equation}
  \begin{aligned}
    & Z_{\vortex,E}\bigl(\{m_{kl}-\sigma_{lj}^+\},
      \{m_{lk}+\sigma_{lj}^+\};\hat{z}_k\bigr)
    \\
    & \quad = (1-\hat{z}_k)^{-\Nc_k-2q_k\Nc_{k-1}-2q_{k+1}\Nc_{k+1}}
    Z_{\vortex,E^\complement}\bigl(\{\I/2-m_{kl}+\sigma_{lj}^+\},
      \{\I/2-m_{lk}-\sigma_{lj}^+\};\hat{z}_k\bigr) \,.
  \end{aligned}
\end{equation}
The signs with which \(\sigma_{lj}^+\) appears in the right-hand side are
inconvenient, as it implies that chiral multiplets which transform under
the fundamental representation of \(U(\Nc_k^\dual)\) also transform in
the fundamental representation of \(U(\Nc_l^\dual)\), and not the
antifundamental representation.  This is fixed by conjugating all
\(U(\Nc_k^\dual)\) charges: \(\hat{z}_k\to\hat{z}_k^{-1}\) and the
vortex partition function becomes a u-channel
(\(\abs{\hat{z}_k^{-1}}\to\infty\)) vortex partition function of SQCD
with \(\Nc_k^\dual\) colors, \(\Nc_{k-1}+\Nc_{k+1}\) flavours, and
\(\hat{z}_k^\dual = \hat{z}_k^{-1}\).  Once this is understood, the
classical contributions (of the electric s-channel vacuum labeled
by~\(E\) and the magnetic u-channel vacuum labeled
by~\(E^\complement\)) are equal up to powers of \(\abs{\hat{z}_k}^2\),
provided \(\hat{z}_k^\dual = \hat{z}_k^{-1}\) and \(\hat{z}_{k\pm
  1}^\dual = \hat{z}_{k\pm 1} \hat{z}_k\).  This is precisely the map
described by the exchange of Toda CFT punctures.

The one-loop determinants, on the other hand, transform non-trivially.
This is expected from the study of Seiberg duality for \(\Nsusy=(2,2)\)
SQCD: the magnetic theory includes mesons whose one-loop determinants
appear in the \(S^2\)~partition function.  There, the mesons are
realized as \(M_{ts} = \anti{\quark}_t \quark_s\) in terms of the electric quarks
and antiquarks \(\quark_s\) and~\(\anti{\quark}_t\), and couple to the magnetic
multiplets through a superpotential term \(\anti{\quark}_t^\dual M_{ts}
\quark_s^\dual\).  In our current setting, the mesons are the four
combinations \(M_{ll'} = \biquark_{lk} \biquark_{kl'}\) in the electric
theory, and couple to the magnetic multiplets through the superpotential
\(\Tr(M_{ll'} \biquark_{l'k}^\dual \biquark_{kl}^\dual)\).  The mesons
\(M_{(k\pm 1)(k\pm 1)}\) transform in the adjoint representations of
\(U(\Nc_{k\pm 1})\), and the mesons \(M_{(k\pm 1)(k\mp 1)}\) in
bifundamental representations of \(U(\Nc_{k+1})\times U(\Nc_{k-1})\).
Since the (electric) superpotential features the term
\(\Tr(M_{(k-1)(k+1)} M_{(k+1)(k-1)})\) for \(\eta_k=-1\), these two
mesons can be integrated out, leaving the term
\(\Tr(\biquark_{(k-1)k}^\dual \biquark_{k(k+1)}^\dual
\biquark_{(k+1)k}^\dual \biquark_{k(k-1)}^\dual)\) in the superpotential
of the magnetic theory.  This term is expected from \(\eta_k^\dual =
-1\).

Next, for each of \(l=k\pm 1\) there are two cases.  If \(\eta_{k\pm 1}
= +1\) then the superpotential term \(\Tr(X_{k\pm 1} M_{(k\pm 1)(k\pm
  1)})\) lets us integrate out both \(X_{k\pm 1}\) and the
meson~\(M_{(k\pm 1)(k\pm 1)}\), leaving the magnetic superpotential
\(\Tr(\biquark_{(k\pm 1)(k\pm 2)} \biquark_{(k\pm 2)(k\pm 1)}
\biquark_{(k\pm 1)k}^\dual \biquark_{k(k\pm 1)}^\dual)\).  This is
expected from \(\eta_{k\pm 1}^\dual = -1\) (multiplets
\(\biquark_{ll'}\) with \(l,l'\neq k\) are not affected by the duality).
If instead \(\eta_{k\pm 1} = -1\), then we integrate out~\(X_{k\pm 1}\), and note the
presence of magnetic superpotential terms \(\Tr(\biquark_{(k\pm 1)(k\pm
  2)} \biquark_{(k\pm 2)(k\pm 1)} M_{(k\pm 1)(k\pm 1)})\) and
\(\Tr(M_{(k\pm 1)(k\pm 1)} \biquark_{(k\pm 1)k}^\dual \biquark_{k(k\pm
  1)}^\dual)\).  Those are expected from \(\eta_{k\pm 1}^\dual = +1\).
In both cases, the change in matter content between the electric and
magnetic theories and the mapping of twisted masses are encoded in the
map \(\eta_{k\pm 1}^\dual = \eta_{k\pm 1} \eta_k\) implied by the
exchange \(\epsilon_{k-1} \leftrightarrow \epsilon_k\).

Combining the classical, one-loop, and vortex contributions yields the
equality of \(S^2\)~partition functions up to powers of
\(\abs{\hat{z}_k}^2\) and \(\abs{1-\hat{z}_k}^2\) when \(\eta_k=-1\).
As for \(\eta_k=+1\), the powers of \(\abs{1-\hat{z}_k}^2\) and
\(\abs{\hat{z}_k}^2\) are an ambiguity of the
\(S^2\)~partition function.
This concludes the proof (for arbitrary~\(\eta\)) that applying Seiberg
duality or the \(\Nsusy=(2,2)^*\) duality to the gauge group factor
\(U(\Nc_k)\) with \(1\leq k<n\) corresponds to transposing the
punctures \(k\) and \(k+1\) in the Toda CFT correlator.  Therefore,
permutations of Toda CFT degenerate punctures encapsulate the mapping of
parameters for arbitrary combinations of dualities which act on the
nodes with \(1\leq k<n\).

\subsubsection{Seiberg Dualities from Momentum Conjugation}
\label{sec:SeibergQ-group}

We now find all theories obtained through dualities acting on any
node, including \(U(\Nc_n)\).

For simplicity, we first consider the theory with \(\eta_n=-1\) and
\(\eta_k=+1\) for \(k<n\), which corresponds to a Toda CFT correlator
where all degenerate punctures are labeled by antisymmetric
representations of~\(A_{\Nf-1}\) (all \(\epsilon_k=-1\)).  Since
\(\eta_n=-1\), the superpotential includes a term \(\Tr X_n^2\) which
lets us integrate out the adjoint chiral multiplet~\(X_n\).  Therefore,
the chiral multiplets which couple to the \(U(\Nc_n)\) vector multiplet
are those of \(\Nsusy=(2,2)\) SQCD with \(\Nc_n\)~colors and
\(\Nf+\Nc_{n-1}\) flavours.  Applying Seiberg duality and charge
conjugation to the node \(U(\Nc_n)\)
yields a similar quiver gauge theory with \(\Nc_n\)
replaced by \(\Nc_n^\dual = \Nf+\Nc_{n-1}-\Nc_n\).
Recall that the Seiberg dual of a theory includes additional multiplets
with charges identical to mesons of the original theory.  Here,
these are \(\Nf^2\)~free chiral multiplets, and \(\Nf\)~fundamental,
\(\Nf\)~antifundamental, and one adjoint of \(U(\Nc_{n-1})\).  The
magnetic theory thus has two adjoints of \(U(\Nc_{n-1})\).  Given the
cubic superpotential which links the bifundamentals of \(U(\Nc_n)\times
U(\Nc_{n-1})\) and the adjoint of \(U(\Nc_{n-1})\) in the electric
theory, the two adjoints of \(U(\Nc_{n-1})\) couple through a quadratic
superpotential term and can thus be integrated out.  Therefore, the two
dual theories are given by the quivers
\begin{equation}
  \quiver[color-group/.append style={minimum size=6ex}, node distance=7.5ex]{
    \node (Nn)  [color-group]                 {$\scriptstyle\Nc_n$};
    \node (Nn-1)[color-group, right of=Nn]    {$\scriptstyle\Nc_{n-1}$};
    \node (dots)[right of=Nn-1]               {$\scriptstyle\cdots$};
    \node (N1)  [color-group, right of=dots]  {$\scriptstyle\Nc_1$};
    \node (f)   [left of=Nn]                  {};
    \node (Nf)  [flavor-group, above=1ex of f]{$\scriptstyle\Nf$};
    \node (Nf') [flavor-group, below=1ex of f]{$\scriptstyle\Nf$};
    \draw[->-=.55] (Nf') -- (Nn);
    \draw[->-=.55] (Nn)  -- (Nf);
    \draw[->-=.55] (Nn)   to [bend left=30] (Nn-1);
    \draw[->-=.55] (Nn-1) to [bend left=30] (dots);
    \draw[->-=.55] (dots) to [bend left=30] (N1);
    \draw[->-=.55] (N1)   to [bend left=30] (dots);
    \draw[->-=.55] (dots) to [bend left=30] (Nn-1);
    \draw[->-=.55] (Nn-1) to [bend left=30] (Nn);
    \draw[->-=.5] (Nn-1)  to [distance=5ex, in=60, out=120, loop] ();
    \draw[->-=.5] (N1)    to [distance=5ex, in=60, out=120, loop] ();
  }
  \:\rightsquigarrow\:
  \quiver[color-group/.append style={minimum size=6ex}, node distance=7.5ex]{%
    \node (Nn)  [color-group]                    {$\scriptstyle\Nc_n^\dual$};
    \node (Nn-1)[color-group, right of=Nn]       {$\scriptstyle\Nc_{n-1}$};
    \node (Nn-2)[color-group, right of=Nn-1]     {$\scriptstyle\Nc_{n-2}$};
    \node (dots)[right of=Nn-2]                  {$\scriptstyle\cdots$};
    \node (N1)  [color-group, right of=dots]     {$\scriptstyle\Nc_1$};
    \node (f)   at ($(Nn)!0.5!(Nn-1)$)           {};
    \node (Nf)  [flavor-group, above=3ex of f]   {$\scriptstyle\Nf$};
    \node (Nf') [flavor-group, below=3ex of f]   {$\scriptstyle\Nf$};
    \draw[->-=.55] (Nf')  to [bend left=90,distance=9ex] (Nf);
    \draw[->-=.55] (Nf)  -- (Nn);
    \draw[->-=.55] (Nn)  -- (Nf');
    \draw[->-=.55] (Nn-1)-- (Nf);
    \draw[->-=.55] (Nf') -- (Nn-1);
    \draw[->-=.55] (Nn)   to [bend left=30] (Nn-1);
    \draw[->-=.55] (Nn-1) to [bend left=30] (Nn-2);
    \draw[->-=.55] (Nn-2) to [bend left=30] (dots);
    \draw[->-=.55] (dots) to [bend left=30] (N1);
    \draw[->-=.55] (N1)   to [bend left=30] (dots);
    \draw[->-=.55] (dots) to [bend left=30] (Nn-2);
    \draw[->-=.55] (Nn-2) to [bend left=30] (Nn-1);
    \draw[->-=.55] (Nn-1) to [bend left=30] (Nn);
    \draw[->-=.5] (Nn-2)  to [distance=5ex, in=60, out=120, loop] ();
    \draw[->-=.5] (N1)    to [distance=5ex, in=60, out=120, loop] ();
  }
\end{equation}
where the superpotential is the sum of all gauge (and flavour) invariant
cubic combinations of bifundamental and adjoint chiral multiplets.  The
complexified FI parameters of the magnetic theory are \(\hat{z}_n^\dual
= \hat{z}_n^{-1}\), \(\hat{z}_{n-1}^\dual = \hat{z}_n \hat{z}_{n-1}\), and
\(\hat{z}_k^\dual = \hat{z}_k\) for \(k\leq n-2\).

The absence of adjoint chiral multiplet of \(U(\Nc_{n-1})\) in the
second theory lets us apply Seiberg duality (and charge conjugation) to
this node of the second quiver.  Once more, the resulting quiver
contains additional matter, including an adjoint of \(U(\Nc_{n-2})\)
which cancels the already present adjoint because of a quadratic
superpotential.  The procedure can thus be continued by acting on
successive nodes from \(U(\Nc_n)\) to \(U(\Nc_1)\).  The resulting
quivers are given in Figure~\ref{fig:zipper}.

\begin{figure}[pt]\centering
  \caption{\label{fig:zipper}%
    Sequence of Seiberg dualities on the quiver with all \(\epsilon_k=-1\).}
  \medskip
  \begin{tabular}{p{0.97\textwidth}}
    \toprule
    The quiver with \(\eta_n=-1\) and \(\eta_k=+1\) for \(k<n\)
    corresponds to a Toda CFT correlator with only antisymmetric
    degenerate operators.  Acting with Seiberg dualities successively
    on all nodes from \(U(\Nc_n)\) to~\(U(\Nc_1)\) yields a quiver of
    the same form, which corresponds to the Toda CFT correlator with
    all momenta conjugated.  The original quiver, the quiver after
    acting on the \(k\)-th node, and the final quiver are drawn here
    without free mesons transforming in the bifundamental
    representation of the flavour group \(S[U(\Nf)\times U(\Nf)]\) to
    avoid clutter.  After acting on the \(k\)-th node,
    the complexified FI parameters are given by
    \((\hat{z}_{n-1},\ldots,\hat{z}_k,
    (\hat{z}_n\hat{z}_{n-1}\cdots\hat{z}_k)^{-1},
    (\hat{z}_n\cdots\hat{z}_{k-1}),\hat{z}_{k-2},\ldots,\hat{z}_1)\)
    in this order.  Dual ranks are
    \(\Nc_j^\dual=(n+1-j)\Nf+\Nc_{j-1}-\Nc_n\).
    \begin{center}
    \quiver[color-group/.append style={minimum size=6ex}, node distance=7.5ex]{%
      \node (LNn)  [color-group]                  {$\scriptstyle\Nc_n$};
      \node (LNn-1)[color-group, right of=LNn]    {$\scriptstyle\Nc_{n-1}$};
      \node (Ldots)[right of=LNn-1]               {$\scriptstyle\cdots$};
      \node (LN1)  [color-group, right of=Ldots]  {$\scriptstyle\Nc_1$};
      \node (Lf)   [left of=LNn]                  {};
      \node (LNf)  [flavor-group, above=1ex of Lf]{$\scriptstyle\Nf$};
      \node (LNf') [flavor-group, below=1ex of Lf]{$\scriptstyle\Nf$};
      \draw[->-=.55] (LNf') -- (LNn);
      \draw[->-=.55] (LNn)  -- (LNf);
      \draw[->-=.55] (LNn)   to [bend left=30] (LNn-1);
      \draw[->-=.55] (LNn-1) to [bend left=30] (Ldots);
      \draw[->-=.55] (Ldots) to [bend left=30] (LN1);
      \draw[->-=.55] (LN1)   to [bend left=30] (Ldots);
      \draw[->-=.55] (Ldots) to [bend left=30] (LNn-1);
      \draw[->-=.55] (LNn-1) to [bend left=30] (LNn);
      \draw[->-=.5] (LNn-1)  to [distance=5ex, in=60, out=120, loop] ();
      \draw[->-=.5] (LN1)    to [distance=5ex, in=60, out=120, loop] ();
      \node (mid)  [right of=LN1] {};
      \node (RNn)  [color-group, right of=mid]       {$\scriptstyle\Nc_n^\dual$};
      \node (Rdots)[right=2ex of RNn]                {$\scriptstyle\cdots$};
      \node (RN2)  [color-group, right=2ex of Rdots] {$\scriptstyle\Nc_2^\dual$};
      \node (RN1)  [color-group, right of=RN2]       {$\scriptstyle\Nc_1^\dual$};
      \node (Rf)   [right of=RN1]                    {};
      \node (RNf)  [flavor-group, above=0ex of Rf]   {$\scriptstyle\Nf$};
      \node (RNf') [flavor-group, below=0ex of Rf]   {$\scriptstyle\Nf$};
      \draw[->-=.55] (RNf)   -- (RN1);
      \draw[->-=.55] (RN1)   -- (RNf');
      \draw[->-=.55] (RNn)   to [bend left=30] (Rdots);
      \draw[->-=.55] (Rdots) to [bend left=30] (RN2);
      \draw[->-=.55] (RN2)   to [bend left=30] (RN1);
      \draw[->-=.55] (RN1)   to [bend left=30] (RN2);
      \draw[->-=.55] (RN2)   to [bend left=30] (Rdots);
      \draw[->-=.55] (Rdots) to [bend left=30] (RNn);
      \draw[->-=.5] (RNn)    to [distance=5ex, in=60, out=120, loop] ();
      \draw[->-=.5] (RN2)    to [distance=5ex, in=60, out=120, loop] ();
      \node (Bf)    [below=10ex of mid]                 {};
      \node (BNk)   [color-group, left=0ex of Bf]       {$\scriptstyle\Nc_k^\dual$};
      \node (BNk+1) [color-group, left of=BNk]          {$\scriptstyle\Nc_{k+1}^\dual$};
      \node (BLdots)[left=2ex of BNk+1]                 {$\scriptstyle\cdots$};
      \node (BNn)   [color-group, left=2ex of BLdots]   {$\scriptstyle\Nc_n^\dual$};
      \node (BNk-1) [color-group, right=0ex of Bf]      {$\scriptstyle\Nc_{k-1}$};
      \node (BNk-2) [color-group, right of=BNk-1]       {$\scriptstyle\Nc_{k-2}$};
      \node (BRdots)[right=2ex of BNk-2]                {$\scriptstyle\cdots$};
      \node (BN1)   [color-group, right=2ex of BRdots]  {$\scriptstyle\Nc_1$};
      \node (BNf)   [flavor-group, above=4.5ex of Bf]   {$\scriptstyle\Nf$};
      \node (BNf')  [flavor-group, below=4.5ex of Bf]   {$\scriptstyle\Nf$};
      \draw[->-=.55] (BNf)   -- (BNk);
      \draw[->-=.55] (BNk)   -- (BNf');
      \draw[->-=.55] (BNf')  -- (BNk-1);
      \draw[->-=.55] (BNk-1) -- (BNf);
      \draw[->-=.55] (BNn)    to [bend left=30] (BLdots);
      \draw[->-=.55] (BLdots) to [bend left=30] (BNk+1);
      \draw[->-=.55] (BNk+1)  to [bend left=30] (BNk);
      \draw[->-=.55] (BNk)    to [bend left=30] (BNk-1);
      \draw[->-=.55] (BNk-1)  to [bend left=30] (BNk-2);
      \draw[->-=.55] (BNk-2)  to [bend left=30] (BRdots);
      \draw[->-=.55] (BRdots) to [bend left=30] (BN1);
      \draw[->-=.55] (BN1)    to [bend left=30] (BRdots);
      \draw[->-=.55] (BRdots) to [bend left=30] (BNk-2);
      \draw[->-=.55] (BNk-2)  to [bend left=30] (BNk-1);
      \draw[->-=.55] (BNk-1)  to [bend left=30] (BNk);
      \draw[->-=.55] (BNk)    to [bend left=30] (BNk+1);
      \draw[->-=.55] (BNk+1)  to [bend left=30] (BLdots);
      \draw[->-=.55] (BLdots) to [bend left=30] (BNn);
      \draw[->-=.5] (BNn)   to [distance=5ex, in=60, out=120, loop] ();
      \draw[->-=.5] (BNk+1) to [distance=5ex, in=60, out=120, loop] ();
      \draw[->-=.5] (BNk-2) to [distance=5ex, in=60, out=120, loop] ();
      \draw[->-=.5] (BN1)   to [distance=5ex, in=60, out=120, loop] ();
      \node at ($(LN1)!0.5!(BNk)$)   {\Huge\rotatebox{-45}{$\rightsquigarrow$}};
      \node at ($(BNk-1)!0.5!(RNn)$) {\Huge\rotatebox {45}{$\rightsquigarrow$}};
    }
    \end{center}
    \\\bottomrule
  \end{tabular}
\end{figure}

We note in particular that the last quiver, obtained after applying
Seiberg duality to all the nodes, has the same form as the original
quiver: only one gauge group factor features fundamental and
antifundamental chiral multiplets.  This quiver gauge theory, or rather
the \(\Nsusy=(2,2)\) surface operator it defines in any class S theory,
has a Toda CFT interpretation as the insertion of some degenerate vertex
operators.  Given the matter content of the gauge theory, all
\(n\)~degenerate vertex operators are labeled by antisymmetric
representations of~\(A_{\Nf-1}\).  The ranks of these representations are
obtained from the number of colors in the dual theory:
\begin{equation}
  K_j^\dual = \Nc_j^\dual - \Nc_{j+1}^\dual
  = \Nf - (\Nc_j - \Nc_{j-1}) = \Nf - K_j
\end{equation}
for \(1\leq j\leq n\), where \(\Nc_{n+1}^\dual=\Nc_0=0\).  The positions
of punctures are obtained from the FI parameters:
\begin{equation}
  x_j^\dual = \prod_{i=1}^j \hat{z}_i^\dual
  = \biggl[ \prod_{i=j}^n \hat{z}_i \biggr]^{-1} = x_j^{-1} \,.
\end{equation}
Both of these maps are reproduced by conjugating all Toda CFT momenta
and applying the conformal transformation \(x\to x^{-1}\) to the
correlator.  This conformal transformation could be avoided by applying
charge conjugation to all nodes of the quiver, mapping all complexified
FI parameters to their inverse in the process.

All in all, Toda CFT conjugation translates to a combination of Seiberg
dualities and charge conjugations.  Here, the precise choice of matter
content of the gauge theory is essential.  On the gauge theory side, it
ensures the absence of adjoint chiral multiplet at each step hence
allows Seiberg duality to be applied.  On the Toda CFT side, the
conjugate of a symmetric representation is neither symmetric nor
antisymmetric, thus momentum conjugation only yields symmetric or
antisymmetric representations if the
original representations were all antisymmetric.  It should be noted
that this choice of signs is identical to that made in
Section~\ref{sec:Quivers-fuse} to fuse degenerate punctures into a
degenerate puncture labeled by an arbitrary representation, hence
conjugating this representation corresponds to a set of Seiberg
dualities on the gauge theory quiver.

We now go back to a quiver given by arbitrary signs~\(\eta_k\), and
determine all dual descriptions obtained through Seiberg and
\(\Nsusy=(2,2)^*\) dualities.  Inspired by the quivers which appeared when
all \(\epsilon_k=-1\), we consider the more general class of quivers
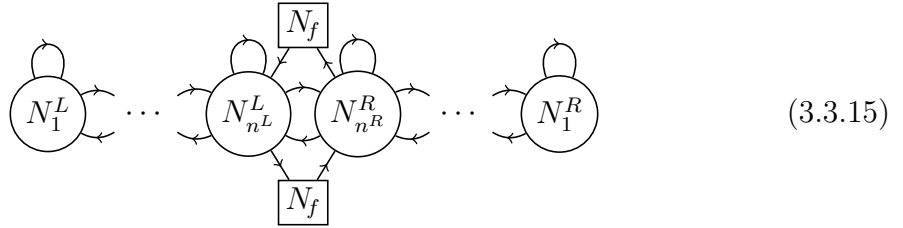
\begin{equation}\label{SeibergQ-biquiver}
  \quiver[color-group/.append style={minimum size=4.5ex}, node distance=8ex]{%
    \node (NL1)   [color-group]                     {$\Nc^L_1$};
    \node (Ldots) [right=2ex of NL1]                {$\cdots$};
    \node (NLn)   [color-group, right=2ex of Ldots] {$\Nc^L_{n^L}$};
    \node (NRn)   [color-group, right of=NLn]       {$\Nc^R_{n^R}$};
    \node (Rdots) [right=2ex of NRn]                {$\cdots$};
    \node (NR1)   [color-group, right=2ex of Rdots] {$\Nc^R_1$};
    \node (f)     at ($(NLn)!0.5!(NRn)$)            {};
    \node (Nf)    [flavor-group, above=4ex of f]    {$\Nf$};
    \node (Nf')   [flavor-group, below=4ex of f]    {$\Nf$};
    \foreach \A/\B in {NL1/Ldots,Ldots/NLn,NLn/NRn,NRn/Rdots,Rdots/NR1}
      {
        \draw[->-=.55] (\A)   to [bend left=30] (\B);
        \draw[->-=.55] (\B)   to [bend left=30] (\A);
      }
    \foreach \A/\B in {NRn/Nf,Nf/NLn,NLn/Nf',Nf'/NRn}
      \draw[->-=.55] (\A) -- (\B);
    \foreach \A in {NL1,NLn,NRn,NR1}
      \draw[->-=.5] (\A) to [distance=3.5ex, in=70, out=110, loop] ();
  }
  \,.
\end{equation}
The multiplets described by this quiver are subject to a superpotential
which depends on some signs \(\eta^L_k\) for \(1\leq k\leq n^L\) and
\(\eta^R_k\) for \(1\leq k\leq n^R\).  Namely, the superpotential is a
sum of \(W_{\eta^L}\)~defined as in~\eqref{Quivers-superpotential} for
fields charged under the \(U(\Nc^L_k)\), \(W_{\eta^R}\)~for fields
charged under the \(U(\Nc^R_k)\), and two cubic terms coupling each
bifundamental of \(U(\Nc^L_{n^L})\times U(\Nc^R_{n^R})\) to multiplets
charged under the flavour groups.  For \(n^L=0\) or \(n^R=0\) we
retrieve the quivers studied throughout this paper.  Whenever
\(\eta^L_k=-1\), the superpotential contains a quadratic term
\(\Tr\bigl((X^L_k)^2\bigr)\) which lets us integrate out the adjoint
chiral multiplet~\(X^L_k\) of \(U(\Nc^L_k)\), and similarly
\(\eta^R_k=-1\) lets us integrate out~\(X^R_k\).

Even though we have not given a Toda CFT interpretation for this class
of quivers, we find analogues of the Toda CFT parameters
\((x_k,\epsilon_k,K_k)\) which are simply transposed under dualities.
Let \(x^L_{n^L+1} = x^R_{n^R+1} = \epsilon^L_{n^L+1} =
\epsilon^R_{n^R+1} = 1\) and
\begin{align}
  x^L_j & = \prod_{i=j}^{n^L} \hat{z}^L_i \,,
  & \epsilon^L_j & = \prod_{i=j}^{n^L} \eta^L_i \,,
  & K^L_j & = \Nc^L_j - \Nc^L_{j-1}
  & & \text{for \(1\leq j\leq n^L\)} \,,
  \\
  x^R_j & = \prod_{i=j}^{n^R} \hat{z}^R_i \,,
  & \epsilon^R_j & = \prod_{i=j}^{n^R} \eta^R_i \,,
  & K^R_j & = \Nc^R_j - \Nc^R_{j-1}
  & & \text{for \(1\leq j\leq n^R\)} \,,
\end{align}
where \(\Nc^L_0 = \Nc^R_0 = 0\).

Acting with Seiberg or the \(\Nsusy=(2,2)^*\) duality (depending
on~\(\eta^L_k\)) on a node \(U(\Nc^L_k)\) with \(k<n^L\) exchanges
\((x^L_k,\epsilon^L_k,K^L_k) \leftrightarrow
(x^L_{k+1},\epsilon^L_{k+1},K^L_{k+1})\).  This is proven through the
same calculations as for the case \(n^R=0\) treated in
Section~\ref{sec:SeibergQ-perm}.  Similarly, acting with a duality on
\(U(\Nc^R_k)\) with \(k<n^R\) exchanges \((x^R_k,\epsilon^R_k,K^R_k)
\leftrightarrow (x^R_{k+1},\epsilon^R_{k+1},K^R_{k+1})\).

Let us now understand how dualities act on \(U(\Nc^R_{n^R})\).  If
\(\epsilon^R_{n^R} = \eta^R_{n^R} = +1\), the fields which couple to the
gauge group factor \(U(\Nc^R_{n^R})\) are those of \(\Nsusy=(2,2)\)
SQCDA, no simplification occurs, and neither Seiberg nor the
\(\Nsusy=(2,2)^*\) duality applies.  However, if \(\epsilon^R_{n^R} =
\eta^R_{n^R} = -1\), we can integrate out the adjoint chiral multiplet
to obtain SQCD with \(\Nc^R_{n^R}\) colors and
\(\Nf+\Nc^L_{n^L}+\Nc^R_{n^R-1}\) flavours, and Seiberg duality yields a
theory with \((\Nc^R_{n^R})^\dual =
\Nf+\Nc^L_{n^L}+\Nc^R_{n^R-1}-\Nc^R_{n^R}\) colors.  The magnetic theory
has the same form~\eqref{SeibergQ-biquiver} as the electric theory, but
it features fundamental and antifundamental chiral multiplets of
\(U\bigl((\Nc^R_{n^R})^\dual\bigr)\) and \(U(\Nc^R_{n^R-1})\) rather
than \(U(\Nc^L_{n^L})\) and \(U(\Nc^R_{n^R})\): in other words, \(n^L
\to n^L + 1\) and \(n^R \to n^R - 1\).  Due to the additional mesons
after Seiberg duality, both \(\eta^L_{n^L}\) and~\(\eta^R_{n^R-1}\)
change signs, thus toggling between the presence or absence of an
adjoint chiral multiplet.  From our previous work on the action of
Seiberg duality on quivers, we also know that FI parameters map as
\(\hat{z}^L_{n^L} \to \hat{z}^L_{n^L} \hat{z}^R_{n^R}\),
\(\hat{z}^R_{n^R} \to (\hat{z}^R_{n^R})^{-1}\) and \(\hat{z}^R_{n^R-1}
\to \hat{z}^R_{n^R-1} \hat{z}^R_{n^R}\).  Translating to the parameters
\((x,\epsilon,K)\), we find that the set
\begin{equation}\label{SeibergQ-set}
  \bigl\{ \bigl( (x^L_j)^{-1} , \epsilon^L_j , \Nf - K^L_j \bigr)
  \bigm| 1\leq j\leq n^L \bigr\}
  \cup
  \bigl\{ (x^R_j, \epsilon^R_j, K^R_j) \bigm| 1\leq j\leq n^R \bigr\}
\end{equation}
is unchanged: the triplet \((x^R_{n^R}, \epsilon^R_{n^R}, K^R_{n^R})\)
is simply moved from the second part of the set (on the right of flavour
nodes) to the first part (on the left).  By symmetry, the discussion
applies to the node \(U(\Nc^L_{n^L})\): if \(\eta^L_{n^L}=+1\) there is
no duality, while if \(\eta^L_{n^L}=-1\) Seiberg duality moves
\(\bigl((x^L_{n^L})^{-1}, \epsilon^L_{n^L}, \Nf-K^L_{n^L}\bigr)\) from
the left part to the right part of~\eqref{SeibergQ-set}.

All in all, Seiberg and \(\Nsusy=(2,2)^*\) dualities acting on any of
the nodes of~\eqref{SeibergQ-biquiver} correspond to transpositions of
\(\bigl((x^L_1)^{-1},\epsilon^L_1,\Nf-K^L_1\bigr)\), \ldots,
\(\bigl((x^L_{n^L})^{-1},\epsilon^L_{n^L},\Nf-K^L_{n^L}\bigr)\),
\(\diamond\), \((x^R_{n^R},\epsilon^R_{n^R},K^R_{n^R})\), \ldots,
\((x^R_1,\epsilon^R_1,K^R_1)\).  The position of~\(\diamond\) indicates
the position of the flavour nodes in the quiver.  Only triplets
\((x,\epsilon,K)\) with \(\epsilon=-1\) can be exchanged
with~\(\diamond\).  Therefore, combinations of
dualities correspond to all permutations which leave triplets with
\(\epsilon^L_j=+1\) to the left of~\(\diamond\) and those
with \(\epsilon^R_j=+1\) to the right
of~\(\diamond\).  Denoting by \(n^L_+\) and~\(n^R_+\) the number of such
triplets, and by \(n_-\)~the total number of triplets with
\(\epsilon=-1\), we conclude that the number of dual descriptions
of the theory~\eqref{SeibergQ-biquiver} is
\begin{equation}
  \sum_{k=0}^{n_-} \binom{n_-}{k} (n^L_+ + k)! (n^R_+ + n_- - k)!
  = \frac{n^L_+! n^R_+! (n^L_+ + n^R_+ + n_- + 1)!} {(n^L_+ + n^R_+ + 1)!}
  \,.
\end{equation}

As a last comment, we propose that the partition function of the
\(S^2\)~surface operator defined by coupling~\eqref{SeibergQ-biquiver}
to \(\Nf^2\)~free hypermultiplets on~\(S^4_b\) should be equal to
\begin{equation}\label{SeibergQ-matching}
  Z_{S^2\subset S^4_b}^{\eqref{SeibergQ-biquiver}}
  = \vev*{\widehat{V}_{\alpha_\infty}(\infty)
    \widehat{V}_{\hat{m}}(1)
    \widehat{V}_{\alpha_0}(0)
    \prod_{j=1}^{n^L}
    \widehat{V}_{-b\Omega(K^L_j,\epsilon^L_j)^\conj} \bigl((x^L_j)^{-1}\bigr)
    \prod_{j=1}^{n^R}
    \widehat{V}_{-b\Omega(K^R_j,\epsilon^R_j)} (x^R_j)
  }
\end{equation}
up to factors that can be absorbed in~\(Z\).  Here,
\(\Omega(K,+1)=Kh_1\) is the highest weight of a symmetric
representation while \(\Omega(K,-1)=\omega_K\) is the highest weight of
an antisymmetric representation.  The proposal is consistent with the
action of dualities as permutations of \((x,\epsilon,K)\) triplets
described above: in particular, an antisymmetric representation with
highest weight \(\omega_K\) can be seen either as part of the left
product (\(\omega_K = \Omega(\Nf-K,-1)^\conj\)) or as part of the right
product (\(\omega_K = \Omega(K,-1)\)), and this choice reproduces the
Seiberg duality map.  On the contrary, the conjugate of a symmetric
representation is neither symmetric nor antisymmetric, so punctures with
\(\epsilon=+1\) belong to a given product and cannot be moved to the
other one.  We have not explored this proposal further, as fusions of
antisymmetric representations are enough to obtain arbitrary representations.

\acknowledgments

We would like to thank Davide Gaiotto, Peter Koroteev, Daniel Park, and
J\"org Teschner for useful discussions.  B.L.F.\@ would like to thank the
Perimeter Institute for hospitality.  Research at Perimeter Institute is
supported by the Government of Canada through Industry Canada and by the
Province of Ontario through the Ministry of Economic Development and
Innovation. J.G.\@ also acknowledges further support from an NSERC
Discovery Grant and from an ERA grant by the Province of Ontario.

\appendix
\addtocontents{toc}{\protect\setcounter{tocdepth}{2}}

\section{Toda CFT}
\label{app:Toda}

This appendix is devoted to the \(A_{\Nf-1}\)~Toda CFT, a generalization
of the Liouville CFT (\(\Nf=2\)), and can be read independently.  It is
split into five topics: we review notations and basic properties
(Appendix \ref{app:Toda-basics}), match products of three-point
functions with gauge theory one-loop determinants (Appendix
\ref{app:1loop-3pt}), derive new braiding matrices useful in the main
text (Appendix \ref{app:braiding}), list known fusion rules and find new
ones (Appendix \ref{app:fusion}), deduce new conformal blocks from the
correspondence (Appendix~\ref{app:blocks}) and finally define some
irregular punctures (Appendix \ref{app:irregular}).

\subsection{Basic Properties}
\label{app:Toda-basics}

We describe here some properties of the \(A_{\Nf-1}\)~Toda CFT, omitting
some details which can be found in~\cite{Fateev:2007ab}.  We introduce
the normalizations \eqref{normalized-V} and~\eqref{normalized-semi} of
vertex operators, which simplify three-point functions hence simplify
constant factors in the main text.

Microscopically, the theory describes a scalar field~\(\phiToda\) in the
Cartan subalgebra of~\(A_{\Nf-1}\), minimally coupled to the metric,
with an exponential potential term.  It depends on a coupling
constant~\(b\), and a cosmological constant~\(\mu\).  We will use the
combination
\begin{equation}
  \hat{\mu} = \bigl[\pi\mu\gamma(b^2)b^{2-2b^2}\bigr]^{1/b}
\end{equation}
where \(\gamma(x)=\Gamma(x)/\Gamma(1-x)\), because the theory is
(non-manifestly) invariant under \((b,\hat{\mu})\to (1/b,\hat{\mu})\).
Besides its local symmetry algebra~\(W_{\Nf}\) (a higher-spin extension of
the Virasoro algebra~\(W_2\)), the theory also possesses a discrete
symmetry \(\phiToda\to\phiToda^\conj\), defined as the \(\bbC\)-linear
map such that
\begin{equation}
  h_s^\conj = -h_{\Nf+1-s}
\end{equation}
for all \(1\leq s\leq\Nf\), where \(h_s\)~are the weights of the
fundamental representation of~\(A_{\Nf-1}\).  These weights form an
overcomplete basis (\(h_1+\cdots+h_{\Nf} = 0\)) of linear forms over the
Cartan subalgebra of~\(A_{\Nf-1}\).  In principle, one should
distinguish the space of linear forms from the Cartan subalgebra, but
the (bilinear) Killing form \(\vev{\ ,\ }\) defined by
\begin{equation}
  \vev{h_s, h_t} = \delta_{st} - \frac{1}{\Nf}
\end{equation}
identifies the two spaces.  Note that the Killing form is invariant
under conjugation, and that \((\phiToda^\conj)^\conj=\phiToda\).
Conjugation maps the highest weight of a representation to the highest
weight of the conjugate representation, hence its name.

Vertex operators \(V_\alpha = e^{\vev{\alpha,\phiToda}}\), labeled by
their momentum~\(\alpha\) in the Cartan subalgebra, are primary
operators for the \(W_{\Nf}\)~symmetry algebra.  The symmetry
\(\phiToda\to\phiToda^\conj\) maps
\(V_\alpha=e^{\vev{\alpha,\phiToda}}\) to
\(e^{\vev{\alpha,\phiToda^\conj}} = e^{\vev{\alpha^\conj,\phiToda}} =
V_{\alpha^\conj}\), and since simple roots are permuted under conjugation,
correlators of vertex operators are invariant
under conjugating all momenta~\(\alpha_i\to\alpha_i^\conj\).

Each vertex operator~\(V_\alpha\) is additionally invariant up to a
constant factor (reflection amplitude) under the Weyl group
of~\(A_{\Nf-1}\), which acts by permuting the \(\Nf\)~components
\(\vev{\alpha-Q,h_s}\).  Here, \(Q=(b+\frac{1}{b})\rho\) is a multiple
of the sum
\begin{equation}
  \rho = \frac{1}{2} \sum_{e>0} e
  = \frac{1}{2} \sum_{s<t}^{\Nf} (h_s-h_t)
  = \sum_s \frac{\Nf+1-2s}{2} h_s
\end{equation}
of all positive roots~\(e=h_s-h_t\), \(1\leq s<t\leq\Nf\),
of~\(A_{\Nf-1}\).  The invariance of~\(V_\alpha\) is confirmed by noting
that its dimension
\begin{equation}
  \dimToda(\alpha) = \frac{1}{2}\vev{\alpha,2Q-\alpha} \,,
\end{equation}
and quantum numbers associated to higher spin generators of~\(W_{\Nf}\),
are invariant under Weyl reflections.  The
normalization~\eqref{normalized-V} later on absorbs reflection
amplitudes.

When decomposing \(n\)-point functions into products of three-point
functions and conformal blocks, we must take into account two-point
functions as well.  Non-zero two-point functions are \(\vev{V_\alpha
  V_{2Q-\alpha}}\) and Weyl reflections thereof.  As a result, the
momenta which appear in two neighboring three-point functions of the
decomposition are related by the map \(\alpha \to 2Q-\alpha\).  For the
Liouville CFT (\(\Nf=2\)), those momenta are Weyl reflections of each
other, hence the distinction is irrelevant.  For the general
\(A_{\Nf-1}\)~Toda CFT, one must include an orientation when labeling
conformal blocks by the various internal momenta, and reversing the
orientation amounts to changing \(\alpha\to 2Q-\alpha\).  External
momenta can also be given an orientation (which must then be retained
for correlators as well as conformal blocks), where an ``incoming''
momentum~\(\alpha\) denotes the presence of the vertex
operator~\(V_\alpha\), and an ``outgoing'' momentum~\(\alpha\)
denotes~\(V_{2Q-\alpha}\).  This incoming/outgoing distinction also
affects the relation between fusion rules and non-zero three-point
functions.

Generically, vertex operators generate irreducible representations
of~\(W_{\Nf}\).  Semi-degenerate vertex operators are defined by the
presence of null-vectors among their \(W_{\Nf}\)~descendants.  In this
paper, all semi-degenerate vertex operators take the form \(V_{\varkappa
  h_1}\) (and~\(V_{-\lambda h_{\Nf}}\)).  The conjugate momentum
\((\varkappa h_1)^\conj = -\varkappa h_{\Nf}\) is in fact mapped by the
Weyl reflection defined by the permutation \((1\,2\cdots\Nf)\) to the
original form
\begin{equation}\label{Toda-conjugate-semi}
  \varkappa^\dual h_1
  = \left[\Nf \left(b+\frac{1}{b}\right)-\varkappa\right] h_1 \,,
\end{equation}
since \(\vev{\varkappa h_1^\conj-Q, h_{\Nf}} = \vev{\varkappa^\dual h_1-Q,
  h_1}\), and \(\vev{\varkappa h_1^\conj - Q, h_{s-1}} =
\vev{\varkappa^\dual h_1-Q, h_s}\) for all \(2\leq s\leq\Nf\).  Thus,
\(V_{\varkappa h_1^\conj}\) and~\(V_{\varkappa^\dual h_1}\) are equal up
to a reflection amplitude, absorbed by the
normalization~\eqref{normalized-semi}.  This equality is crucial to
obtain dualities in Section~\ref{sec:Seiberg} and
Section~\ref{sec:Kutasov} as conjugation of momenta.

Fully degenerate momenta \(\alpha=-b\omega-\frac{1}{b}\omega'\) are
labeled by pairs \((\omega,\omega')\) of highest weights
of~\(A_{\Nf-1}\) representations.  We only consider in this work
degenerate momenta of the form \(\alpha=-b\omega\), and mapping
\(b\to\frac{1}{b}\) would probe degenerate momenta
\(\alpha=-\frac{1}{b}\omega\), but the mixed case with non-zero
\(\omega\) and~\(\omega'\) is hard to access.  We denote the
representation of~\(A_{\Nf-1}\) with highest weight~\(\omega\) by
\(\repr(\omega)\).  In particular, the fundamental representation
\(\repr(h_1)\) has weights~\(h_s\) for \(1\leq s\leq\Nf\), and highest
weight~\(h_1\).  The \(\Nc\)-th antisymmetric power of~\(\repr(h_1)\) is
\(\repr(\omega_\Nc)\), where \(\omega_\Nc = \sum_{j=1}^{\Nc} h_j\); it
has weights \(h_{\{p\}} = \sum_{j=1}^\Nc h_{p_j}\) for \(1\leq
p_1<\cdots<p_\Nc\leq\Nf\).  The \(\Nc\)-th symmetric power \(\repr(\Nc
h_1)\) has weights \(\sum_{j=1}^\Nc h_{p_j}\) for \(1\leq
p_1\leq\cdots\leq p_\Nc\leq\Nf\), or equivalently \(h_{[n]} = \sum_{s=1}^{\Nf} n_s
h_s\) for (non-negative) integers \(n_1+\cdots+n_{\Nf}=\Nc\).  We also
consider quasi-rectangular Young diagrams: for \(0\leq j<l\) and \(0\leq
k<\Nf\), the highest weight \(\omega_{kl+j,l} = l\omega_k+jh_{k+1}\)
corresponds to a Young diagram with \(kl+j\) boxes, organized as
\(k\)~rows of \(l\)~boxes, followed by a \(j\)-box row.  This reproduces
the antisymmetric case \(\omega_{\Nc,1} = \omega_\Nc\) for \(l=1\), and
the symmetric case \(\omega_{\Nc,l} = \Nc h_1\) for \(l\geq\Nc\).

In view of the matching of parameters with gauge theory, we write
generic momenta as \(\alpha = Q - \I a\).  The dimension is
\(\dimToda(Q-\I a)=\frac{1}{2}\vev{Q,Q}-\frac{1}{2}\vev{\I a,\I a}\).
Weyl reflections act by permuting the \(\vev{a,h_s}\).  In terms of the
Upsilon function~\eqref{Upsilon-shift} below, we introduce the
normalization
\begin{equation}\label{normalized-V}
  \widehat{V}_{Q-\I a}
  = \frac{\hat{\mu}^{-\vev{\I a,\rho}}}
  {\prod_{s<t}^{\Nf} \Upsilon(\vev{\I a,h_s-h_t})}
  V_{Q-\I a}
  \,,
\end{equation}
where the product ranges over positive roots \(e=h_s-h_t\).  The
normalization factor is invariant under conjugation, hence does not
spoil this symmetry of Toda CFT correlators involving generic
operators~\(\widehat{V}_\alpha\).  The three-point function
\(\vev{\widehat{V}_{\alpha} \widehat{V}_{\alpha'} \widehat{V}_{\varkappa
    h_1}}\) given in~\eqref{FL-1-39} is invariant under Weyl reflections
permuting the \(\vev{a,h_s}\), hence the normalized
operator~\(\widehat{V}_{\alpha}\) is Weyl invariant.  To further
simplify three-point functions, we also provide a normalization for
semi-degenerate operators and fully degenerate operators,
\begin{equation}\label{normalized-semi}
  \widehat{V}_{\varkappa h_1}
  = \frac{\hat{\mu}^{\vev{\varkappa h_1,\rho}}}
  {\bigl(\Upsilon(b)\bigr)^{\Nf-1} \Upsilon(\varkappa)}
  V_{\varkappa h_1}
  \,, \qquad
  \widehat{V}_{-b\omega}
  = \bigl[\hat{\mu}b^{2(b+\frac{1}{b})}\bigr]^{\vev{-b\omega, \rho}}
  V_{-b\omega}
  \,.
\end{equation}
The normalizations of generic and semi-degenerate operators are
invariant under \(b\to \frac{1}{b}\).

The Upsilon function appearing above depends implicitly on the coupling
constant~\(b\) (it is invariant under \(b\to \frac{1}{b}\)), and for
generic real~\(b\) it is a holomorphic function, uniquely determined
by its normalization
\(\Upsilon\bigl(\frac{1}{2}(b+\frac{1}{b})\bigr)=1\) and by shift
relations
\begin{equation}\label{Upsilon-shift}
  \Upsilon(x+b) = \gamma(bx) b^{1-2bx} \Upsilon(x)\,,
  \qquad
  \Upsilon(x+1/b)
  = \gamma(x/b) b^{2x/b-1} \Upsilon(x) \,.
\end{equation}
Also, \(\Upsilon(b+\frac{1}{b}-x)=\Upsilon(x)\) and the function has
zeros at \(-mb-n\frac{1}{b}\) and \((m+1)b+(n+1)\frac{1}{b}\) for
integers \(m,n\geq 0\), and no poles.  As \(x\to\pm\I\infty\), one has
\begin{gather}
  \label{Upsilon-asymptotic-ratio}
  \frac{\Upsilon(x+a)}{\Upsilon(x)}
  \sim \biggl(\frac{-x^2}{e^2}\biggr)^{ax} \abs{x}^{a(a-b-1/b)}
  \sim \bigl(\gamma(bx)b^{1-2bx}\bigr)^{a/b} \abs{x}^{a(a-b)}
  \,,
  \\
  \label{Upsilon-asymptotic-prod}
  \prod_{s=1}^{\Nf} \frac{\Upsilon(x+\vev{\alpha,h_s})}{\Upsilon(x)}
  \sim \abs{x}^{\vev{\alpha,\alpha}}
  \,,
\end{gather}
for any~\(a\) and any momentum~\(\alpha\).  The gamma function
\(\gamma(x) = \Gamma(x) / \Gamma(1-x)\) obeys by construction
\(\gamma(1-x) = 1 / \gamma(x)\) and also appears in one-loop
determinants of chiral multiplets.  Vortex partition functions involve
Pochhammer symbols
\begin{equation}
  (x)_k = \frac{\Gamma(x+k)}{\Gamma(x)}
  = (-1)^k \frac{\Gamma(1-x)}{\Gamma(1-x-k)}
  = \frac{(-1)^k}{(1-x)_{-k}} \,.
\end{equation}
This equality is shown using the Euler identity \(\Gamma(x) \Gamma(1-x)
= \pi/\sin\pi x\).

\subsection{Three-Point Functions}
\label{app:1loop-3pt}

We check in this appendix that the one-loop determinants which appear in
Higgs branch expansions of \(S^2\)~partition functions, considered in
the main text, are equal to products of three-point functions which
appear in s-~and u-channel decompositions of the corresponding Toda CFT
correlators.  This relies on the three-point functions provided
by~\cite{Fateev:2007ab}, equations (1.39), (1.51), and~(1.56), which we
first translate to our normalization.

The three-point function of two generic operators \(\widehat{V}_{Q-\I
  a_1}\) and~\(\widehat{V}_{Q-\I a_2}\) and of a semi-degenerate
operator~\(\widehat{V}_{\varkappa h_1}\) is expressed as the
normalizations \eqref{normalized-V} and~\eqref{normalized-semi}
multiplied by equation~(1.39) of~\cite{Fateev:2007ab} with all momenta
conjugated:
\begin{equation}\label{FL-1-39}
  \begin{aligned}
    \widehat{C}(Q-\I a_1,Q-\I a_2,\varkappa h_1)
    & = \frac
      {\hat{\mu}^{-\vev{\I a_1+\I a_2-\varkappa h_1,\rho}}
        C(Q-\I a_1^\conj,Q-\I a_2^\conj,\varkappa h_1^\conj)}
      {\bigl(\Upsilon(b)\bigr)^{\Nf-1} \Upsilon(\varkappa)
        \prod_{s<t}^{\Nf} \Upsilon(\vev{\I a_1,h_s-h_t})
        \Upsilon(\vev{\I a_2,h_s-h_t})}
    \\
    & =
    \frac{1}
    {\prod_{s,t=1}^{\Nf} \Upsilon\bigl(\frac{\varkappa}{\Nf}
        +\vev{\I a_1,h_s}+\vev{\I a_2,h_t}\bigr)} \,.
  \end{aligned}
\end{equation}
The three-point function is invariant under Weyl transformations of each
\(\widehat{V}_{Q-\I a_i}\), which permute the~\(\vev{\I a_i,h_s}\),
hence the normalized~\(\widehat{V}_{Q-\I a}\) are Weyl invariant, as
claimed earlier.  The three-point function is also invariant under
conjugation of all momenta, followed by the Weyl transformation~\eqref{Toda-conjugate-semi} which
maps \((\varkappa h_1)^\conj \to \varkappa^\dual h_1 = (\Nf(b+1/b)-\varkappa)h_1\): indeed,
\(\vev{\I a_i,h_s}\to -\vev{\I a_i,h_s}\) and \(\varkappa/\Nf\to
(b+1/b)-\varkappa/\Nf\) under this transformation, and we know that
\(\Upsilon(b+1/b-x)=\Upsilon(x)\).

Besides this three-point function, we also need some three-point
functions involving a degenerate operator~\(\widehat{V}_{-b\omega}\).
The OPE of this operator with a generic~\(\widehat{V}_{Q-\I a}\) is
\begin{equation}
  \widehat{V}_{-b\omega} \widehat{V}_{Q-\I a}
  = \sum_{h\in\repr(\omega)}
  \widehat{C}_{-b\omega,Q-\I a}^{Q-\I a-bh}
  \bigl[\widehat{V}_{Q-\I a-bh}\bigr] \,,
\end{equation}
where the sum runs over weights of \(\repr(\omega)\) and the brackets
denote the contribution from \(W_{\Nf}\)~descendants (see
Appendix~\ref{app:fusion} for a description of which momenta can appear
in various OPE\@).  The structure constants \(\widehat{C}_{-b\omega,Q-\I
  a}^{Q-\I a-bh}\) are equal to their analogues given
in~\cite{Fateev:2007ab} for usual vertex operators, multiplied by the
normalization factors of \(\widehat{V}_{-b\omega}\)
and~\(\widehat{V}_{Q-\I a}\), and divided by the normalization
of~\(\widehat{V}_{Q-\I a-bh}\), namely
\begin{equation}\label{normalized-Cdeg}
  \widehat{C}_{-b\omega,Q-\I a}^{Q-\I a-bh}
  =
  \hat{\mu}^{\vev{bh-b\omega, \rho}} b^{2\vev{-b\omega,Q}}
  \prod_{s<t}^{\Nf} \biggl[
  \frac{\Upsilon(\vev{\I a+bh,h_s-h_t})}{\Upsilon(\vev{\I a,h_s-h_t})}
  \biggr]
  C_{-b\omega,Q-\I a}^{Q-\I a-bh} \,.
\end{equation}
The structure constants are also closely related to three-point
functions:
\begin{equation}
  \widehat{C}_{-b\omega,Q-\I a}^{Q-\I a-bh}
  = \prod_{s\neq t}^{\Nf} \Bigl[ \Upsilon(\vev{\I a+bh,h_s-h_t}) \Bigr]
  \widehat{C}(-b\omega,Q-\I a,Q+\I a+bh) \,.
\end{equation}
The change \(Q-\I a-bh\to Q+\I a+bh\) and the Upsilon functions both
come from the non-zero two-point functions
\(\vev{\widehat{V}_\alpha(z,\bar{z}) \widehat{V}_{2Q-\alpha}(0)} =
\abs{z}^{-4\dimToda(\alpha)} \bigm/ \prod_{s\neq t}^{\Nf}
\Upsilon(\vev{Q-\alpha,h_s-h_t})\).

Equation~(1.51) of~\cite{Fateev:2007ab} covers the case of a degenerate
field~\(\widehat{V}_{-b\omega_\Nc}\) labeled by the antisymmetric
representation~\(\repr(\omega_\Nc)\), whose weights \(h = h_{p_1}
+ \cdots + h_{p_\Nc}\) are labeled by \(\Nc\)-element subsets of
\(\{1,\ldots,\Nf\}\) without repetition.  With our
normalization~\eqref{normalized-Cdeg}, all Upsilon functions cancel
through the shift relation~\eqref{Upsilon-shift}, and leave only gamma
functions:
\begin{equation}\label{3pt-antisym}
  \widehat{C}_{-b\omega_\Nc,Q-\I a}^{Q-\I a-bh}
  =
  b^{-\Nf \vev{2\I a+bh, bh}}
  \prod_{s\not\in\{p\}}^{\Nf} \prod_{t\in\{p\}} \gamma(b\vev{\I a,h_t-h_s})
  \,.
\end{equation}
When matching with the \(S^2\)~partition function of SQCDA, we need
three-point functions involving~\(\widehat{V}_{-\Nc bh_1}\).  Weights of
the \(\Nc\)-th symmetric representation~\(\repr(\Nc h_1)\) are \(h
= \sum_{s=1}^{\Nf} n_s h_s\) for a choice of \(\Nf\)~integers \(n_s \geq
0\) with \(n_1+\cdots+n_{\Nf} = \Nc\).  The three-point function can be
derived from~\eqref{FL-1-39} by setting \(\varkappa = -\Nc b\), taking
into account the normalization, and extracting the residue at \(\I a_1 =
\I a\) and \(\I a_2 = - \I a - b h\).  This yields
\begin{equation}\label{3pt-sym}
  \widehat{C}_{-\Nc bh_1, Q-\I a}^{Q-\I a-bh}
  =
  \frac{b^{-\Nf\vev{2\I a+bh,bh}}}{\prod_{\nu=1}^\Nc \gamma(-\nu b^2)}
  \prod_{s,t=1}^{\Nf} \prod_{\nu=0}^{n_t-1}
  \gamma(b\vev{\I a,h_t-h_s}+(\nu-n_s)b^2)
  \,.
\end{equation}
Taking \(\Nc = 1\) in either \eqref{3pt-antisym} or~\eqref{3pt-sym}
yields the (same) expression for the case of a fundamental degenerate
field,
\begin{equation}
  \widehat{C}_{-bh_1, Q-\I a}^{Q-\I a-bh_p}
  =
  b^{-\Nf \vev{2\I a+bh_p,bh_p}}
  \prod_{s\neq p}^{\Nf} \gamma(b\vev{\I a,h_p-h_s}) \,.
\end{equation}
For theories with a superpotential, we also use some three-point
functions with a degenerate \(\widehat{V}_{-be_0}\) labeled by the
highest weight \(e_0 = h_1-h_{\Nf}\) of the adjoint representation.
Because the weight~\(0\) has multiplicity in this representation,
\(\widehat{C}_{-b(h_1-h_{\Nf}),\alpha}^\alpha\) is not a ratio of Gamma
functions.  We will focus on other weights \(h=h_i-h_j\), which have no
multiplicity.  From equation~(1.56) of~\cite{Fateev:2007ab},
\begin{equation}
  \widehat{C}_{-be_0, Q-\I a}^{Q-\I a-bh}
  =
  b^{-\Nf\vev{2\I a+bh,bh}}
  \frac{\gamma(b\vev{\I a,h_i-h_j}+b^2)}{\gamma(b\vev{\I a,h_i-h_j})}
  \prod_{s\neq i}^{\Nf} \! \gamma(b\vev{\I a,h_i-h_s})
  \prod_{s\neq j}^{\Nf} \! \gamma(b\vev{\I a,h_s-h_j})
  \,.
\end{equation}

We are now ready to consider the products of three-point functions
appearing in s-~and u-channel decompositions of Toda CFT correlators of
interest.  Our first computation concerns the s-channel
decomposition~\eqref{SQED-conf-s} of a four-point function with the
degenerate insertion~\(\widehat{V}_{-bh_1}\), which corresponds to the
\(S^2\)~partition function of SQED, multiplied by the
contribution~\(Z_{S^4_b}^\free =
\widehat{C}(\alpha_\infty,\alpha_0,\varkappa h_1)\) of free
hypermultiplets.  We set \(\alpha_\infty = Q - \I a_\infty\) and
\(\alpha_0 = Q - \I a_0\), and evaluate:
\begin{align}\label{1loop-3pt-SQED}
  & \widehat{C}_{-bh_1,\alpha_0}^{\alpha_0-bh_p}
  \widehat{C}(\alpha_\infty, \alpha_0-bh_p,(\varkappa+b)h_1)
  \bigm/
  \widehat{C}(\alpha_\infty,\alpha_0,\varkappa h_1)
  \\\nonumber
  & \quad =
  \widehat{C}_{-bh_1,\alpha_0}^{\alpha_0-bh_p}
  \prod_{s,t=1}^{\Nf} \biggl[
  \frac{\Upsilon\bigl(\frac{\varkappa}{\Nf}
    +\vev{\I a_0,h_s}+\vev{\I a_\infty,h_t}\bigr)}
  {\Upsilon\bigl(\frac{\varkappa}{\Nf}
    +\vev{\I a_0,h_s}+\vev{\I a_\infty,h_t}+b\delta_{ps}\bigr)}
  \biggr]
  \\\nonumber
  & \quad =
  \underbrace{b^{2b\varkappa-\Nf(1+b^2)+b^2}}_{A^{-1}}
  \frac{\prod_{s\neq p}^{\Nf} \gamma(b\vev{\I a_0,h_p-h_s})}
    {\prod_{t=1}^{\Nf} \gamma(\frac{b\varkappa}{\Nf}
      +b\vev{\I a_0,h_p}+b\vev{\I a_\infty,h_t})}
  \,.
\end{align}
The numerator gamma function is \(\gamma(\I m_p-\I m_s)\) in terms of
gauge theory twisted masses, and the denominator is \(\gamma(1+\I m_p
+\I\anti{m}_t)\).  We thus recognize the one-loop determinant appearing
in the s-channel decomposition~\eqref{SQED-Zs} of~\(Z\), divided by
the constant~\(A\) given
in~\eqref{SQED-A}.  Since \(A\) is invariant under the exchange of
\(\alpha_0\) and~\(\alpha_\infty\), which amounts to exchanging the s-
and u-channels, the matching of three-point functions and one-loop
contributions also occurs in the u-channel.

Next is the case of SQCD\@.  The corresponding Toda four-point function
involves the degenerate insertion~\(\widehat{V}_{-b\omega_\Nc}\), the
semi-degenerate~\(\widehat{V}_{(\varkappa+\Nc b)h_1}\), and two generic
momenta \(\alpha_0 = Q - \I a_0\) and \(\alpha_\infty = Q - \I
a_\infty\).  We factor out the \(S^4_b\)~contribution
\(\widehat{C}(\alpha_\infty,\alpha_0,\varkappa h_1)\).  For a given
weight \(h = h_{p_1} + \cdots + h_{p_\Nc}\)
of~\(\repr(\omega_\Nc)\), we find
\begin{align}\label{1loop-3pt-SQCD}
  & \widehat{C}_{-b\omega_\Nc,\alpha_0}^{\alpha_0-bh}
  \widehat{C}(\alpha_\infty, \alpha_0-bh, (\varkappa+\Nc b)h_1)
  \bigm/
  \widehat{C}(\alpha_\infty, \alpha_0, \varkappa h_1)
  \\\nonumber
  & \quad =
  \widehat{C}_{-b\omega_\Nc,\alpha_0}^{\alpha_0-bh}
  \prod_{s,t=1}^{\Nf} \biggl[
  \frac{\Upsilon\bigl(\frac{\varkappa}{\Nf}
    +\vev{\I a_0,h_t}+\vev{\I a_\infty,h_s}\bigr)}
  {\Upsilon\bigl(\frac{\varkappa}{\Nf}
    +\vev{\I a_0,h_t}+\vev{\I a_\infty,h_s}+b\delta_{t\in\{p\}}\bigr)}
  \biggr]
  \\\nonumber
  & \quad =
  \underbrace{b^{2\Nc b\varkappa-\Nc\Nf(1+b^2)+\Nc^2 b^2}}_{A^{-1}}
  \prod_{t\in\{p\}} \biggl[
  \frac{\prod_{s\not\in\{p\}}^{\Nf} \gamma(-b\vev{\I a_0,h_s - h_t})}
  {\prod_{s=1}^{\Nf} \gamma(\frac{b\varkappa}{\Nf}
    +b\vev{\I a_0,h_t}+b\vev{\I a_\infty,h_s})}
  \biggr]
  \,.
\end{align}
Again, we recognize the ratio of \(\gamma(-\I m_s+\I m_t)\) and
\(\gamma(1+\I\anti{m}_s+\I m_t)\) as being the one-loop
determinants~\eqref{SQCD-Zs} of SQCD\@.  This fixes the constant~\(A\) to
be~\eqref{SQCD-A} in the matching with SQCD\@.  Since \(A\)~is invariant
under the exchange of \(\alpha_0\) and~\(\alpha_\infty\), the u-channel
three-point functions and one-loop determinant match up to the same
constant.

Our last four-point function corresponds to the \(S^2\)~partition
function of SQCDA, and involves the degenerate field~\(\widehat{V}_{-\Nc
  bh_1}\).  With notations as above, and for \(h = \sum_s n_s h_s\), we
compute
\begin{align}\label{1loop-3pt-SQCDA}
  &
  \widehat{C}_{-\Nc bh_1,\alpha_0}^{\alpha_0-bh}
  \widehat{C}(\alpha_\infty, \alpha_0-bh,(\varkappa+\Nc b)h_1)
  \bigm/ \widehat{C}(\alpha_\infty, \alpha_0, \varkappa h_1)
  \\\nonumber
  & \quad =
  \underbrace{
    \frac{b^{2\Nc b\varkappa-\Nc\Nf(1+b^2)+\Nc^2 b^2}}
    {\prod_{\nu=1}^\Nc \gamma(-\nu b^2)}
  }_{A^{-1}}
  \prod_{s,t=1}^{\Nf} \prod_{\nu=0}^{n_t-1} \biggl[
  \frac{\gamma(b\vev{\I a_0,h_t-h_s}+(\nu-n_s)b^2)}
  {\gamma(\frac{b\varkappa}{\Nf}
    +b\vev{\I a_0,h_t}+b\vev{\I a_\infty,h_s}+\nu b^2)}
  \biggr]
  \,.
\end{align}
In terms of gauge theory variables, the numerator gamma functions have
arguments \(\I m_t + \nu b^2 - \I m_s - n_s b^2\), while the denominator
have arguments \(1 + \I\anti{m}_s + \I m_t + \nu b^2\), hence we obtain
the one-loop determinant~\eqref{SQCDA-Z1l}, divided by the
constant~\(A\) given in~\eqref{SQCDA-A}.  Once more,
\(A\)~is invariant under \(\alpha_0\leftrightarrow\alpha_\infty\), hence
u-channel three-point functions and one-loop determinants match.

Perhaps interestingly, the power of~\(b\) appearing in~\(A\) can be recast as
\begin{equation}
  b^{\Nc\Nf(1+b^2)-\Nc^2 b^2-2\Nc b\varkappa}
  = b^{-\varkappa \left[\Nf\left(b+\frac{1}{b}\right)-\varkappa\right]}
  \Big/
  b^{-(\varkappa+\Nc b)\left[\Nf\left(b+\frac{1}{b}\right)-\varkappa-\Nc b\right]} \,.
\end{equation}
We do not absorb these powers of~\(b\) into the normalization
of~\(\widehat{V}_{\varkappa h_1}\) and \(\widehat{V}_{(\varkappa+\Nc
  b)h_1}\), because they would spoil the \(b\to\frac{1}{b}\) symmetry
which~\eqref{normalized-semi} enjoys.  Note that these powers are
invariant under conjugation, which maps
\(\varkappa\to\Nf\bigl(b+\frac{1}{b}\bigr)-\varkappa\) and
\(\varkappa+\Nc b\) similarly.

\subsection{Braiding Matrices}
\label{app:braiding}

In this appendix, we compute the braiding matrix of the antisymmetric
degenerate operator~\(V_{-b\omega_\Nc}\) around the semi-degenerate
operator~\(V_{\varkappa h_1}\), as well as its gauge theory analogue,
and check that they are equal.

\subsubsection{Gauge Theory Transfer Matrices}

Let us start on the gauge theory side: namely, we find the matrix
relating Higgs branch decompositions near \(z=0\) and near \(z=\infty\)
of the partition function~\eqref{SQCD-Zs} of SQCD\@.

First, focus on the case of SQED (\(\Nc=1\)), which uses the same
techniques as Appendix~B of~\cite{Gomis:2010kv}.  Recall the Higgs
branch decomposition~\eqref{SQED-Zs} near~\(0\),
\begin{equation}
  Z = \sum_{p=1}^{\Nf} \Biggl\{
  (x\bar{x})^{-\I m_p}
  \frac{\prod_{s\neq p}^{\Nf}\gamma(-\I m_s+\I m_p)}
    {\prod_{s=1}^{\Nf}\gamma(1+\I\anti{m}_s+\I m_p)}
  f^\sch_p(m,\anti{m},x) f^\sch_p(m,\anti{m},\bar{x}) \Biggl\} \,,
\end{equation}
where the series~\eqref{SQED-fs} defining \(f^\sch_p(x)\) converges for
\(\abs{x}<1\).  Similarly, the Higgs branch decomposition
near~\(\infty\) involves series which converge for \(\abs{x}>1\).  We
wish to relate the two sets of holomorphic factors, or rather, their
analytic continuation to a common domain.  This is done through the
integral representation~\eqref{SQED-fs-integral} also given
in~\eqref{braid-SQED-f} below, which converges away from the positive
real axis.  For \(\abs{x}\lessgtr 1\) we close the contour integral
towards \(\kappa\to\pm\infty\), enclosing either the poles at
\(\kappa+\I m_p \in \bbZ_{\geq 0}\) or the \(\Nf\)~families of poles at
\(\kappa-\I\anti{m}_s \in \bbZ_{\leq 0}\) labeled by a flavour~\(s\).
The first choice yields a single s-channel factor, while the second
yields a sum of \(\Nf\) u-channel factors:
\begin{align}
  \nonumber
  & (-x)^{-\I m_p} f^\sch_p(x)
  \\\nonumber
  & \quad \stackrel{\text{cont}}{=}
  \prod_{s=1}^{\Nf}\biggl[
    \frac{\Gamma(1+\I m_s-\I m_p)}{\Gamma(-\I\anti{m}_s-\I m_p)}
  \biggr]
  \int_{-\I\infty}^{\I\infty} \frac{\dd{\kappa}}{2\pi\I}
  \frac{\prod_{s=1}^{\Nf}\Gamma(-\I\anti{m}_s+\kappa)}
    {\prod_{s\neq p}^{\Nf}\Gamma(1+\I m_s+\kappa)}
  \Gamma(-\kappa-\I m_p) (-x)^{\kappa}
  \\\label{braid-SQED-f}
  & \quad \stackrel{\text{cont}}{=}
  \sum_{s=1}^{\Nf} B_{ps}^0
  (-x)^{\I\anti{m}_s} f^\uch_s(x)
  =
  \sum_{s=1}^{\Nf} D_p \check{B}_{ps}^0 \anti{D}_s
  (-x)^{\I\anti{m}_s} f^\uch_s(x) \,.
\end{align}
The transfer matrix~\(B_{ps}^0\) is the product of simpler matrices
\(D\), \(\check{B}^0\) and~\(\anti{D}\) given in~\eqref{braid-SQED-BDD}.
It is also convenient to work with the s-channel factors \(x^{-\I m_p}
f^\sch_p(x)\), analytically continued with branch cuts on \((-\infty,0]
\cup [1,+\infty)\), and the u-channel factors \(x^{\I\anti{m}_s}
f^\uch_s(x)\), with branch cuts along \((-\infty,0] \cup [0,1]\), rather
than with the factors appearing in~\eqref{braid-SQED-f}, which all have
branch cuts along \([0,1] \cup [1,+\infty)\).  Using \((-x)^\lambda =
e^{-\I\pi\epsilon\lambda} x^\lambda\) for \(\epsilon=\operatorname{sign}(\Im x)\), we
obtain
\begin{equation}\label{braid-SQED-f-eps}
  x^{-\I m_p} f^\sch_p(x)
  \stackrel{\text{cont}}{=}
  \sum_{s=1}^{\Nf} B^\epsilon_{ps} x^{\I\anti{m}_s} f^\uch_s(x)
  =
  \sum_{s=1}^{\Nf} D_p \check{B}^\epsilon_{ps} \anti{D}_s
  x^{\I\anti{m}_s} f^\uch_s(x) \,,
\end{equation}
which only differs from~\eqref{braid-SQED-f} by a phase
in~\(\check{B}^\epsilon\):
\begin{align}\label{braid-SQED-BDD}
  \check{B}^\epsilon_{ps} & =
  \frac{\pi e^{\pi\epsilon(m_p+\anti{m}_s)}}{\sin\pi(-\I\anti{m}_s-\I m_p)}
  \,,
  &
  \begin{aligned}
    D_p & = \prod_{t=1}^{\Nf}
    \frac{\Gamma(1+\I m_t-\I m_p)}{\Gamma(-\I\anti{m}_t-\I m_p)} \,,
    \\
    \anti{D}_s & =
    \frac{\prod_{t\neq s}^{\Nf} \Gamma(-\I\anti{m}_t+\I\anti{m}_s)}
    {\prod_{t=1}^{\Nf} \Gamma(1+\I m_t+\I\anti{m}_s)} \,.
  \end{aligned}
\end{align}
Through the matching of parameters~\eqref{SQED-alpha}, the matrix
\(D\check{B}^\epsilon\anti{D}\) reproduces the appropriate braiding
matrix (B.11) of~\cite{Gomis:2010kv}.  This is expected since conformal
blocks and vortex partition functions are already known explicitly to
match.

The monodromy matrix around~\(1\) is a product \(M_1 = B^+ (B^-)^{-1}\)
of braiding matrices.  Since all \(\check{B}^+_{ps} - \check{B}^-_{ps} =
2\pi\I\), the matrix \(M_1 - \id = D (\check{B}^+ - \check{B}^-)
\anti{D} (B^-)^{-1}\) has rank~\(1\).  Thus, \(M_1\)~has the
eigenvalue~\(1\) with multiplicity~\(\Nf-1\).  This was used
below~\eqref{SQED-Zt}.

Next, recall that the partition function of SQCD can be expressed
as~\eqref{SQCD-Z-from-SQED} in terms of derivatives of a product of SQED
partition functions.  This also holds for s-channel (and u-channel)
holomorphic factors~\eqref{SQCD-fs-from-SQED}, and we can analytically
continue each SQED factor using~\eqref{braid-SQED-f-eps}:
\begin{align}
  & x^{-\sum_{j=1}^\Nc \I m_{p_j}} f^\sch_{\{p\}}(x)
  =
  \Biggl[\prod_{i<j}
  \frac{x_i \partial_{x_i} - x_j \partial_{x_j}}
    {-\I m_{p_i} + \I m_{p_j}}
  \prod_{j=1}^\Nc \Bigl[ x_j^{-\I m_{p_j}} f^\sch_{p_j}(x_j) \Bigr]
  \Biggr]_{x_j=x}
  \\
  & \quad \stackrel{\mathclap{\text{cont}}}{=}
  \Biggl[\prod_{i<j}
  \frac{x_i \partial_{x_i} - x_j \partial_{x_j}}
    {-\I m_{p_i} + \I m_{p_j}}
  \prod_{j=1}^\Nc \sum_{s_j=1}^{\Nf} \Bigl[
  D_{p_j} \check{B}^\epsilon_{p_js_j} \anti{D}_{s_j} \,
  x_j^{\I\anti{m}_{s_j}} f^\uch_{s_j}(x_j)
  \Bigr]
  \Biggr]_{\mathrlap{x_j=x}}
  \\\label{braid-why-s-neq}
  & \quad =
  \sum_{s_1\neq\cdots\neq s_\Nc}
  \Biggl[
  \prod_{j=1}^\Nc \Bigl[ D_{p_j} \check{B}^\epsilon_{p_js_j} \anti{D}_{s_j} \Bigr]
  \prod_{i<j} \biggl[
  \frac{\I\anti{m}_{s_i}-\I\anti{m}_{s_j}}
    {-\I m_{p_i} + \I m_{p_j}}
  \biggr]
  x^{\sum_{j=1}^\Nc \I\anti{m}_{s_j}} f^\uch_{\{s\}}(x)
  \Biggr]
  \\\label{braid-SQCD}
  & \quad =
  \sum_{1\leq s_1<\cdots<s_\Nc\leq\Nf} B^\epsilon_{\{p\}\{s\}}
  x^{\sum_{j=1}^\Nc \I\anti{m}_{s_j}} f^\uch_{\{s\}}(x) \,,
\end{align}
where \(B^\epsilon_{\{p\}\{s\}} = D_{\{p\}}
\check{B}^\epsilon_{\{p\}\{s\}} \anti{D}_{\{s\}}\) in terms of matrices
given below, and another form of~\(B^\epsilon_{\{p\}\{s\}}\) is
in~\eqref{braid-SQCD-B}.  To get~\eqref{braid-why-s-neq}, we note that
if \(s_i=s_j\) for some \(i\neq j\), the differential operators
\(x_i \partial_{x_i}\) and \(x_j \partial_{x_j}\) act identically on the
product of SQED factors (once \(x_i\) and~\(x_j\) are set to~\(x\)),
hence the term does not contribute.  After restricting ourselves to
terms with all \(s_i\) distinct, we can safely extract the product of
\(\I\anti{m}_{s_i}-\I\anti{m}_{s_j}\) to convert SQED u-channel factors
to the SQCD one.  The last step sums over permutations of the~\(s_i\),
to collect terms with the same factor, labeled by the set~\(\{s\}\).
The various ingredients are two diagonal matrices,
\begin{align}
  D_{\{p\}} &=
  \frac{\prod_{j=1}^\Nc D_{p_j}}{\prod_{i<j}(-\I m_{p_i} + \I m_{p_j})} \,,
  &
  \anti{D}_{\{s\}} &=
  \prod_{i<j} (\I\anti{m}_{s_i}-\I\anti{m}_{s_j})
  \prod_{j=1}^\Nc \anti{D}_{s_j} \,,
\end{align}
and the \(\Nc\)-th wedge power \(\check{B}^\epsilon_{\{p\}\{s\}}\) of
the SQED matrix~\(\check{B}^\epsilon_{ps}\):
\begin{align}\label{braid-SQCD-Bdet}
  & \check{B}^\epsilon_{\{p\}\{s\}}
  = \sum_{\sigma\in S_\Nc} (-1)^\sigma
  \prod_{j=1}^\Nc \check{B}^\epsilon_{p_j s_{\sigma(j)}}
  = \sum_{\sigma\in S_\Nc} (-1)^\sigma
  \prod_{j=1}^\Nc
  \frac{\pi e^{\pi\epsilon(m_{p_j}+\anti{m}_{s_{\sigma(j)}})}}
    {\sin\pi(-\I\anti{m}_{s_{\sigma(j)}}-\I m_{p_j})}
  \\\label{braid-SQCD-Bint}
  & \quad = \int \frac{\dd{\kappa_1}}{2\I}\cdots
  \frac{\dd{\kappa_\Nc}}{2\I}
  \frac{\prod_{i<j} \sin\pi(\kappa_i-\kappa_j)
    \sin\pi(\I\anti{m}_{s_i}-\I\anti{m}_{s_j})}
  {\prod_{i,j=1}^\Nc \sin\pi(\kappa_j+\I\anti{m}_{s_i})}
  \prod_{j=1}^\Nc \frac{\pi e^{\pi\epsilon(m_{p_j}+\anti{m}_{s_j})}}
  {\sin\pi(\kappa_j-\I m_{p_j})}
  \\
  & \quad = \frac{\pi^\Nc e^{\pi\epsilon\sum_{j=1}^\Nc(m_{p_j}+\anti{m}_{s_j})}
    \prod_{i<j} \sin\pi(\I\anti{m}_{s_i}-\I\anti{m}_{s_j})
    \prod_{i<j} \sin\pi(\I m_{p_i}-\I m_{p_j})}
  {\prod_{i,j} \sin\pi(-\I\anti{m}_{s_i}-\I m_{p_j})} \,.
\end{align}
The \(\dd{\kappa_j}\) contours in~\eqref{braid-SQCD-Bint} are each a
pair of vertical lines \(\frac{1}{2}-\I\infty\to\frac{1}{2}+\I\infty\)
and \(\I\infty\to-\I\infty\), surrounding poles at \(\kappa_j =
-\I\anti{m}_{s_{\sigma(j)}}\).  Convergence is guaranteed since the
integrand decreases exponentially as \(\Im\kappa\to\pm\infty\) (for
\(-1\leq\epsilon\leq 1\)).  If two \(\sigma(j)\) are equal, the
numerator sines lead to a vanishing residue.  Otherwise, the first
fraction completely cancels and we retrieve~\eqref{braid-SQCD-Bdet}.
Next, we note that the integrand has period~\(1\), hence the contour can
be replaced by \(-\frac{1}{2}-\I\infty\to-\frac{1}{2}+\I\infty\) and
\(\I\infty\to-\I\infty\), which surrounds poles at \(\kappa_j = \I
m_{p_j}\), with a factor of \((-1)^\Nc\) to account for the orientation
of the contour.  This yields the last expression.

All in all, the matrix relating s-channel and u-channel factors
in~\eqref{braid-SQCD} is
\begin{equation}\label{braid-SQCD-B}
  B^\epsilon_{\{p\}\{s\}}
  =
  \prod_{\mathclap{p\in\{p\}}}
  \frac{e^{\pi\epsilon m_p}\prod_{t\not\in\{p\}}^{\Nf} \Gamma(1+\I m_t-\I m_p)}
  {\prod_{u\not\in\{s\}}^{\Nf} \Gamma(-\I m_p-\I\anti{m}_u)}
  \prod_{\mathclap{s\in\{s\}}}
  \frac{e^{\pi\epsilon\anti{m}_s}
    \prod_{u\not\in\{s\}}^{\Nf} \Gamma(\I\anti{m}_s-\I\anti{m}_u)}
  {\prod_{t\not\in\{p\}}^{\Nf} \Gamma(1+\I m_t+\I\anti{m}_s)} \,.
\end{equation}

\subsubsection{Toda CFT Braiding Matrices}

So far in this appendix, we have manipulated gauge theory factors only.
For the gauge theory/Toda CFT correspondence to hold, those should be
equal to conformal blocks multiplied by the factor
\(z^{\gamma_0}(1-z)^{\gamma_1}\) appearing in~\eqref{SQCD-matching}.  We
will show that the braiding matrix relating s-channel and u-channel
conformal blocks is \(\Braiding{\epsilon}{PS}{} =
e^{\I\pi\epsilon\gamma_1} B^\epsilon_{PS}\), where we denote \(P=\{p\}\)
and \(S=\{s\}\).  This implies in particular that all monodromy matrices
on the gauge theory side and the Toda CFT side match, thus establishes
the correspondence for SQCD, up to a factor fixed in
Appendix~\ref{app:1loop-3pt}.

The braiding matrix \(\Braiding{\epsilon}{PS}{}\) is defined by
\begin{equation}\label{braid-Toda-anti}
  \Fblock{s}{\alpha_0-bh_P}{
    \hat{m}&-b\omega_\Nc\\
    \alpha_\infty&\alpha_0}{x}
  =
  \sum_{\substack{S\subseteq\intset{1}{\Nf}\\\#S=\Nc}}
  \Braiding{\epsilon}{PS}{
    \hat{m}&-b\omega_\Nc\\\alpha_\infty&\alpha_0}
  \Fblock{u}{\alpha_\infty-bh_S}{
    \hat{m}&-b\omega_\Nc\\
    \alpha_\infty&\alpha_0}{x}
\end{equation}
where \(\hat{m}=(\varkappa+\Nc b)h_1\), and we will often decompose
\(\alpha_0=Q-\I a_0\), \(\alpha_\infty=Q-\I a_\infty\).  Using the
dictionary \(\gamma_1 = \frac{\Nc}{\Nf}(b\varkappa+\Nc b)-\Nc(1+b^2)\),
\(a_0 = \frac{1}{b}\sum_{t=1}^{\Nf} m_t h_t\), \(a_\infty =
\frac{1}{b}\sum_{t=1}^{\Nf} \anti{m}_t h_t\), and \(\varkappa =
\frac{1}{b}\sum_{t=1}^{\Nf} (1+m_t+\anti{m}_t)\) of~\eqref{SQCD-alpha}, we
wish to prove that
\begin{equation}\label{braid-Toda-anti-B}
  \begin{aligned}
    & \Braiding{\epsilon}{PS}{
      (\varkappa+\Nc b)h_1&-b\omega_\Nc\\Q-\I a_\infty&Q-\I a_0}
    =
    e^{\I\pi\epsilon\gamma_1}
    B^\epsilon_{PS}
    \\
    & \quad =
    \begin{aligned}[t]
      e^{-\I\pi\epsilon\frac{\Nc(\Nf-\Nc)}{\Nf}b^2}
      & \prod_{p\in P}
      \frac{e^{\pi\epsilon b\vev{a_0,h_p}}
        \prod_{t\not\in P}^{\Nf} \Gamma(1+b\vev{\I a_0,h_t-h_p})}
      {\prod_{u\not\in S}^{\Nf} \Gamma(1-\frac{b\varkappa}{\Nf}
        -b\vev{\I a_0,h_p}-b\vev{\I a_\infty,h_u})}
      \\
      \cdot
      & \prod_{s\in S}
      \frac{e^{\pi\epsilon b\vev{a_\infty,h_s}}
        \prod_{u\not\in S}^{\Nf} \Gamma(b\vev{\I a_\infty,h_s-h_u})}
      {\prod_{t\not\in P}^{\Nf} \Gamma(\frac{b\varkappa}{\Nf}
        +b\vev{\I a_0,h_t}+b\vev{\I a_\infty,h_s})} \,.
    \end{aligned}
  \end{aligned}
\end{equation}
We proceed by induction on~\(\Nc\).  For \(\Nc=1\), the Toda CFT
braiding matrix is known, as mentioned below~\eqref{braid-SQED-BDD},
and matches with the gauge theory transfer matrix, thus
\eqref{braid-Toda-anti-B}~holds.  From here on, we
assume~\eqref{braid-Toda-anti-B} for a given~\(\Nc\).  In particular,
the s-channel conformal blocks are given by the gauge theory holomorphic
factors~\eqref{blocks-SQCD} for that value of~\(\Nc\), because conformal
blocks are uniquely determined by their monodromy exponents at
\(\{0,1,\infty\}\) and the braiding matrix around~\(1\).

%\bigskip

We first deduce the fusion of \(V_{-bh_1}\) and~\(V_{-b\omega_\Nc}\)
into~\(V_{-b\omega_{\Nc+1}}\),
\begin{equation}
  \label{fusion-fun-anti}
  \Fblock{t}{-b\omega_{\Nc+1}}{
    -bh_1&\!\!\!-b\omega_\Nc\\
    2Q-\alpha_0+bh_P&\alpha_0}{}
  =
  \sum_{p\in P}
  \Fusionname_{p,P}[\alpha_0]
  \Fblock{s}{\alpha_0-bh_{P\setminus\{p\}}}{
    -bh_1&\!\!\!-b\omega_\Nc\\
    2Q-\alpha_0+bh_P&\alpha_0}{}
  \,,
\end{equation}
which must have the monodromy
\(e^{2\pi\I[\dimToda(-b\omega_\Nc) + \dimToda(-bh_1) - \dimToda(-b\omega_{\Nc+1})]}
= e^{-2\pi\I[\Nc (b^2+1) + b^2 \Nc/\Nf]}\)
around \(x=1\).  We shall prove that the fusion coefficients
\begin{equation}
  \label{fusion-coefs}
  \Fusionname_{p,P}[\alpha_0]
  =
  \frac{\Gamma\bigl((\Nc+1)(1+b^2)\bigr)}{\Gamma(1+b^2)}
  \prod_{t\in P\setminus\{p\}} \biggl[
  \frac{\Gamma(b\vev{Q-\alpha_0, h_t - h_p})}
  {\Gamma(1+b^2+b\vev{Q-\alpha_0, h_t - h_p})} \biggr]
\end{equation}
give this monodromy, and are normalized so that the dominant power of
\(1-x\) has a coefficient~\(1\).

Braid \(V_{-b\omega_\Nc}\) and~\(V_{-bh_1}\) in the right-hand side of~\eqref{fusion-fun-anti}
using~\eqref{braid-Toda-anti-B} with \(P\to P\setminus\{p\}\),
\(\varkappa \to -(\Nc+1)b\), \(\I a_\infty \to -\I a_0 - bh_P\) and \(S
\to P\setminus\{s\}\) for some \(s\in P\) (\(h_P-h_S\) must be a weight
of the fundamental representation, because of \(V_{-bh_1}\)):
\begin{align}
  \nonumber
  & \sum_{p\in P}
  \Fusionname_{p,P}[\alpha_0]
  \Braiding{\epsilon}{P\setminus\{p\},P\setminus\{s\}}{
    -bh_1&-b\omega_\Nc\\2Q-\alpha_0+bh_P&\alpha_0}
  \\
  & \quad = \begin{aligned}[t]
    &
    e^{-\I\pi\epsilon\frac{\Nc}{\Nf}b^2}
    \sum_{p\in P}
    e^{\pi\epsilon b\vev{a_0,h_s-h_p}}
    \frac{\prod_{t\in P\setminus\{s\}} \sin\pi(1+b^2+b\vev{\I a_0,h_t-h_p})}
    {\prod_{t\in P\setminus\{p\}} \sin\pi(b\vev{\I a_0,h_t-h_p})}
    \\
    & \cdot
    \frac{\Gamma\bigl((\Nc+1)(1+b^2)\bigr)}{\Gamma(1+b^2)}
    \prod_{t\in P\setminus\{s\}} \biggl[
    \frac{\Gamma(b\vev{\I a_0,h_s-h_t})}
    {\Gamma(1+b^2+b\vev{\I a_0,h_s-h_t})} \biggr]
  \end{aligned}
  \\\label{braid-half}
  & \quad =
  e^{-\I\pi\epsilon \bigl[\frac{\Nc}{\Nf}b^2 + \Nc(1+b^2)\bigr]}
  \Fusionname_{s,P}[2Q-\alpha_0+bh_P] \,.
\end{align}
We have used
\begin{equation}
  \begin{aligned}
    & \sum_{p\in P}
    e^{\pi\epsilon b\vev{a_0,h_s-h_p}}
    \frac{\prod_{t\in P\setminus\{s\}} \sin\pi(1+b^2+b\vev{\I a_0,h_t-h_p})}
    {\prod_{t\in P\setminus\{p\}} \sin\pi(b\vev{\I a_0,h_t-h_p})}
    \\
    & \quad =
    \int \frac{\dd{\kappa}}{2\I}
    \frac{e^{\pi\epsilon (b\vev{a_0,h_s}-\I\kappa)}
      \prod_{t\in P\setminus\{s\}} \sin\pi(1+b^2+\I m_t+\kappa)}
    {\prod_{t\in P} \sin\pi(\I m_t+\kappa)}
    =
    e^{-\I\pi\epsilon\Nc(1+b^2)} \,,
  \end{aligned}
\end{equation}
where the contour surrounds the rectangle \(\Re\kappa\in[0,1]\),
\(\Im\kappa\in(-\infty,\infty)\).  Summing over poles yields the sum
over \(p\in P\) in the first line.  The integrals over the lines
\(1-\I\infty \to 1+\I\infty\) and \(\I\infty \to -\I\infty\) cancel
because the integrand is \(1\)-periodic, and the integrals over
\(1+\I\infty \to \I\infty\) and \(-\I\infty \to 1-\I\infty\) yield \(0\)
and~\(e^{-\I\pi\epsilon\Nc(1+b^2)}\) in some order.

In~\eqref{braid-half}, we have only done one braiding move, not a full
monodromy (two braiding moves).  However, the combination of u-channel
conformal blocks is identical to~\eqref{fusion-fun-anti} after changing
\(\I a_0\to \I a_\infty=-\I a_0-bh_P\), thus, by symmetry, braiding once
more to reach the s-channel yields the same phase factor.  Therefore,
\eqref{fusion-fun-anti}~has the announced monodromy around \(x=1\).

There remains to fix the normalization.  We evaluate at \(x=1\) the
explicit expression~\eqref{blocks-SQCD} of s-channel conformal blocks
which appear in~\eqref{fusion-fun-anti}, after removing a power
of~\((1-x)\),
\begin{equation}
  \begin{aligned}
    & \biggl[ (1-x)^{-\Nc(b^2+1) - \frac{\Nc}{\Nf} b^2}
    \Fblock{s}{\alpha_0-bh_{P\setminus\{p\}}}{
      -bh_1&\!\!\!-b\omega_\Nc\\
      2Q-\alpha_0+bh_P&\alpha_0}{x}
    \biggr]_{x=1}
    \\
    & \qquad\qquad
    = \sum_{\substack{\mathsemiclap{6}{k\colon P\to\bbZ_{\geq 0}}\\k_p=0}}
    (-1)^{\sum_{s\in P} k_s}
    \prod_{s,t\in P}
    \frac
    {(1+b^2+b\vev{\I a_0,h_t-h_s})_{k_s}}
    {(b\vev{\I a_0,h_t-h_s}-k_t+\delta_{tp})_{k_s}}
    \,.
  \end{aligned}
\end{equation}
This only depends on the \(\vev{\I a_0,h_t}\) with \(t\in P\), and does
not depend on~\(\Nf\).  We can thus take \(\Nf=\Nc+1\), in which case
\(-b\omega_\Nc = bh_{\Nf}\) and the fusion is a special case of
equation~(B.14) of~\cite{Gomis:2010kv}, where the normalization is known
to be~\eqref{fusion-coefs}.

%\bigskip

We are now ready to find the braiding matrix of \(V_{-b\omega_{\Nc+1}}\)
with~\(V_{\hat{m}}\) (where \(\hat{m}=(\varkappa+(\Nc+1)b)h_1\)).  This
braiding, followed by writing \(V_{-b\omega_{\Nc+1}}\) as the fusion
of~\(V_{-bh_1}\) and~\(V_{-b\omega_\Nc}\), is equivalent to performing
the fusion step first, then braiding each of \(V_{-bh_1}\)
and~\(V_{-b\omega_\Nc}\) in turn around the semi-degenerate operator.
The equivalence is encoded as a pentagon identity: for any
\((\Nc+1)\)-element sets of flavours \(P\) and~\(S\), and for \(s\in
S\),
\begin{equation}
  \begin{aligned}
    &
    \Braiding{\epsilon}{PS}{
      \hat{m}&-b\omega_{\Nc+1}\\\alpha_\infty&\alpha_0}
    \Fusionname_{s,S}[2Q-\alpha_\infty]
    \\
    & \quad =
    \sum_{p\in P}
    \Fusionname_{p,P}[\alpha_0]
    \Braiding{\epsilon}{ps}{
      \hat{m}&-bh_1\\
      \alpha_\infty&\alpha_0-bh_{P\setminus\{p\}}}
    \Braiding{\epsilon}{P\setminus\{p\},S\setminus\{s\}}{
      \hat{m}&-b\omega_\Nc\\\alpha_\infty-bh_s&\alpha_0}
    \,.
  \end{aligned}
\end{equation}
As a consistency check, we compute a slightly more general right-hand
side, with \(S\setminus\{s\}\) replaced by any \(\Nc\)-element
subset~\(S'\) of \(\intset{1}{\Nf}\).  This altered right-hand side must
vanish whenever \(s\in S'\).  After extracting factors independent
of~\(p\) in~\eqref{braid-collect} below, we will obtain a sum over~\(p\)
of products of sines, which is a sum of residues:
\begin{equation}\label{braid-sum-vanishes}
  \begin{aligned}
    & \sum_{p\in P}
    \frac{\prod_{u\in S'} \frac{1}{\pi}\sin\pi(\frac{b\varkappa}{\Nf}
      +b\vev{\I a_0,h_p}+b\vev{\I a_\infty,h_u}+b^2\delta_{us})}
    {\frac{1}{\pi}\sin\pi(\frac{b\varkappa}{\Nf}+b^2
      +b\vev{\I a_0,h_p}+b\vev{\I a_\infty,h_s})
      \prod_{t\in P\setminus\{p\}}
      \frac{1}{\pi}\sin\pi(b\vev{\I a_0,h_t-h_p})}
    \\
    & \quad =
    - \sum_{p\in P}
    \res_{\kappa = \I m_p}
    \frac{\prod_{u\in S'}
      \frac{1}{\pi}\sin\pi(1+b^2\delta_{us}+\kappa+\I\anti{m}_u)}
    {\frac{1}{\pi}\sin\pi(1+b^2+\kappa+\I\anti{m}_s)
      \prod_{t\in P} \frac{1}{\pi}\sin\pi(\I m_t-\kappa)}
    \\
    & \quad =
    \frac{\prod_{u\in S'}
      \frac{1}{\pi}\sin\pi(b^2\delta_{us}+\I\anti{m}_u-b^2-\I\anti{m}_s)}
    {\prod_{t\in P} \frac{1}{\pi}\sin\pi(\I m_t+\I\anti{m}_s+1+b^2)} \,.
  \end{aligned}
\end{equation}
This sum of residues is equal to the residue at
\(\kappa=-1-b^2-\I\anti{m}_s\) written in the last line, because the
function of~\(\kappa\) is \(1\)-periodic and vanishes at
\(\kappa\to\pm\I\infty\), hence the integral over the boundary of
\([0,1]\times(-\infty,\infty)\) vanishes.  As expected, the result
is~\(0\) when \(s\in S'\) (take \(u=s\)).  It is otherwise a product of
sines, and we get in that case the last equality below (denoting
\(S=S'\cup\{s\}\)):
\begin{align}
  \nonumber
  & \sum_{p\in P}
  \Fusionname_{p,P}[\alpha_0]
  \Braiding{\epsilon}{ps}{
    \hat{m}&-bh_1\\
    \alpha_\infty&\alpha_0-bh_{P\setminus\{p\}}}
  \Braiding{\epsilon}{P\setminus\{p\},S'}{
    \hat{m}&-b\omega_\Nc\\\alpha_\infty-bh_s&\alpha_0}
  \\\label{braid-collect}
  & = \begin{aligned}[t]
    &
    \frac{
      e^{-\I\pi\epsilon\frac{(\Nc+1)(\Nf-\Nc-1)}{\Nf}b^2
        -\I\pi\epsilon\delta_{s\in S'} b^2
        +\pi\epsilon b\vev{a_0,h_P}
        +\pi\epsilon b\vev{a_\infty,h_s+h_{S'}}}
      \prod_{u\neq s}^{\Nf} \Gamma(b\vev{\I a_\infty,h_s-h_u})}
    {\prod_{t=1}^{\Nf} \Gamma(\frac{b\varkappa}{\Nf}+b^2\delta_{t\in P}
      +b\vev{\I a_0,h_t}+b\vev{\I a_\infty,h_s})}
    \\
    & \cdot
    \prod_{t\in P}
    \frac{\prod_{v\not\in P}^{\Nf} \Gamma(1+b\vev{\I a_0,h_v-h_t})}
    {\prod_{w\not\in S'}^{\Nf} \Gamma(1-\frac{b\varkappa}{\Nf}
      -b\vev{\I a_0,h_t}-b\vev{\I a_\infty,h_w}-b^2\delta_{sw})}
    \\
    & \cdot
    \frac{\Gamma\bigl((\Nc+1)(1+b^2)\bigr)}{\Gamma(1+b^2)}
    \prod_{u\in S'}
    \frac{
      \prod_{w\not\in S'}^{\Nf}
      \Gamma(b\vev{\I a_\infty,h_u-h_w}+b^2\delta_{su}-b^2\delta_{sw})}
    {\prod_{v\not\in P}^{\Nf} \Gamma(\frac{b\varkappa}{\Nf}
      +b\vev{\I a_0,h_v}+b\vev{\I a_\infty,h_u}+b^2\delta_{us})}
    \\
    & \cdot
    \sum_{p\in P}
    \frac{\prod_{u\in S'} \frac{1}{\pi}\sin\pi(\frac{b\varkappa}{\Nf}
      +b\vev{\I a_0,h_p}+b\vev{\I a_\infty,h_u}+b^2\delta_{us})}
    {\frac{1}{\pi}\sin\pi(\frac{b\varkappa}{\Nf}+b^2
      +b\vev{\I a_0,h_p}+b\vev{\I a_\infty,h_s})
      \prod_{t\in P\setminus\{p\}}
      \frac{1}{\pi}\sin\pi(b\vev{\I a_0,h_t-h_p})}
  \end{aligned}
  \\\label{braid-BF}
  & \stackrel{\mathclap{s\not\in S'}}{=}
  \begin{aligned}[t]
    &
    e^{-\I\pi\epsilon\frac{(\Nc+1)(\Nf-\Nc-1)}{\Nf}b^2}
    \prod_{t\in P}
    \frac{e^{\pi\epsilon b\vev{a_0,h_t}}
      \prod_{v\not\in P}^{\Nf} \Gamma(1+b\vev{\I a_0,h_v-h_t})}
    {\prod_{w\not\in S}^{\Nf} \Gamma(1-\frac{b\varkappa}{\Nf}
      -b\vev{\I a_0,h_t}-b\vev{\I a_\infty,h_w})}
    \\
    & \cdot
    \prod_{u\in S}
    \frac{e^{\pi\epsilon b\vev{a_\infty,h_u}}
      \prod_{w\not\in S}^{\Nf} \Gamma(b\vev{\I a_\infty,h_u-h_w})}
    {\prod_{v\not\in P}^{\Nf} \Gamma(\frac{b\varkappa}{\Nf}
      +b\vev{\I a_0,h_v}+b\vev{\I a_\infty,h_u})}
    \\
    & \cdot
    \frac{\Gamma\bigl((\Nc+1)(1+b^2)\bigr)}{\Gamma(1+b^2)}
    \prod_{u\in S\setminus\{s\}}
    \frac{\Gamma(b\vev{\I a_\infty,h_s-h_u})}
      {\Gamma(1+b^2+b\vev{\I a_\infty,h_s-h_u})} \,.
  \end{aligned}
\end{align}
We recognize in the last line the fusion coefficient
\(\Fusionname_{s,S}[2Q-\alpha_\infty]\).  What remains is the braiding
matrix of \(V_{-b\omega_{\Nc+1}}\) with~\(V_{\hat{m}}\), which we check
to be~\eqref{braid-Toda-anti} with \(\Nc\to\Nc+1\).  This concludes the
induction, and the proof of the relation between conformal blocks and
vortex partition functions for SQCD\@.  Together with the
equality~\eqref{1loop-3pt-SQCD} of constant factors, checked in
Appendix~\ref{app:1loop-3pt}, this establishes the
relation~\eqref{SQCD-matching} between the partition function of an SQCD
surface operator and the appropriate correlator in the Toda CFT\@.

\subsection{Fusion Rules}
\label{app:fusion}

We provide here the fusion rules between various pairs of vertex
operators, in particular the fusion~\eqref{fusion-sh1-sh1} of two
semi-degenerate operators, and the fusion~\eqref{fusion-kappa-deg} of a
semi-degenerate operator with a fully-degenerate operator labeled by an
arbitrary representation of~\(A_{\Nf-1}\).  We propose that operators
resulting from the latter fusion appear with
multiplicity~\eqref{fusion-mult-almost} in the fusion of two generic
operators.

Null vectors among \(W_{\Nf}\)~descendants of a fully degenerate vertex
operator~\(V_{-b\omega-\omega'/b}\) constrain its three-point function
with arbitrary vertex operators \(V_\alpha\) and~\(V_\beta\).  Namely,
the three point function vanishes unless \(\alpha + \beta = 2Q + bh +
h'/b\) for some weights \(h\) of~\(\repr(\omega)\) and \(h'\)
of~\(\repr(\omega')\).  This results in the fusion rule
\begin{equation}\label{fusion-generic-degenerate}
  V_\alpha \times V_{-b\omega-\omega'/b}
  = \sum_{h\in\repr(\omega)} \sum_{h'\in\repr(\omega')}
  V_{\alpha - b h - h'/b} \,,
\end{equation}
with outgoing momenta \(\alpha-bh-h'/b=2Q-\beta\): the degenerate
operator shifts the incoming momentum by \(-bh-h'/b\).  Each
operator~\(V_{\alpha-bh-h'/b}\) appears
in~\eqref{fusion-generic-degenerate} with a multiplicity equal to the
product of the multiplicity of~\(h\) in \(\repr(\omega)\) and that
of~\(h'\) in \(\repr(\omega')\).  Henceforth, we take
\(\omega'=0\), thus \(h'=0\).

Later in this appendix, we find that the fusion~\eqref{fusion-kappa-deg}
of a semi-degenerate operator \(V_{\varkappa h_1}\)
with~\(V_{-b\omega}\) only allows some of the shifts \(-bh\)
of~\eqref{fusion-generic-degenerate}.  Let us first describe the case
\(\omega=h_1\) based on~\cite[Appendix~B]{Gomis:2010kv}: the fusion of
\(-\varkappa h_{\Nf}\) and \(-bh_1\) yields the momenta
\(-(\varkappa+b)h_{\Nf}\) and \(-\varkappa h_{\Nf}-bh_1\).  After the Weyl
rotation \((1\,2\cdots\Nf)\), we get
\begin{equation}\label{fusion-kappa-bh1}
  V_{\varkappa h_1} \times V_{-bh_1}
  = V_{\varkappa h_1 - b h_1} + V_{\varkappa h_1 - b h_2} \,.
\end{equation}
The s-channel expansion of \(\vev{V_{\alpha_\infty}(\infty) V_{\varkappa
    h_1}(1) V_{-bh_1}(x,\bar{x})V_{\alpha_0}(0)}\) involves
\(\Nf\)~products of holomorphic and antiholomorphic conformal blocks.
The t-channel expansion only features two
momenta~\eqref{fusion-kappa-bh1}, and takes the form
\begin{equation}
  \abs{1-z}^{2[\dimToda(\varkappa h_1-bh_1)-\dimToda(\varkappa h_1)-\dimToda(-bh_1)]}
  ({\cdots})
  +
  \abs{1-z}^{2[\dimToda(\varkappa h_1-bh_2)-\dimToda(\varkappa h_1)-\dimToda(-bh_1)]}
  ({\cdots})
\end{equation}
where \(({\cdots})\) are series in powers of \((1-z)\) and
\((1-\bar{z})\).  The first series factorizes as the product of a
holomorphic and an anti-holomorphic conformal blocks, multiplied by
\(C_{-bh_1,\varkappa h_1}^{\varkappa h_1-bh_1}
C(\alpha_0,\alpha_\infty,(\varkappa-b)h_1)\).  The second does not, but
can be written non-canonically as a sum of \(\Nf-1\) products of the
same form.  This multiplicity implies that the fusion \(V_{\alpha_0}\)
and~\(V_{\alpha_\infty}\) includes \(\Nf-1\) copies of the
representations of the \(W_{\Nf}\)~algebra generated by~\(V_{\varkappa h_1
  - bh_2}\), while it only includes one copy of the representation
generated by any semi-degenerate operator~\(V_{\varkappa h_1}\).  We
generalize the statement to all momenta of the form \(\varkappa h_1 -
b\omega\) in~\eqref{fusion-mult-almost}.

\subsubsection{Fusion of Two Semi-Degenerate Operators}
\label{app:fusion-semi}

To reach more complicated degenerate operators, we first find which
momenta result from the fusion of two semi-degenerate momenta
\(-\varkappa h_{\Nf}\) and~\(\lambda h_1\).  In principle, one could write
null vectors descending from \(V_{-\varkappa h_{\Nf}}\) and~\(V_{\lambda
  h_1}\) for a given~\(\Nf\) and, through those, constrain the momenta
which arise in the OPE\@.  Such constraints are polynomial in the momenta,
and any constraint shown for generic \((b,\varkappa,\lambda)\) must hold
for all \((b,\varkappa,\lambda)\) by continuity: in other words, fusion
rules for more specific momenta can only become more restrictive.  We
are thus free to assume that \((b,\varkappa,\lambda)\) is generic.

Since null vectors are very difficult to write down for general~\(\Nf\),
we use a different route: the braiding matrix relating the s-channel
(\(x\to 0\)) and u-channel (\(x\to\infty\)) conformal blocks of
\(\vev{V_{\lambda h_1}(\infty) V_{-\varkappa h_{\Nf}}(1)
  V_{-bh_1}(x,\bar{x}) V_{\alpha_0}(0)}\) should only lead to u-channel
conformal blocks with internal momenta \(\lambda h_1-bh_1\) and
\(\lambda h_1-bh_2\), and all other components must vanish.
Specifically, we take \(\alpha_2=2Q-\lambda h_1\), \(\hat{m}=-\varkappa
h_{\Nf}\), \(\mu=-bh_1\) and \(\alpha_1=2Q-\alpha+bh_l\) in
equation~(B.12) of~\cite{Gomis:2010kv}:
\begin{equation}
  \begin{aligned}
    \Fblock{s}{2Q-\alpha}
      {-\varkappa h_{\Nf}&-bh_1\\\lambda h_1&2Q-\alpha+bh_l}{x}
    =
    \sum_{k=1}^{\Nf}
    e^{i\pi\epsilon(\phi_{kl} - b\varkappa/\Nf)}
    \prod_{j\neq l}
    \frac{\Gamma(1+b^2+b\vev{\alpha-Q,h_j-h_l})}
    {\Gamma(1+b^2-\phi_{kj})}
    \\
    \cdot
    \prod_{j\neq k}
    \frac{\Gamma(b\vev{\lambda h_1-Q, h_j- h_k})}
    {\Gamma(\phi_{jl})}
    \Fblock{u}{\lambda h_1-bh_k}
    {-\varkappa h_{\Nf}&-bh_1\\\lambda h_1&2Q-\alpha+bh_l}{x}
  \end{aligned}
\end{equation}
where
\begin{equation}
  \phi_{st} = b\vev{-\varkappa h_{\Nf},h_1}
  + b\vev{\lambda h_1-Q,h_s} - b\vev{\alpha-Q,h_t} \,.
\end{equation}
The coefficient must vanish for all \(k\not\in\{1,2\}\) and all~\(l\),
hence one of the denominator Gamma functions must have a non-positive
integer argument:
\begin{equation}\label{fusion-semi-condition}
  \forall k\in\intset{3}{\Nf} \;
  \forall l\in\intset{1}{\Nf} \;
  -\phi_{jl}\in\bbZ_{\geq 0}
  \ \text{or} \
  \phi_{kj}-1-b^2=\phi_{(k-1)j}\in\bbZ_{\geq 0} \,.
\end{equation}
If for each \(1\leq s\leq\Nf\) one had \(\phi_{p_s s}=n_s\) for some
integers \(1\leq p_s\leq\Nf\) and~\(n_s\), then summing over~\(s\) would
yield
\begin{equation}
  0 = \sum_{s=1}^{\Nf} b\vev{\alpha-Q,h_s}
  = \sum_{s=1}^{\Nf} \biggl(\frac{b\varkappa}{\Nf}
  + b\vev{\lambda h_1-Q,h_{p_s}} - n_s\biggr)
  = b\varkappa + k_1 b\lambda + k_2 b^2 + k_3
\end{equation}
for some integers \(k_i\): this cannot happen for generic
\((b,\varkappa,\lambda)\).  Thus there exists \(1\leq u\leq\Nf\) such
that none of the \(\phi_{pu}\) are integers.  The
condition~\eqref{fusion-semi-condition} for \(l=u\) then implies that
for each \(3\leq k\leq\Nf\), \(\phi_{(k-1)t_k}\in\bbZ_{\geq 0}\) for
some \(1\leq t_k\leq\Nf\).  No two~\(t_k\) can be equal, because
\(\phi_{(k-1)t}-\phi_{(l-1)t} = (k-l)(b^2+1)\) is non-integer for
\(k\neq l\).  We can thus permute the components of \(\alpha-Q\) through
a Weyl transformation so that \(t_k=k-1\):
\begin{equation}
  b \vev{\alpha-Q,h_{k-1}}
  = b\vev{-\varkappa h_{\Nf},h_1} + b \vev{\lambda h_1-Q,h_{k-1}} - n_k
\end{equation}
for all \(3\leq k\leq\Nf\), where \(n_k\geq 0\)~are some integers.  We
deduce that
\begin{equation}\label{fusion-kap-lambda}
  \alpha = (\lambda+\nu) h_1 - (\varkappa+\nu) h_{\Nf}
  + \frac{1}{b} \sum_{k=3}^{\Nf} n_k (h_1-h_{k-1}) \,.
\end{equation}
The same considerations applied to the braiding of \(-\frac{1}{b}h_1\)
and \(-\varkappa h_{\Nf}\) yield the constraint above with \(\frac{1}{b}\)
replaced by~\(b\).  We can thus restrict to
momenta~\eqref{fusion-kap-lambda} which also have, up to a Weyl
transformation, the \(b\to\frac{1}{b}\) form.  All in all, the fusion of
two semi-degenerate operators can only allow a one-parameter set of
momenta, and some isolated momenta
\begin{equation}
  \begin{aligned}
    V_{-\varkappa h_{\Nf}} \times V_{\lambda h_1}
    & = \int\dd{\nu} V_{(\lambda+\nu) h_1 - (\varkappa+\nu) h_{\Nf}}
    \\
    & \quad + \sum_{n\in\bbZ} \sum_{n'\in\bbZ} \sum_{k=1}^{\Nf}
    V_{(\lambda-\varkappa) h_1 + (n/b) (h_1-h_k) + [n'b - (\Nf-k)/b] (h_1-h_{\Nf})}
    \,.
  \end{aligned}
\end{equation}
In the case \(\Nf=3\), we wrote down explicitly null vectors descending
from \(V_{-\varkappa h_{\Nf}}\) and~\(V_{\lambda h_1}\) (higher
\(W_{\Nf}\)~algebras are intractable), and found that the isolated momenta
are in fact not allowed.  We propose that this holds for
general~\(\Nf\).  After performing some Weyl reflections of momenta on
the left and right-hand side and redefining~\(\nu\), we deduce the
fusion rules
\begin{align}
  V_{-\varkappa h_{\Nf}} \times V_{\lambda h_1}
  & = \int\dd{\nu} V_{-\varkappa h_{\Nf} + \lambda h_1 + \nu (h_1 - h_{\Nf})}
  \\
  \label{fusion-sh1-sh1}
  V_{\varkappa h_1} \times V_{\lambda h_1}
  & = \int\dd{\nu} V_{\varkappa h_1 + \lambda h_1 + \nu (h_1 - h_2)}
  \\
  V_{-\varkappa h_{\Nf}} \times V_{-\lambda h_{\Nf}}
  & = \int\dd{\nu} V_{-\varkappa h_{\Nf} - \lambda h_{\Nf} + \nu (h_{\Nf-1}-h_{\Nf})}
  \,.
\end{align}

For completeness, we find the corresponding structure constant as the
main residue of \(C(\alpha_1,\alpha_2,\varkappa h_1)\) at
\(\alpha_1=\lambda h_1\) and
\(\alpha_2=2Q-(\varkappa+\lambda+\nu)h_1+\nu h_2\), after removing our
normalization from~\eqref{FL-1-39}, and recognize a Liouville CFT
three-point function:
\begin{equation}
  \begin{aligned}
    C_{\varkappa h_1,\lambda h_1}^{(\varkappa+\lambda+\nu)h_1-\nu h_2}
    & =
    \frac
    {\hat{\mu}^\nu \Upsilon(b)^{\Nf-1} \Upsilon(\varkappa)
      \Upsilon(\lambda) \Upsilon(\varkappa+\lambda+2\nu-b-1/b)}
    {\Upsilon(-\nu) \Upsilon(\varkappa+\nu) \Upsilon(\lambda+\nu)
      \Upsilon(\varkappa+\lambda+\nu-b-1/b)}
    \\
    & = \Upsilon(b)^{\Nf-2} \,
    C_{\text{Liouville}}\biggl(\frac{\varkappa}{2}, \frac{\lambda}{2},
    b+\frac{1}{b}-\frac{\varkappa+\lambda}{2}-\nu\biggr) \,.
  \end{aligned}
\end{equation}
The equality is true by construction for \(\Nf=2\), as a Liouville
momentum of \(\varkappa/2\) corresponds in the Toda CFT language to a
momentum of \((\varkappa/2)(h_1-h_2) = \varkappa h_1\).  More generally,
the equality may hint to a deeper relation between Toda CFTs for
different values of~\(\Nf\).

\subsubsection{Fusion of Semi-Degenerate and Degenerate Operators}
\label{app:fusion-deg-semi}

We are now ready to tackle the fusion of other degenerate vertex
operators~\(V_{-b\omega}\) with semi-degenerate operators~\(V_{\varkappa
  h_1}\).

For \(\omega = \Nc h_1\) the fusion is a special case
of~\eqref{fusion-sh1-sh1} with \(\lambda = -\Nc b\), hence only allows
the momenta \((\varkappa-\Nc b)h_1 + \nu (h_1-h_2)\).  Given the fusion
rule~\eqref{fusion-generic-degenerate} of a degenerate operator,
\((\Nc-\nu/b)h_1 + (\nu/b) h_2\) must be a weight of \(\repr(\Nc
h_1)\) hence \(\nu=nb\) with \(0\leq n\leq\Nc\), and
\begin{equation}\label{fusion-kap-sym}
  V_{\varkappa h_1} \times V_{-\Nc bh_1}
  =
  \sum_{n=0}^\Nc V_{(\varkappa-(\Nc-n)b)h_1-nbh_2}
\end{equation}
with no multiplicity since the weight \((\Nc-n)h_1+nh_2\) of
\(\repr(\Nc h_1)\) has no multiplicity.  Through the Weyl rotation
\((\Nf\cdots 2\,1)\), an equivalent statement is that the fusion of
\(-\varkappa h_{\Nf}\) and \(-\Nc bh_1\) yields the momenta
\(-nbh_1-(\varkappa+(\Nc-n)b)h_{\Nf}\).

The correlator \(\vev{V_{\alpha_\infty}(\infty) V_{(\varkappa'+lb)h_1}(1)
  V_{-lbh_1}(x,\bar{x}) V_{\alpha_0}(0)}\) has
\(\dim\bigl(\repr(lh_1)\bigr)\) s-channel conformal blocks, and
must have the same number of t-channel conformal blocks.  The
fusion~\eqref{fusion-kap-sym} allows the t-channel internal momenta
\(\varkappa'h_1+nb(h_1-h_2)\) for \(0\leq n\leq l\), with no
multiplicity, hence any multiplicity is due to the fusion of
\(V_{\alpha_0}\) and~\(V_{\alpha_\infty}\).  The number of t-channel
conformal blocks is thus
\begin{equation}
  \sum_{n=0}^l N_{\alpha_0,\,\alpha_\infty}^{\varkappa'h_1+nb(h_1-h_2)}
  = \dim\repr(lh_1) = \binom{\Nf+l-1}{l}
\end{equation}
where \(V_\beta\) appears \(N_{\alpha_0,\alpha_\infty}^\beta\) times in the
fusion of \(V_{\alpha_0}\) and~\(V_{\alpha_\infty}\).  Solving, we find
\begin{equation}\label{fusion-mult-ksym}
  \begin{aligned}
    N_{\alpha_0,\,\alpha_\infty}^{\varkappa'h_1+nb(h_1-h_2)}
    & = \dim\repr\bigl(nh_1\bigr)
    - \dim\repr\bigl((n-1)h_1\bigr)
    \\
    & = \binom{\Nf+n-1}{n} - \binom{\Nf+n-2}{n-1}
    = \binom{\Nf+n-2}{n} \,.
  \end{aligned}
\end{equation}
None of these multiplicities vanish, so all \(\Nc+1\) momenta
of~\eqref{fusion-kap-sym} do appear in the fusion.

Restricting the fusion rule~\eqref{fusion-kap-sym} to \(\varkappa h_1 =
-Kbh_1\) with \(K\geq\Nc\), we retrieve the decomposition into
irreducible representations of the tensor product of two symmetric
representations, given by the Littlewood--Richardson rule:
\begin{equation}\label{Litt-Rich-sym}
  \repr(Kh_1) \otimes \repr(\Nc h_1)
  = \bigoplus_{n=0}^\Nc \repr\bigl((K+\Nc-n)h_1 + nh_2\bigr) \,.
\end{equation}
One could go in the other direction: the
decomposition~\eqref{Litt-Rich-sym} for \(K\geq\Nc\) implies that the
fusion of \(V_{-Kbh_1}\) with~\(V_{-\Nc bh_1}\) yields the momenta
\(-(K+\Nc-n)bh_1 - nbh_2\).  This set of \(\Nc+1\) momenta only involves
\(-Kbh_1\) as an overall constant part, hence the natural generalization
from \(V_{-Kbh_1}\) to~\(V_{\varkappa h_1}\) is~\eqref{fusion-kap-sym}.
We will apply a similar reasoning\footnote{In principle, one could go
  further, and guess the fusion rule~\eqref{fusion-sh1-sh1} for two
  semi-degenerate operators by replacing \(-\Nc bh_1 \to \lambda h_1\)
  and allowing shifts by continuous multiples of \(h_2-h_1\).  It could
  be interesting to obtain a continuous analogue of the
  Littlewood--Richardson rule along those lines.} to guess the fusion of
other degenerate operators with a semi-degenerate operator.

The tensor product of an antisymmetric and a symmetric representations
of~\(A_{\Nf-1}\) is the sum of two irreducible representations,
\begin{equation}\label{Litt-Rich-anti}
  \repr(Kh_1) \otimes \repr(\omega_\Nc)
  = \repr(Kh_1 + \omega_\Nc)
  \oplus \repr\bigl((K-1)h_1 + \omega_{\Nc+1}\bigr) \,.
\end{equation}
This naturally generalizes to the fusion rule
\begin{equation}\label{fusion-kap-anti}
  V_{\varkappa h_1} \times V_{-b\omega_\Nc}
  =
  V_{\varkappa h_1 - b\omega_\Nc}
  + V_{\varkappa h_1 - b(\omega_{\Nc+1}-h_1)} \,.
\end{equation}
For completeness, a Weyl reflection yields the fusion of \(-\varkappa
h_{\Nf}\) and \(-b\omega_\Nc\), which features the momenta
\(-(\varkappa+b)h_{\Nf}-b\omega_{\Nc-1}\) and \(-\varkappa
h_{\Nf}-b\omega_\Nc\).

We show in Section~\ref{sec:SQCD}, together with appendices
\ref{app:braiding} and~\ref{app:1loop-3pt} which do not depend on this
fusion rule, that the Toda CFT correlator of two generic operators with
\(V_{\varkappa h_1}\) and~\(V_{-b\omega_\Nc}\) is equal to the partition
function of a surface operator.  At the end of
Section~\ref{sec:SQCD-expand} we expand the partition function in a
limit which corresponds to the fusion of \(V_{\varkappa h_1}\)
and~\(V_{-b\omega_\Nc}\).  The exponents found there prove the fusion
rule~\eqref{fusion-kap-anti}.  Once more, the number of t-channel and
s-channel conformal blocks must be equal:
\begin{equation}
  N_{\alpha_0,\,\alpha_\infty}^{\varkappa h_1 - b\omega_\Nc}
  + N_{\alpha_0,\,\alpha_\infty}^{(\varkappa+b) h_1 - b\omega_{\Nc+1}}
  = \binom{\Nf}{\Nc} \,.
\end{equation}
We deduce for each \(n\geq 0\) that for all~\(\varkappa\),
\begin{equation}\label{fusion-mult-kanti}
  N_{\alpha_0,\,\alpha_\infty}^{\varkappa h_1 - b\omega_{n+1}}
  = \binom{\Nf-1}{n} \,.
\end{equation}
This is consistent with multiplicities of the two powers of
\(\abs{1-x}^2\) found at the end of Section~\ref{sec:SQCD-expand}, and
matches with~\eqref{fusion-mult-ksym} for \(n=1\) and \(n=0\).

Consider now an arbitrary highest weight \(\omega = \sum_{j=1}^{\Nf-1}
n_j \omega_j\) of~\(A_{\Nf-1}\).  For each \(j\) from \(\Nf-1\)
to~\(1\), its Young diagram has \(n_j\)~columns with \(j\)~boxes.
Through the Littlewood--Richardson rule, we find a decomposition valid
for \(K\geq \sum_{j=1}^{\Nf-1} n_j\),
\begin{equation}\label{Litt-Rich-gen}
  \repr(Kh_1) \otimes \repr(\omega)
  = \bigoplus_{k_{\Nf-1}=0}^{n_{\Nf-1}} \cdots \bigoplus_{k_1=0}^{n_1}
  \repr\biggl(Kh_1+\omega
  +\sum_{j=1}^{\Nf-1}\Bigl[k_j(h_{j+1}-h_1)\Bigr]\biggr)
\end{equation}
into \(\prod_{j=1}^{\Nf-1} (n_j+1)\) irreducible representations.  We
thus propose the fusion rule
\begin{equation}
  \label{fusion-kappa-deg}
  V_{\varkappa h_1} \times V_{-b\sum_{j=1}^{\Nf-1} n_j\omega_j}
  =
  \sum_{k_1=0}^{n_1} \cdots \sum_{k_{\Nf-1}=0}^{n_{\Nf-1}}
  V_{\varkappa h_1 - b \sum_{j=1}^{\Nf-1}[n_j\omega_j + k_j(h_{j+1}-h_1)]} \,.
\end{equation}
As a natural generalization of \eqref{fusion-mult-ksym}
and~\eqref{fusion-mult-kanti}, we propose that vertex operators with a
momentum \(\varkappa h_1 - b \sum_{j=1}^{\Nf-1} l_j \omega_j\) appear
with multiplicity
\begin{equation}\label{fusion-mult-almost}
  N_{\alpha_0,\,\alpha_\infty}^{\varkappa h_1 - b \sum_{j=1}^{\Nf-1} l_j \omega_j}
  = \dim \repr_{A_{\Nf-2}}
    \Biggl( \sum_{j=1}^{\Nf-1} l_j \omega_{j-1} \Biggr) \,,
\end{equation}
where the \(j=1\) term can be absorbed in a shift of~\(\varkappa\), and
the right-hand side is the dimension of the representation
of~\(A_{\Nf-2}\) whose Young diagram is obtained from that of
\(\repr\bigl(\sum_{j=1}^{\Nf-1} l_j \omega_j\bigr)\) by removing
the first row: \(h_1\to 0\) and \(h_i \to h_{i-1}\).  Besides
reproducing the correct multiplicities for the symmetric and
antisymmetric case, the proposal~\eqref{fusion-mult-almost} correctly
leads to equally many s-channel and t-channel conformal blocks in the
four-point function \(\vev{V_{\alpha_\infty}(\infty) V_{\varkappa
    h_1}(1) V_{-b\omega}(x,\bar{x}) V_{\alpha_0}(0)}\) since
\begin{equation}
  \dim\repr\Biggl(\sum_{j=1}^{\Nf-1} n_j\omega_j\Biggr)
  =
  \sum_{k_1=0}^{n_1} \cdots \sum_{k_{\Nf-1}=0}^{n_{\Nf-1}}
  \dim \repr_{A_{\Nf-2}}\Biggl(
  \sum_{j=1}^{\Nf-1}\Bigl[n_j\omega_{j-1} + k_j h_j\Bigr]
  \Biggr)
  \,.
\end{equation}
The equality holds because the representations on the right-hand side
are the decomposition of \(\repr(\omega)\) into irreducible
representations of the subalgebra \(A_{\Nf-2}\) of~\(A_{\Nf-1}\).

\subsection{Conformal Blocks}
\label{app:blocks}

In this section we give explicit expressions of conformal blocks which
are labeled by two generic momenta \(\alpha_\infty=Q-\I a_\infty\) and
\(\alpha_0=Q-\I a_0\) at \(\infty\) and~\(0\), one semi-degenerate
momentum \(\hat{m}=\lambda h_1\) at~\(1\), and some degenerate momenta
\(-b\Omega_j\) inserted at the positions \((x_j,\bar{x}_j)\) for \(1\leq
j\leq n\).  The expressions are direct translations of the gauge theory
vortex partition functions through the correspondence described in the
main text.  We only consider conformal blocks in the s-channel (the
region \(1>\abs{x_n}>\cdots>\abs{x_1}>0\)), which are series in powers
of \(x_n\), \(x_{n-1}/x_n\), \ldots{}, \(x_1/x_2\).

First comes the case of a single degenerate momentum \(-b\omega_\Nc\)
labeled by the \(\Nc\)-th antisymmetric representation
of~\(A_{\Nf-1}\).  The four-point conformal blocks are equal, up to
powers of \(x\) and \(1-x\), to the vortex partition
functions~\eqref{SQCD-fs} of SQCD,
\begin{equation}\label{blocks-SQCD}
  \begin{aligned}
    &
    \Fblock{s}{Q-\I a_0-bh_{\{p\}}}
      {\lambda h_1&-b\omega_\Nc\\Q-\I a_\infty&Q-\I a_0}{x}
    \\
    & =
    x^{-b\vev{\I a_0,h_{\{p\}}}+\frac{\Nc(\Nf-\Nc)}{2}(b^2+1)}
    (1-x)^{\Nc(b^2+1-b\lambda/\Nf)}
    \\
    & \quad \!\!
    \sum_{k_1,\ldots,k_\Nc\geq 0}
    \frac{x^{\sum_{j=1}^\Nc k_j}}{\prod_{j=1}^\Nc k_j!}
    \frac{\prod_{j=1}^\Nc \prod_{s=1}^{\Nf}
      (1 - b(\lambda-\Nc b)/\Nf
      - b\vev{\I a_0,h_{p_j}}-b\vev{\I a_\infty,h_s})_{k_j}}
      {\prod_{i\neq j}^\Nc (b\vev{\I a_0, h_{p_i}-h_{p_j}}-k_i)_{k_j}
        \prod_{j=1}^\Nc \prod_{s\not\in\{p\}}^{\Nf}
        (1+b\vev{\I a_0,h_s-h_{p_j}})_{k_j}}
    \,.
  \end{aligned}
\end{equation}
The internal momentum \(Q-\I a_0-bh_{\{p\}}\) is labeled by a weight
\(h_{\{p\}}=h_{p_1}+\cdots+h_{p_\Nc}\) of \(\repr(\omega_\Nc)\), where
\(1\leq p_1<\cdots<p_\Nc\leq\Nf\).  While the
expression~\eqref{blocks-SQCD} is established since we provide a proof
of the correspondence in this case, the conformal blocks below are not.
However, they are supported by the evidence we gave for the
correspondence in the main text.

Next, s-channel conformal blocks with the degenerate momentum \(-\Nc
bh_1\) have an internal momentum \(Q-\I a_0-bh\) labeled by a
weight~\(h\) of the \(\Nc\)-th symmetric representation \(\repr(\Nc
h_1)\) of~\(A_{\Nf-1}\).  We let \(h=h_{[n]}=\sum_{s=1}^{\Nf} n_s
h_s\) for \(\sum_{s=1}^{\Nf} n_s = \Nc\), and \(I=\{(s,\mu)\mid 1\leq
s\leq\Nf, 0\leq\mu<n_s\}\). Conformal blocks are vortex partition
functions~\eqref{SQCDA-Zv} up to factors \(x^{-\gamma_0}
(1-x)^{-\gamma_1}\) from the correspondence~\eqref{SQCDA-matching}:
\begin{equation}\label{blocks-SQCDA}
  \begin{aligned}
    &
    \Fblock{s}{Q-\I a_0-bh_{[n]}}
      {\lambda h_1&-\Nc bh_1\\Q-\I a_\infty&Q-\I a_0}{x}
    \\
    & =
    x^{\dimToda(Q-\I a_0-bh_{[n]})-\dimToda(Q-\I a_0)-\dimToda(-\Nc bh_1)}
    (1-x)^{\dimToda(\lambda h_1-\Nc bh_2)
      - \dimToda(\lambda h_1) - \dimToda(-\Nc bh_1)}
    \\
    & \quad
    \sum_{k:I\to\bbZ_{\geq 0}}
    \prod_{(s,\mu) \in I} \biggl[
    x^{k_{s\mu}}
    \prod_{t=1}^{\Nf}
    \frac
    {(1 - b(\lambda-\Nc b)/\Nf
      - b\vev{\I a_0,h_s} - b\vev{\I a_\infty, h_t} - \mu b^2)_{k_{s\mu}}}
    {(1 + b\vev{\I a_0, h_t - h_s} + (n_t - \mu) b^2)_{k_{s\mu}}}
    \\
    & \qquad \qquad \qquad \cdot
    \frac
    {\prod_{t=1}^{\Nf} (1 + b\vev{\I a_0, h_t - h_s} + (n_t - \mu) b^2
      + k_{s\mu} - k_{t(n_t - 1)})_{k_{t(n_t - 1)}}}
    {\prod_{(t,\nu)\in I}
      (1 + b\vev{\I a_0, h_t - h_s} + (\nu - \mu) b^2
      + k_{s\mu} - k_{t\nu})_{k_{t\nu} - k_{t(\nu - 1)}}}
    \biggr]
    \,.
  \end{aligned}
\end{equation}

We now come to the case of \((n+3)\)-point conformal blocks with two
generic, one semi-degenerate, and \(n\)~degenerate momenta
\(-b\Omega_j = -b\Omega(K_j,\epsilon_j)\), where
\(\Omega(K,-1)=\omega_K\) and \(\Omega(K,+1)=Kh_1\).
In the s-channel \(1>\abs{x_n}>\ldots>\abs{x_1}\), the internal momentum
running between the punctures at \(x_j\) and~\(x_{j+1}\) (here
\(x_{n+1}=1\)) has the form \(\alpha_0 - b h_{[n^j]} = \alpha_0 - b
\sum_{t=1}^{\Nf} n_t^j h_t\), for some integers \(n_t^j\geq 0\).  These
integers must be such that \(h_{[n^j]}-h_{[n^{j-1}]}\) is a weight of
\(\repr(\Omega_j)\) for each \(1\leq j\leq n\) (here \(n_t^0=0\)).
Explicitly, \(\sum_{t=1}^{\Nf} (n_t^j-n_t^{j-1}) = K_j\), and
\(n_t^j-n_t^{j-1}\) is in \(\bbZ_{\geq 0}\) if \(\epsilon_j=+1\) and in
\(\{0,1\}\) if \(\epsilon_j=-1\).

In Section~\ref{sec:Quivers} we find
a quiver gauge theory whose vacua are labeled by the same data, and
perform various checks that its partition function is equal to the Toda
correlator we are now considering.  Up to simple factors, the conformal
blocks are thus equal to the vortex partition functions, themselves a
sum of residues in the Coulomb branch representation of the partition
function.  Let us introduce the sets \(I_j = \bigl\{(s,\mu) \bigm| 0\leq
\nu<n_s^j , 1 \leq s\leq\Nf \bigr\}\) (\(I_0\)~is empty), the notation
\(\I a_{s,\mu} = \vev{\I a_0, h_s} + \mu b\), and the parameters
\(q_{n+1} = b^2/2\) and \(q_j = \epsilon_j (b^2/2+1/4) - 1/4\).
We find
\begin{equation}\label{blocks-Quivers}
  \begin{aligned}
    &
    \Fblock{s}{}
      {
        \mathtikz[x=1em,y=1ex,thick]{
          \draw[->-=.55] (0,0) -- (2,0);
          \draw[->-=.55] (20,0) -- (18,0);
          \draw[->-=.55] (18,0) -- (12,0);
          \draw          (12,0) -- (8,0);
          \draw[->-=.55] (8,0) -- (2,0);
          \draw[->-=.55] (2,4) -- (2,0);
          \draw[->-=.55] (8,4) -- (8,0);
          \draw[->-=.55] (12,4) -- (12,0);
          \draw[->-=.55] (18,4) -- (18,0);
          \node at (0,1.5) {$\alpha_\infty$};
          \node at (20,1.5) {$\alpha_0$};
          \node at (15,1.5) {$\alpha_0-bh_{[n^1]}$};
          \node at (5,1.5) {$\alpha_0-bh_{[n^n]}$};
          \node at (18,5.5) {$-b\Omega_1$};
          \node at (12,5.5) {$-b\Omega_2$};
          \node at (10,2.5) {$\cdots$};
          \node at (8,5.5) {$-b\Omega_n$};
          \node at (2,5.5) {$\lambda h_1$};
        }
      }{x}
    \\
    & =
    \prod_{j=1}^n \biggl[ x_j^{\dimToda(\alpha_0-bh_{[n^j]})
      - \dimToda(\alpha_0-bh_{[n^{j-1}]})- \dimToda(-b\Omega_j)}
    [1-x_j]^{(1+b^2-\frac{b\lambda}{\Nf})K_j} \biggr]
    \prod_{i<j}^n \biggl[1-\frac{x_i}{x_j}
    \biggr]^{(1+2q_j+b^2 \frac{K_j}{\Nf}) K_i}
    \\
    & \quad \cdot
    \sum_{\{k_{j,s,\mu}\geq 0\}} \Biggl\{
    \begin{aligned}[t]
      & \prod_{j=1}^n \prod_{(s,\mu)\in I_j} \Biggl[
      \begin{aligned}[t]
        & \biggl[ \frac{x_j}{x_{j+1}} \biggr]^{k_{j,s,\mu}}
        \prod_{(t,\nu)\in I_j}
        \frac{(1+b\I a_{s,\mu}-b\I a_{t,\nu})_{k_{j,t,\nu}-k_{j,s,\mu}}}
        {(1+q_j+q_{j+1}+b\I a_{s,\mu}-b\I a_{t,\nu})_{k_{j,t,\nu}-k_{j,s,\mu}}}
        \\
        & \cdot
        \prod_{(t,\nu)\in I_{j-1}}
        \frac{(1+2q_j+b\I a_{s,\mu}-b\I a_{t,\nu})_{k_{j-1,t,\nu}-k_{j,s,\mu}}}
        {(1+b\I a_{s,\mu}-b\I a_{t,\nu})_{k_{j-1,t,\nu}-k_{j,s,\mu}}}
        \Biggr]
      \end{aligned}
      \\
      & \cdot
      \prod_{s=1}^{\Nf} \prod_{(t,\nu)\in I_n}
      \frac{\bigl(1-b(\lambda-b\sum_{j=1}^n K_j)/\Nf
        -b\vev{\I a_\infty,h_s}-b\I a_{t,\nu}\bigr)_{k_{n,t,\nu}}}
      {(1+b\vev{\I a_0,h_s}-b\I a_{t,\nu})_{k_{n,t,\nu}}}
      \Biggr\} \,.
    \end{aligned}
  \end{aligned}
\end{equation}

As discussed in the main text, when all \(\epsilon_j=-1\), placing all
degenerate punctures at the same position \(x_j=x\) yields the conformal
block for one particular fusion of the degenerate momenta, which turns
out to be
\begin{equation}
  -b\Omega = -b\sum_{j=1}^n \Omega_j =-b\sum_{j=1}^n \omega_{K_j} \,.
\end{equation}
This provides an explicit expression for the four-point conformal block
of two generic and one semi-degenerate momentum, and one degenerate
momentum labeled by an arbitrary representation of~\(A_{\Nf-1}\).
Fusing degenerate punctures in several sets gives conformal
blocks with several arbitrary degenerate momenta \(-b\Omega\), but these
quickly become unwieldy.

\subsection{Irregular Punctures}
\label{app:irregular}

We study irregular punctures obtained as collision limits of vertex
operators in the Toda CFT\@.  Such collisions were studied for Virasoro
primaries in~\cite{Gaiotto:2012sf}, and extended to other algebras
in~\cite{Kanno:2013vi,Gaiotto:2013rk}.  We give evidence that the
limit%
\begin{equation}\label{irregular-rank-K-def}
  \mathbb{V}_{c_0;c_1,\bar{c}_1;\cdots;c_K,\bar{c}_K} (w,\bar{w})
  = \lim_{(w_I,\bar{w}_I)\to(w,\bar{w})}
  \prod_{I<J} \abs{w_J-w_I}^{2\vev{\alpha_J,\alpha_I}}
  \prod_{I=0}^K V_{\alpha_I}(w_I,\bar{w}_I)
\end{equation}
exists, provided that the momenta~\(\alpha_I\) of vertex operators, and
their position~\((w_J,\bar{w}_J)\), vary in such a way that
\begin{align}
  C_j & = \sum_{I=0}^K (w_I-w)^j \alpha_I
  \to c_j
  &
  \bar{C}_j & = \sum_{I=0}^K (\bar{w}_I-\bar{w})^j \alpha_I
  \to \bar{c}_j
\end{align}
for all~\(j\geq 0\).  Not every choice of \(c_j\) and~\(\bar{c}_j\) can
appear (for a given rank~\(K\)).  Firstly, \(\bar{c}_0 = c_0\).
Secondly, \(\bar{c}_j = c_j = 0\) for all \(j>K\).  Indeed, any \(C_j\)
with \(j>K\) is a linear combination \(C_j = \sum_{k=0}^K
P_{j,k}(\{w_I-w\}) C_k\) whose coefficients \(P_{j,k}\) are homogeneous
polynomial of degree \(j-k\geq 1\) in the variables \(w_I-w\), and such
polynomials vanish as \(w_I\to w\).  The limits of \(C_j\)
and~\(\bar{C}_j\) are thus described by the \(2K+1\) momenta
\((c_0;c_1,\bar{c}_1;\cdots;c_K,\bar{c}_K)\), as indicated by the
notation in~\eqref{irregular-rank-K-def}.

There is (at least) one other condition on the \(c_j\)
and~\(\bar{c}_j\): for each \(0\leq m\leq K\) the vectors \(\{c_n,
\bar{c}_n\mid m\leq n\leq K\}\) must span a space of dimension at most
\(K - m + 1\), for instance \(c_K\) and~\(\bar{c}_K\) must be collinear.
This third restriction relies on
\begin{equation}\label{irregular-C-alpha}
  \sum_{j=0}^n \biggl(
  C_j \sum_{\substack{S\in\intset{0}{n-1}\\\#S=n-j}} \prod_{I\in S} (w-w_I)
  \biggr)
  = \sum_{J=n}^K \prod_{I=0}^{n-1} (w_J-w_I)^n \alpha_J \,,
\end{equation}
whose left-hand side goes to~\(c_n\) in our limit, and on its analogue
for~\(\bar{c}_n\).  Since rank is lower semicontinuous, the rank of the
space spanned by \(\{c_n,\bar{c}_n\mid m\leq n\leq K\}\) is at most that
of the space spanned by~\eqref{irregular-C-alpha} and by their
antiholomorphic counterparts (for \(m\leq n\leq K\)).  This second space
lies within the span of \(\{\alpha_J \mid m\leq J\leq K\}\), which has
rank at most \(K - m + 1\).

\subsubsection{OPE with the Stress-Energy Tensor}

Our first piece of evidence is to write the OPE of the stress-energy
tensor with
\begin{equation}
  \mathbf{V}_{\{\alpha_I\}}\bigl(\{w_I,\bar{w}_I\}\bigr)
  =
  \prod_{I<J} \abs{w_J-w_I}^{2\vev{\alpha_J,\alpha_I}}
  \prod_{I=0}^K V_{\alpha_I}(w_I,\bar{w}_I)
\end{equation}
in the limit which defines \(\mathbb{V}_{c_0;\cdots;c_K,\bar{c}_K}\).
The operators~\(V_{\alpha_I}\) are primary, hence
\begin{align}
  \nonumber
  & T(z) \mathbf{V}_{\{\alpha_I\}}\bigl(\{w_I,\bar{w}_I\}\bigr)
  \\
  & \quad \sim
  \prod_{I<J} \abs{w_J-w_I}^{2\vev{\alpha_J,\alpha_I}}
  \sum_{I=0}^K \biggl( \frac{\dimToda(\alpha_I)}{(z-w_I)^2}
  + \frac{1}{z-w_I} \partial_{w_I} \biggr)
  \prod_{I=0}^K V_{\alpha_I}(w_I,\bar{w}_I)
  \\
  & \quad = \sum_{I=0}^K \Biggl( \frac{\dimToda(\alpha_I)}{(z-w_I)^2}
  + \frac{1}{z-w_I} \biggl(\partial_{w_I}
    + \sum_{J\neq I} \frac{\vev{\alpha_I,\alpha_J}}{w_J-w_I} \biggr)
  \Biggr) \mathbf{V}_{\{\alpha_I\}}\bigl(\{w_I,\bar{w}_I\}\bigr)
  \\
  & \quad = \Biggl(
  \vev{Q, \partial_z\partial_z \phiToda_{\text{sing}}}
  - \frac{1}{2} \vev{\partial_z \phiToda_{\text{sing}},
    \partial_z \phiToda_{\text{sing}}}
  + \sum_{I=0}^K \frac{\partial_{w_I}}{z-w_I}
  \Biggr) \mathbf{V}_{\{\alpha_I\}}\bigl(\{w_I,\bar{w}_I\}\bigr)
\end{align}
where in the last line we use \(\dimToda(\alpha_I) = \vev{Q,\alpha_I} -
\frac{1}{2}\vev{\alpha_I,\alpha_I}\) to express all but the
\(\partial_{w_I}\) piece in terms of
\begin{equation}
  \partial_z \phiToda_{\text{sing}}
  = \sum_{I=0}^K \frac{-\alpha_I}{z-w_I}
  = \sum_{n\geq 0} \frac{-\sum_{I=0}^K (w_I-w)^n\alpha_I}{(z-w)^{n+1}}
  = \sum_{n\geq 0} \frac{-C_n}{(z-w)^{n+1}}
  \to \sum_{n=0}^{K} \frac{-c_n}{(z-w)^{n+1}} \,.
\end{equation}
In the domain where all \(\abs{w_I-w}<\abs{z-w}\), we can expand the
derivative term as
\begin{equation}
  \sum_{I=0}^K (z-w_I)^{-1} \partial_{w_I}
  = \sum_{n\geq -1} (z-w)^{-n-2} \sum_{I=0}^K (w_I-w)^{n+1} \partial_{w_I} \,.
\end{equation}
The term with \(n=-1\) is \(\sum_{I=0}^K \partial_{w_I}\), which translates
all vertex operators, hence its limit is~\(\partial_w\).
The other terms do not have such a simple geometrical
interpretation.  Instead, let us write their action on~\(C_m\) for
\(0\leq m\leq K\):
\begin{align}
  \sum_{I=0}^K (w_I-w)^{n+1} \partial_{w_I} C_m
  & = \sum_{I=0}^K (w_I-w)^{n+1} \partial_{w_I}
  \sum_{J=0}^K (w_J-w)^m \alpha_J
  \\
  & = \sum_{I=0}^K m (w_I-w)^{n+m} \alpha_I = m C_{n+m} \,.
\end{align}
The limit of \(\sum_{I=0}^K (w_I-w)^{n+1} \partial_{w_I}\) must thus be
a differential operator which maps \(c_m \to m c_{n+m}\) for all~\(0\leq
m\leq K-n\) and \(c_m \to 0\) for \(K-n<m\leq K\).  This is naturally
realized by
\begin{equation}
  \sum_{I=0}^K (w_I-w)^{n+1} \partial_{w_I}
  \to \sum_{j=1}^{K-n} j \vev{c_{n+j}, \partial_{c_j}} \,.
\end{equation}

All in all, the OPE of \(T(z)\) with \(\mathbb{V} =
\mathbb{V}_{c_0;\cdots;c_K,\bar{c}_K}(w,\bar{w})\) is
\begin{align}
  \nonumber
  T(z) \mathbb{V}
  & \sim
  \Biggl(
  \vev*{Q, \partial_z \sum_{n=0}^K \frac{-c_n}{(z-w)^{n+1}}}
  - \frac{1}{2} \vev*{\sum_{j=0}^K \frac{-c_j}{(z-w)^{j+1}},
    \sum_{l=0}^K \frac{-c_l}{(z-w)^{l+1}}}
  \\
  & \quad \qquad
  + \frac{\partial_w}{z-w}
  + \sum_{n\geq 0} \frac{1}{(z-w)^{n+2}}
  \sum_{j=1}^{K-n} j \vev{c_{n+j}, \partial_{c_j}}
  \Biggr)
  \mathbb{V}
  \\\label{irregular-T-OPE}
  & = \Biggl(
  \frac{1}{z-w} \partial_w
  + \sum_{n=0}^{2K} \frac{
    (n+1)\vev{Q, c_n}
    - \frac{1}{2} \sum_{j=0}^n \vev{c_j,c_{n-j}}
    + \sum_{j=1}^{K-n} j \vev{c_{n+j}, \partial_{c_j}}
  }{(z-w)^{n+2}}
  \Biggr) \mathbb{V}
\end{align}
where we recall that~\(c_n=0\) for \(n>K\).  The presence of
singularities up to \((z-w)^{-2K-2}\) in this OPE implies that the
Virasoro generators~\(L_n\) act non-trivially on the state~\(\lvert
c\rangle = \mathbb{V}_{c_0;\cdots;c_K,\bar{c}_K}(0)\lvert 0\rangle\) for
\(n\leq 2K\).  More precisely,
\begin{equation}\label{irregular-Virasoro}
  L_n \lvert c\rangle
  =
  \biggl(
  (n+1)\vev{Q, c_n}
  - \frac{1}{2} \sum_{j=0}^n \vev{c_j,c_{n-j}}
  + \sum_{j=1}^{K-n} j \vev{c_{n+j}, \partial_{c_j}}
  \biggr)
  \lvert c\rangle
\end{equation}
for \(0\leq n\leq 2K\), while \(L_{-1}\) translates~\(w\), and
\(L_n\lvert c\rangle = 0\) for \(n>2K\).  This is the natural
generalization of equation~(2.7) of~\cite{Gaiotto:2012sf}.

In the rank~\(1\) case (\(c_n=0\) for \(n>1\)), we can exponentiate
explicitly the action of the Virasoro generators~\(L_n\) to find how
large conformal transformations act.  From above, we know that \(L_n
\lvert c\rangle = 0\) for \(n>2\), that \(L_{-1}\)~acts
like~\(\partial_w\), and that
\begin{align}
  L_2 \lvert c\rangle
  & = - \frac{1}{2} \vev{c_1,c_1} \lvert c\rangle \,,
  \\
  L_1 \lvert c\rangle
  & = \vev{2Q-c_0, c_1} \lvert c\rangle \,,
  \\
  L_0 \lvert c\rangle
  & = \bigl(\dimToda(c_0)+\vev{c_1,\partial_{c_1}}\bigr) \lvert c\rangle \,,
\end{align}
where as usual \(\dimToda(c_0) = \vev{Q, c_0} - \vev{c_0,c_0}/2\).
Omitting the parameters \(\bar{z}\), \(\bar{w}\) and~\(\bar{c}_n\) which play no role
for holomorphic transformations, we claim that
\begin{equation}\label{irregular-conformal-transfo}
  \mathbb{V}_{c_0,c_1}(z)
  = \bigl(\partial_z w\bigr)^{\dimToda(c_0)}
  \exp\Biggl(
    \frac{\vev{2Q-c_0,c_1}}{2} \frac{\partial_z^2 w}{\partial_z w}
    - \frac{\vev{c_1,c_1}}{12} \biggl[
    \frac{\partial_z^3 w}{\partial_z w}
    - \frac{3}{2} \frac{(\partial_z^2 w)^2}{(\partial_z w)^2}
    \biggr]
  \Biggr) \mathbb{V}_{c_0,(\partial_z w)c_1}(w)
\end{equation}
under a conformal map \(z\to w(z)\).  Indeed, this transformation is
transitive and has the correct infinitesimal behavior: for \(\partial_z
w=1+\epsilon\),
\begin{equation}
  \mathbb{V}_{c_0,c_1}(z)
  =
  \biggl(1
  + \epsilon \bigl(\dimToda(c_0) + \vev{c_1,\partial_{c_1}}\bigr)
  + \frac{\vev{2Q-c_0,c_1}}{2} \partial_z\epsilon
  - \frac{\vev{c_1,c_1}}{12} \partial_z^2 \epsilon
  + O(\epsilon^2)
  \biggr)
  \mathbb{V}_{c_0,c_1}(w) \,.
\end{equation}

\subsubsection{Free Field Realization}

Our derivation of~\eqref{irregular-T-OPE} only relies on the OPE of
\(T(z)\) with vertex operators~\(V_\alpha\).  This OPE has a free field
realization as the OPE of \(T^\free_Q = \vev{Q,\partial\partial\phiToda}
- \frac{1}{2} \nop{\vev{\partial\phiToda,\partial\phiToda}}\) with
\(V^\free_\alpha = \nop{e^{\vev{\alpha,\phiToda}}}\).  We
rederive~\eqref{irregular-T-OPE} more directly by first building the
collision limit~\(\mathbb{V}^\free\) of vertex
operators~\(V^\free_\alpha\), then computing its OPE with~\(T^\free_Q\).
We then go further and consider the OPE of higher spin currents of the
\(W_{\Nf}\)~algebra with~\(\mathbb{V}^\free\).

First, \(\nop{e^{\vev{\alpha,\phiToda(z,\bar{z})}}}
\nop{e^{\vev{\beta,\phiToda(w,\bar{w})}}} =
\abs{z-w}^{-2\vev{\alpha,\beta}}
\nop{e^{\vev{\alpha,\phiToda(z,\bar{z})}+\vev{\beta,\phiToda(w,\bar{w})}}}\)
implies by induction
\begin{equation}
  \prod_{I<J} \abs{w_I-w_J}^{2\vev{\alpha_I,\alpha_J}}
  \prod_{I=0}^K \nop{e^{\vev{\alpha_I,\phiToda(w_I,\bar{w}_I)}}}
  =
  \nop{e^{\sum_{I=0}^K \vev{\alpha_I,\phiToda(w_I,\bar{w}_I)}}} \,.
\end{equation}
Expanding \(\phiToda(w_I, \bar{w}_I) = \phiToda(w, \bar{w}) +
\sum_{n\geq 1} \frac{1}{n!} \bigl( (w_I-w)^n \partial^n\phiToda(w)
+(\bar{w}_I-\bar{w})^n \bar{\partial}^n\phiToda(\bar{w}) \bigr)\) thanks
to \(\partial\bar{\partial}\phiToda = 0\) and using the limit
\(\sum_{I=0}^K (w_I-w)^n \vev{\alpha_I,\partial^n \phiToda} \to
\vev{c_n,\partial^n\phiToda}\) and its anti\-holomorphic counterpart
yields the free field collision limit
\begin{equation}\label{irregular-free-field}
  \mathbb{V}^\free_{c_0;\cdots;c_K,\bar{c}_K}(w,\bar{w})
  = \nop{\exp\biggl(\vev{c_0,\phiToda(w,\bar{w})}
    + \sum_{n=1}^K \frac{1}{n!} \Bigl( \vev{c_n,\partial^n\phiToda(w)}
      + \vev{\bar{c}_n,\bar{\partial}^n\phiToda(\bar{w})} \Bigr)
    \biggr)} \,.
\end{equation}

The stress-energy tensor~\(T^\free_Q(z)\) and higher spin currents are
polynomials in \(\partial\phiToda(z)\) and its derivatives.  We thus
evaluate
\begin{align}
  \partial\phiToda(z) \mathbb{V}^\free_{c_0;\cdots}(w,\bar{w})
  & = \nop{\biggl( \partial\phiToda(z)
    + \sum_{n\geq 0} \frac{c_n}{n!} \partial_w^n \frac{-1}{z-w} \biggr)
    \mathbb{V}^\free_{c_0;\cdots}(w,\bar{w})}
  \\
  & = \nop{\biggl(
    \sum_{n\geq 1} (z-w)^{n-1} n \partial_{c_n}
    - \sum_{n\geq 0} \frac{c_n}{(z-w)^{n+1}}
    \biggr)
    \mathbb{V}^\free_{c_0;\cdots}(w,\bar{w})}
\end{align}
where the first equality relies on
\(\contraction{\partial}{\phiToda}{(z)}{\phiToda}
  \partial\phiToda(z)\phiToda(w,\bar{w}) = -1/(z-w)\), and the
second on the Taylor expansion of \(\partial\phiToda(z)\) and on
\(\partial^n\phiToda\mathbb{V}^\free_{c_0;\cdots} =
n!\partial_{c_n}\mathbb{V}^\free_{c_0;\cdots}\).
The OPE of \(\mathbb{V}^\free_{c_0;\cdots}(w,\bar{w})\)
with any polynomial in derivatives of \(\partial\phiToda(z)\) is
thus obtained by replacing all
\begin{equation}\label{irregular-OPE-replacement}
  \partial_z^{l+1}\phiToda(z)
  \to
  \partial_z^l \biggl(
  \sum_{n\geq 1} (z-w)^{n-1} n \partial_{c_n}
  - \sum_{n\geq 0} \frac{c_n}{(z-w)^{n+1}}
  \biggr)
  =
  -\partial_z^l \sum_{n\in\bbZ}
  \frac{\theta_n c_n + \theta_{-n} n \partial_{c_{-n}}}{(z-w)^{n+1}}
\end{equation}
where \(\theta_n=1\) if \(n\geq 0\) and \(0\)~if \(n<0\), then dropping
terms that are regular as \(z\to w\).

In particular,
\begin{align}
  T^\free_Q(z)
  \mathbb{V}^\free
  & = \Bigl(\vev{Q,\partial\partial\phiToda(z)}
  -\frac{1}{2}\nop{\vev{\partial\phiToda(z),\partial\phiToda(z)}}\Bigr)
  \mathbb{V}^\free_{c_0;\cdots;c_K,\bar{c}_K}(w,\bar{w})
  \\\nonumber
  & \sim
  \Biggl(
  \vev*{Q, \sum_{n\geq 0} \frac{(n+1)c_n}{(z-w)^{n+2}}}
  - \frac{1}{2} \vev*{\sum_{i\geq 0} \frac{c_i}{(z-w)^{i+1}},
    \sum_{j\geq 0} \frac{c_j}{(z-w)^{j+1}}}
  \\
  & \qquad\quad
  + \vev*{\sum_{i\geq 0} \frac{c_i}{(z-w)^{i+1}},
    \sum_{j\geq 1} (z-w)^{j-1} j \partial_{c_j}}
  \Biggr)
  \mathbb{V}^\free
  \displaybreak[0]\\
  & =
  \Biggl( \sum_{n=0}^{2K}
  \frac{(n+1)\vev{Q,c_n}
    -\frac{1}{2}\sum_{i=0}^n \vev{c_i,c_{n-i}}}{(z-w)^{n+2}}
  + \sum_{n=-1}^{K-1} \frac{\sum_{j=1}^{K-n}
    \vev{c_{j+n},j\partial_{c_j}}}{(z-w)^{n+2}}
  \Biggr)
  \mathbb{V}^\free \,.
\end{align}
Upper bounds could be omitted since \(c_m=0\) for \(m\geq K\).  Note the
presence of~\(\partial_{c_{K+1}}\) in the last term for \(n=-1\) and
\(j=K+1\).  This derivative is inconvenient as it involves irregular
punctures with a rank higher than~\(\mathbb{V}^\free\).  It turns out
that the terms with \(n=-1\) combine nicely into
\begin{equation}
  \sum_{j=1}^{K+1} \frac{\vev{c_{j-1},\partial^j\phiToda(w)}}{(j-1)!}
  \mathbb{V}^\free_{c_0;\cdots;c_K,\bar{c}_K}(w,\bar{w})
  = \partial_w \mathbb{V}^\free_{c_0;\cdots;c_K,\bar{c}_K}(w,\bar{w}) \,.
\end{equation}
As expected, the free field OPE reproduces the
OPE~\eqref{irregular-T-OPE}.

We are ready to consider higher spin currents.  A basis of those
currents is obtained via the Miura transform
\begin{equation}\label{irregular-Miura}
  \prod_{s=\Nf}^1 \bigl(q\partial_z+\vev{h_s,\partial_z\phiToda(z)}\bigr)
  = \sum_{p=0}^{\Nf} W^p(z) \bigl(q\partial_z\bigr)^{\Nf-p}
\end{equation}
where \(q=b+\frac{1}{b}\).  In particular, \(W^0(z)=1\), \(W^1(z)=0\),
and \(W^2(z) = T^\free_Q(z)\).
The prescription~\eqref{irregular-OPE-replacement} then yields the OPE
of \(W^p(z)\) with the
irregular~\(\mathbb{V}^\free_{c_0;\cdots}(w,\bar{w})\), but expressions
quickly become very unwieldy.  However, we can get valuable information
by applying the prescription~\eqref{irregular-OPE-replacement} directly
to the Miura transform~\eqref{irregular-Miura}:
\begin{equation}
  \sum_{p=0}^{\Nf} W^p(z)
  \mathbb{V}^\free_{c_0;\cdots}(w,\bar{w})
  \bigl(q\partial_z\bigr)^{\Nf-p}
  =
  \prod_{s=\Nf}^1 \biggl(q\partial_z
  + \sum_{n\in\bbZ} \frac{\vev*{h_s, - \theta_n c_n
      - \theta_{-n} n \partial_{c_{-n}}}}{(z-w)^{n+1}}
  \biggr)
  \mathbb{V}^\free_{c_0;\cdots}(w,\bar{w})
\end{equation}
where \(\partial_{c_j}\) only acts on~\(\mathbb{V}^\free\) and not on
intervening~\(c_j\), and where \(\theta_n=1\) if \(n\geq 0\) and
\(0\)~otherwise.  The sums over~\(n\) actually truncate to \(n\leq K\)
for rank~\(K\) punctures, thus only a finite number of negative powers
of \((z-w)\) appear in the OPE\@.

Let us find out the most singular terms of the OPE of a given \(W^p(z)\)
with~\(\mathbb{V}^\free\) as \(z\to w\).  Thanks to the mode expansion
\(W^p(z) = \sum_{n\in\bbZ} W^p_n(w) (z-w)^{-n-p}\), the \((z-w)^{-n-p}\) term
in the OPE encodes the action of~\(W^p_n\) on the rank~\(K\) puncture
\(\lvert c\rangle =
\mathbb{V}^\free_{c_0;\cdots;c_K,\bar{c}_K}(w,\bar{w})\lvert 0\rangle\).
Terms where \(0\leq m<p\) of
the \(q\partial_z\) act on some \(\vev{h_s,\ldots}\) are at most of
order \(O\bigl((z-w)^{-(K+1)(p-m)-m}\bigr)\).  Those
involving \(\partial_{c_j}\)~derivatives are of order
\(O\bigl((z-w)^{-(K+1)(p-m-1)-m}\bigr)\) or more regular.  Thus,
\(W^p_n \lvert c\rangle = 0\) for \(n>pK\),
\begin{equation}
  \begin{aligned}
    W^p_n \lvert c\rangle
    = (-1)^p & \sum_{1\leq s_1<\cdots<s_p\leq\Nf} \Biggl[
    \sum_{k_1+\cdots+k_p = pK-n} \prod_{i=1}^p \vev{h_{s_i}, c_{K-k_i}}
    \\
    & \qquad + \delta_{n,(p-1)K} (K+1) q
    \sum_{j=1}^p \biggl( (j-1) \prod_{i\neq j}^p \vev{h_{s_i}, c_K} \biggr)
    \Biggr]
    \lvert c\rangle
  \end{aligned}
\end{equation}
for \((p-1)K\leq n\leq pK\),
and lower components of \(W^p(z)\) act with
\(\partial_{c_j}\)~derivatives.  This is consistent with the
action~\eqref{irregular-Virasoro} of the Virasoro algebra for \(p=2\).

For \(n<(p-1)K\), the action of \(W^p_n\) on \(\lvert c\rangle\)
involves derivatives~\(\partial_{c_j}\) for each \(1\leq j\leq
(p-1)K-n\).  In particular, if \(n<(p-2)K\), derivatives with \(j>K\)
appear: the set of rank~\(K\) irregular punctures is not stable under
those components~\(W^p_n\).  One exception is that
\(L_{-1}=W^2_{-1}\) involves derivatives up to~\(\partial_{c_{K+1}}\)
but turns out to be identical to an infinitesimal translation.  The set
of all (finite, integer) rank irregular punctures is stable under
all~\(W^p_n\).

Before closing this appendix, we go back to the Toda CFT and compute
various two-point functions of vertex operators with rank \(K=1\)
irregular punctures as a test that the collision limit is finite.

\subsubsection{Two-Point Functions}
\label{app:irregular-2pt}

Irregular punctures only arise in Section~\ref{sec:irreg} as the
collision of a semi-degenerate and a generic vertex operators.  We
compute here the two-point function of the resulting rank~\(1\) puncture
with any generic vertex operator~\eqref{irregular-2pt} in a useful
normalization~\eqref{irregular-normalized-V}.

The collision limits of interest are a special case of the general
collision limit~\eqref{irregular-rank-K-def} which defines rank~\(K\)
irregular punctures.  Using notations close to the main text,
\begin{equation}
  \mathbb{V}_{c_0;-(x/b)h_1,(\bar{x}/b)h_1}(0)
  = \lim_{\Lambda\to\infty} \left[
    \abs*{\frac{x}{\Lambda}}^{2\vev{\varkappa h_1, c_0 - \varkappa h_1}}
    V_{\varkappa h_1}
    \biggl(\frac{x}{-\I\Lambda}, \frac{\bar{x}}{\I\Lambda}\biggr)
    V_{c_0-\varkappa h_1}(0)
  \right]_{\varkappa = \I\Lambda/b+O(1)}
\end{equation}
where \(\Lambda\in\bbR\)~is the gauge theory cutoff scale, \(c_0\),
\(b\), \(x\) and~\(\bar{x}\) are various physical parameters, and only
the leading behavior of \(\varkappa\) in~\(\Lambda\) affects the limit.
We also introduce the normalization
\begin{align}\label{irregular-normalized-V}
  & \widehat{\mathbb{V}}_{c_0;-(x/b)h_1,(\bar{x}/b)h_1}(0)
  =
  \frac{\hat{\mu}^{\vev{c_0-Q,\rho}}\mathbb{V}_{c_0;-(x/b)h_1,(\bar{x}/b)h_1}(0)}
  {\Upsilon(b)^{\Nf-1} \prod_{2\leq s<t\leq\Nf}\Upsilon(\vev{Q-c_0,h_s-h_t})}
  \\\nonumber
  & \: = \lim_{\Lambda\to\infty} \left[
    \frac{\Upsilon\bigl(\varkappa+\vev{Q-c_0,h_1}\bigr)^{\Nf}}
      {\abs{\Lambda/b}^{2\dimToda(c_0)-\vev{Q,Q}}}
    \abs*{\frac{x}{\Lambda}}^{2\vev{\varkappa h_1, c_0 - \varkappa h_1}}
    \widehat{V}_{\varkappa h_1}
    \biggl(\frac{x}{-\I\Lambda}, \frac{\bar{x}}{\I\Lambda}\biggr)
    \widehat{V}_{c_0-\varkappa h_1}(0)
  \right]_{\varkappa = \I\Lambda/b+O(1)}
\end{align}
where the second line is obtained by combining the factors
\eqref{normalized-V} and~\eqref{normalized-semi} which relate
\(\widehat{V}\) and~\(V\) with those relating \(\widehat{\mathbb{V}}\)
and~\(\mathbb{V}\).  The only non-trivial step is that the
asymptotics~\eqref{Upsilon-asymptotic-prod} of the Upsilon function
simplify \(\prod_{t=1}^{\Nf} \Upsilon(\varkappa+\vev{Q-c_0,h_1-h_t})\) to
\(\Upsilon(\varkappa+\vev{Q-c_0,h_1})^{\Nf}
\abs{\Lambda/b}^{\vev{Q,Q}-2\dimToda(c_0)}\).

Let us compute the two-point function of the irregular
puncture~\eqref{irregular-normalized-V} with a generic vertex
operator~\(\widehat{V}_{\alpha_0}\).  Throughout the calculation,
\(\varkappa = \I\Lambda/b+O(1)\).  Scale covariance and the explicit
form~\eqref{FL-1-39} of the three-point function give
\begin{align}
  \nonumber
  &
  \frac{\Upsilon\bigl(\varkappa+\vev{Q-c_0,h_1}\bigr)^{\Nf}}
    {\abs{\Lambda/b}^{2\dimToda(c_0)-\vev{Q,Q}}}
  \abs*{\frac{x}{\Lambda}}^{2\vev{\varkappa h_1, c_0 - \varkappa h_1}}
  \vev*{\widehat{V}_{\alpha_0}(\infty)
    \widehat{V}_{\varkappa h_1}
    \biggl(\frac{x}{-\I\Lambda}, \frac{\bar{x}}{\I\Lambda}\biggr)
    \widehat{V}_{c_0-\varkappa h_1}(0)
  }
  \\
  & =
  \frac{
    \abs{x/\Lambda}^{
      2\vev{\varkappa h_1, c_0 - \varkappa h_1}
      - 2\dimToda(\varkappa h_1) - 2\dimToda(c_0-\varkappa h_1)
      + 2\dimToda(\alpha_0)}
    \Upsilon(\varkappa+\vev{Q-c_0,h_1})^{\Nf}}
  {\abs{\Lambda/b}^{2\dimToda(c_0)-\vev{Q,Q}}
    \prod_{s,t=1}^{\Nf} \Upsilon\bigl(
    \frac{\varkappa}{\Nf}
    +\vev{Q-c_0+\varkappa h_1,h_s}+\vev{Q-\alpha_0,h_t}\bigr)}
  \\
  & \sim
  \frac{\abs{x/\Lambda}^{2\dimToda(\alpha_0) - 2\dimToda(c_0)}
    \abs{\Lambda/b}^{2\dimToda(\alpha_0)-\vev{Q,Q}}}
  {\abs{\Lambda/b}^{2\dimToda(c_0)-\vev{Q,Q}}
    \prod_{s=2}^{\Nf} \prod_{t=1}^{\Nf} \Upsilon\bigl(
    \vev{Q-c_0,h_s}+\vev{Q-\alpha_0,h_t}\bigr)} \,.
\end{align}
All powers of~\(\Lambda\) cancel, and we deduce that
\begin{equation}\label{irregular-2pt}
  \vev*{\widehat{V}_{\alpha_0}(\infty)
    \widehat{\mathbb{V}}_{c_0;-(x/b)h_1,(\bar{x}/b)h_1}(0)}
  =
  \frac{\abs{x/b}^{2\dimToda(\alpha_0) - 2\dimToda(c_0)}}
  {\prod_{s=2}^{\Nf} \prod_{t=1}^{\Nf} \Upsilon\bigl(
    \vev{Q-c_0,h_s}+\vev{Q-\alpha_0,h_t}\bigr)} \,.
\end{equation}
Note that the dependence on \(\abs{x/b}\) is as expected from the
transformation~\eqref{irregular-conformal-transfo} of rank~\(1\)
irregular punctures under a scaling.  Both the OPE with \(W_{\Nf}\)
currents, and the two-point function we have just computed, are finite,
and independent of details such as the precise value of~\(\varkappa\) in
the limit~\eqref{irregular-normalized-V}.  This gives credence to our
claim that collision limits \(\mathbb{V}_{c_0;\cdots;c_K,\bar{c}_K}\)
are finite and only depend on the \(c_j\) and~\(\bar{c}_j\).

A similar calculation (not used in the main text) is the two-point
function of a rank~\(1\) irregular puncture (with \(c_1\) and
\(\bar{c}_1\) collinear) and a semidegenerate operator:
\begin{equation}\label{irregular-2pt-full}
  \begin{aligned}
    & \vev*{\widehat{V}_{\varkappa h_1}(x,\bar{x})
      \mathbb{V}_{c_0;c_1,\bar{c}_1}(y,\bar{y})}
    =
    \frac{\exp\bigl(\frac{\vev{c_0-2Q,c_1}}{y-x}+\frac{\vev{c_0-2Q,\bar{c}_1}}{\bar{y}-\bar{x}}\bigr)}
    {\abs{x-y}^{4\dimToda(\varkappa h_1)}
      \hat{\mu}^{\vev{c_0-2Q,\rho}}
      \prod_{s=1}^{\Nf} \Upsilon\bigl(\frac{\varkappa}{\Nf}+\vev{2Q-c_0,h_s}\bigr)}
    \\
    & \qquad\qquad\cdot
    \prod_{s<t} \bigl(-\vev{c_1,h_s-h_t}\vev{\bar{c}_1,h_s-h_t}\bigr)^{(\varkappa/\Nf-\vev{c_0-2Q,h_t})(b+1/b-\varkappa/\Nf+\vev{c_0-2Q,h_s})}
    \,.
  \end{aligned}
\end{equation}
It is instructive to note how this expression is consistent with
transformation properties~\eqref{irregular-conformal-transfo} of
rank~\(1\) irregular punctures.  Under a special
conformal transformation \(z\to w(z)\) which keeps \(x\) and~\(y\)
invariant, \(c_1\)~is scaled by \(\partial_z w\) at~\(y\).  Since
at the fixed point~\(y\) one has
\(\frac{\partial_z w - 1}{y-x} = \frac{\partial_z^2 w}{2\partial_z w}\),
the exponential in~\eqref{irregular-2pt-full} is shifted by
\(\frac{1}{2}\vev{2Q-c_0,c_1} \partial_z^2 w/\partial_z w\),
as required by~\eqref{irregular-conformal-transfo}.

\section{Vortex Partition Function Dualities}
\label{app:Proof}

We prove here that the vortex partition functions of some dual theories
are equal up to simple factors.  The equalities are most easily seen
through the matching with Toda CFT correlators, as done in the main
text.  However, the matching is not proven in all cases, so we proceed
to establish the equalities directly using integral representations of
the vortex partition functions.  We cover the case of Seiberg duality
for \(\Nsusy=(2,2)\) SQCD in Appendix~\ref{app:Proof-SQCD}.  We then add
adjoint matter and a superpotential in Appendix~\ref{app:Proof-SQCDAW}:
this includes as special cases the Seiberg duality for
\(\Nsusy=(2,2)^*\) SQCD, and the Kutasov--Schwimmer duality.  The two
appendices use similar ideas but are independent.

\subsection{SQCD Vortex Partition Functions}
\label{app:Proof-SQCD}

We focus first on the \(S^2\)~partition function of an \(\Nsusy=(2,2)\)
theory of a \(U(\Nc)\) vector multiplet coupled to \(\Nf\)~fundamental
and \(\anti{\Nf}\)~antifundamental chiral multiplets.  Its expression
can be decomposed as~\eqref{SQCD-Zs} into vortex partition
functions~\cite{Benini:2012ui,Doroud:2012xw}.  By symmetry we can assume
that \(\anti{\Nf}<\Nf\), or that \(\anti{\Nf}=\Nf\) and \(\abs{z}<1\).
The relevant vortex partition functions are then labeled by
\(\Nc\)-element subsets of \(\intset{1}{\Nf}\) and take the form
\begin{equation}
  Z_{\vortex,\{p\}}(m,\anti{m},x)
  = \sum_{k=0}^\infty x^k Z_{k,\{p\}}(m,\anti{m}) \,,
\end{equation}
where $x=(-1)^{\Nf+\Nc-1}z$ and the \(k\)-vortex partition function is
\begin{equation}
  Z_{k,\{p\}}(m,\anti{m}) = \sum_{k_1+\cdots+k_\Nc=k}
  \prod_{j=1}^\Nc \biggl[
  \frac{1}{k_j!}
  \frac{\prod_{s=1}^{\anti{\Nf}}
    (-\I\anti{m}_s-\I m_{p_j})_{k_j}}
  {\prod_{i\neq j}^\Nc (\I m_{p_i}-\I m_{p_j}-k_i)_{k_j}
    \prod_{s\not\in\{p\}}^{\Nf} (1+\I m_s-\I m_{p_j})_{k_j}}
  \biggr]
  \,.
\end{equation}
We prove that the vortex partition function is invariant under the
Seiberg duality map \(\Nc^\dual = \Nf-\Nc\), \(\{p\}^\dual\to\{p\}^\complement\)
(the set complement), \(m_s^\dual = \frac{\I}{2} - m_s\), \(\anti{m}_s^\dual =
\frac{\I}{2} - \anti{m}_s\), \(z^\dual = (-1)^{\anti{\Nf}} z\), up to a
simple overall factor.  This is based on the proof~\cite{Benini:2012ui}
that for \(\anti{\Nf}\leq\Nf-2\) the \(k\)-vortex partition function
is invariant.  Since \(Z_{k,\{p\}}\) depends analytically on the \(m_s\)
and~\(\anti{m}_s\), we only need to prove the equality when
\(R\)-charges \(\Re(-2\I m_s)\) and \(\Re(-2\I\anti{m}_s)\) are between
\(0\) and~\(1\); the same is then true of the \(R\)-charges in the dual
theory.

Consider a closed contour~\(C_k^+\) which lies in the half-plane
\(\Re(\varphi)>-\frac{1}{2}\) and surrounds with a positive orientation
all points \(-\I m_s + \nu\) and \(\frac{1}{2}+\I m_s+\nu\) for \(1\leq
s\leq\Nf\) and integer \(0\leq\nu<\kappa\).  This set of points, which
all have positive real part, is invariant under the duality map
\(-\I m_s^\dual = \frac{1}{2}+\I m_s\).  The contour
\(C_k^-=-\frac{1}{2}-C_k^+\) lies in the half-plane \(\Re(\varphi)<0\)
and surrounds with a positive orientation all points \(-1-\I m_s-\nu\)
and \(-\frac{1}{2}+\I m_s-\nu\) for \(1\leq s\leq\Nf\) and integer
\(0\leq\nu<\kappa\).  Define the contour integrals
\begin{equation}\label{Proof-vortex-integral}
  I_{k,\{p\}}^\pm(m,\anti{m})
  =
  \frac{1}{k!}
  \int_{(C_k^\pm)^k}\frac{\dd[^k]{\varphi}}{(2\pi\I)^k}
  \prod_{\kappa\neq\lambda}^k
  \frac{\varphi_\kappa-\varphi_\lambda}{\varphi_\kappa-\varphi_\lambda-1}
  \prod_{\kappa=1}^k
  \frac{\prod_{s=1}^{\anti{\Nf}}(\varphi_\kappa-\I\anti{m}_s)}
    {\prod_{s=1}^{\Nf}(\varphi_\kappa+\I m_s+\delta_{s\not\in\{p\}})} \,.
\end{equation}
As we will see shortly, \(I^\pm\)~are essentially \(k\)-vortex partition
functions of Seiberg dual theories~\eqref{Proof-ZkZkpm}.  Given our
choice of contours, the change of variables
\(\varphi\to\varphi^\dual=-\frac{1}{2}-\varphi\) maps \(C_k^\pm\to
C_k^\mp\), and we find
\begin{equation}\label{Proof-contour-duals}
  I_{k,\{p\}^\complement}^\pm(m^\dual,\anti{m}^\dual)
  = (-1)^{(1+\anti{\Nf}+\Nf)k} I_{k,\{p\}}^\mp(m,\anti{m}) \,,
\end{equation}
where the sign comes from \(\dd{\varphi^\dual}=-\dd{\varphi}\),
\(\varphi^\dual-\I\anti{m}^\dual = -(\varphi-\I\anti{m})\), and \(\varphi^\dual+\I
m_s^\dual+\delta_{s\in\{p\}} = -(\varphi+\I m_s+\delta_{s\not\in\{p\}})\).

Poles of the integrand for which all
\(\Re(\varphi_\kappa)>-\frac{1}{2}\) are labeled by choices of
\(\Nc\)~integers \(k_s\geq 0\) with \(\sum_{s\in\{p\}} k_s=k\), such
that the \(\varphi_\kappa\) are given in some order by
\begin{equation}
  \{\varphi_\kappa\}
  = \bigl\{ -\I m_s+\nu \bigm| s\in\{p\}, 0\leq\nu<k_s \bigr\} \,,
\end{equation}
hence \((C_k^+)^k\)~surrounds precisely those poles.  Similarly, poles
with \(\Re(\varphi_\kappa)<0\) are
\begin{equation}
  \{\varphi_\kappa\}
  = \bigl\{ -1-\I m_s-\nu \bigm| s\not\in\{p\}, 0\leq\nu<k_s \bigr\} \,,
\end{equation}
labeled by \(\Nf-\Nc\) integers \(k_s\geq 0\) for \(s\not\in\{p\}\),
summing to~\(k\), and \((C_k^-)^k\)~surrounds precisely those poles.
For a given choice of \(k_1+\cdots+k_\Nc=k\), the residue at each of the
\(k!\)~points \(\{\varphi_\kappa\}=\{-\I m_{p_j}+\nu\mid 1\leq j\leq\Nc,
0\leq\nu<k_j\}\) reproduces the corresponding term in the \(k\)-vortex
partition function (the factor~\(1/k!\) cancels the choice of ordering
of \(\varphi_\kappa\)), hence the \(k\)-vortex partition functions are
\begin{equation}\label{Proof-ZkZkpm}
  \begin{aligned}
    Z_{k,\{p\}}(m,\anti{m})
    & = I_{k,\{p\}}^+(m,\anti{m})
    \\
    Z_{k,\{p\}^\complement}(m^\dual,\anti{m}^\dual)
    & = (-1)^{(1+\anti{\Nf}+\Nf)k} I_{k,\{p\}}^-(m,\anti{m}) \,,
  \end{aligned}
\end{equation}
where the dual relation derives from~\eqref{Proof-contour-duals} or from
summing residues at poles surrounded by \((C_k^-)^k\).

\subsubsection{SQCD with \(\anti{\Nf}<\Nf\)}

As long as \(\anti{\Nf}\leq\Nf-2\), the integrand
in~\eqref{Proof-vortex-integral} is regular at infinity, hence we can
choose \(C^+\)~along \(-\frac{1}{4}+\I\bbR\), from \(\I\infty\) to
\(-\I\infty\): then \(C^-=-\frac{1}{2}-C^+\) has the opposite
orientation, and \(I_{k,\{p\}}^-(m,\anti{m}) = (-1)^k
I_{k,\{p\}}^+(m,\anti{m})\).  Therefore
\begin{equation}\label{Proof-Seiberg-Zk-1}
  Z_{k,\{p\}^\complement}(m^\dual,\anti{m}^\dual)
  = (-1)^{(\anti{\Nf}+\Nf)k} Z_{k,\{p\}}(m,\anti{m}) \,,
\end{equation}
hence vortex partition functions are equal:
\begin{equation}\label{Proof-Seiberg-1}
  Z_{\vortex,\{p\}^\complement}(m^\dual,\anti{m}^\dual,x^\dual)
  = Z_{\vortex,\{p\}}(m,\anti{m},x) \,,
\end{equation}
where \(x^\dual=(-1)^{\Nf+\anti{\Nf}}x\) hence \(z^\dual = (-1)^{\anti{\Nf}} z\).  This result strongly relies on
our ability to reverse contours, that is, on the absence of poles at
infinity for \(\anti{\Nf}\leq\Nf-2\).  For \(\anti{\Nf}=\Nf-1\) or
\(\anti{\Nf}=\Nf\), we must take into account the contribution
from infinity.

Consider first the case \(\anti{\Nf}=\Nf-1\).  We shift the pole at
infinity to a finite position through the regulating factor \(\I
M/(\varphi_\kappa+\I M)\).  This is equivalent to adding a fundamental
chiral multiplet with twisted mass~\(M\) in
the strip \(0<\Re(-2\I M)<1\).  In the limit \(\abs{M}\to\infty\), the
contours \((C_k^\pm)^k\) only surround poles of the original integral,
which are independent of~\(M\), and the regulator does not affect
residues.  Therefore,
\begin{equation}
  \begin{aligned}
    I_{k,\{p\}}^\pm(m,\anti{m})
    =
    \lim_{\abs{M}\to\infty}
    \frac{1}{k!}
    & \int_{(C_k^\pm)^k}\frac{\dd[^k]{\varphi}}{(2\pi\I)^k}
    \Biggl\{
    \prod_{\kappa\neq\lambda}^k
    \frac{\varphi_\kappa-\varphi_\lambda}{\varphi_\kappa-\varphi_\lambda-1}
    \\
    & \qquad \cdot
    \prod_{\kappa=1}^k \Biggl[
    \frac{\prod_{s=1}^{\Nf-1}(\varphi_\kappa-\I\anti{m}_s)}
      {\prod_{s=1}^{\Nf}(\varphi_\kappa+\I m_s+\delta_{s\not\in\{p\}})}
    \frac{\I M}{\varphi_\kappa+\I M}
    \Biggr]
    \Biggr\}
    \,.
  \end{aligned}
\end{equation}
Poles of the integrand above with all \(\Re(\varphi_\kappa)<0\) are
identical to those of the non-regulated integral, hence integrating
along the contour \(-\frac{1}{4}+\I\bbR\) yields
\(I_{k,\{p\}}^-(m,\anti{m})\) by closing the contour towards
\(-\infty\).  Closing the contour instead towards \(+\infty\) surrounds
poles at
\begin{equation}
  \{\varphi_\kappa\}
  = \bigl\{ -\I m_s+\nu \bigm| s\in\{p\}, 0\leq\nu<k_s \bigr\}
  \cup \bigl\{ -\I M+\nu \bigm| 0\leq\nu<l \bigr\} \,,
\end{equation}
for each choice of non-negative integers~\(k_s\) for \(s\in\{p\}\),
and~\(l\), such that \(l+\sum_{s\in\{p\}} k_s=k\).  The residue at such
a point is (factors of \(\I M\) cancel out)
\begin{equation}
  \frac{(-1)^l}{l!}
  \res_{\{\varphi_\kappa\mid 1\leq\kappa\leq k-l\}} \Biggl[
  \prod_{\kappa\neq\lambda}^{k-l}
  \frac{\varphi_\kappa-\varphi_\lambda}{\varphi_\kappa-\varphi_\lambda-1}
  \prod_{\kappa=1}^{k-l}
  \frac{\prod_{s=1}^{\Nf-1}(\varphi_\kappa-\I\anti{m}_s)}
    {\prod_{s=1}^{\Nf}(\varphi_\kappa+\I m_s+\delta_{s\not\in\{p\}})}
  \Biggr] \,,
\end{equation}
where the residue is precisely one of the contributions to
\(I_{k-l,\{p\}}^+(m,\anti{m})\).  The contributions for a fixed~\(l\)
combine into the full \((k-l)\)-vortex partition function.  All in all,
using~\eqref{Proof-ZkZkpm} \(I_{k,\{p\}}^-(m,\anti{m}) =
Z_{k,\{p\}^\complement}(m^\dual,\anti{m}^\dual)\) and \(I_{k,\{p\}}^+ =
Z_{k,\{p\}}\) when \(\anti{\Nf}=\Nf-1\),
\begin{align}
  Z_{k,\{p\}^\complement}(m^\dual,\anti{m}^\dual)
  & = (-1)^k \sum_{l=0}^k \frac{(-1)^l}{l!} Z_{k-l,\{p\}}(m,\anti{m}) \,,
  \\
  \label{Proof-Seiberg-exp}
  Z_{\{p\}^\complement}(m^\dual,\anti{m}^\dual,x^\dual)
  & = e^{-x} Z_{\{p\}}(m,\anti{m},x) \,.
\end{align}

Alternatively, the factor~\(e^x\) can be obtained from the case
\(\Nf=\anti{\Nf}+2\) (where there is no factor) by decoupling one of the
fundamental chiral multiplets through the limit
\(\abs{m_{\Nf}}\to\infty\).  For an arbitrary \(\Nf>\anti{\Nf}\),
\begin{equation}\label{Proof-fs-limit}
  Z^{\Nf,\anti{\Nf}}_{k,\{p\}}
  \sim
  \begin{cases}
    \displaystyle
    (\I m_{\Nf})^{-k} \sum_{l=0}^k (-1)^l \binom{k}{l}
    (-\I m_{\Nf})^{l(\anti{\Nf}+2-\Nf)}
    Z^{\Nf-1,\anti{\Nf}}_{k-l,\{p\}}
    & \text{if \(\Nf\in\{p\}\),}
    \\
    \displaystyle\vphantom{\sum_l^k}
    (\I m_{\Nf})^{-k}
    Z^{\Nf-1,\anti{\Nf}}_{k,\{p\}}
    & \text{if \(\Nf\not\in\{p\}\).}
  \end{cases}
\end{equation}
If \(\Nf\geq\anti{\Nf}+3\), terms other than \(l=0\) in the sum are of a
lower order, thus \(Z^{\Nf,\anti{\Nf}}_{k,\{p\}} \sim (\I m_{\Nf})^{-k}
Z^{\Nf-1,\anti{\Nf}}_{k,\{p\}}\), consistent with the
equality~\eqref{Proof-Seiberg-Zk-1} of Seiberg-dual vortex partition
functions in those cases.  If \(\Nf=\anti{\Nf}+2\), we find
\begin{equation}
  Z^{\anti{\Nf}+2,\anti{\Nf}}_{\{p\}}(\I m_{\Nf} x)
  \sim
  e^{-x\delta_{\Nf\in\{p\}}}
  Z^{\anti{\Nf}+1,\anti{\Nf}}_{\{p\}}(x) \,.
\end{equation}
Exactly one of two Seiberg-dual vortex partition functions exhibits this
exponential factor, and with opposite signs since \(\I m_{\Nf}^\dual \sim -\I
m_{\Nf}\).  Starting from the Seiberg duality
relation~\eqref{Proof-Seiberg-1} for \(\Nf\geq\anti{\Nf}+2\), we thus
obtain the exponential factor in~\eqref{Proof-Seiberg-exp} for
\(\Nf=\anti{\Nf}+1\).  Unfortunately, the same technique fails to reach
the case \(\Nf=\anti{\Nf}\), because terms beyond~\eqref{Proof-fs-limit}
contribute to the limit \(\abs{m_{\Nf}}\to\infty\) (with \(x/m_{\Nf}\) kept
constant).  We avoid this issue in the contour integral approach by
introducing different parameters for each occurrence of~\(m_{\Nf}\), as we
now see.

\subsubsection{SQCD with \(\anti{\Nf}=\Nf\)}

When \(\anti{\Nf}=\Nf\), we regulate using \(\prod_{\kappa=1}^k\bigl[
-(\I M_\kappa)^2 \bigm/ \bigl(\varphi_\kappa^2-(\I M_\kappa)^2\bigr)
\bigr]\) with \(M_\kappa\) real for simplicity.  This factor is similar
to the contribution from two fundamental chiral multiplets with opposite
twisted masses, but importantly we let the parameter~\(M_\kappa\) depend
on~\(\kappa\).  In fact, we will consider the limit where masses have
different scales, \(1\ll\abs{M_1}\ll\cdots\ll\abs{M_k}\), as this
simplifies the expansion of residues.  For large
enough~\(\abs{M_\kappa}\), the additional poles lie outside the contours
\((C_k^\pm)^k\), and the regulating factor tends to~\(1\) when evaluated
on the contour (or at poles it encloses), thus
\begin{equation}
  \begin{aligned}
    I_{k,\{p\}}^\pm(m,\anti{m})
    =
    \lim_{\abs{M_\kappa}\to\infty}
    \frac{1}{k!}
    & \int_{(C_k^\pm)^k}\frac{\dd[^k]{\varphi}}{(2\pi\I)^k} \Biggl\{
    \prod_{\kappa=1}^k \biggl[
    \frac{-(\I M_\kappa)^2}{\varphi_\kappa^2-(\I M_\kappa)^2}
    \biggr]
    \\
    & \qquad \cdot
    \prod_{\kappa\neq\lambda}^k \biggl[
    \frac{\varphi_\kappa-\varphi_\lambda}{\varphi_\kappa-\varphi_\lambda-1}
    \biggr]
    \prod_{\kappa=1}^k \prod_{s=1}^{\Nf} \biggl[
    \frac{\varphi_\kappa-\I\anti{m}_s}
      {\varphi_\kappa+\I m_s+\delta_{s\not\in\{p\}}}
    \biggr]
    \Biggr\}
    \,.
  \end{aligned}
\end{equation}
Poles of the integrand above with all
\(\Re(\varphi_\kappa)\leq-\frac{1}{4}\) are identical to those of the
non-regulated integral, hence integrating along the contour
\(-\frac{1}{4}+\I\bbR\) yields \(Z_{k,\{p\}}^-(m,\anti{m})\) by closing
the contour towards \(-\infty\).

Closing the contour instead towards \(+\infty\) surrounds numerous
poles:
\begin{equation}\label{Proof-Nf-reg-poles}
  \{\varphi_\kappa\}
  = \bigl\{ -\I m_s+\mu \bigm| s\in\{p\}, 0\leq\mu<k_s \bigr\}
  \cup \bigl\{ \epsilon_\kappa\I M_\kappa+\nu \bigm|
  \kappa\in K, 0\leq\nu<l_\kappa \bigr\} \,,
\end{equation}
where \(K\)~is the set of \(1\leq\kappa\leq k\) such that
\(\varphi_\kappa = \epsilon_\kappa \I M_\kappa\) for some sign
\(\epsilon_\kappa = \pm 1\), and where the integers \(k_s\geq 0\) for
\(s\in\{p\}\) and \(l_\kappa > 0\) for \(\kappa\in K\) sum to~\(k\).  To
specify a pole completely, one needs to know \(\{K,
\epsilon_\kappa,l_\kappa,k_s\}\), but also which component of
\(\varphi\) is equal to each \(-\I m_s+\mu\) and each
\(\epsilon_\kappa\I M_\kappa+\nu\).  This is encoded in maps \(\sigma\)
and~\(\tau\) such that
\begin{equation}
  \varphi_{\sigma(s,\mu)} = -\I m_s + \mu
  \quad \text{and} \quad
  \varphi_{\tau(\kappa,\nu)} = \epsilon_\kappa \I M_\kappa + \nu \,.
\end{equation}
Note that \(\tau(\kappa,\nu)=\kappa\) if and only if \(\nu=0\).

We expand the residue at the pole defined by
\(\{K,\epsilon_\kappa,l_\kappa,k_s,\sigma,\tau\}\) in the limit
\(1\ll\abs{M_1}\ll\cdots\ll\abs{M_k}\):
\begin{equation}\label{Proof-residue}
  \begin{aligned}
    & \frac{1}{k!}
    \prod_{\kappa\in K} \Biggl[
    \frac{-\epsilon_\kappa\I M_\kappa}{2}
    \prod_{\nu=1}^{l_\kappa-1} \biggl[
    \frac{-(\I M_{\tau(\kappa,\nu)})^2}
      {(\epsilon_\kappa \I M_\kappa+\nu)^2-(\I M_{\tau(\kappa,\nu)})^2}
    \biggr]
    \Biggr]
    \prod_{s\in\{p\}} \prod_{\mu=0}^{k_s-1} \biggl[
    1 + O\biggl(\frac{1}{M_{\sigma(s,\mu)}^2}\biggr)\biggr]
    \\
    & \quad \cdot
    \prod_{\kappa\in K} \biggl[
    1 - \frac{l_\kappa \Sigma}{\epsilon_\kappa\I M_\kappa}
    + O\biggl(\frac{1}{M_\kappa^2}\biggr)
    \biggr]
    \prod_{\kappa\in K} \biggl[
    1
    + O\biggl(\frac{1}{M_\kappa^2}\biggr)
    \biggr]
    \prod_{\kappa\in K} \biggl[ \frac{1}{l_\kappa} \biggr]
    \\
    & \quad \cdot
    \res_{\varphi_{\sigma(s,\mu)} = -\I m_s+\mu}
    \Biggl\{
    \prod_{\kappa\neq\lambda\in\{\sigma(s,\mu)\}} \biggl[
    \frac{\varphi_\kappa-\varphi_\lambda}{\varphi_\kappa-\varphi_\lambda-1}
    \biggr]
    \prod_{\kappa\in\{\sigma(s,\mu)\}} \prod_{t=1}^{\Nf} \biggl[
    \frac{\varphi_\kappa-\I\anti{m}_t}
      {\varphi_\kappa+\I m_t+\delta_{t\not\in\{p\}}}
    \biggr]
    \Biggr\}
    \,.
  \end{aligned}
\end{equation}
The first line consists of all factors coming from the regulator; the
next factor comes from \((\varphi_{\tau(\dots)}-\I\anti{m}_s) /
(\varphi_{\tau(\dots)}+\I m_s+\delta_{s\not\in\{p\}})\) and involves
\begin{equation}
  \Sigma = \sum_{s=1}^{\Nf} (\I\anti{m}_s+\I m_s+\delta_{s\not\in\{p\}}) \,;
\end{equation}
the following two factors come from the ratio
\((\varphi-\varphi)/(\varphi-\varphi-1)\) where either one or both
components of~\(\varphi\) take the form~\(\varphi_{\tau(\kappa,\mu)}\);
the last line consists of all finite factors, independent of
the~\(M_\kappa\), which organize themselves into a residue along the
components~\(\varphi_{\sigma(s,\mu)}\).  A useful simplification is
\begin{equation}\label{Proof-tau-kappa-limit}
  \frac{-(\I M_{\tau(\kappa,\nu)})^2}
    {(\epsilon_\kappa\I M_\kappa+\nu)^2-(\I M_{\tau(\kappa,\nu)})^2}
  \sim \begin{cases}
    - M_{\tau(\kappa,\nu)}^2 M_\kappa^{-2}
    & \text{if \(\tau(\kappa,\nu)<\kappa\)} \,, \\
    1 & \text{if \(\tau(\kappa,\nu)>\kappa\)} \,.
  \end{cases}
\end{equation}

On its own, the residue~\eqref{Proof-residue} grows like
\(\prod_\kappa(-\epsilon_\kappa \I M_\kappa)\), but we will see that the
sum over all possible choices of the signs~\(\epsilon_\kappa\) (keeping
\(\{K,l_\kappa,k_s,\sigma,\tau\}\) fixed) has a finite limit.  More
precisely, starting from \(\lambda=k\), and all the way down to
\(\lambda=1\), we sum over \(\epsilon_\lambda=\pm 1\) (if \(\lambda\in
K\)) and take the limit \(\abs{M_\lambda}\to\infty\).  At each step
there are three cases.  If \(\lambda=\sigma(s,\mu)\), the twisted mass
appears only in a factor \(1+O(1/M_\lambda^2)\), which thus drops out.
If \(\lambda=\tau(\kappa,\nu)>\kappa\), then the
factor~\eqref{Proof-tau-kappa-limit} containing~\(M_\lambda\) drops out.
The case \(\lambda=\tau(\kappa,\nu)<\kappa\) does not appear, as we see
shortly.  Finally, if \(\lambda\in K\), several factors
contain~\(M_\lambda\):
\begin{equation}
  \frac{-\epsilon_\lambda\I M_\lambda}{2}
  \prod_{\substack{1\leq\nu<l_\lambda\\\tau(\lambda,\nu)<\lambda}} \biggl[
  \frac{-(\I M_{\tau(\lambda,\nu)})^2}
    {(\epsilon_\lambda \I M_\lambda+\nu)^2-(\I M_{\tau(\lambda,\nu)})^2}
  \biggr]
  \biggl[
  1 - \frac{l_\lambda \Sigma}{\epsilon_\lambda\I M_\lambda}
  + O\biggl(\frac{1}{M_\lambda^2}\biggr)
  \biggr]
  \biggl[
  1 + O\biggl(\frac{1}{M_\lambda^2}\biggr)
  \biggr] \,.
\end{equation}
This expression vanishes in the limit \(\abs{M_\lambda}\to\infty\) if
any \(\tau(\lambda,\nu)<\lambda\), thus only poles for which all
\(\tau(\lambda,\nu)\geq\lambda\) contribute in the limit we consider.
Otherwise, the expression above is \(\frac{1}{2}\bigl(
-\epsilon_\lambda\I M_\lambda + l_\lambda \Sigma + O(1/M_\lambda)
\bigr)\), whose sum over \(\epsilon_\lambda=\pm 1\) is the finite result
\(l_\lambda \Sigma\).  All in all, the sum over all choices of
signs~\(\epsilon\) of the residue at the pole defined by
\(\{K,\epsilon_\kappa,l_\kappa,k_s,\sigma,\tau\}\) has a finite limit
\begin{equation}\label{Proof-k-Sigma-res}
  \frac{1}{k!}
  \Sigma^{\#K}
  \res_{\varphi_{\sigma(s,\mu)} = -\I m_s+\mu}
  \Biggl\{
  \prod_{\kappa\neq\lambda\in\{\sigma(s,\mu)\}} \biggl[
  \frac{\varphi_\kappa-\varphi_\lambda}{\varphi_\kappa-\varphi_\lambda-1}
  \biggr]
  \prod_{\kappa\in\{\sigma(s,\mu)\}} \prod_{t=1}^{\Nf} \biggl[
  \frac{\varphi_\kappa-\I\anti{m}_t}
    {\varphi_\kappa+\I m_t+\delta_{t\not\in\{p\}}}
  \biggr]
  \Biggr\}
  \,,
\end{equation}
which turns out to only depends on the number~\(\#K\) of elements
in~\(K\) and on the~\(k_s\).

We must now sum this expression over all choices of sets~\(K\), of
integers \(l_\kappa>0\) and~\(k_s\geq 0\), and of indices
\(\sigma(s,\mu)\) and \(\tau(\kappa,\nu)>\kappa\).  The choice of
\(\{K,l_\kappa,k_s,\sigma,\tau\}\) can be split into a choice of
\(\{K,l_\kappa,\tau\}\) followed by a choice of integers \(k_s\geq 0\)
summing to \(k-l\), where \(l = \sum_{\kappa\in K} l_\kappa\), and
finally a choice of~\(\sigma\) labeling the complement of
\(T=\{\tau({\cdot},{\cdot})\}\) by pairs~\((s,\mu)\).  This last choice
does not affect the residue, hence contributes a factor of \((k-l)!\).
The sum over \(\{k_s\}\) (summing to \(k-l\)) of the residue
in~\eqref{Proof-k-Sigma-res} yields the \((k-l)\)-vortex partition
function.  Thus,
\begin{equation}
  Z_{k,\{p\}}^-(m,\anti{m})
  =
  (-1)^k
  \sum_{l=0}^\infty \Biggl[
  \frac{(k-l)!}{k!}
  \sum_{T\mid\#T=l} \sum_{K\subseteq T} \sum_{\{l_\kappa\geq 1\}} \sum_\tau
  \bigl[ \Sigma^{\#K} \bigr]
  Z_{k-l,\{p\}}^+(m,\anti{m})
  \Biggr] \,.
\end{equation}
The number of choices of \(\{K,l_\kappa,\tau\}\) with a given~\(\#K\)
only depends on the size \(l=\#T\), thus the choice of~\(T\) contributes
a factor \(k!/[l!(k-l)!]\).  At this point, we could conclude by noting
that we expressed \(Z_{k,\{p\}}^-(m,\anti{m})\) in terms of the
\(Z_{k-l,\{p\}}^+(m,\anti{m})\) with coefficients depending only on
\(l\)~and the combination~\(\Sigma\) of twisted masses, and neither
on~\(\Nf\) nor on~\(\Nc\).  The coefficients can thus be obtained
through the special case \(\anti{\Nf}=\Nf=1\), \(\Nc=0\), for which
computations are elementary, leading to a Seiberg duality relation valid
for arbitrary \(\anti{\Nf}=\Nf\) and~\(\Nc\).

For completeness, we go through the combinatorical exercise.  Since only
\(l=\#T\) affects the counting, we can fix \(T=\intset{1}{l}\) to
simplify the discussion.  Define the map~\(\upsilon:T\to T\) such that
for each \(\kappa\in K\), \(\upsilon(\kappa)=\kappa\) and
\(\upsilon(\tau(\kappa,\nu)) = \max \{\tau(\kappa,\mu)\mid 0\leq\mu<\nu,
\tau(\kappa,\mu)<\tau(\kappa,\nu)\}\) for \(\nu>0\).  The data of
\(K\subseteq T=\intset{1}{l}\) and \(\upsilon:T\to T\) with
\(\upsilon(\kappa)=\kappa\) for \(\kappa\in K\) and
\(\upsilon(\lambda)<\lambda\) for \(\lambda\in T\setminus K\) is in fact
equivalent to that of \(\{K,l_\kappa,\tau\}\).  There are
\(\prod_{\lambda\in T\setminus K} (\lambda-1)\) maps~\(\upsilon\), hence
\begin{equation}
  \begin{aligned}
    Z_{k,\{p\}}^-(m,\anti{m})
    & =
    (-1)^k
    \sum_{l=0}^\infty \Biggl\{ \frac{1}{l!} \Biggl(
    \sum_{K\subseteq\intset{1}{l}} \Sigma^{\#K}
    \prod_{\lambda\in\intset{1}{l}\setminus K} (\lambda-1)
    \Biggr)
    Z_{k-l,\{p\}}^+(m,\anti{m})
    \Biggr\}
    \\
    & =
    (-1)^k
    \sum_{l=0}^\infty \frac{(\Sigma)_l}{l!}
    Z_{k-l,\{p\}}^+(m,\anti{m})
    \,,
  \end{aligned}
\end{equation}
where \((\Sigma)_l = \Sigma \cdots (\Sigma+l-1)\) is the Pochhammer
symbol.  From this, we can finally deduce the Seiberg duality relation
\begin{equation}\label{Proof-vortex-relation}
  Z_{\{p\}^\complement}(m^\dual,\anti{m}^\dual,x^\dual)
  = (1-x)^{-\Sigma} Z_{\{p\}}(m,\anti{m},x) \,,
\end{equation}
with \(x^\dual=x\) hence \(z^\dual=(-1)^{\anti{\Nf}}z\), and where we recall \(\Sigma = \sum_{s=1}^{\Nf}
(\I\anti{m}_s+\I m_s)+\Nf-\Nc\).  This relation precisely matches that
obtained in the main text as Toda conjugation, in particular the
exponent~\eqref{Seiberg-delta1}.

\subsection{SQCDA Vortex Partition Functions}
\label{app:Proof-SQCDAW}

We now adapt the proof to \(\Nsusy=(2,2)\) SQCDA theories with a
superpotential.  The field content consists of a vector multiplet
coupled to one adjoint chiral multiplet~\(X\), \(\Nf\)~fundamental
chiral multiplets~\(\quark_s\), and \(\Nf\)~antifundamental chiral
multiplets~\(\anti{\quark}_s\).  As in Section~\ref{sec:SQCDAW} we
consider two cases: the superpotential \(W = \sum_{t=1}^{\Nf}
\anti{\quark}_t X^{l_t} \quark_t\) and the superpotential \(W = \Tr
X^{l+1}\) for integers \(l_t,l\geq 0\).  Both choices exhibit common
features, with \(l_t\) replaced by~\(l\) for the second superpotential.

In sections \ref{sec:SeibergW-cross} and~\ref{sec:Kutasov}, we find that
pairs of such theories with gauge groups \(U(\Nc)\) and \(U(\Nc^\dual)\) are
dual, using symmetries of Toda CFT correlators.  Parameters are mapped
as follows: \(m_X^\dual = m_X\), \(\Nc^\dual = L - \Nc\) with \(L =
\sum_{t=1}^{\Nf} l_t\), and
\begin{align}
  m_t^\dual & = m_t \,,
  & \anti{m}_t^\dual & = m_t^\dual \,,
  & z^\dual & = z^{-1}
  & \text{for } W & = \sum_{t=1}^{\Nf} \anti{\quark}_t X^{l_t} \quark_t
  \\
  m_t^\dual & = m_X - m_t \,,
  & \anti{m}_t^\dual & = m_X - \anti{m}_t \,,
  & z^\dual & = z
  & \text{for } W & = \Tr X^{l+1} \,.
\end{align}
Higgs branch vacua of the \(U(\Nc)\) theory are labeled by integers
\(0\leq n_t\leq l_t\) with sum~\(\Nc\).  Those are in a natural
bijection \(n_t^\dual = l_t-n_t\) to integers \(0\leq n_t^\dual\leq l_t\) with
sum \(L-\Nc\), which label Higgs branch vacua of the dual theory.  We
compare classical and one-loop contributions
in sections \ref{sec:SeibergW-cross} and~\ref{sec:Kutasov}.  We now
prove the relations \eqref{Proof-SQCDAW-Z=Z}
and~\eqref{Proof-SQCDAW-Z=Z-Kutasov} between the vortex partition
functions of the \(U(\Nc)\) theory in the vacuum~\(\{n_t\}\) and of the
\(U(L-\Nc)\) theory in the vacuum \(\{l_t-n_t\}\).  As in the main text,
\(y=(-1)^{\Nf}z\).

\subsubsection{Preliminary Result for \(\Nf=1\)}
\label{app:Proof-SQCDAW-Zvn}

Later on, we prove that dual vortex partition functions are equal up to
some factor which only depends on very little data.  To fix this factor,
we will use the simple case of \(\Nf=1\) SQCDA with \(1+\I
m_1+\I\anti{m}_1+\Nc\I m_X = 0\), which we consider now.  Its unique
vacuum has \(n_1=\Nc\), and we prove that
\(Z_{\vortex,\{\Nc\}}(y)=(1-y)^{-\Nc(1+\I m_X)}\).  By analyticity, it
is enough to show this when \(\Re(\I\anti{m}_1) < 0 < \Re(-\I m_1) <
\Re(-\I m_X)\).

The vortex partition function, given by the series~\eqref{SQCDA-Zv}, has
a Mellin--Barnes integral representation
\begin{align}
  \label{Proof-SQCDAW-Zvn-series}
  & Z_{\vortex,\{\Nc\}}(y)
  = \sum_{\{k_\mu\geq 0\}} y^{\sum k_\mu}
  \prod_{\mu,\nu=0}^{\Nc-1}
  \frac{((\nu-\mu-1)\I m_X-k_\nu)_{k_\mu}}
    {((\nu-\mu)\I m_X-k_\nu)_{k_\mu}}
  \\
  \label{Proof-SQCDAW-Zvn-MB}
  & =
  (-y)^{\Nc\I m_1+\frac{1}{2}(\Nc-1)\Nc\I m_X}
  \prod_{\mu=1}^\Nc\frac{\sin\pi(-\mu\I m_X)}{\pi}
  \\\nonumber
  & \quad \cdot \frac{1}{\Nc!}
  \int_{\bbR^\Nc} \frac{\dd[^\Nc]{\sigma}}{(2\pi)^\Nc}
  (-y)^{\Tr\I\sigma}
  \prod_{j=1}^\Nc \Bigl[ \Gamma(-\I m_1-\I\sigma_j)
  \Gamma(-\I\anti{m}_1+\I\sigma_j)\Bigr]
  \frac{\prod_{i,j=1}^\Nc \Gamma(\I\sigma_i-\I\sigma_j-\I m_X)}
  {\prod_{i\neq j}^\Nc \Gamma(\I\sigma_i-\I\sigma_j)}
\end{align}
which analytically continues \(Z_{\vortex,\{\Nc\}}(y)\) from the unit
disc to \(y\not\in\bbR_{\geq 0}\).  Closing contours towards
\(\I\infty\) yields a similar relation for \(\abs{y}>1\), with
\(m_1\leftrightarrow\anti{m}_1\) and \(y\to y^{-1}\).  Hence, the
analytic continuations obey
\begin{equation}
  Z_{\vortex,\{\Nc\}}(y)
  = (-y)^{-\Nc(1+\I m_X)} Z_{\vortex,\{\Nc\}}(y^{-1}) \,.
\end{equation}
The function \((1-y)^{\Nc(1+\I m_X)} Z_{\vortex,\{\Nc\}}(y)\) is thus
analytic on the Riemann sphere away from \(y=1\).  Furthermore, we can
bound it by a power of \(\abs{1-y}\) in two pairs of angular sectors
centered at \(y=1\), whose union is a neighborhood of \(y=1\).

The first angular sector is defined by \(\abs{1-y} < M(1-\abs{y})\) for
some \(M>0\) and is contained in the open unit disc.  The coefficients
in the series~\eqref{Proof-SQCDAW-Zvn-series} grow at most polynomially
in the exponent \(\sum_\mu k_\mu\) of~\(y\), and the number of terms
contributing for a given power of~\(y\) also grows polynomially.  Hence,
\begin{equation}
  \abs{Z_{\vortex,\{\Nc\}}(y)}
  \leq \sum_{k\geq 0} C_1 (k+1)_{C_2} \abs{y}^k
  = C_2! C_1 (1-\abs{y})^{-1-C_2}
\end{equation}
for some \(C_1,C_2>0\) which do not depend on~\(y\).  Thus
\(\abs{1-y}^{1+C_2} Z_{\vortex,\{\Nc\}}(y)\) is bounded in each sector
\(\abs{1-y} < M(1-\abs{y})\).  By the symmetry \(y\to y^{-1}\), the
function is also bounded in a similar sector \(\abs{y-1} <
M(\abs{y}-1)\).  We have thus probed the function away from the unit
circle.

The next pair of sectors is probed using the Mellin--Barnes
representation~\eqref{Proof-SQCDAW-Zvn-MB}, which converges away from
the real axis.  Set \(y = r e^{\epsilon\I\theta}\) with
\(\frac{1}{2}<r<2\) (to avoid \(\{0,\infty\}\)), \(\epsilon=\pm 1\), and
\(0<\theta<\pi\) (that is, \(y\not\in\bbR\)).  Then
\begin{equation}
  \abs{(-y)^{\I\sigma_j}}
  = e^{\epsilon(\pi-\theta)\sigma_j}
  \leq e^{(\pi-\theta)\abs{\sigma}} \,.
\end{equation}
For some large enough \(C_1,C_2>0\) which depend on the twisted masses,
we have
\begin{equation}
  \abs*{\frac{\Gamma(-\I m_1-\I\sigma_j)}
    {\Gamma(1+\I\anti{m}_1-\I\sigma_j)}}
  < C_1 \bigl(\abs{\sigma} + 1\bigr)^{\Nc\Re(\I m_X)}
  \:,\:
  \abs*{\frac{\Gamma(\I\sigma_i-\I\sigma_j-\I m_X)}
    {\Gamma(\I\sigma_i-\I\sigma_j)}}
  < C_2 \bigl(\abs{\sigma} + 1\bigr)^{\Re(-\I m_X)}
\end{equation}
for all~\(\sigma\), where \(\abs{\sigma} = \bigl(\sum_{i=1}^\Nc
\abs{\sigma_i}^2\bigr)^{1/2}\) is larger than all \(\abs{\sigma_j}\) and
all \(\abs{\sigma_i-\sigma_j}\).  The inequalities rely on the
asymptotics \(\Gamma(a+\I\upsilon) / \Gamma(b+\I\upsilon) \sim
(\I\upsilon)^{a-b}\) as \(\upsilon\to\pm\infty\), and the continuity of
both ratios of Gamma functions.  Since \(0<\Im(\anti{m}_1)<1\), we
also have
\begin{equation}
  \abs{\Gamma(1+\I\anti{m}_1-\I\sigma_j)
    \Gamma(-\I\anti{m}_1+\I\sigma_j)}
  \leq \frac{2\pi e^{-\pi\abs{\sigma_j-\Re(\anti{m}_1)}}}
    {\abs{\sin(\pi\Im(\anti{m}_1))}}
  < C_3 e^{-\pi\abs{\sigma}}
\end{equation}
for some \(\anti{m}_1\)-dependent \(C_3 > 0\).  Combining the bounds
into~\eqref{Proof-SQCDAW-Zvn-MB} yields
\begin{equation}
  \abs{Z_{\vortex,\{\Nc\}}(y)}
  \leq
  C_4
  \int_{\bbR^\Nc} \dd[^\Nc]{\sigma}
  e^{-\Nc\theta\abs{\sigma}}
  \bigl(\abs{\sigma} + 1\bigr)^{\Nc\Re(\I m_X)}
\end{equation}
for some \(C_4 > 0\).  Switching to polar coordinates, letting \(\tau =
\theta(\abs{\sigma}+1)\), and bounding
\((\tau-\theta)^{\Nc-1}<\tau^{\Nc-1}\) leads to
\begin{equation}
  \abs{\theta^{\Nc(1+\I m_X)} Z_{\vortex,\{\Nc\}}(y)}
  \leq C_5 \int_\theta^\infty \dd{\tau} e^{-\Nc\tau} \tau^{\Nc\Re(\I m_X)+\Nc-1}
  \leq C_6
\end{equation}
for some \(C_5,C_6>0\).  In any angular sector centered at \(y=1\) and
away from the real axis, \(\abs{1-y}\) is bounded by some multiple of
\(\theta=\arg(y)\), hence \((1-y)^{\Nc(1+\I m_X)}
Z_{\vortex,\{\Nc\}}(y)\) is bounded both above and below the real axis.

We have bounded the function \((1-y)^{\Nc(1+\I m_X)}
Z_{\vortex,\{\Nc\}}(y)\) by a power of \(\abs{1-y}\) in a neighborhood
of \(y=1\).  Since the function is analytic away from~\(1\), it takes
the form \(P(y)/(1-y)^n\), where \(P(y)\) is a polynomial of degree at
most \(n\geq 0\).  In the second pair of sectors, we found that the
function is bounded as \(y\to 1\), thus \(n=0\) and the function is the
constant \((1-y)^{\Nc(1+\I m_X)} Z_{\vortex,\{\Nc\}}(y) =
Z_{\vortex,\{\Nc\}}(0) = 1\).

\subsubsection{Proof for SQCDA}

Let us move on to the proof per se.  We start with the vortex partition
function~\eqref{SQCDA-Zv} of the \(U(\Nc)\) SQCDA theory in a given
Higgs branch vacuum~\(\{n_s\}\).  The terms of this series in
\(y=(-1)^{\Nf}z\) are labeled by integer vorticities
\(k_{s\mu}\geq 0\) for \(1\leq s\leq\Nf\) and \(0\leq\mu<n_s\):
\begin{equation}
  Z_{\vortex,\{n_s\}}(y)
  = \sum_{k\geq 0} y^k Z_{\vortex,\{n_s\},k}
  = \sum_{k\geq 0} y^k \sum_{\sum k_{s\mu} = k} V_{\{n_s\}}^{\{k_{s\mu}\}} \,.
\end{equation}
The contribution \(V_{\{n_s\}}^{\{k_{s\mu}\}}\) for a given choice of
vorticities is a ratio of Pochhammer symbols, which we massage using the
identity \((1-x-k)_{k-j}=(-1)^{k-j}(x)_k/(x)_j\) into
\begin{equation}\label{Proof-SQCDAW-V}
  V_{\{n_s\}}^{\{k_{s\mu}\}}
  =
  \prod_{(s,\mu)\in I} \Biggl[
  \prod_{t=1}^{\Nf}
  \frac
  {(-\I\anti{m}_t - \I m_{s\mu})_{k_{s\mu}}}
  {(1+\I m_t + n_t \I m_X - \I m_{s\mu})_{k_{s\mu}}}
  \prod_{(t,\nu)\in I}
  \frac{(\I m_{t\nu}-\I m_{s\mu} - \I m_X - k_{t\nu})_{k_{s\mu}}}
  {(\I m_{t\nu}-\I m_{s\mu} - k_{t\nu})_{k_{s\mu}}}
  \Biggr] \,.
\end{equation}
Here, \(m_s\), \(\anti{m}_s\), and \(m_X\) are complexified twisted
masses of the chiral multiplets, we denote \(m_{s\mu} = m_s+\mu m_X\),
the products range over \(I=\{(s,\mu)\mid 1\leq
s\leq\Nf,0\leq\mu<n_s\}\), and we have swapped \((s,\mu)\leftrightarrow
(t,\nu)\) compared to~\eqref{SQCDA-Zv}.  Using that
\begin{equation}
  \prod_{(s,\mu)\in I} \prod_{(t,\nu)\in I}
  \frac{(\I m_{t\nu}-\I m_{s\mu}-A-k_{t\nu})_{k_{s\mu}}}
    {(\I m_{t\nu}-\I m_{s\mu}-A)_{k_{s\mu}}}
  = \mspace{-5mu}
  \prod_{\substack{(s,\mu)\in I\\0\leq i<k_{s\mu}}}
  \prod_{\substack{(t,\nu)\in I\\0\leq j<k_{t\nu}}}
  \frac{\I m_{t\nu}-j-\I m_{s\mu}+i-A-1}
    {\I m_{t\nu}-j-\I m_{s\mu}+i-A}
\end{equation}
for a generic \(A\in\bbC\), we can
express~\(V_{\{n_s\}}^{\{k_{s\mu}\}}\) in terms of the combinations
\(-\I m_{s\mu}+i\) for \((s,\mu)\in I\) and \(0\leq i<k_{s\mu}\).  We
find that \((-1)^k V_{\{n_s\}}^{\{k_{s\mu}\}}\) is the residue at
\(\{\varphi_\kappa\}=\{-\I m_{s\mu}+i\mid 0\leq i<k_{s\mu}\}\) of the
integrand in~\eqref{Proof-SQCDAW-Ik} below, after
\(\abs{M_\kappa}\to\infty\).

The discussion above leads us to the contour integral (\(I_0=1\) is an
empty product)%
\begin{equation}\label{Proof-SQCDAW-Ik}
  \begin{aligned}
    & I_k
    =
    \lim_{\abs{M_1}\to\infty}
    \cdots
    \lim_{\abs{M_k}\to\infty}
    \frac{1}{k!}
    \prod_{\kappa=1}^k \Biggl[
    \int_{-\I\infty}^{\I\infty} \frac{\dd{\varphi_\kappa}}{2\pi\I} \Biggr]
    \Biggl\{
    \prod_{\kappa=1}^k
    \frac{-(\I M_\kappa)^2}
      {(\varphi_\kappa-\frac{1}{2})^2-(\I M_\kappa)^2}
    \prod_{\kappa\neq\lambda}^k
    \frac{\varphi_\kappa-\varphi_\lambda}
      {\varphi_\kappa-\varphi_\lambda-1}
    \\
    & \cdot
    \prod_{\kappa,\lambda=1}^k
    \frac{\varphi_\kappa-\varphi_\lambda-1-\I m_X}
      {\varphi_\kappa-\varphi_\lambda-\I m_X}
    \prod_{\kappa=1}^k \prod_{s=1}^{\Nf} \biggl[
    \frac{\varphi_\kappa-\I\anti{m}_s}
      {\varphi_\kappa+1+\I m_s+n_s\I m_X}
    \frac{\varphi_\kappa+\I m_s-\I m_X}
      {\varphi_\kappa+\I m_s+(n_s-1)\I m_X}
    \biggr]
    \Biggr\}
  \end{aligned}
\end{equation}
whose residues include all contributions to the \(k\)-vortex partition
function~\(Z_{\vortex,\{n_s\},k}\).  As in the SQCD case, we move the
pole at infinity to a finite value through a regulating factor, which
depends on large real parameters with
\(1\ll\abs{M_1}\ll\cdots\ll\abs{M_k}\).  The small shift
by~\(\frac{1}{2}\) moves poles away from the imaginary axis.  We assume
that the complex parameters \(m_s\) and~\(m_X\) are in the ranges
\begin{equation}
  0 < \Re(\I m_X) < 1 \,,
  \qquad
  (n_s-1) \Re(\I m_X) < \Re(-\I m_s) < n_s \Re(\I m_X) \,.
\end{equation}
This constraint is eventually lifted since the relation we will deduce
between vortex partition functions is analytic in \(m_s\) and~\(m_X\).

Close the contours of~\eqref{Proof-SQCDAW-Ik} towards \(+\infty\) first.
Because of the factors \(1/(\varphi_\kappa-\varphi_\lambda-1)\) and
\(1/(\varphi_\kappa-\varphi_\lambda-\I m_X)\), the surrounded poles are
such that the~\(\varphi_\lambda\) are organized into groups of
components with related values:
\begin{align}
  \{\varphi_\lambda\mid\lambda\in T\}
  & = \coprod_{\kappa\in K}
  \bigl\{ \tfrac{1}{2}+\epsilon_\kappa\I M_\kappa+\nu\I m_X+j
  \bigm| 0\leq\nu<\hat{n}_\kappa, 0\leq j<\hat{k}_{\kappa\nu} \bigr\} \,,
  \\
  \{\varphi_\lambda\mid\lambda\in S\}
  & = \coprod_{\mathclap{1\leq s\leq\Nf}}
  \bigl\{ -\I m_s+(1-n_s+\mu)\I m_X+i \bigm|
  0\leq\mu<n'_s, 0\leq i<k'_{s\mu} \bigr\}
\end{align}
where~\(K\) is the set of indices for which \(\varphi_\kappa =
\frac{1}{2}+\epsilon_\kappa\I M_\kappa\), and \(\intset{1}{k}=S\sqcup
T\).  Note that all \(n'_s\leq n_s\), otherwise the numerator factor
\(\prod_\lambda(\varphi_\lambda+\I m_s-\I m_X)\) would vanish.
Introducing if necessary \(k'_{sn'_s}=\cdots=k'_{s(n_s-1)}=0\), we set
\(n'_s = n_s\), then define \(k_{s\mu} = k'_{s(n_s-1-\mu)}\).

The pole is uniquely determined by the partition \(\intset{1}{k}=S\sqcup
T\), the set~\(K\subseteq T\), the signs~\(\epsilon_\kappa=\pm 1\), the
non-negative integers \(n_s\) (fixed when defining~\(I_k\)),
\(k_{s\mu}\), \(\hat{n}_\kappa\), and~\(\hat{k}_{\kappa\nu}\), and the
maps \(\sigma\) and~\(\tau\) defined by
\begin{equation}\label{Proof-SQCDAW-positive}
  \varphi_{\sigma(s,\mu,i)} = -\I m_s-\mu\I m_X+i
  \quad\text{and}\quad
  \varphi_{\tau(\kappa,\nu,j)}
  = \frac{1}{2}+\epsilon_\kappa\I M_\kappa+\nu\I m_X+j
\end{equation}
for \(1\leq s\leq\Nf\), \(0\leq\mu<n_s\), \(0\leq i<k_{s\mu}\), and for
\(\kappa\in K\), \(0\leq\mu<\hat{n}_\kappa\), \(0\leq
j<\hat{k}_{\kappa\mu}\).  This data is constrained: \(\sigma\)~is a
bijection from \(\{(s,\mu,i)\mid 0\leq i<k_{s\mu}\}\) to~\(S\), hence
\(\sum k_{s\mu} = \#S\), and \(\tau\)~is a bijection from
\(\{(\kappa,\nu,j)\mid 0\leq j<\hat{k}_{\kappa\nu}\}\) to~\(T\), hence
\(\sum \hat{k}_{\kappa\nu} = \#T\).  Also, \(\tau(\kappa,0,0)=\kappa\)
for all \(\kappa\in K\).

Let \(t=\#T\).  It is convenient to parametrize poles in terms of the
data \(t\), \(T\), \((K,\hat{n}_\kappa,\hat{k}_{\kappa\nu},\tau)\),
\((k_{s\mu},\sigma)\), and~\(\epsilon_\kappa\).  When summing residues
of~\(I_k\) at such poles, we will first sum over choices of
signs~\(\epsilon_\kappa\) and take the limits
\(\abs{M_\kappa}\to\infty\).  The result is independent of~\(\sigma\),
which thus contributes only a combinatorical factor.  Then follows a sum
over choices of~\(k_{s\mu}\), whose only constraint is \(\sum k_{s\mu} =
k - t\).  Since the residue of~\(I_k\) involves the vortex contribution
\(V_{\{n_s\}}^{\{k_{s\mu}\}}\), the sum over~\(k_{s\mu}\) yields the
\((k-t)\)-vortex partition function.  Summing over the remaining data,
we find that \(I_k\)~is a linear combination of \((k-t)\)-vortex
partition functions for \(0\leq t\leq k\), whose coefficients only
depend on \(t\), \(\I m_X\), and a single combination~\(\Sigma\) of the
twisted masses.  This allows us to fix the coefficients by considering a
simple case.

Let us proceed.  The residue at~\eqref{Proof-SQCDAW-positive}
of~\eqref{Proof-SQCDAW-Ik} has the following asymptotics:
\begin{equation}\label{Proof-SQCDAW-residue}
  \begin{aligned}
    &
    \prod_{\kappa=1}^k
    \biggl[1+O\biggl(\frac{1}{M_\kappa^2}\biggl)\biggr]
    \prod_{\kappa<\lambda}^k
    \biggl[1+O\biggl(\frac{M_\kappa^2}{M_\lambda^2}\biggr)\biggr]
    \prod_{\tau(\kappa,\mu,j)<\kappa}
    \biggl[O\biggl(\frac{M_{\tau(\kappa,\mu,j)}^2}{M_\kappa^2}\biggr)\biggr]
    \\
    & \cdot
    \frac{(-1)^k}{k!}
    \prod_{\kappa\in K}
    f_{\{\hat{n}_\kappa\},\{\hat{k}_{\kappa\nu}\}}(\I m_X)
    \prod_{\kappa\in K} \biggl[
    \frac{-\epsilon_\kappa\I M_\kappa}{2}
    + \frac{\Sigma}{2} \sum_{\nu=0}^{\hat{n}_\kappa-1} \hat{k}_{\kappa\nu}
    + O\biggl(\frac{1}{M_\kappa}\biggr)
    \biggr]
    V_{\{n_s\}}^{\{k_{s\mu}\}}
  \end{aligned}
\end{equation}
where \(f\)~is a rational function of~\(\I m_X\) with integer
coefficients, and
\begin{equation}
  \Sigma = 2\Nc\I m_X+\sum_{s=1}^{\Nf}(1+\I m_s+\I\anti{m}_s) \,.
\end{equation}

We expect the divergent piece \(-\epsilon_\kappa\I M_\kappa/2\) of the
residue to cancel when summing over signs~\(\epsilon_\kappa\).  Let us
take limits \(\abs{M_\lambda}\to\infty\) from \(\lambda=k\) down to
\(\lambda=1\) carefully.  At each step there are two cases.  If
\(\lambda\in K\), then the limit vanishes whenever any
\(\tau(\lambda,\mu,j)<\lambda\).  Hence, only poles with all
\(\tau(\lambda,\mu,j)\geq\lambda\) contribute and we can focus on those.
The \(M_\lambda\)-dependent terms are then of the form
\(-\epsilon_\lambda\I M_\lambda/2\) plus a finite part.  Summing
over~\(\epsilon_\lambda=\pm 1\) only leaves the finite part.  On the
other hand, if \(\lambda\not\in K\), then taking the limit
\(\abs{M_\lambda}\to\infty\) is trivial as \(M_\lambda\)~only appears in
factors \([1+O(1/M_\lambda^2)]\) and \([1+O(M_\kappa^2/M_\lambda^2)]\)
for \(\kappa<\lambda\) (importantly, we have already taken the limits
\(\abs{M_\kappa}\to\infty\) for all \(\kappa>\lambda\)).

All in all, we are left with a non-divergent expression for~\(I\):
\begin{equation}
  I_k
  =
  \frac{1}{k!}
  \sum_t
  \sum_T
  \sum_{K,\{\hat{n}_\kappa\},\{\hat{k}_{\kappa\nu}\},\tau}
  \sum_{\{k_{s\mu}\},\sigma}
  \Sigma^{\#K}
  \prod_{\kappa\in K} \biggl[
  f_{\{\hat{n}_\kappa\},\{\hat{k}_{\kappa\nu}\}}(\I m_X)
  \sum_{\nu=0}^{\hat{n}_\kappa-1} \hat{k}_{\kappa\nu}
  \biggr]
  V_{\{n_s\}}^{\{k_{s\mu}\}} \,.
\end{equation}
The summand is independent of~\(\sigma\), and there are \((k-t)!\)
maps~\(\sigma\).  Summing \(V_{\{n_s\}}^{\{k_{s\mu}\}}\) over
\(k_{s\mu}\) with \(\sum k_{s\mu} = k-t\) yields
\(Z_{\vortex,\{n_s\},k-t}\).  The sum over \(K\), \(\hat{n}_\kappa\),
\(\hat{k}_{\kappa\nu}\), \(\tau\) does not depend on the precise
set~\(T\), but only on \(t=\#T\).  The choice of~\(T\) thus simply
contributes a factor \(k!/[t!(k-t)!]\), which cancels the overall
\(1/k!\), and \((k-t)!\) coming from the choice of~\(\sigma\).  For a
fixed \(j=\#K\), the remaining sums yield a rational function of~\(\I
m_X\) which can only depend on the two integers \(0\leq j\leq t\leq k\):
\begin{equation}
  I_k = \sum_{t=0}^k \sum_{j=0}^t f_{tj}(\I m_X) \Sigma^j
  Z_{\vortex,\{n_s\},k-t} \,.
\end{equation}
Since the~\(f_{tj}\) do not depend on~\(k\), summing over~\(k\) yields
\begin{equation}\label{Proof-SQCDAW-Ik-Zv}
  \sum_{k\geq 0} y^k I_k
  = \sum_{t\geq 0} \sum_{j=0}^t \bigl[ y^t f_{tj}(\I m_X) \Sigma^j \bigr]
  Z_{\vortex,\{n_s\}}(y)
  = f\bigl(\I m_X,\Sigma; y\bigr) Z_{\vortex,\{n_s\}}(y) \,.
\end{equation}

In Appendix~\ref{app:Proof-SQCDAW-Zvn}, we consider the case \(\Nf=1\),
\(n_1=\Nc\), \(1+\I m_1+\I\anti{m}_1+\Nc\I m_X=0\), for which
\(\Sigma=\Nc\I m_X\), and find that
\begin{equation}
  Z_{\vortex,\{\Nc\}}\bigl(1+\I m_1+\I\anti{m}_1+\Nc\I m_X=0; y\bigr)
  = (1-y)^{-\Nc(\I m_X+1)}
  = (1-y)^{-[1+1/(\I m_X)]\Sigma}
  \,.
\end{equation}
On the other hand, since the factors \(\varphi_\kappa-\I\anti{m}_1\) and
\(\varphi_\kappa+1+\I m_1+n_1\I m_X\) in~\eqref{Proof-SQCDAW-Ik} cancel,
the integrand of~\(I_k\) has no pole with \(\Re(\varphi_\kappa)<0\),
thus \(I_k = \delta_{k0}\).  As a result,
\begin{equation}\label{Proof-SQCDAW-f}
  f\bigl(\I m_X,\Sigma; y\bigr) = (1-y)^{[1+1/(\I m_X)]\Sigma}
\end{equation}
for all \(\Sigma=\Nc\I m_X\).  This fixes each polynomial \(\sum_{j=0}^t
f_{tj}(\I m_X) \Sigma^j\) at an infinite set of values, hence
determines~\(f\) completely.

At last, we are ready to wrap up, by showing that~\(I_k\) is the
\(k\)-vortex partition function of the dual theory.  Close contours
of~\eqref{Proof-SQCDAW-Ik} towards \(-\infty\).  The surrounded poles
are labeled by non-negative integers \(n'_t\geq 0\) and \(k'_{t\nu}\geq
0\) for \(1\leq t\leq\Nf\) and \(0\leq\nu<n'_t\):
\begin{equation}
  \{\varphi_\kappa\} = \bigl\{ -1-\I m_t-n_t\I m_X-\nu\I m_X-j \bigm|
  0\leq\nu<n'_t, 0\leq j<k'_{t\nu} \bigr\} \,.
\end{equation}
For the choice of superpotential \(W=\sum_{t=1}^{\Nf} \anti{\quark}_t
X^{l_t} \quark_t\), the constraint \(1+\I m_t+\I\anti{m}_t+l_t\I m_X=0\)
implies that the numerator factor \(\prod_\kappa
(\varphi_\kappa-\I\anti{m}_t)\) vanishes unless all \(k'_{t\nu} = 0\)
for \(\nu \geq l_t-n_t\).  For the choice of superpotential \(W=\Tr
X^{l+1}\), the constraint \(1+(l+1)\I m_X=0\) implies that
\(\prod_\kappa (\varphi_\kappa+\I m_t-\I m_X)\) vanishes unless all
\(k'_{t\nu} = 0\) for \(\nu \geq l-n_t\).  We can thus take \(n'_t = l_t
- n_t\) in both cases, and let \(k_{t\nu} = k'_{t(l_t-n_t-1-\nu)}\) so
that
\begin{equation}
  \{\varphi_\kappa\} = \bigl\{ -1-\I m_t-(l_t-1-\nu)\I m_X-j \bigm|
  0\leq\nu<l_t-n_t, 0\leq j<k_{t\nu} \bigr\} \,.
\end{equation}
Summing over residues yields, after some massaging,
\begin{equation}
  \begin{aligned}
    & I_k
    =
    \sum_{\{k_{t\nu}\geq 0 \mid 0\leq\nu<l_t-n_t\}}
    \prod_{(s,\mu)} \Biggl[ \prod_{(t,\nu)}
    \frac
      {(\I m_s-\I m_t+(l_s-l_t+\nu-\mu-1)\I m_X-k_{t\nu})_{k_{s\mu}}}
      {(\I m_s-\I m_t+(l_s-l_t+\nu-\mu)\I m_X-k_{t\nu})_{k_{s\mu}}}
    \\
    & \cdot
    \prod_{t=1}^{\Nf} \biggl[
    \frac
      {(1+\I\anti{m}_t+\I m_s+(l_s-1-\mu)\I m_X)_{k_{s\mu}}}
      {(\I m_s-\I m_t+(l_s-l_t-1-\mu)\I m_X)_{k_{s\mu}}}
    \frac
      {(\I m_s-\I m_t+(l_s-\mu)\I m_X+1)_{k_{s\mu}}}
      {(\I m_s-\I m_t+(l_s-n_t-\mu)\I m_X+1)_{k_{s\mu}}}
    \biggr] \Biggr] \,.
  \end{aligned}
\end{equation}

For \(W=\sum_{t=1}^{\Nf} \anti{\quark}_t X^{l_t} \quark_t\), the summand
takes the general form~\eqref{Proof-SQCDAW-V}
of~\(V_{\{l_t-n_t\}}^{\{k\}}\), with \(m_t\leftrightarrow\anti{m}_t\)
since \(1+\I\anti{m}_t+\I m_s+(l_s-1-\mu)\I m_X = \I m_s - \I m_t +
(l_s-l_t-1-\mu)\I m_X\).  Thus, \(I_k\)~is the \(k\)-vortex partition
function of the SQCDA theory with \(\Nf\)~flavour, \(L-\Nc\) colors, the
superpotential \(W=\sum_t \anti{\quark}_t X^{l_t} \quark_t\),
interchanged twisted masses \(m_t\leftrightarrow\anti{m}_t\) compared to
the \(U(\Nc)\) theory, and the same value of~\(y\).  Charge conjugation
maps twisted masses back to those of the \(U(\Nc)\) theory, and maps
\(y\to y^\dual = y^{-1}\) hence \(z^\dual=z^{-1}\).  Summing \(y^k I_k\) then yields the u-channel
vortex partition function of the \(U(L-\Nc)\) theory (that is, a series
in powers of \((y^\dual)^{-1}\)).  We finally combine the
relation~\eqref{Proof-SQCDAW-Ik-Zv} and the explicit
factor~\eqref{Proof-SQCDAW-f} with \(\Sigma = (2\Nc-L)\I m_X\) to get
\begin{equation}\label{Proof-SQCDAW-Z=Z}
  Z^{U(L-\Nc)}_{\vortex,\{l_s-n_s\}} \bigl((y^\dual)^{-1}\bigr)
  = (1-y)^{(2\Nc-L)(1+\I m_X)} Z^{U(\Nc)}_{\vortex,\{n_s\}}(y)
  \quad\text{for } W=\sum_{t=1}^{\Nf} \anti{\quark}_t X^{l_t} \quark_t\,.
\end{equation}

For \(W=\Tr X^{l+1}\), we have \(\I m_s-\I m_t + (l_s-l_t-1-\mu)\I m_X =
\I m_s-\I m_t + (l_s-\mu)\I m_X+1\), and again the summand takes the
form of~\(V_{\{l_t-n_t\}}^{\{k\}}\), with \(\I m_s \to \I m_s^\dual = \I
m_X-\I m_s\) and \(\I\anti{m}_t \to \I\anti{m}_t^\dual = \I
m_X-\I\anti{m}_t\).  Combining the relation~\eqref{Proof-SQCDAW-Ik-Zv}
and the explicit factor~\eqref{Proof-SQCDAW-f} with \(\Sigma = 2\Nc\I
m_X+\sum_{t=1}^{\Nf} (1+\I m_t+\I\anti{m}_t)\) and \(1+1/(\I m_X)=-l\)
yields the Kutasov--Schwimmer duality relation
\begin{equation}\label{Proof-SQCDAW-Z=Z-Kutasov}
  Z^{U(l\Nf-\Nc)}_{\vortex,\{l-n_s\}}
  \bigl(m_t^\dual, \anti{m}_t^\dual ; y^\dual\bigr)
  = (1-y)^{-\delta_1} Z^{U(\Nc)}_{\vortex,\{n_s\}}
  \bigl(m_t , \anti{m}_t ; y\bigr)
  \quad \text{for } W = \Tr X^{l+1}
\end{equation}
with \(\delta_1 = - \frac{2l}{l+1}\Nc + l\sum_{t=1}^{\Nf} (1+\I m_t +
\I\anti{m}_t)\), as obtained in~\eqref{Kutasov-delta} through the
relation with conjugation of momenta in the Toda CFT\@.  Since \(y^\dual =
y\), we get \(z^\dual = z\).  In Section~\ref{sec:Kutasov}, we extend
the Kutasov--Schwimmer duality to theories with a different number of
fundamental and antifundamental chiral multiplets.

\addtocontents{toc}{\protect\setcounter{tocdepth}{3}}
\clearpage
\phantomsection
\addcontentsline{toc}{section}{References}
\bibliography{refs}

\providecommand{\href}[2]{#2}\begingroup\raggedright\begin{thebibliography}{10}

\bibitem{Wilson:1974sk}
K.~G. Wilson, \emph{{Confinement of quarks}},
  \href{http://dx.doi.org/10.1103/PhysRevD.10.2445}{\emph{Phys. Rev.} {\bf D10}
  (1974) 2445--2459}.

\bibitem{Hooft:1977hy}
G.~'t~Hooft, \emph{{On the Phase Transition Towards Permanent Quark
  Confinement}},
  \href{http://dx.doi.org/10.1016/0550-3213(78)90153-0}{\emph{Nucl.Phys.} {\bf
  B138} (1978) 1}.

\bibitem{Gukov-Witten-surface}
S.~Gukov and E.~Witten, \emph{{Gauge theory, ramification, and the geometric
  Langlands program}},  \href{http://arxiv.org/abs/hep-th/0612073}{{\tt
  hep-th/0612073}}.

\bibitem{Gukov:2013zka}
S.~Gukov and A.~Kapustin, \emph{{Topological Quantum Field Theory, Nonlocal
  Operators, and Gapped Phases of Gauge Theories}},
  \href{http://arxiv.org/abs/1307.4793}{{\tt 1307.4793}}.

\bibitem{Gomis-bubbling-surface}
J.~Gomis and S.~Matsuura, \emph{{Bubbling surface operators and S-duality}},
  \href{http://dx.doi.org/10.1088/1126-6708/2007/06/025}{\emph{JHEP} {\bf 06}
  (2007) 025}, [\href{http://arxiv.org/abs/0704.1657}{{\tt 0704.1657}}].

\bibitem{Gukov:2007ck}
S.~Gukov, \emph{{Gauge theory and knot homologies}},
  \href{http://dx.doi.org/10.1002/prop.200610385}{\emph{Fortsch.Phys.} {\bf 55}
  (2007) 473--490}, [\href{http://arxiv.org/abs/0706.2369}{{\tt 0706.2369}}].

\bibitem{Witten:2007td}
E.~Witten, \emph{{Gauge theory and wild ramification}},
  \href{http://arxiv.org/abs/0710.0631}{{\tt 0710.0631}}.

\bibitem{Buchbinder:2007ar}
E.~I. Buchbinder, J.~Gomis and F.~Passerini, \emph{{Holographic gauge theories
  in background fields and surface operators}},
  \href{http://dx.doi.org/10.1088/1126-6708/2007/12/101}{\emph{JHEP} {\bf 12}
  (2007) 101}, [\href{http://arxiv.org/abs/0710.5170}{{\tt 0710.5170}}].

\bibitem{Beasley:2008dc}
C.~Beasley, J.~J. Heckman and C.~Vafa, \emph{{GUTs and Exceptional Branes in
  F-theory - I}},
  \href{http://dx.doi.org/10.1088/1126-6708/2009/01/058}{\emph{JHEP} {\bf 0901}
  (2009) 058}, [\href{http://arxiv.org/abs/0802.3391}{{\tt 0802.3391}}].

\bibitem{Gukov:2008sn}
S.~Gukov and E.~Witten, \emph{{Rigid Surface Operators}},
  {\emph{Adv.Theor.Math.Phys.} {\bf 14} (2010) },
  [\href{http://arxiv.org/abs/0804.1561}{{\tt 0804.1561}}].

\bibitem{Dru-surface}
N.~Drukker, J.~Gomis and S.~Matsuura, \emph{{Probing ${\cal N}=4$ SYM with
  surface operators}},
  \href{http://dx.doi.org/10.1088/1126-6708/2008/10/048}{\emph{JHEP} {\bf 10}
  (2008) 048}, [\href{http://arxiv.org/abs/0805.4199}{{\tt 0805.4199}}].

\bibitem{Gaiotto:2009fs}
D.~Gaiotto, \emph{{Surface Operators in ${\cal N}=2$ 4d Gauge Theories}},
  \href{http://arxiv.org/abs/0911.1316}{{\tt 0911.1316}}.

\bibitem{Gaiotto:2009we}
D.~Gaiotto, \emph{{${\cal N}=2$ dualities}},
  \href{http://arxiv.org/abs/0904.2715}{{\tt 0904.2715}}.

\bibitem{Pestun:2007rz}
V.~Pestun, \emph{{Localization of gauge theory on a four-sphere and
  supersymmetric Wilson loops}},  \href{http://arxiv.org/abs/0712.2824}{{\tt
  0712.2824}}.

\bibitem{Hama:2012bg}
N.~Hama and K.~Hosomichi, \emph{{Seiberg-Witten Theories on Ellipsoids}},
  \href{http://dx.doi.org/10.1007/JHEP09(2012)033,
  10.1007/JHEP10(2012)051}{\emph{JHEP} {\bf 1209} (2012) 033},
  [\href{http://arxiv.org/abs/1206.6359}{{\tt 1206.6359}}].

\bibitem{Gomis:2011pf}
J.~Gomis, T.~Okuda and V.~Pestun, \emph{{Exact Results for 't Hooft Loops in
  Gauge Theories on $S^4$}}, {\emph{JHEP} {\bf 1205} (2012) 141},
  [\href{http://arxiv.org/abs/1105.2568}{{\tt 1105.2568}}].

\bibitem{Benini:2012ui}
F.~Benini and S.~Cremonesi, \emph{{Partition functions of $N=(2,2)$ gauge
  theories on $S^2$ and vortices}},  \href{http://arxiv.org/abs/1206.2356}{{\tt
  1206.2356}}.

\bibitem{Doroud:2012xw}
N.~Doroud, J.~Gomis, B.~Le~Floch and S.~Lee, \emph{{Exact Results in D=2
  Supersymmetric Gauge Theories}},  \href{http://arxiv.org/abs/1206.2606}{{\tt
  1206.2606}}.

\bibitem{Gomis:2012wy}
J.~Gomis and S.~Lee, \emph{{Exact Kahler Potential from Gauge Theory and Mirror
  Symmetry}}, \href{http://dx.doi.org/10.1007/JHEP04(2013)019}{\emph{JHEP} {\bf
  1304} (2013) 019}, [\href{http://arxiv.org/abs/1210.6022}{{\tt 1210.6022}}].

\bibitem{Doroud:2013pka}
N.~Doroud and J.~Gomis, \emph{{Gauge Theory Dynamics and Kahler Potential for
  Calabi-Yau Complex Moduli}},  \href{http://arxiv.org/abs/1309.2305}{{\tt
  1309.2305}}.

\bibitem{AGT}
L.~F. Alday, D.~Gaiotto and Y.~Tachikawa, \emph{{Liouville Correlation
  Functions from Four-dimensional Gauge Theories}},
  \href{http://dx.doi.org/10.1007/s11005-010-0369-5}{\emph{Lett. Math. Phys.}
  {\bf 91} (2010) 167--197}, [\href{http://arxiv.org/abs/0906.3219}{{\tt
  0906.3219}}].

\bibitem{Wyllard:2009hg}
N.~Wyllard, \emph{{$A_{N-1}$ conformal Toda field theory correlation functions
  from conformal ${\cal N}=2$ $SU(N)$ quiver gauge theories}},
  \href{http://dx.doi.org/10.1088/1126-6708/2009/11/002}{\emph{JHEP} {\bf 11}
  (2009) 002}, [\href{http://arxiv.org/abs/0907.2189}{{\tt 0907.2189}}].

\bibitem{Alday:2009fs}
L.~F. Alday, D.~Gaiotto, S.~Gukov, Y.~Tachikawa and H.~Verlinde, \emph{{Loop
  and surface operators in ${\cal N}=2$ gauge theory and Liouville modular
  geometry}}, \href{http://dx.doi.org/10.1007/JHEP01(2010)113}{\emph{JHEP} {\bf
  01} (2010) 113}, [\href{http://arxiv.org/abs/0909.0945}{{\tt 0909.0945}}].

\bibitem{Drukker:2009id}
N.~Drukker, J.~Gomis, T.~Okuda and J.~Teschner, \emph{{Gauge Theory Loop
  Operators and Liouville Theory}},
  \href{http://dx.doi.org/10.1007/JHEP02(2010)057}{\emph{JHEP} {\bf 02} (2010)
  057}, [\href{http://arxiv.org/abs/0909.1105}{{\tt 0909.1105}}].

\bibitem{Drukker:2010jp}
N.~Drukker, D.~Gaiotto and J.~Gomis, \emph{{The Virtue of Defects in 4D Gauge
  Theories and 2D CFTs}},
  \href{http://dx.doi.org/10.1007/JHEP06(2011)025}{\emph{JHEP} {\bf 1106}
  (2011) 025}, [\href{http://arxiv.org/abs/1003.1112}{{\tt 1003.1112}}].

\bibitem{Passerini:2010pr}
F.~Passerini, \emph{{Gauge Theory Wilson Loops and Conformal Toda Field
  Theory}}, \href{http://dx.doi.org/10.1007/JHEP03(2010)125}{\emph{JHEP} {\bf
  03} (2010) 125}, [\href{http://arxiv.org/abs/1003.1151}{{\tt 1003.1151}}].

\bibitem{Gomis:2010kv}
J.~Gomis and B.~Le~Floch, \emph{{'t Hooft Operators in Gauge Theory from Toda
  CFT}}, \href{http://dx.doi.org/10.1007/JHEP11(2011)114}{\emph{JHEP} {\bf
  1111} (2011) 114}, [\href{http://arxiv.org/abs/1008.4139}{{\tt 1008.4139}}].

\bibitem{Bullimore:2013xsa}
M.~Bullimore, \emph{{Defect Networks and Supersymmetric Loop Operators}},
  \href{http://arxiv.org/abs/1312.5001}{{\tt 1312.5001}}.

\bibitem{Dimofte:2010tz}
T.~Dimofte, S.~Gukov and L.~Hollands, \emph{{Vortex Counting and Lagrangian
  3-manifolds}},
  \href{http://dx.doi.org/10.1007/s11005-011-0531-8}{\emph{Lett.Math.Phys.}
  {\bf 98} (2011) 225--287}, [\href{http://arxiv.org/abs/1006.0977}{{\tt
  1006.0977}}].

\bibitem{Taki:2010bj}
M.~Taki, \emph{{Surface Operator, Bubbling Calabi-Yau and AGT Relation}},
  \href{http://dx.doi.org/10.1007/JHEP07(2011)047}{\emph{JHEP} {\bf 1107}
  (2011) 047}, [\href{http://arxiv.org/abs/1007.2524}{{\tt 1007.2524}}].

\bibitem{Bonelli:2011fq}
G.~Bonelli, A.~Tanzini and J.~Zhao, \emph{{Vertices, Vortices and Interacting
  Surface Operators}},  \href{http://arxiv.org/abs/1102.0184}{{\tt 1102.0184}}.

\bibitem{Bonelli:2011wx}
G.~Bonelli, A.~Tanzini and J.~Zhao, \emph{{The Liouville side of the Vortex}},
  {\emph{JHEP} {\bf 1109} (2011) 096},
  [\href{http://arxiv.org/abs/1107.2787}{{\tt 1107.2787}}].

\bibitem{Braverman:2010ef}
A.~Braverman, B.~Feigin, M.~Finkelberg and L.~Rybnikov, \emph{{A Finite analog
  of the AGT relation I: Finite $W$-algebras and quasimaps' spaces}},
  \href{http://dx.doi.org/10.1007/s00220-011-1300-3}{\emph{Commun.Math.Phys.}
  {\bf 308} (2011) 457--478}, [\href{http://arxiv.org/abs/1008.3655}{{\tt
  1008.3655}}].

\bibitem{Alday:2010vg}
L.~F. Alday and Y.~Tachikawa, \emph{{Affine SL(2) conformal blocks from 4d
  gauge theories}},
  \href{http://dx.doi.org/10.1007/s11005-010-0422-4}{\emph{Lett.Math.Phys.}
  {\bf 94} (2010) 87--114}, [\href{http://arxiv.org/abs/1005.4469}{{\tt
  1005.4469}}].

\bibitem{Kozcaz:2010yp}
C.~Kozcaz, S.~Pasquetti, F.~Passerini and N.~Wyllard, \emph{{Affine sl(N)
  conformal blocks from N=2 SU(N) gauge theories}},
  \href{http://dx.doi.org/10.1007/JHEP01(2011)045}{\emph{JHEP} {\bf 1101}
  (2011) 045}, [\href{http://arxiv.org/abs/1008.1412}{{\tt 1008.1412}}].

\bibitem{Wyllard:2010rp}
N.~Wyllard, \emph{{W-algebras and surface operators in N=2 gauge theories}},
  \href{http://dx.doi.org/10.1088/1751-8113/44/15/155401}{\emph{J.Phys.} {\bf
  A44} (2011) 155401}, [\href{http://arxiv.org/abs/1011.0289}{{\tt
  1011.0289}}].

\bibitem{Wyllard:2010vi}
N.~Wyllard, \emph{{Instanton partition functions in N=2 SU(N) gauge theories
  with a general surface operator, and their W-algebra duals}},
  \href{http://dx.doi.org/10.1007/JHEP02(2011)114}{\emph{JHEP} {\bf 1102}
  (2011) 114}, [\href{http://arxiv.org/abs/1012.1355}{{\tt 1012.1355}}].

\bibitem{Tachikawa:2011dz}
Y.~Tachikawa, \emph{{On W-algebras and the symmetries of defects of 6d N=(2,0)
  theory}}, \href{http://dx.doi.org/10.1007/JHEP03(2011)043}{\emph{JHEP} {\bf
  1103} (2011) 043}, [\href{http://arxiv.org/abs/1102.0076}{{\tt 1102.0076}}].

\bibitem{Kanno:2011fw}
H.~Kanno and Y.~Tachikawa, \emph{{Instanton counting with a surface operator
  and the chain-saw quiver}},
  \href{http://dx.doi.org/10.1007/JHEP06(2011)119}{\emph{JHEP} {\bf 1106}
  (2011) 119}, [\href{http://arxiv.org/abs/1105.0357}{{\tt 1105.0357}}].

\bibitem{Belavin:2012qh}
V.~Belavin, \emph{{Conformal blocks of Chiral fields in N=2 SUSY CFT and Affine
  Laumon Spaces}}, \href{http://dx.doi.org/10.1007/JHEP10(2012)156}{\emph{JHEP}
  {\bf 1210} (2012) 156}, [\href{http://arxiv.org/abs/1209.2992}{{\tt
  1209.2992}}].

\bibitem{Tan:2013tq}
M.-C. Tan, \emph{{M-Theoretic Derivations of 4d-2d Dualities: From a Geometric
  Langlands Duality for Surfaces, to the AGT Correspondence, to Integrable
  Systems}}, \href{http://dx.doi.org/10.1007/JHEP07(2013)171}{\emph{JHEP} {\bf
  1307} (2013) 171}, [\href{http://arxiv.org/abs/1301.1977}{{\tt 1301.1977}}].

\bibitem{Gadde:2013dda}
A.~Gadde and S.~Gukov, \emph{{2d Index and Surface operators}},
  \href{http://dx.doi.org/10.1007/JHEP03(2014)080}{\emph{JHEP} {\bf 1403}
  (2014) 080}, [\href{http://arxiv.org/abs/1305.0266}{{\tt 1305.0266}}].

\bibitem{Nakayama:2011pa}
Y.~Nakayama, \emph{{4D and 2D superconformal index with surface operator}},
  \href{http://dx.doi.org/10.1007/JHEP08(2011)084}{\emph{JHEP} {\bf 1108}
  (2011) 084}, [\href{http://arxiv.org/abs/1105.4883}{{\tt 1105.4883}}].

\bibitem{Gaiotto:2012xa}
D.~Gaiotto, L.~Rastelli and S.~S. Razamat, \emph{{Bootstrapping the
  superconformal index with surface defects}},
  \href{http://arxiv.org/abs/1207.3577}{{\tt 1207.3577}}.

\bibitem{Alday:2013kda}
L.~F. Alday, M.~Bullimore, M.~Fluder and L.~Hollands, \emph{{Surface defects,
  the superconformal index and q-deformed Yang-Mills}},
  \href{http://dx.doi.org/10.1007/JHEP10(2013)018}{\emph{JHEP} {\bf 1310}
  (2013) 018}, [\href{http://arxiv.org/abs/1303.4460}{{\tt 1303.4460}}].

\bibitem{Bullimore:2014nla}
M.~Bullimore, M.~Fluder, L.~Hollands and P.~Richmond, \emph{{The superconformal
  index and an elliptic algebra of surface defects}},
  \href{http://arxiv.org/abs/1401.3379}{{\tt 1401.3379}}.

\bibitem{Witten:1993yc}
E.~Witten, \emph{{Phases of N = 2 theories in two dimensions}},
  \href{http://dx.doi.org/10.1016/0550-3213(93)90033-L}{\emph{Nucl. Phys.} {\bf
  B403} (1993) 159--222}, [\href{http://arxiv.org/abs/hep-th/9301042}{{\tt
  hep-th/9301042}}].

\bibitem{Benini:2014mia}
F.~Benini, D.~S. Park and P.~Zhao, \emph{{Cluster algebras from dualities of 2d
  N=(2,2) quiver gauge theories}},  \href{http://arxiv.org/abs/1406.2699}{{\tt
  1406.2699}}.

\bibitem{Hori:2006dk}
K.~Hori and D.~Tong, \emph{{Aspects of Non-Abelian Gauge Dynamics in
  Two-Dimensional N=(2,2) Theories}},
  \href{http://dx.doi.org/10.1088/1126-6708/2007/05/079}{\emph{JHEP} {\bf 0705}
  (2007) 079}, [\href{http://arxiv.org/abs/hep-th/0609032}{{\tt
  hep-th/0609032}}].

\bibitem{Seiberg:1994pq}
N.~Seiberg, \emph{{Electric - magnetic duality in supersymmetric nonAbelian
  gauge theories}},
  \href{http://dx.doi.org/10.1016/0550-3213(94)00023-8}{\emph{Nucl.Phys.} {\bf
  B435} (1995) 129--146}, [\href{http://arxiv.org/abs/hep-th/9411149}{{\tt
  hep-th/9411149}}].

\bibitem{Gaiotto:2012sf}
D.~Gaiotto and J.~Teschner, \emph{{Irregular singularities in Liouville theory
  and Argyres-Douglas type gauge theories, I}},
  \href{http://arxiv.org/abs/1203.1052}{{\tt 1203.1052}}.

\bibitem{Kutasov:1995ve}
D.~Kutasov, \emph{{A Comment on duality in N=1 supersymmetric nonAbelian gauge
  theories}},
  \href{http://dx.doi.org/10.1016/0370-2693(95)00392-X}{\emph{Phys.Lett.} {\bf
  B351} (1995) 230--234}, [\href{http://arxiv.org/abs/hep-th/9503086}{{\tt
  hep-th/9503086}}].

\bibitem{Kutasov:1995np}
D.~Kutasov and A.~Schwimmer, \emph{{On duality in supersymmetric Yang-Mills
  theory}},
  \href{http://dx.doi.org/10.1016/0370-2693(95)00676-C}{\emph{Phys.Lett.} {\bf
  B354} (1995) 315--321}, [\href{http://arxiv.org/abs/hep-th/9505004}{{\tt
  hep-th/9505004}}].

\bibitem{Hori:2013ika}
K.~Hori and M.~Romo, \emph{{Exact Results In Two-Dimensional (2,2)
  Supersymmetric Gauge Theories With Boundary}},
  \href{http://arxiv.org/abs/1308.2438}{{\tt 1308.2438}}.

\bibitem{Honda:2013uca}
D.~Honda and T.~Okuda, \emph{{Exact results for boundaries and domain walls in
  2d supersymmetric theories}},  \href{http://arxiv.org/abs/1308.2217}{{\tt
  1308.2217}}.

\bibitem{Closset:2014pda}
C.~Closset and S.~Cremonesi, \emph{{Comments on N=(2,2) Supersymmetry on
  Two-Manifolds}},  \href{http://arxiv.org/abs/1404.2636}{{\tt 1404.2636}}.

\bibitem{Gerchkovitz:2014gta}
E.~Gerchkovitz, J.~Gomis and Z.~Komargodski, \emph{{Sphere Partition Functions
  and the Zamolodchikov Metric}},  \href{http://arxiv.org/abs/1405.7271}{{\tt
  1405.7271}}.

\bibitem{Gomis:2016ljm}
J.~Gomis, B.~Le~Floch, Y.~Pan and W.~Peelaers, \emph{{Intersecting Surface
  Defects and Two-Dimensional CFT}},
  \href{http://arxiv.org/abs/1610.03501}{{\tt 1610.03501}}.

\bibitem{Fateev:2007ab}
V.~A. Fateev and A.~V. Litvinov, \emph{{Correlation functions in conformal Toda
  field theory I}},
  \href{http://dx.doi.org/10.1088/1126-6708/2007/11/002}{\emph{JHEP} {\bf 11}
  (2007) 002}, [\href{http://arxiv.org/abs/0709.3806}{{\tt 0709.3806}}].

\bibitem{Orlando:2010uu}
D.~Orlando and S.~Reffert, \emph{{Relating Gauge Theories via Gauge/Bethe
  Correspondence}},
  \href{http://dx.doi.org/10.1007/JHEP10(2010)071}{\emph{JHEP} {\bf 1010}
  (2010) 071}, [\href{http://arxiv.org/abs/1005.4445}{{\tt 1005.4445}}].

\bibitem{Kanno:2013vi}
H.~Kanno, K.~Maruyoshi, S.~Shiba and M.~Taki, \emph{{$W_3$ irregular states and
  isolated N=2 superconformal field theories}},
  \href{http://dx.doi.org/10.1007/JHEP03(2013)147}{\emph{JHEP} {\bf 1303}
  (2013) 147}, [\href{http://arxiv.org/abs/1301.0721}{{\tt 1301.0721}}].

\bibitem{Gaiotto:2013rk}
D.~Gaiotto and J.~Lamy-Poirier, \emph{{Irregular Singularities in the $H_3^+$
  WZW Model}},  \href{http://arxiv.org/abs/1301.5342}{{\tt 1301.5342}}.

\end{thebibliography}\endgroup

\end{document}